\documentclass[10pt,twoside]{article} 
\usepackage{graphicx}
\usepackage{verbatim}
\usepackage{framed}
\usepackage{hyperref}
\usepackage{listings}
\usepackage{amsmath}
\usepackage{morefloats}

\begin{document}

\title{Sensitivity of the Static Earthquake Triggering Mechanism to Elastic Heterogeneity and Main Event Slip}
\author{Musa Maharramov \href{musa@sep.stanford.edu}{musa@sep.stanford.edu}}
\date{\today}

\maketitle

\setcounter{secnumdepth}{2}

\abstract{This paper has evolved out of our previous work on static stress transfer, where we used the full-space elastostatic Green's tensor to compute the Coulomb stress transfer impact of the Landers earthquake on the Hector Mine event. In this work, we use the elastostatic Green's tensor for an arbitrary layered Earth model with free-surface boundary conditions to study the impact of elastic heterogeneity as well as source-fault slip and geometry on the stress transfer mechanism. Slip distribution and fault geometry of the source have a significant impact on the stress transfer, especially in case of spatially extended triggered events. Maximization of the Coulomb stress transfer function for known aftershocks provides a mechanism for inverting for the source event slip. Heterogeneity of the elastic earth parameters is shown to have a sizeable, but lower-magnitude, impact on the static stress transfer in 3D. The analysis is applied to Landers/Hector Mine and 100 small ``aftershocks'' of the Landers event. A computational toolkit is provided for the study of static stress transfer for arbitrary source and receiver faults in layered Earth.}

\section{Introduction}

In this work we analyze the sensitivity of the static stress transfer mechanism (\cite{King}) to the heterogeneity of the elastic earth model, source fault geometry and slip. Our approach to computing the change in the Coulomb Failure Function (CFF -- see \cite{RG}) due to a triggered stress change is similar to e.g. \cite{King} and \cite{Stein}, and the one used in \cite{Parsons}. The results are applied to analyzing a possible Coulomb stress transfer impact of the 1992 {\bf M}=7.3 Landers earthquake on the {\bf M}=7.1 1999 Hector Mine event and 100 smaller regional ``aftershock'' events. A connection between Landers and Hector Mine was conjectured soon after the occurrence of the second event; a static-stress impact causality analysis was presented in \cite{Parsons} and will be compared to our results.

We numerically compute the elastostatic Green's tensor for an arbitrary layered Earth model, and use the obtained tensor in the double-couple representation of displacement due to slip on a \emph{finite fault} (\cite{PS}).  As in our previous work, the obtained displacements are used to compute the induced change in the stress field and CFF. To validate our approach, first we test our heterogeneous displacement modeling algorithm on a series of source events with varying geometry and magnitude.

In the application to Landers and Hector Mine, we use three differently oriented receiver fault planes and a multi-fault source model. Maps of CFF, normal and shear stress change at all points of the receiver faults are generated, and the impact of input parameters on the receiver slip likelihood is studied.

Results of both stress transfer modeling and slip inversion algorithms are sensitive to the accuracy of input data for the source and receiver events (e.g. fault geometry). The focal mechanism inversion data for the Landers and Hector Mine events, provided by the Southern California Earthquake Data Center (see \cite{FPCAT}), fails a simple sense check: e.g., the data indicates that the Landers ``fault'' has the down-dip azimuth of 165$^\circ$ -- this disagrees with the known geometric features of the Johnson Valley, Landers, Homestead Valley, Emerson and Camp Rock faults that experienced temporally and spatially distributed slips during the Landers earthquake (see \cite{WALD}). The reason for such discrepancies is that the focal mechanism inversion ``reconstructs'' each event as a slip on a single fault based on early arrivals. More accurate fault data can be obtained from alternative inversion/observation methods that take into account regional displacement measurements. In this study, we use the fault orientation and earthquake nucleation data for the Landers event presented in \cite{WALD}. Likewise for the Hector Mine event, rather than using the SCED-provided single-fault inversion parameters, we use the parameters estimated in \cite{KAV}, \cite{JON} and \cite{HM}. In particular, we compute the Coulomb stress transfer for \emph{three} fault orientations/locations associated with the \emph{nucleation} of the Hector Mine event (see \cite{KAV}). The 100 small Landers ``aftershock'' events used in the analysis are taken from the SCED small-event database (\cite{Hardenbeck}). Along with using the original fault parameters from the database, we allow for the uncertainty in the parameters and create three ``aftershock'' datasets made up of events local to Landers, spread around the entire rupture, and a decimated version of the latter. The direct and inverse stress transfer analysis is performed for all three datasets and Hector Mine.

The entire heterogeneous displacement modeling/stress transfer framework has been implemented using {\tt MATLAB} for easy reproducibility. A description and brief user manual for the provided tools are included in Appendix A.

\section{Method}

We model the source earthquake as a series of \emph{spatially-distributed} slips on \emph{finite} faults in 3D. In the modeling part of this exercise, the source events are specified by 3D fault geometry (coordinates of the fault center, fault length, width, strike, dip and rake) and moment magnitude $M_w$ \emph{for each fault}. Once the elastostatic Green's tensor is obtained, the 3D \emph{displacement field} due to the slip on the source fault is modeled using the \emph{Volterra equation} (see e.g. \cite{PS}):
\begin{equation}
u_k(\mathbf{x})\;=\;\int\limits_{\cup F} {\mu s_i \left[\frac{\partial g^i_k(\mathbf{x},\mathbf{y})}{\partial y_j}+\frac{\partial g^j_k(\mathbf{x},\mathbf{y})}{\partial y_i}\right] n_j d S_y }.
\label{eq:VOLT}
\end{equation}
In (\ref{eq:VOLT}), we have omitted the tensile (diagonal) components of the tensor in the square brackets as we assume that the slip is parallel to the fault surface, the integration is carried out over all faults $\cup F$, and $g^i_k(\mathbf{x},\mathbf{y})$ is elastostatic Green's tensor -- i.e., the displacement along axis $k$ at point $\mathbf{x}$ due to a unit force applied along axis $i$ at point $\mathbf{y}$ (see e.g. \cite{PS}). Variables $\mu,s_i,n_j$ have the usual meaning of the shear modulus, fault slip and \emph{external} normal vector components. Note that the tensile components of stress can be easily included in (\ref{eq:VOLT}) if required for e.g. modeling tensile cracks.

For a homogeneous half-space earth model with the free-surface boundary conditions, the elastostatic tensor was given by Mindlin in \cite{M36}. In case of a fixed and known beforehand number of elastic layers, the Green's function can be found analytically using e.g. the \emph{method of images} (\cite{PS}, \cite{Kausel}), or the \emph{analytic} form of the \emph{propagation/stiffness matrix} method (\cite{Kausel}). If the number of layers are unknown, the numerical form of the propagator matrix may be used (e.g. \cite{GilbertBackus}). However, in this paper we use a different approach based on numerically solving the system of elastostatic equations with the free-surface boundary conditions at $z=0$ and assuming zero \emph{co-seismic} displacement at some maximum depth (20{\tt km} in our tests). More specifically, we solve the elastostatic wave equation in the Navier form
 \begin{equation}
\mu(x_3) \Delta u^i+ \frac{\mu(x_3)}{1-2\nu(x_3)} \frac{ \partial}{\partial x_i} \frac{ \partial u_k}{\partial x_k}=0,\;i=1,2,3.
\label{eq:ewe}
\end{equation}
After Fourier-transforming (\ref{eq:ewe}) with respect to the two horizontal variables we obtain
\begin{align}
\mu(x_3)\left[ (-k_x^2-k_y^2)u^1+\frac{\partial^2 u^1}{\partial z^2}\right]+(\lambda+\mu)\left[ -k_x^2 u^1-k_x k_y u^2+ik_x\frac{\partial u^3}{\partial z}\right] =& 0, \nonumber \\
\mu(x_3)\left[ (-k_x^2-k_y^2)u^2+\frac{\partial^2 u^2}{\partial z^2}\right]+(\lambda+\mu))\left[ -k_x k_y u^1-k_y^2 u^2+ik_y\frac{\partial u^3}{\partial z}\right] =& 0, \nonumber \\
\mu(x_3)\left[ (-k_x^2-k_y^2)u^3+\frac{\partial^2 u^3}{\partial z^2}\right]+(\lambda+\mu))\left[ i k_x \frac{\partial u^1}{\partial z}+i k_y \frac{\partial u^2}{\partial z} + \frac{\partial^2 u^3}{\partial z^2}\right] =& 0,
\label{eq:fewe}
\end{align}
where $\lambda=\lambda(x_3)=2\mu\nu/(1-2\nu)$ is the Lam\'{e} coefficient. Rather than solving (\ref{eq:fewe}) for both displacements and stresses, we discretize in $z$ and complement the resulting system with the condition of continuity of normal tractions (\cite{PS}) between discretization intervals.  Once the elastostatic tensor is computed, the regional stress change is computed from forward-modeled displacements (\ref{eq:VOLT}) as
\begin{equation}
\Delta \sigma^i_k\;=\;\mu \left(\frac{\partial u_i}{\partial x_k}+\frac{\partial u_k}{\partial x_i}\right)+\frac{2\mu\nu}{1-2\nu} \mathbf{div}\, \mathbf{u} \delta^i_k,
\label{eq:DSIGMA}
\end{equation}
where \emph{tensile stress} is positive. Coulomb stress transfer ($\Delta \mathrm{CFF}$) at the receiver fault is then computed form (\ref{eq:DSIGMA}), taking into account the stress sign conventions, as follows (see \cite{RG}):
\begin{align}
\Delta \sigma_n \; & = \; \mathbf{n}^T_r \Delta \sigma \mathbf{n}_r , \nonumber \\
\Delta \tau \; & = \; \mathbf{s}^T_r \Delta \sigma \mathbf{n}_r, \nonumber \\
\Delta \mathrm{CFF}\;& =\;\Delta \tau + \mu_f \left( \Delta \sigma_n - \Delta P_p \right)\;=\;\Delta \tau + \mu_f \Delta \sigma_n \left( 1-B \right),
\label{eq:CFF}
\end{align}
where $\mathbf{n}_r,\,\mathbf{s}_r$ are the \emph{external} unit normal and unit slip (rake) vectors for the receiver fault, $\Delta \sigma_n$ is the \emph{normal stress} change (positive in tension), and $\Delta \tau$ is the shear stress change\footnote{in the direction of the receiver slip (rake)} induced by the source events. The Skempton coefficient $B$ relates the change in the confining stress to the resulting change in pore pressure (\cite{LINPORO}) and is a source of uncertainty in our analysis. In this work we assume $B\approx 0.5$ based on the fact the range of values for solid rocks is $\approx$0.5-0.8 (\cite{LINPORO}). The coefficient of fault friction $\mu_f$ typically ranges between $0.5-0.9$ and is depends on gouging/brecciation within the receiver fault. We attempt to estimate this parameter based on plausible stress transfer scenarios for the analyzed source and receiver faults.

Equations (\ref{eq:VOLT},\ref{eq:CFF}) provide a mechanism for mapping known slips on the source fault to the stress change on the receiver fault, and form the basis of the static stress triggering analysis. We additionally suggest to solve an \emph{inverse problem} for identifying the slips on the source fault from the \emph{receiver fault parameters}. In our proposed inversion method we do not use values of the stress change at the receiver faults but rather attempt to estimate the source slip from the \emph{assumption} that the user-specified spatially-distributed receiver faults \emph{have been encouraged to slip} by the source events. More specifically, if $\mathbf{A},\mathbf{B},\mathbf{C}$ are \emph{linear operators} that map source slips to $\Delta \sigma_n,\Delta \tau, \Delta \mathrm{CFF}$
\begin{align}
\Delta \sigma_n \; & =\; \mathbf{A}\mathbf{s}      \nonumber \\
\Delta \tau \; & =\;  \mathbf{B}\mathbf{s}         \nonumber \\
\Delta \mathrm{CFF} \; & =\; \mathbf{C}\mathbf{s},
\label{eq:STM} 
\end{align}
then we formulate the  following \emph{linear programming} problem for computing $\mathbf{s}$:
\begin{align}
& \overbrace{[1 1 \ldots 1 1]}^\text{all grid points of all receiver faults} \mathbf{C}\mathbf{s} \rightarrow \mathrm{max}, \nonumber \\
& [0  \ldots 0 \overbrace{1 1 \ldots 1 1}^{\text{grid of receiver fault \#}\,k} 0 \ldots  0 ] \mathbf{A}\mathbf{s}\; \ge \; 0, \nonumber \\
& [0  \ldots 0 \overbrace{1 1 \ldots 1 1}^{\text{grid of receiver fault \#}\,k} 0 \ldots  0 ] \mathbf{B}\mathbf{s}\; \ge \; 0, \nonumber \\
& [0  \ldots 0 \overbrace{1 1 \ldots 1 1}^{\text{grid of receiver fault \#}\,k} 0 \ldots  0 ] \mathbf{C}\mathbf{s}\; \ge \; 0, \nonumber \\
& \int_{F_l}\mu(s^l_i n_j+s^l_j n_i)dS\;=\;M^l_{i j}, \nonumber \\
& \mathbf{s}\ge0,
\label{eq:INV}
\end{align}
 where $k$ ranges across all receiver fault indices, $l$ ranges across all source faults $F_l$, $\mathbf{s}$ is a vector made up of non-negative slip along respective rake direction at each grid point of each source fault, $M^l_{i j}$ is the moment tensor associated with the source event (slipping fault) No $l$. Equations (\ref{eq:INV}) mean that we maximize the total change of the Coulomb Friction Function $\sum \Delta \mathrm{CFF}$ for all receiver faults under the constraints that the total  $\Delta \sigma_n,\Delta \tau, \Delta \mathrm{CFF}$ \emph{for each receiver fault}\footnote{integrated over the fault surface} is positive. Additionally, the slips are constrained by the known moment tensor and the obvious requirement that the slips are non-negative in the rake direction. The three stress-related constrains can actually be reduced, depending on whichever slip-encouragement mechanism is likely relevant (e.g., shear advancement or un-clamping). This is especially useful with uncertain friction and Skempton coefficients.

\section{Implementation}

The elastostatic tensor is computed on a 3D grid for a number of source depths\footnote{The Green's tensor for a layered Earth is invariant with respect to planar translations of both source and receiver hence the source can always be assumed to have $x=0,y=0$; however, source and receiver depths must be independent.}. In all of our experiments we use the \emph{geological coordinate system} with the axis $x$ pointing North, axis $y$ pointing East and $z$ pointing down (depth). Negative vertical displacements correspond to uplift while positive ones, to subsidence. The Green's tensor is computed for two modeling cases: 

\begin{enumerate}
\item $128\times128\times20$ {\tt km} on a $64\times674\times20$ grid, with source depths (index of $z$) ranging from 2 to 20;
\item $256\times80\times20$ {\tt km} on a $256\times80\times20$ grid, with source depths ranging from 2 to 20.
\end{enumerate}
The delta function was modeled as a Gaussian with the standard deviation of 3 and $1.5$ {\tt km}, respectively. The second grid has been chosen so as to encompass just the Landers/Hector Mine geography (see e.g. \cite{WALD}). Local fault grids have $x$ pointing in the strike direction, $y$ down dip and $z$ into the foot wall\footnote{e.g. for a strike-slip fault striking North the local $z$ axis in the fault coordinate system points West}.

\begin{figure}[htbp]
\begin{center}
\begin{tabular}{ccc}
\includegraphics[width=.33\textwidth]{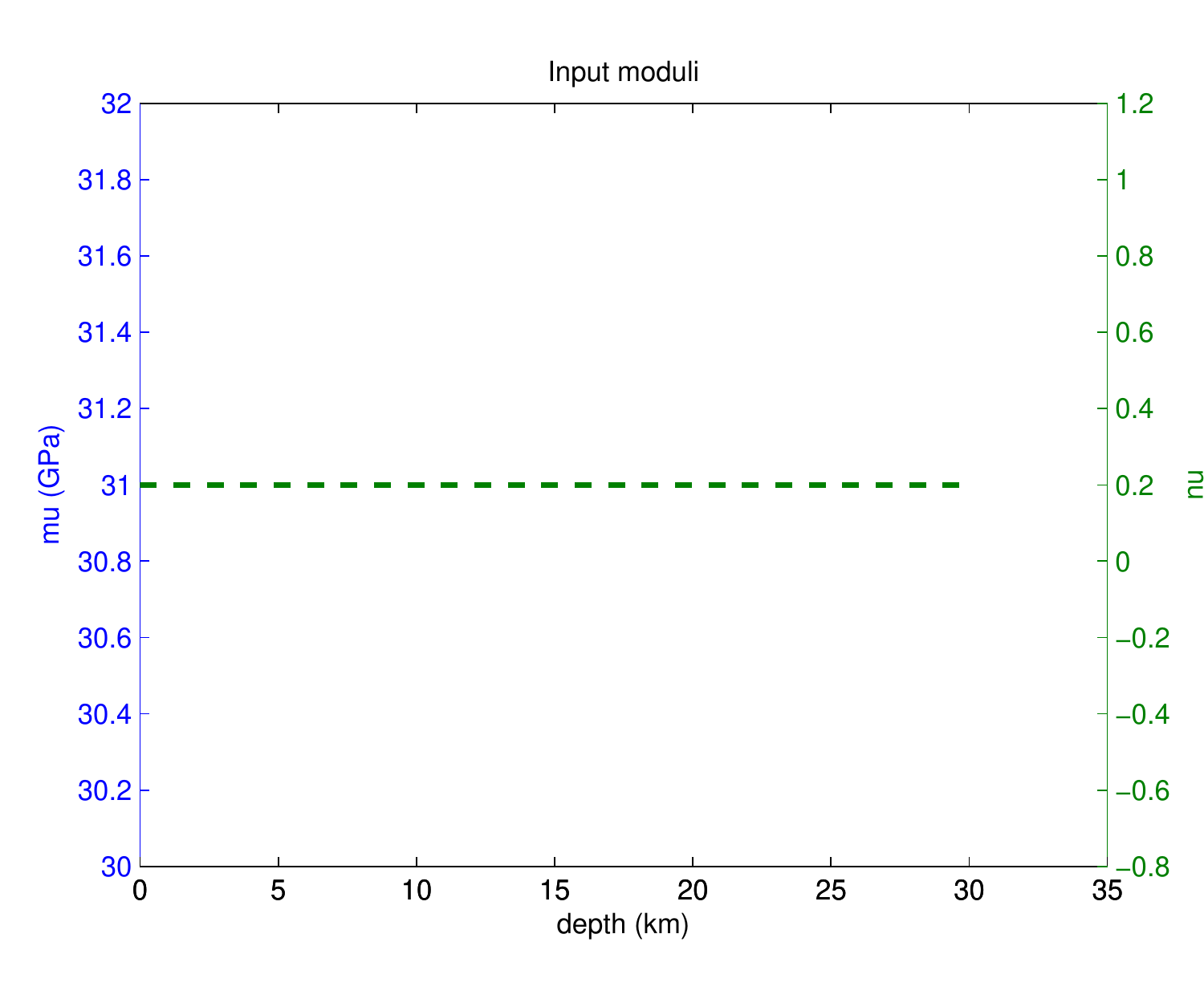} &
\includegraphics[width=.33\textwidth]{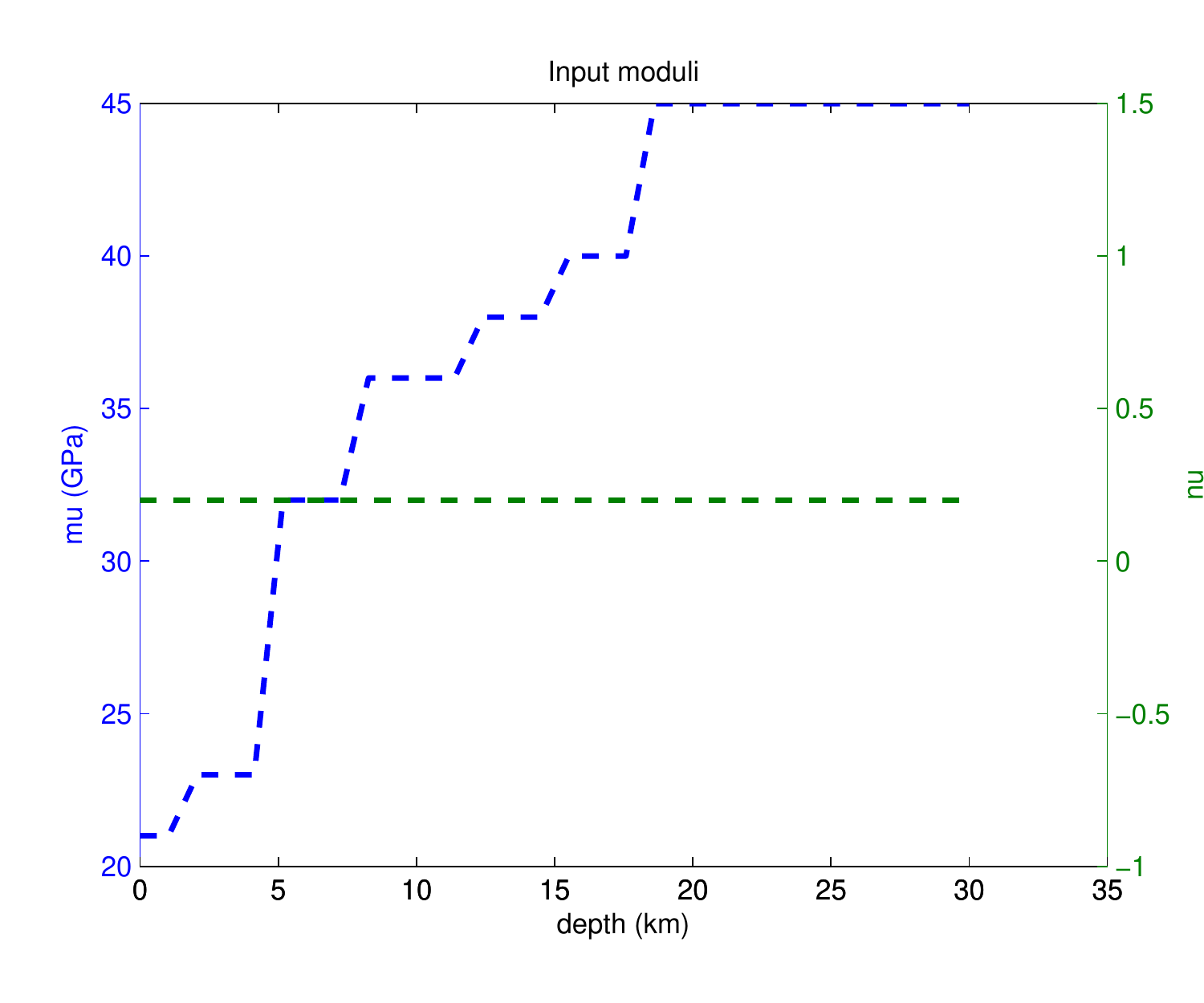} &
\includegraphics[width=.33\textwidth]{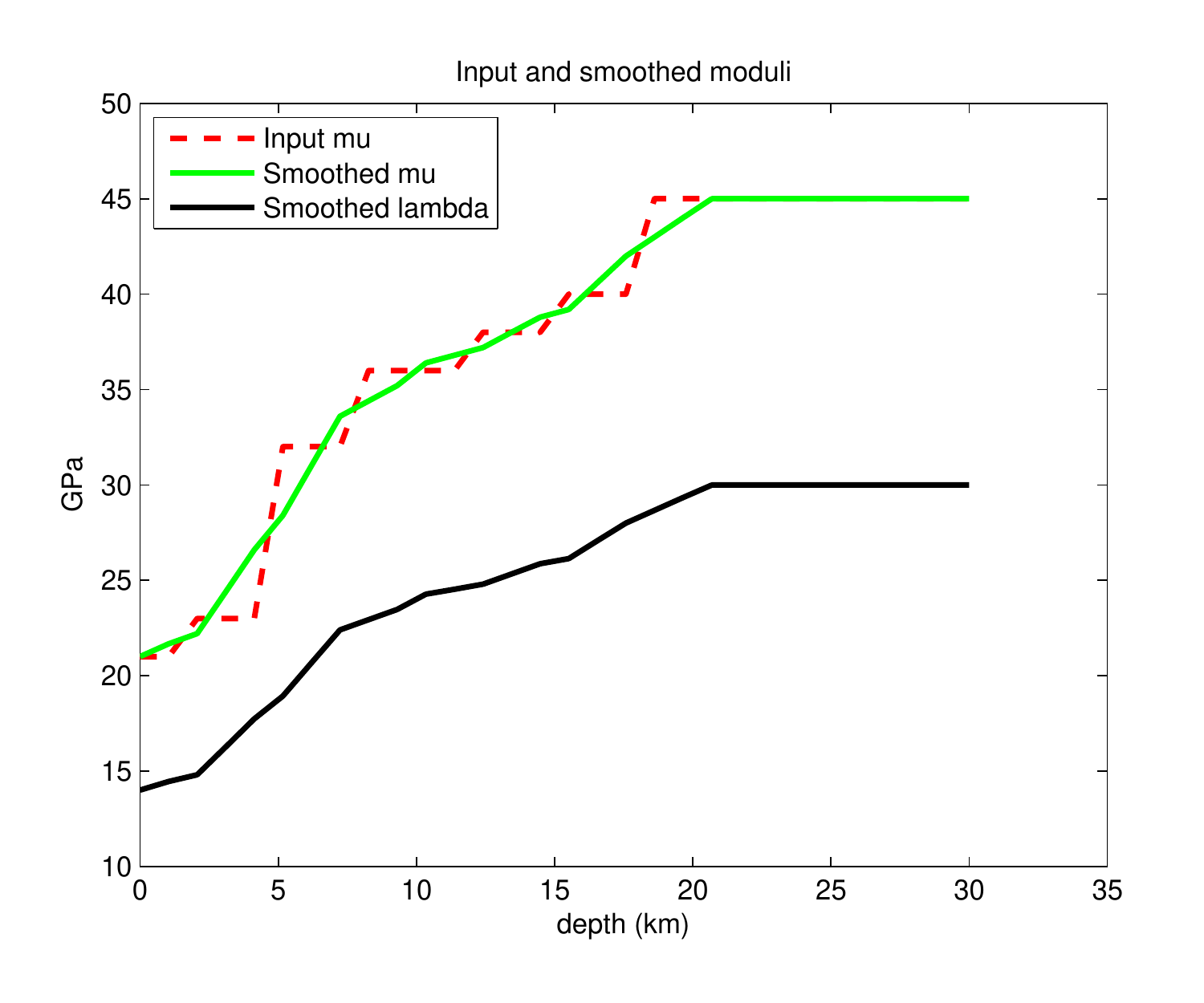}
\end{tabular}
\end{center}
\vspace{-5 mm}
\caption{Homogeneous and layered Earth model. Smoothed model.}
\label{fig:moduli}
\end{figure}

In our analysis of sensitivity to elastic heterogeneity, we use two earth models: homogeneous $\mu=31$ {\tt GPa} and a horizontally layered one with more compliant top layers but a similar average shear modulus over seismogenic depths of $\approx$15 {\tt km} (see fig \ref{fig:moduli}). Poisson ratio is assumed constant in both cases, equal to $0.2$. The shear modulus and (computed) Lam\'{e} constant are smoothed with a 4-{\tt km} rectangular filter before solving (\ref{eq:fewe}) -- see fig \ref{fig:moduli}. 

The elastostatic Green's tensor computation is performed by the \newline
 {\tt elastostatic\_green.m} {\tt MATLAB} module, described in Appendix A.

\begin{figure}[htbp]
\begin{center}
\begin{tabular}{cc}
\includegraphics[width=.5\textwidth]{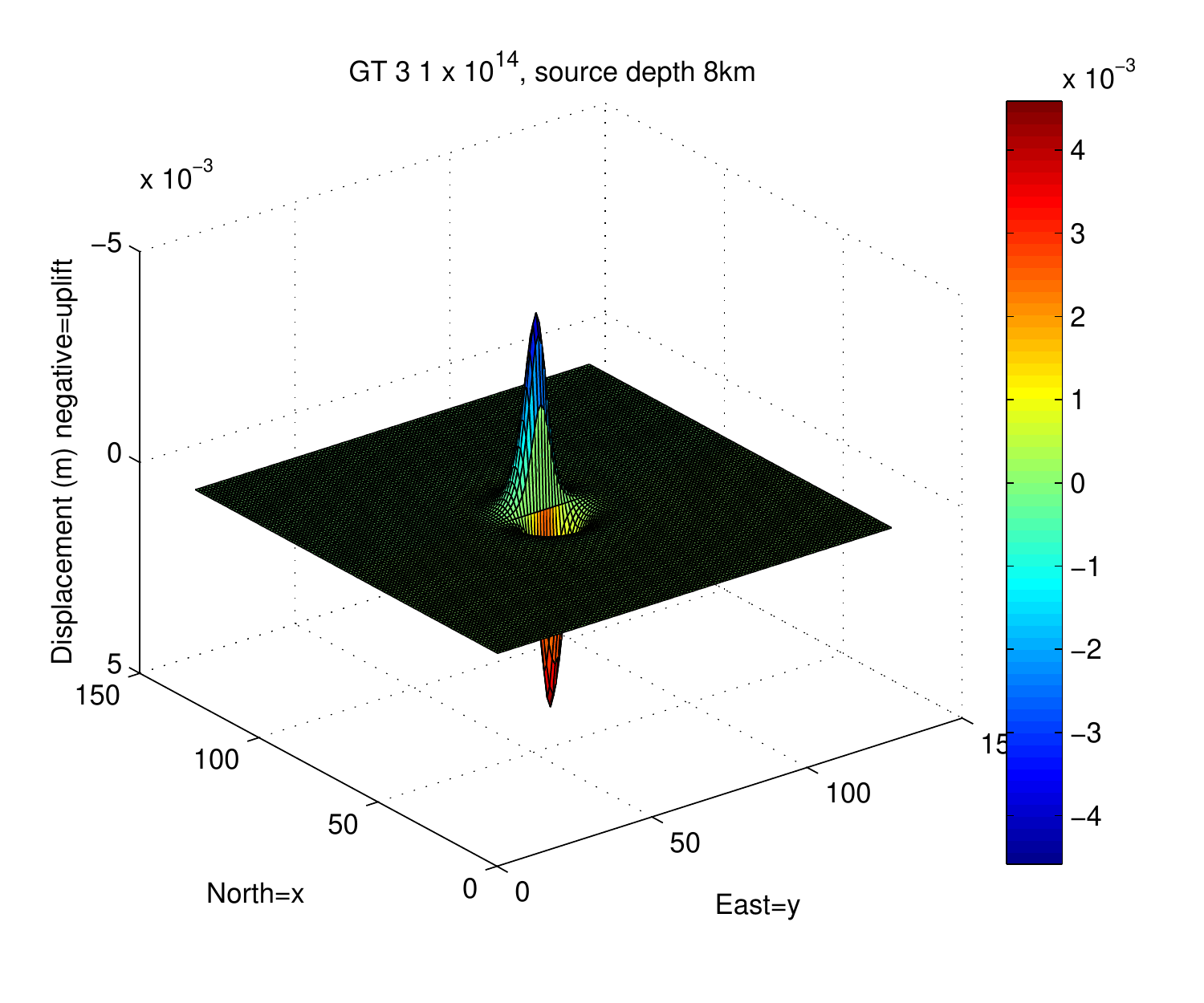} &
\includegraphics[width=.5\textwidth]{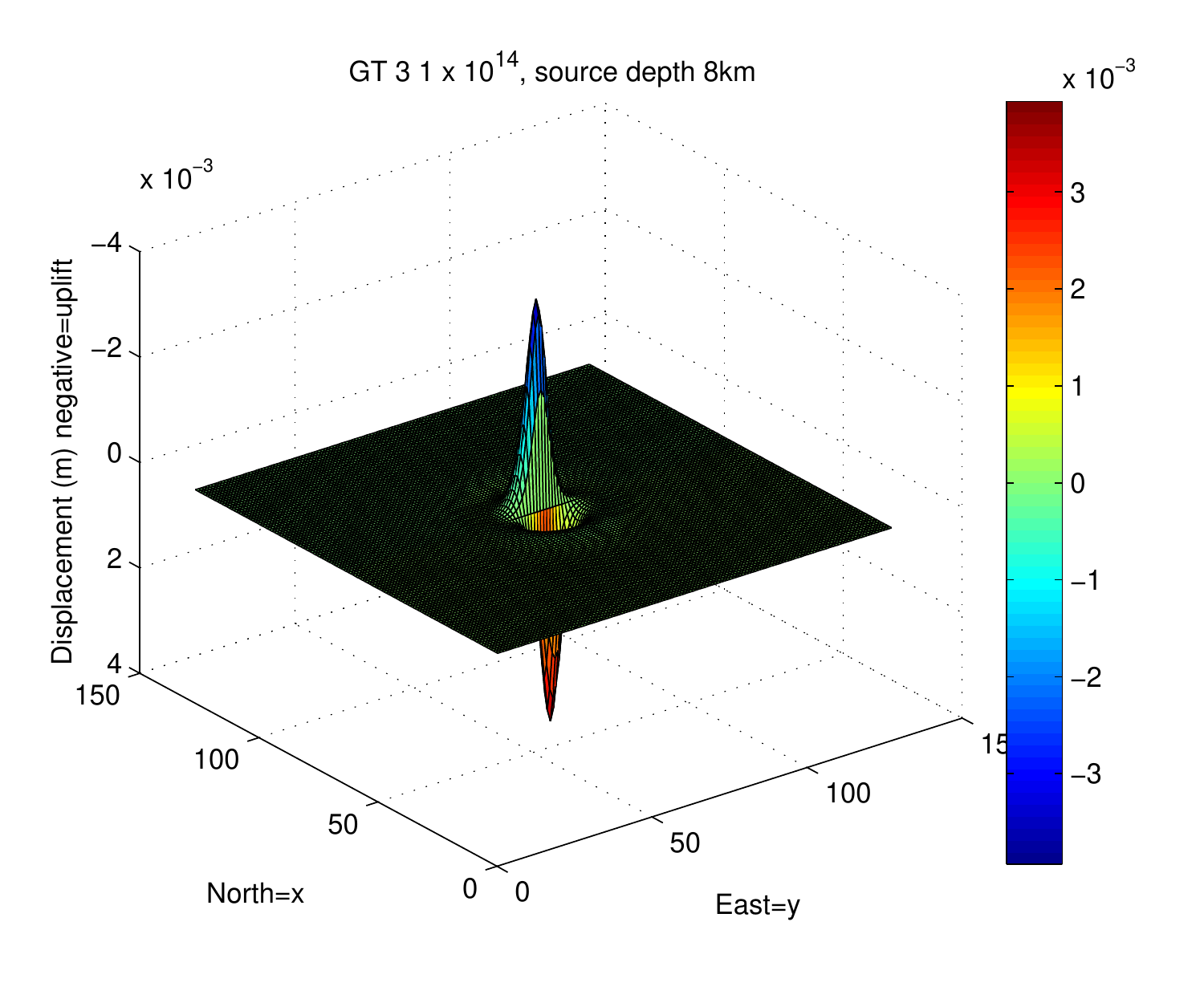}
\end{tabular}
\end{center}
\vspace{-10 mm}
\caption{Green's Tensor Component: vertical displacement due to unit force in Northern direction, E 90$^\circ$ strike, homogeneous vs layered Earth.}
\label{fig:green}
\end{figure}

\begin{figure}[htbp]
\begin{center}
\begin{tabular}{cc}
\includegraphics[width=.5\textwidth]{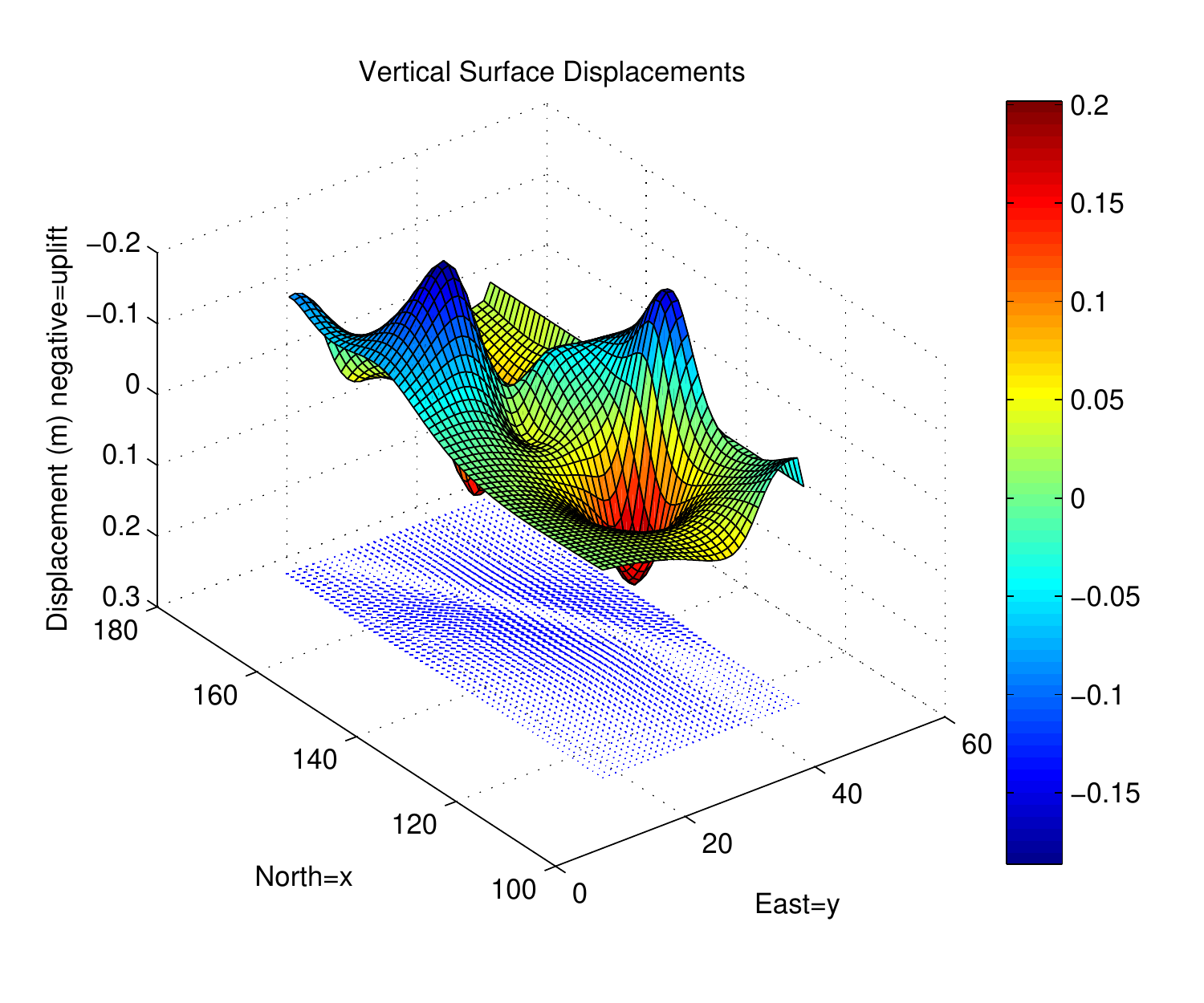} &
\includegraphics[width=.5\textwidth]{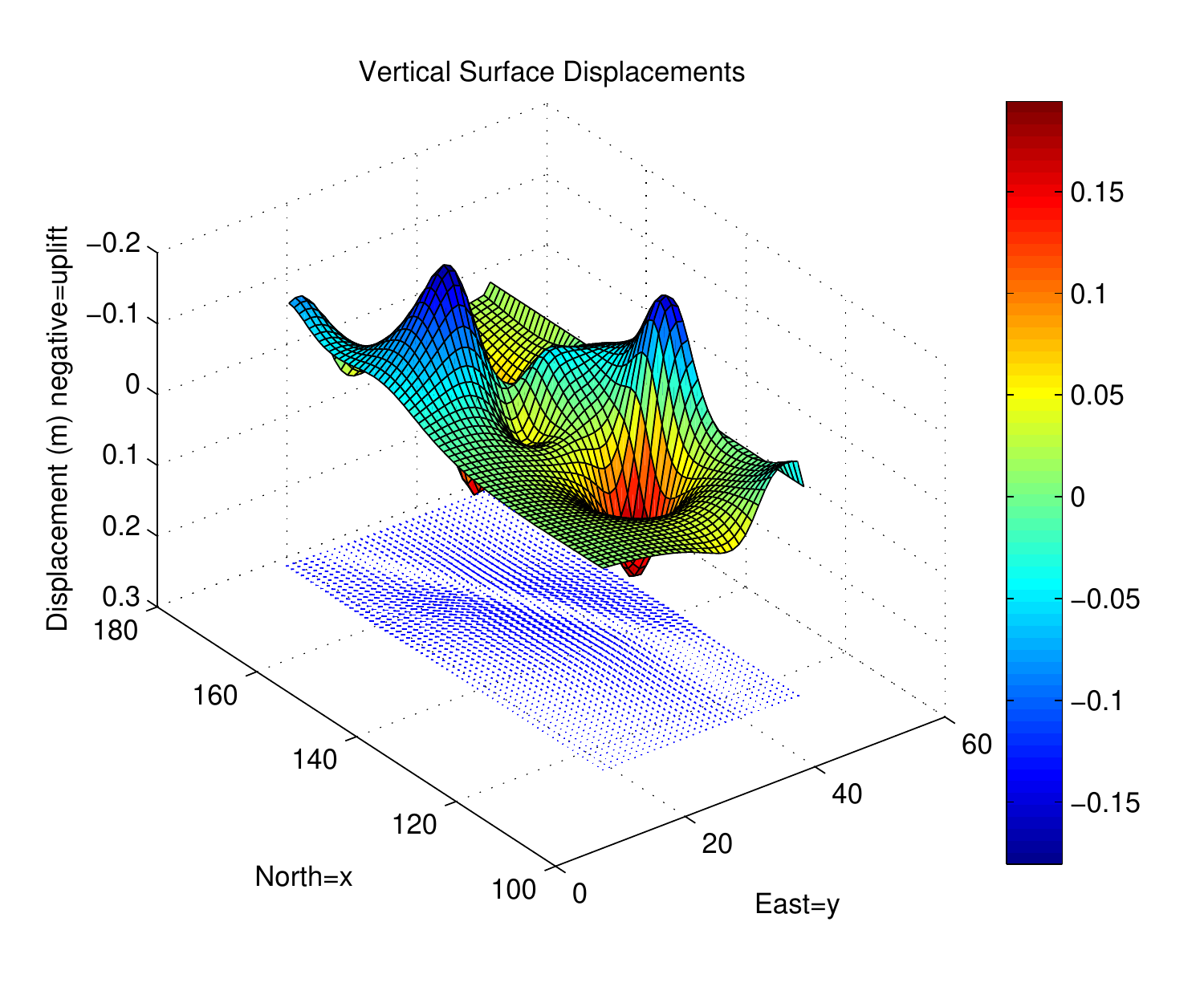}
\end{tabular}
\end{center}
\vspace{-10 mm}
\caption{Landers Forward Model, vertical displacement, homogeneous vs layered Earth.}
\label{fig:landersvert}
\end{figure}

\begin{figure}[htbp]
\begin{center}
\begin{tabular}{cc}
\includegraphics[width=.5\textwidth]{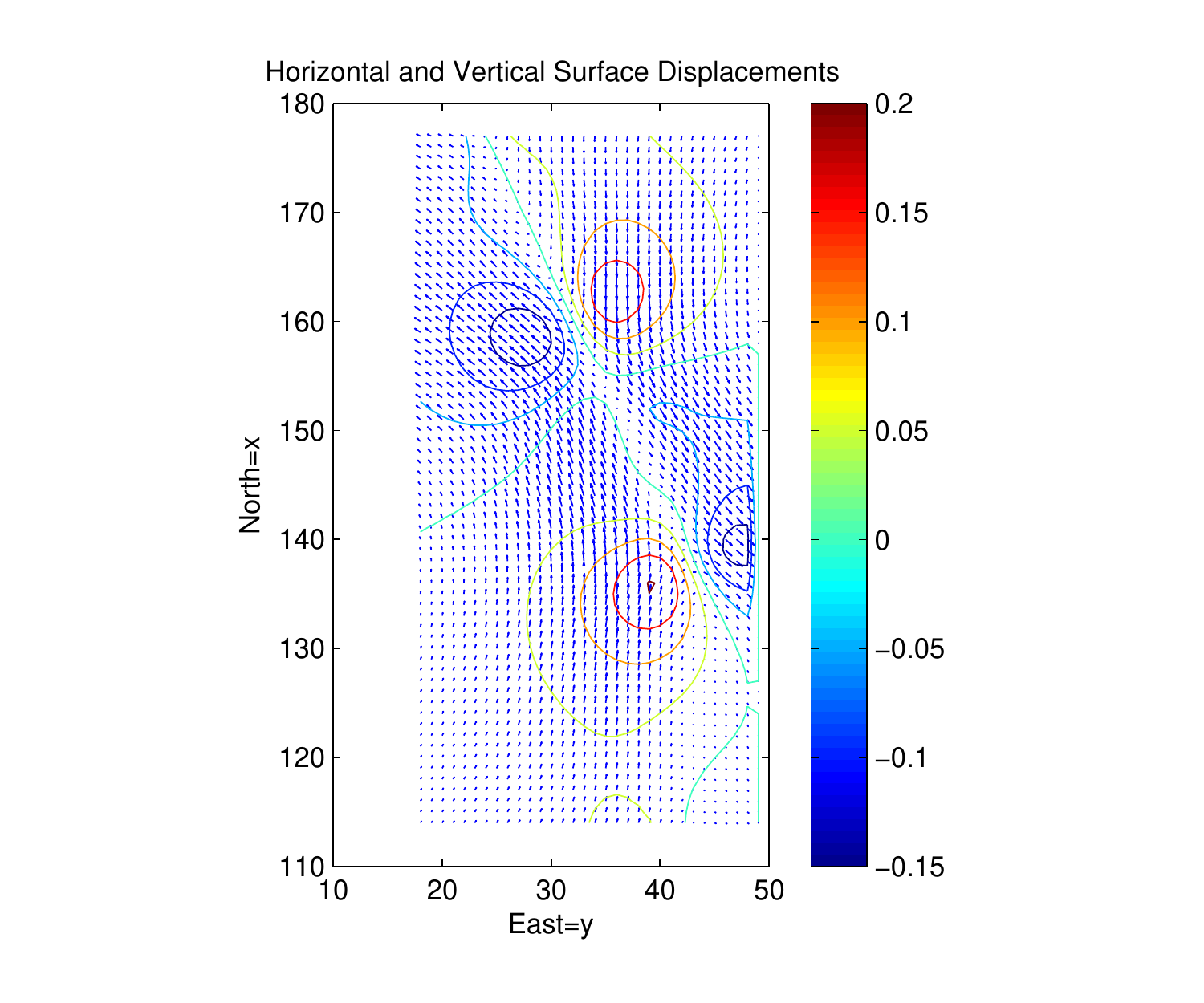} &
\includegraphics[width=.5\textwidth]{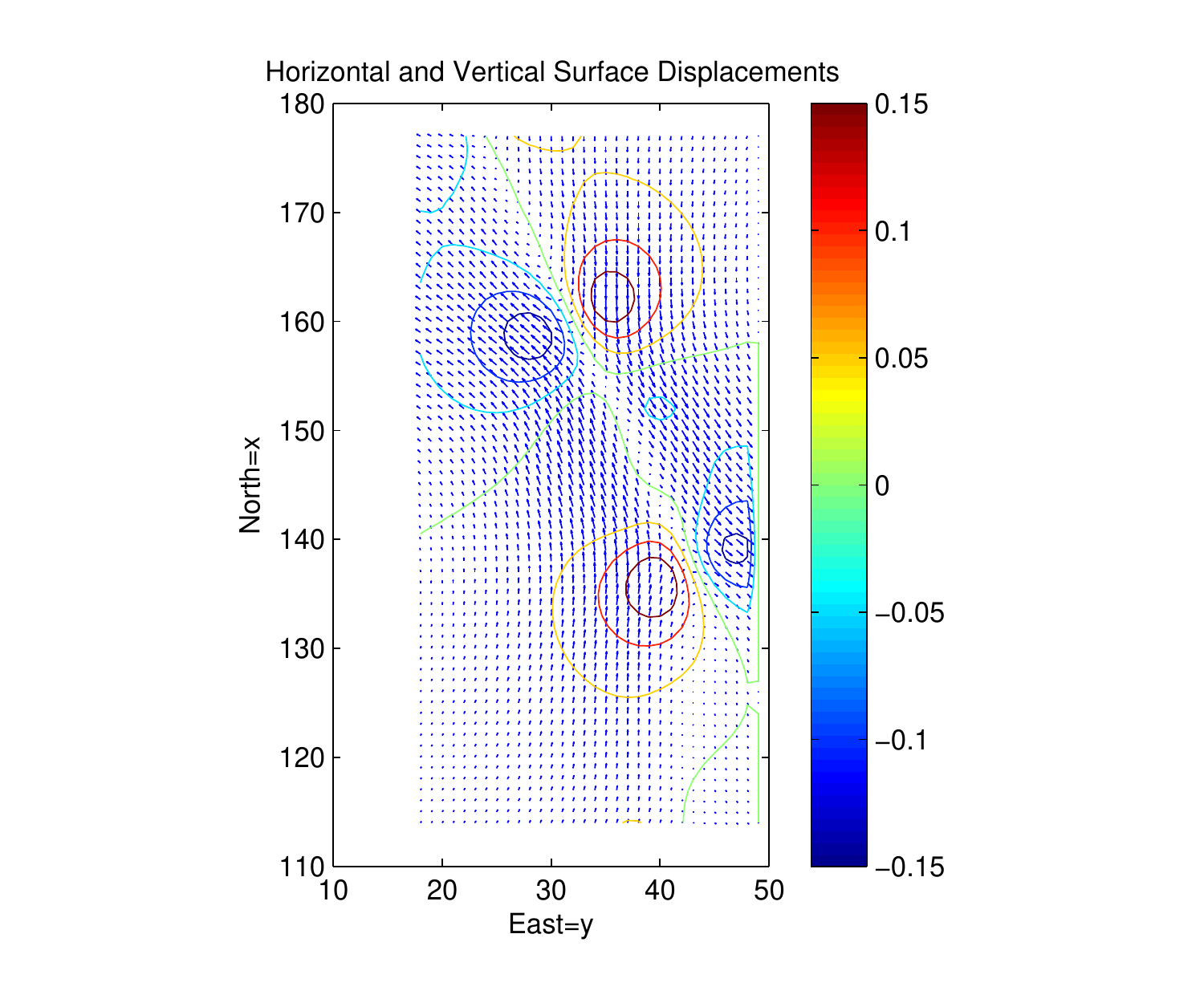}
\end{tabular}
\end{center}
\vspace{-10 mm}
\caption{Landers Forward Model, vertical displacement contour, homogeneous vs layered Earth.}
\label{fig:landerscon}
\end{figure}

\begin{figure}[htbp]
\begin{center}
\begin{tabular}{cc}
\includegraphics[width=.5\textwidth]{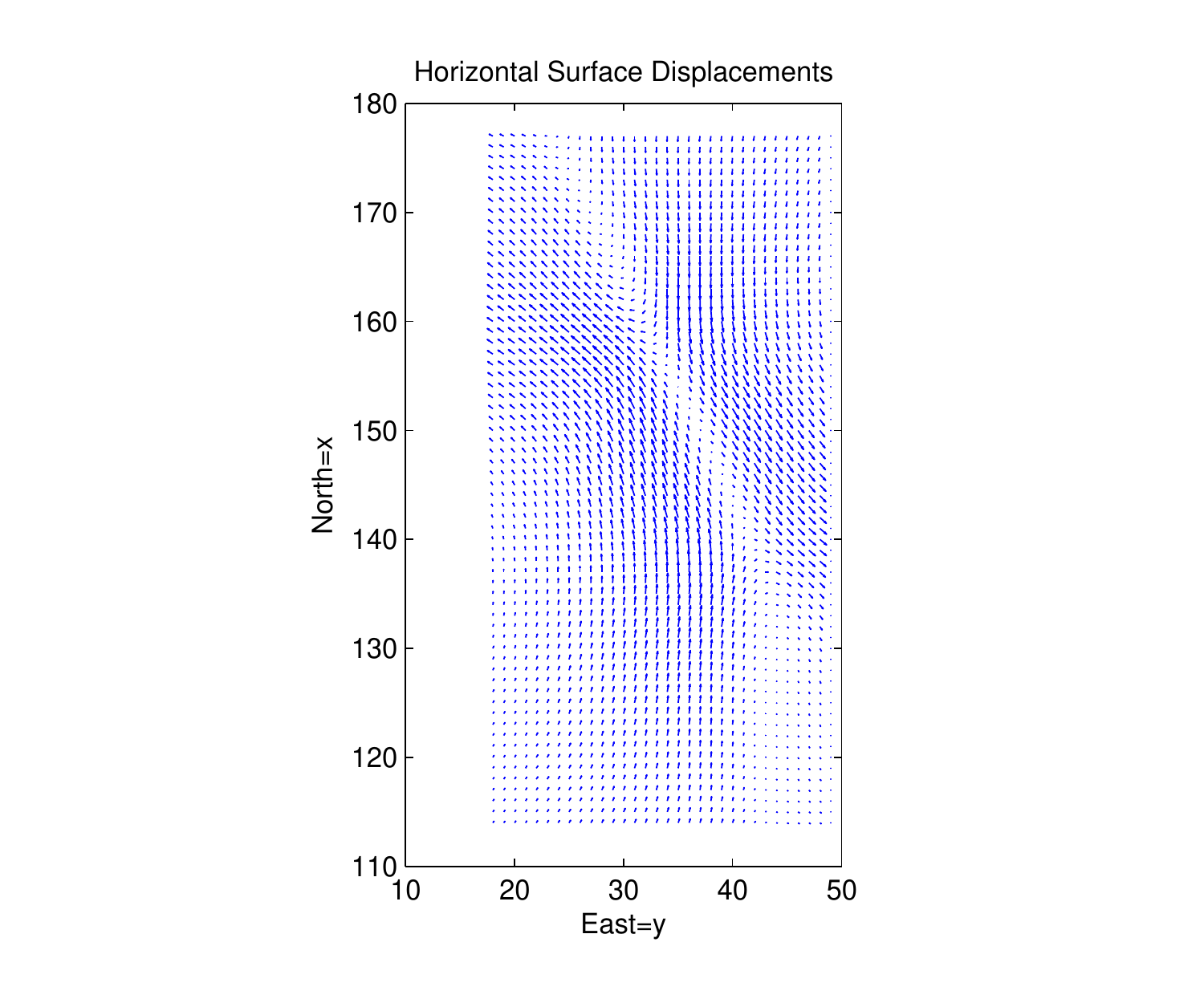} &
\includegraphics[width=.5\textwidth]{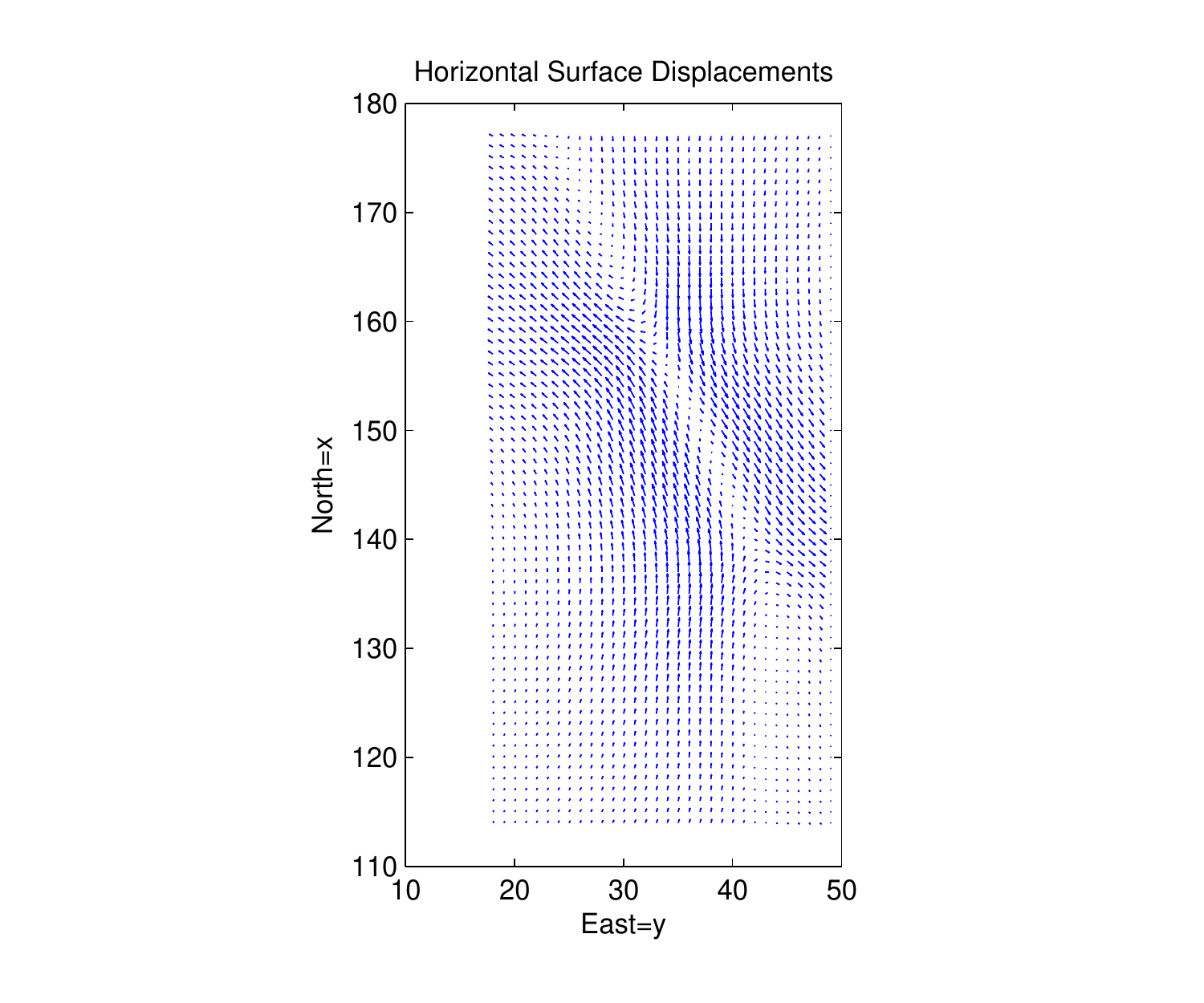}
\end{tabular}
\end{center}
\vspace{-10 mm}
\caption{Landers Forward Model, horizontal displacement directions, homogeneous vs layered Earth.}
\label{fig:landersxy}
\end{figure}

\begin{figure}[htbp]
\begin{center}
\begin{tabular}{cc}
\includegraphics[width=.5\textwidth]{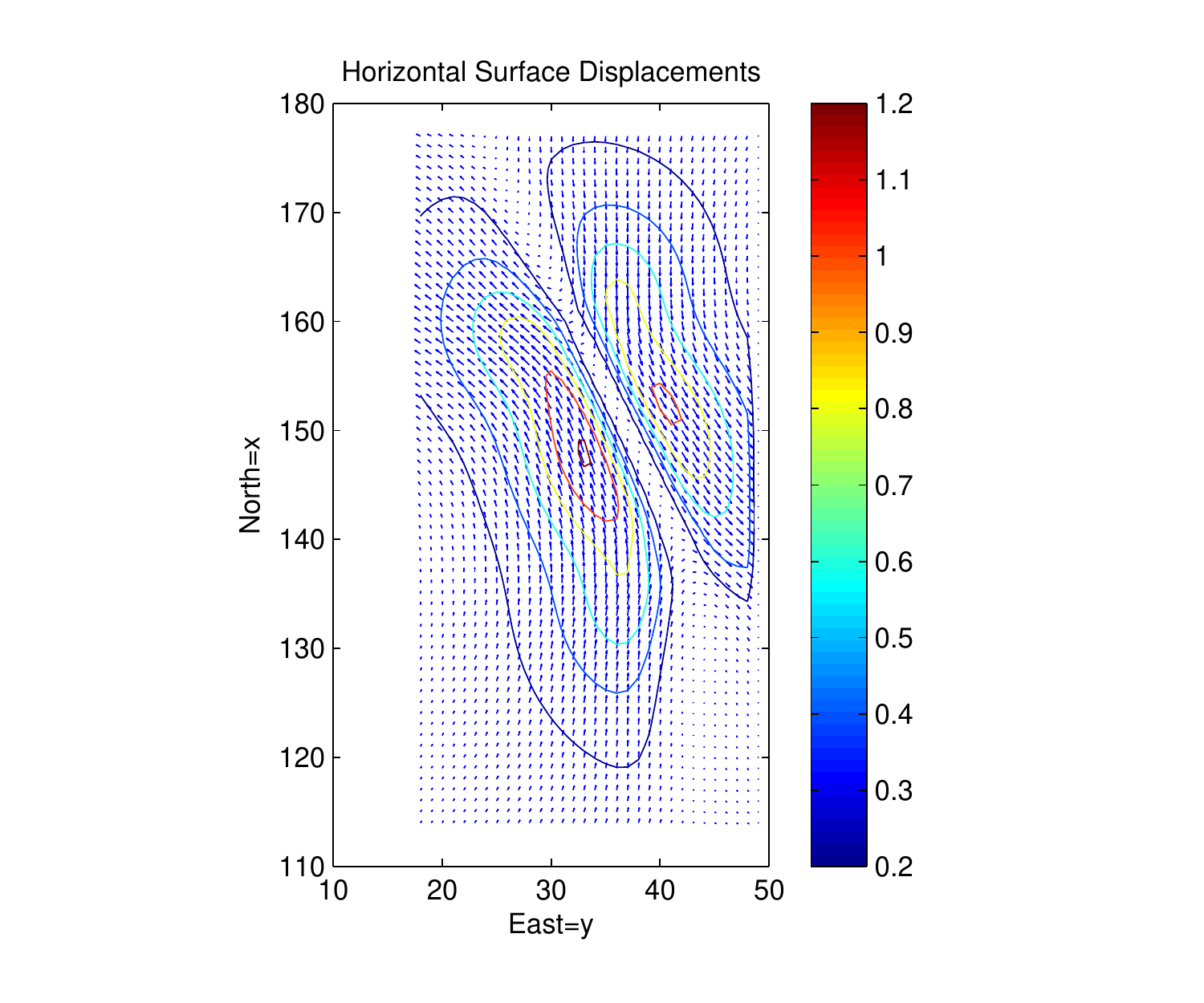} &
\includegraphics[width=.5\textwidth]{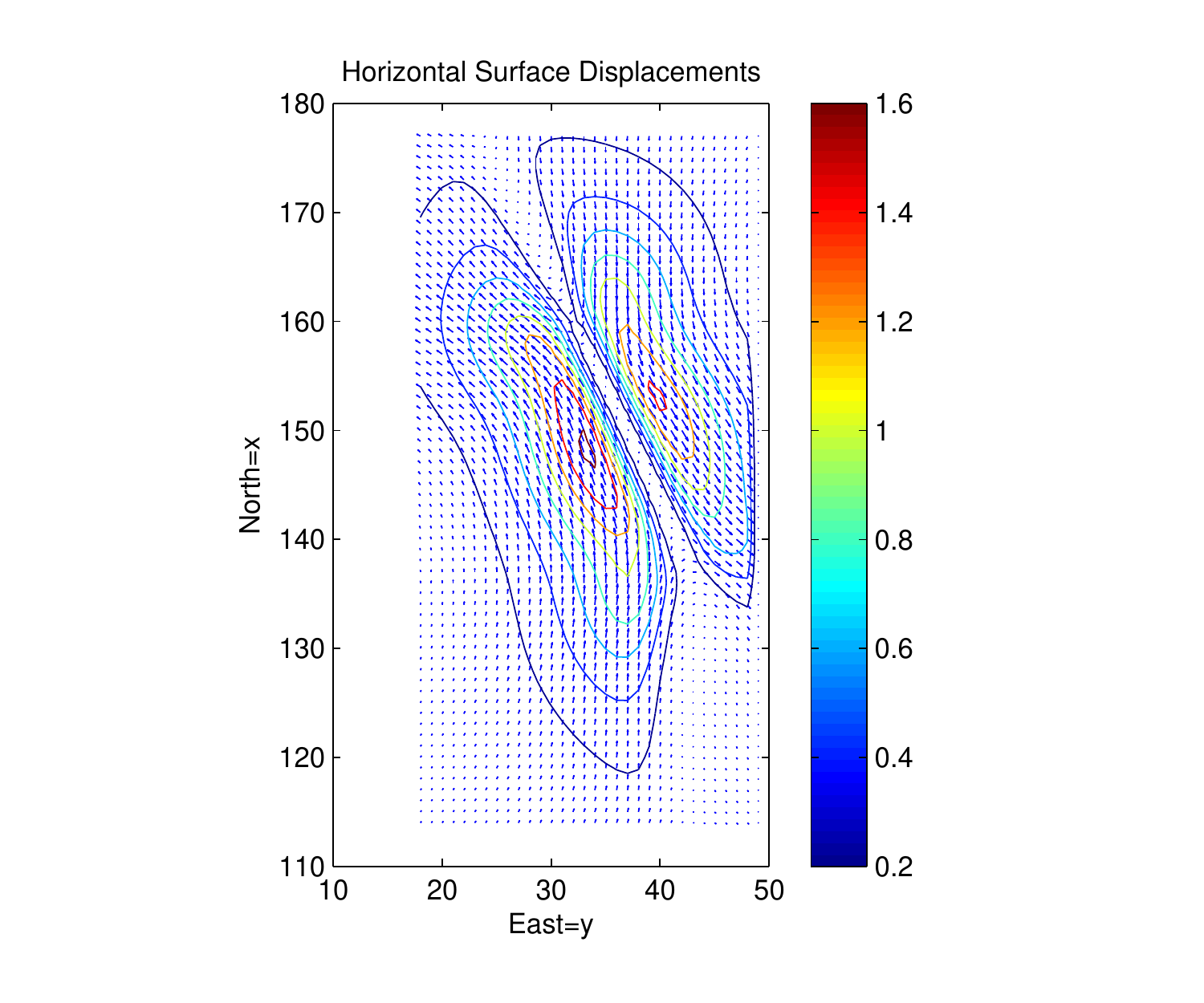}
\end{tabular}
\end{center}
\vspace{-10 mm}
\caption{Landers Forward Model, horizontal displacement, homogeneous vs layered Earth.}
\label{fig:landersxycon}
\end{figure}

A component of the Green's tensor computed on the square grid is shown on fig~\ref{fig:green}. Note that the more compliant top layers resulted in a smaller uplift in the compressive area to the north of the source. Figures \ref{fig:landersvert},~\ref{fig:landerscon},~\ref{fig:landersxy},~\ref{fig:landersxycon} present the results of forward-modeling surface displacements due to the Landers earthquake using the elastostatic Green's tensor computed on the $256\times80\times20$ grid. The Landers is modeled as three slips along 
\begin{itemize}
\item the combined Johnson Valley/Landers fault (centered at $34.225^\circ$N, $116.425^\circ$W, depth $9$ {\tt km}, strike -5$^\circ$, length $30$ {\tt km}, width $14$ {\tt km});
\item Hempsted Valley fault (centered at $34.425^\circ$N, $116.475^\circ$W, depth $9$ {\tt km}, strike -26$^\circ$, length $25$ {\tt km}, width $14$ {\tt km});
\item Emerson/Camp Rock fault (centered at $34.575^\circ$N, $116.575^\circ$W, depth $9$ {\tt km}, strike -40$^\circ$, length $30$ {\tt km}, width $14$ {\tt km}),
\end{itemize}
with the constant dip of 90$^\circ$ and rake of 180$^\circ$. The total Landers moment is distributed as $18\%,55\%,27\%$ among the three ``finite-plane'' events according to the estimated spatial distribution of the slip over fault surfaces (see Fig 5 of \cite{WALD}).

Note that since the event is modeled as a strike slip close (2 {\tt km}) but below the surface, the resulting behavior is consistent with the presence of more compliant top layers (see fig~\ref{fig:landerscon},~\ref{fig:landersxycon}): a smaller shear modulus on the fault near the top means greater slips that translate (across the thin 2 {\tt km} top layer) to greater horizontal displacements on the surface fig~\ref{fig:landersxycon}. And a greater compliance of the medium near the surface results in a smaller uplift/subsidence (fig~\ref{fig:landerscon}). Note the uplift in the compressional areas and subsidence in the tensile regions (fig~\ref{fig:landersvert},\ref{fig:landersxy}). Additional experiments for simple strike-slip and thrust faults indicate good qualitative agreement of the results of our modeling algorithm with the expected medium behavior (\cite{PS}). Surface displacement modeling is performed by the {\tt model\_surface\_displacements.m} {\tt MATLAB} module, described in Appendix A. Note that this module is not required for the stress transfer computations and is provided simply for validating the elastostatic tensor.

Both forward and inverse stress transfer problems (\ref{eq:CFF}) and (\ref{eq:INV}) are solved by the {\tt transfer\_stress.m} module.

\section{Stress Transfer}

Fig~\ref{fig:cat} shows a regional stress map with some pre-Landers events indicated by dots. Fault and nucleation data for the Landers event (\cite{WALD}) and Hector Mine (\cite{KAV}) place the source and receiver faults in \emph{block 72} of the map, with the Hector Mine faults situated to the North East of Landers.

Stress tensor inversion studies for this region indicate the principal maximum regional stress directions \emph{near} the block 72 ranging from -10$^\circ$ to 5$^\circ$ measured from North. The indicated stress directions are consistent with the Hector Mine faults being in a state of failure equilibrium (see \cite{RG}), as the estimated strike directions of the Hector Mine fault (-35$^\circ$ and -15$^\circ$ -- see \cite{KAV}) position those faults within $\approx 25-30^\circ$ of the maximum stress axis. Assuming an approximately strike-slip faulting regime, this range of angles corresponds to values of the friction coefficient between $0.6\approx 1/\tan {2 \times 30^\circ}$ and $0.8\approx 1/\tan{2\times 25^\circ}$. 

\begin{figure}[htbp]
\begin{center}
\begin{tabular}{cc}
\includegraphics[width=.5\textwidth]{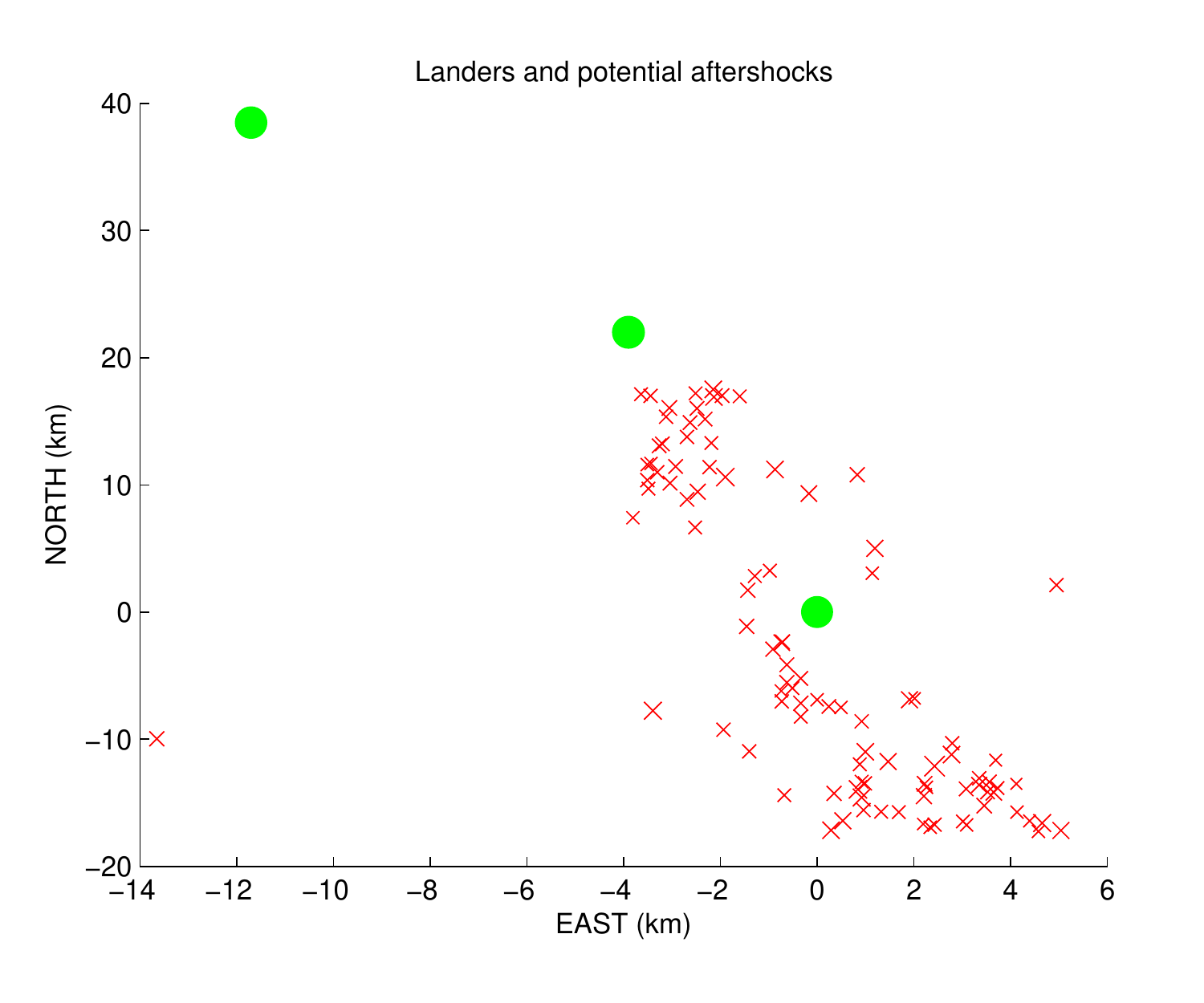} &
\includegraphics[width=.5\textwidth]{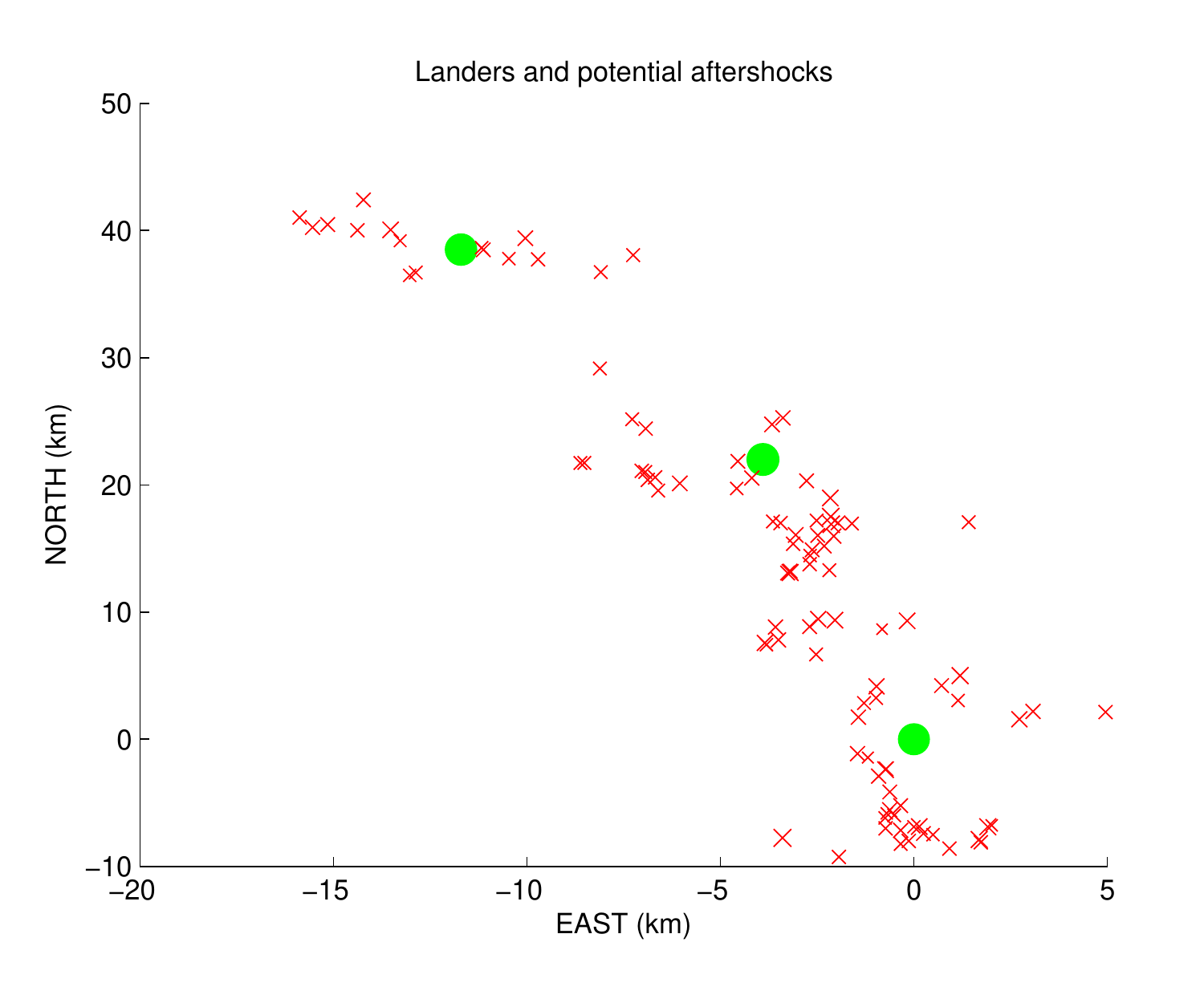} 
\end{tabular}
\end{center}
\vspace{-10 mm}
\caption{Possible Landers aftershocks used in the inversion: (a) clustered around Landers (b) more uniformly distributed around the entire rupture. Centers of Landers/Johnson Valley, Homestead Valley, Emerson/Camprock segments of the rupture marked.}
\label{fig:aftershocks}
\end{figure}

We have randomly selected 100 events from the SCED database (\cite{Hardenbeck}) within 18 {\tt km} from the center of the combined Landers/Johnson Valley faults (`local dataset') and spread uniformly around the entire Landers rupture (`regional dataset') -- see fig~\ref{fig:aftershocks}. Additionally, we solve the direct and inverse stress transfer problems for a decimated regional dataset consisting of 20 randomly picked events. The results of stress transfer computation for the local dataset are shown on fig~\ref{fig:dcffaftershockslocal},\ref{fig:dshearaftershockslocal},\ref{fig:dnormaftershockslocal}. The corresponding plots for the complete and decimated regional datasets are on fig~\ref{fig:dcffspreadaftershocks},\ref{fig:dshearspreadaftershocks},\ref{fig:dnormspreadaftershocks} and fig~\ref{fig:dcffaftershocks},\ref{fig:dshearaftershocks},\ref{fig:dnormaftershocks},respectively. In all cases majority of faults appear to be encouraged to slip through the advancement in shear and/or un-clamping. Green circles indicate that the corresponding quantity advances the slip, while red crosses denote the opposite\footnote{symbol sizes are proportional to the corresponding values}. Due to the proximity of  these aftershocks to Landers, elastic heterogeneity (e.g. compliance of the top layer) does not have a significant qualitative effect on stress transfer. However, elastic heterogeneity becomes important at greater distances when the stress change is $\ll 10$ {\tt Bar}. We will demonstrate the effect of heterogeneity by computing the stress transfer on the receiver faults for Hector Mine.

\begin{figure}[htb]
\begin{center}
\includegraphics[width=1.\textwidth]{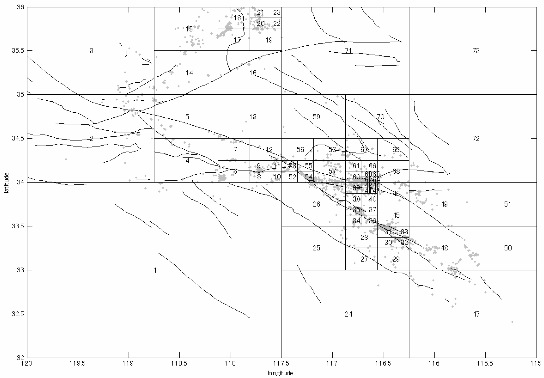}
\end{center}
\caption{Regional Fault Map.}
\label{fig:cat}
\end{figure}

\begin{figure}[htb]
\begin{center}
\includegraphics[width=1.\textwidth]{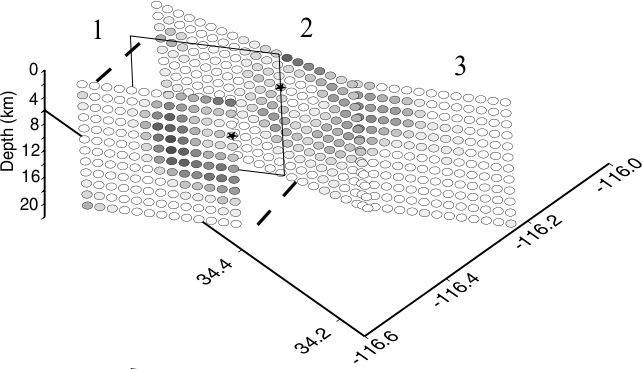}
\end{center}
\caption{Hector Mine faults with the previously estimated slips, from \cite{KAV}.}
\label{fig:kav}
\end{figure}

The Hector Mine is included in a separate dataset and modeled, following \cite{KAV}, as three faults shown on fig~\ref{fig:kav} from \cite{KAV}, using the following parameters:
\begin{verbatim}
x    y        z     length  width     strike   dip   rake  mu   B
50   5.85     10    26       16       -35      77    180   0.7   0.5
44   13.65    10    38       16       -15      77    180   0.7   0.5
16.5  23.4    10    26       16       -35      77    180   0.7   0.5
\end{verbatim}
where $x,y,z$ are the coordinates of respective fault centers with center of the Landers/Johnson Valley fault at the origin. Note the nucleation into faults 1 and 2 (see fig~\ref{fig:kav}) that appear to spatially overlap if viewed e.g. in the eastern direction. 

Results of the stress transfer calculations indicate that un-clamping appears to be the dominant slip-encouraging effect of Landers on Hector Mine (see fig~\ref{fig:dcffhm7},\ref{fig:dcffhm8},\ref{fig:dnormhm7},\ref{fig:dnormhm8}). Stress transfer is computed on local fault grids of $7\times8,10\times8$ and $7\times8$ points. Note that due to the spatial overlap of faults 1 and 2 (see fig~\ref{fig:kav}) and relative sizes of the two faults (see the parameters above) the northern half of fault 2 (i.e., the right half along the strike axis) can be expected to exhibit roughly the same properties as the entire fault 1. As can be seen from e.g. fig~\ref{fig:dcffhm7},\ref{fig:dshearhm7},\ref{fig:dnormhm7} this is indeed the case, with the difference simply due to the different strikes (15$^\circ$ and 24$^\circ$). The shear stress appears to drop throughout the Hector Mine rupture (fig~\ref{fig:dshearhm7},\ref{fig:dshearhm8}), while a significant un-clamping is predicted on faults 1 and 2 (fig~\ref{fig:dnormhm7},\ref{fig:dnormhm8}). Coulomb transfer appears negative almost everywhere except at the ``junction'' of faults 1 and 2 -- i.e., where the maximum slip has been predicted along the Hector Mine rupture (see fig~\ref{fig:kav} and \cite{KAV}). It is plausible that Landers statically triggered Hector Mine by stress transfer to this point that happens to coincide with the hypocenter (\cite{KAV},\cite{Parsons}). A slightly higher friction $0.8$ results in a greater slip encouragement, however, does not alter the qualitative picture. Note that our results are in a good qualitative agreement with \cite{Parsons} (compare with plate 2 of \cite{Parsons}) and our previous estimates in HW7.

The results indicate a significant sensitivity to the elastic heterogeneity of the medium. Stress plots appear to be shifted down for the heterogeneous test, consistent with equal average moduli but more compliant top layers in the heterogeneous case. Note the resulting quantitative impact e.g. normal stress drop on fig~\ref{fig:dnormhm7},\ref{fig:dnormhm8}. This indicates the importance of accounting for elastic heterogeneity in stress transfer analysis.

\begin{figure}[htbp]
\begin{center}
\begin{tabular}{cc}
\includegraphics[width=.5\textwidth]{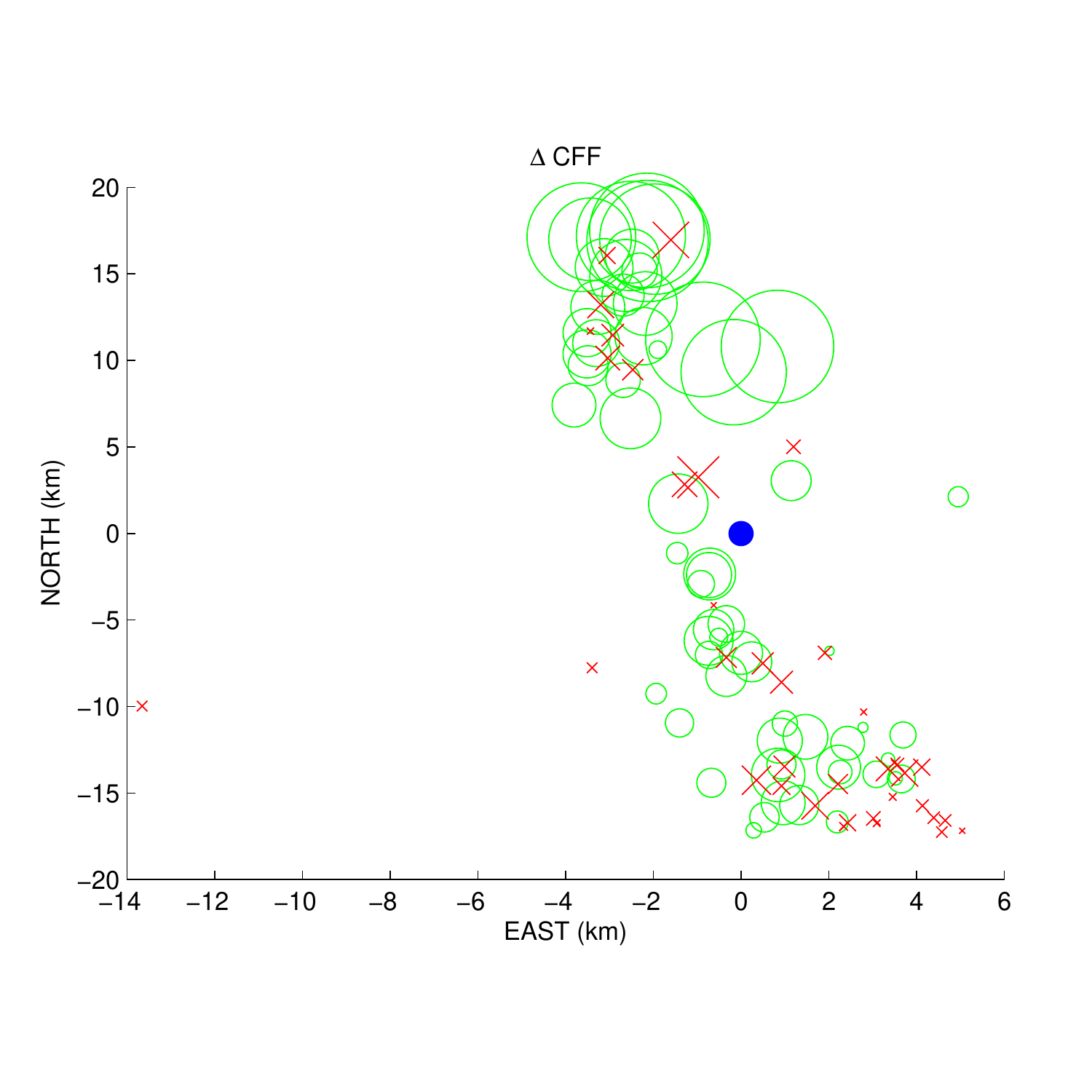} &
\includegraphics[width=.5\textwidth]{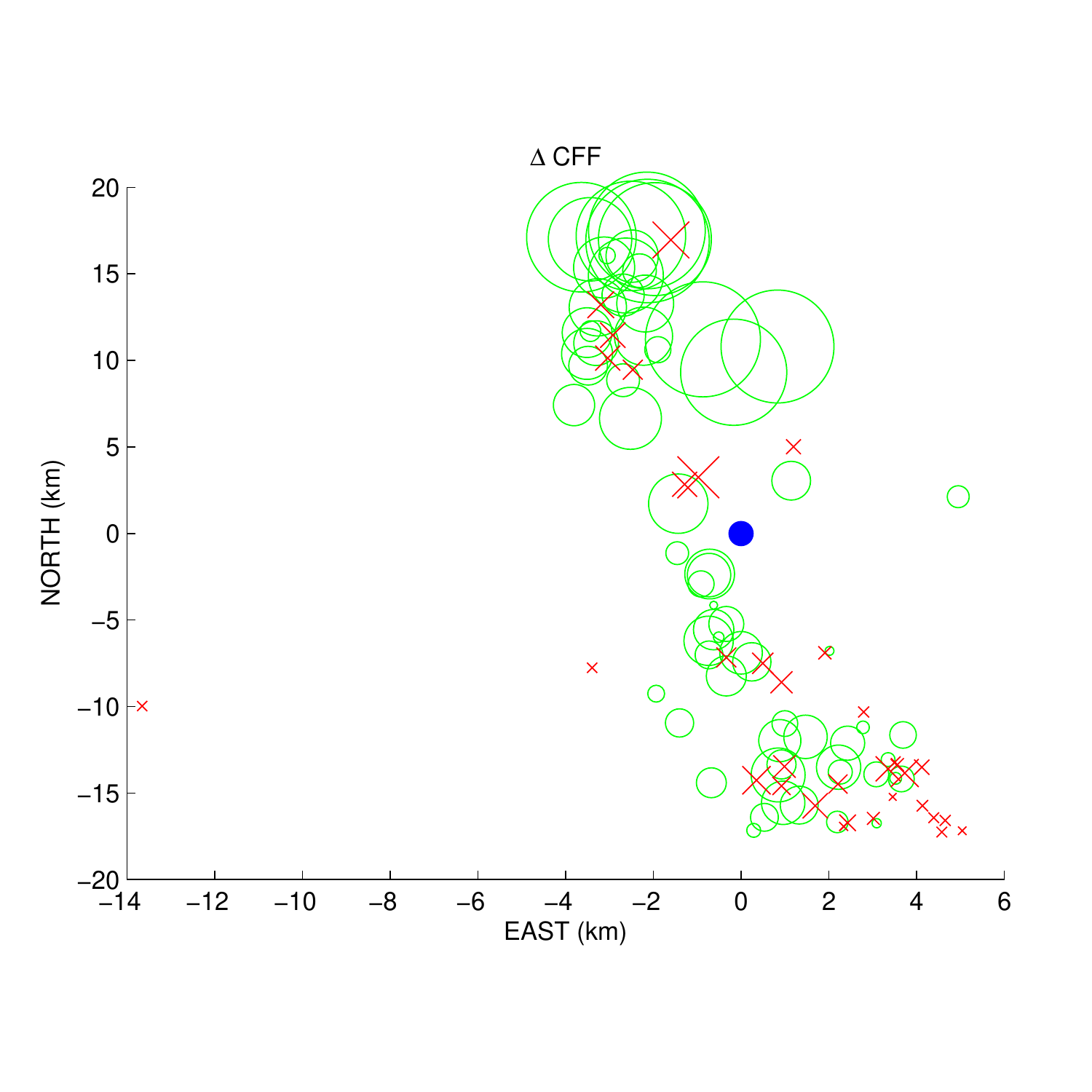} 
\end{tabular}
\end{center}
\vspace{-10 mm}
\caption{Coulomb transfer for the local dataset of Landers aftershocks (20 events), homogeneous vs layered Earth.}
\label{fig:dcffaftershockslocal}
\end{figure}

\begin{figure}[htbp]
\begin{center}
\begin{tabular}{cc}
\includegraphics[width=.5\textwidth]{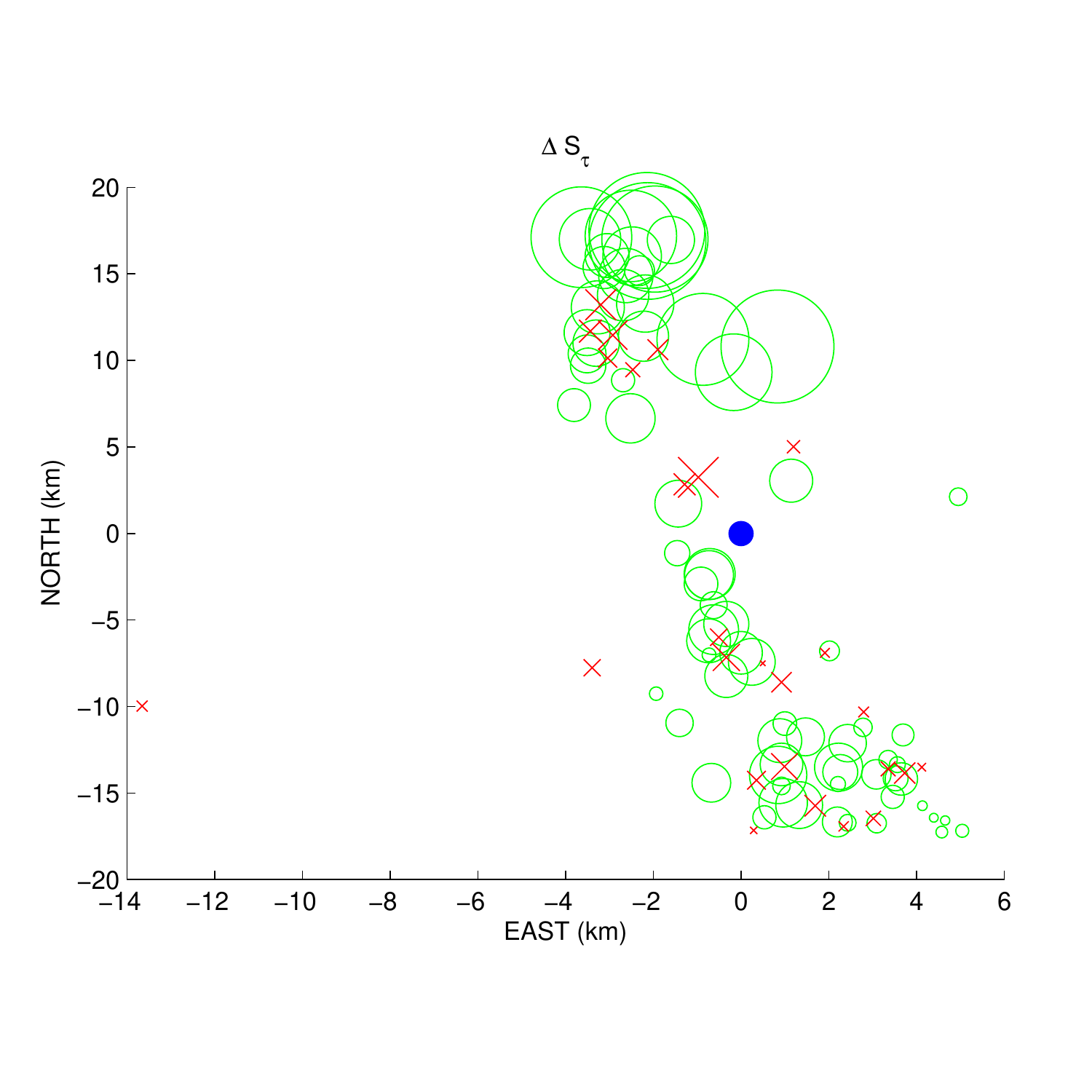} &
\includegraphics[width=.5\textwidth]{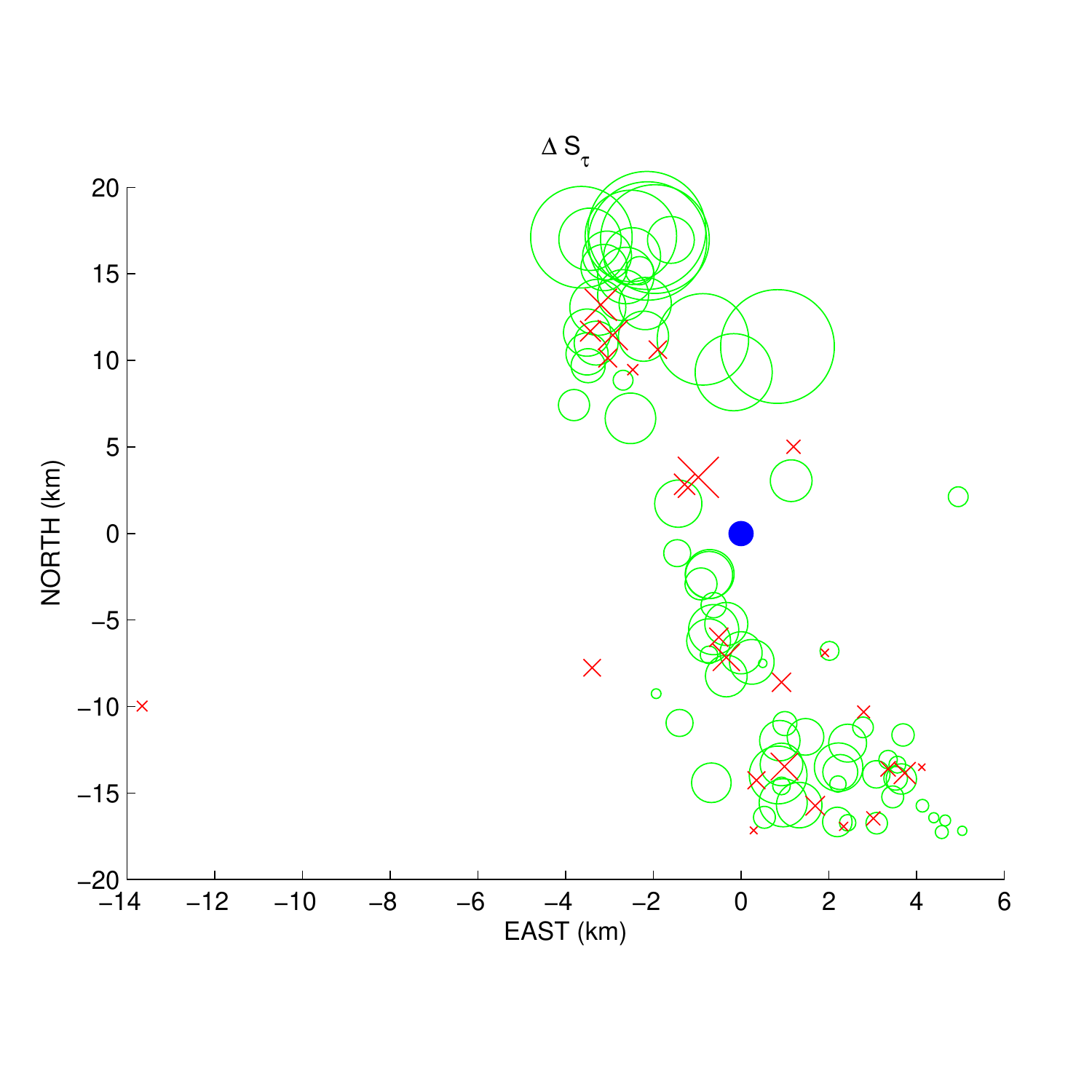} 
\end{tabular}
\end{center}
\vspace{-10 mm}
\caption{Advancement in shear for the local dataset of Landers aftershocks, homogeneous vs layered Earth.}
\label{fig:dshearaftershockslocal}
\end{figure}

\begin{figure}[htbp]
\begin{center}
\begin{tabular}{cc}
\includegraphics[width=.5\textwidth]{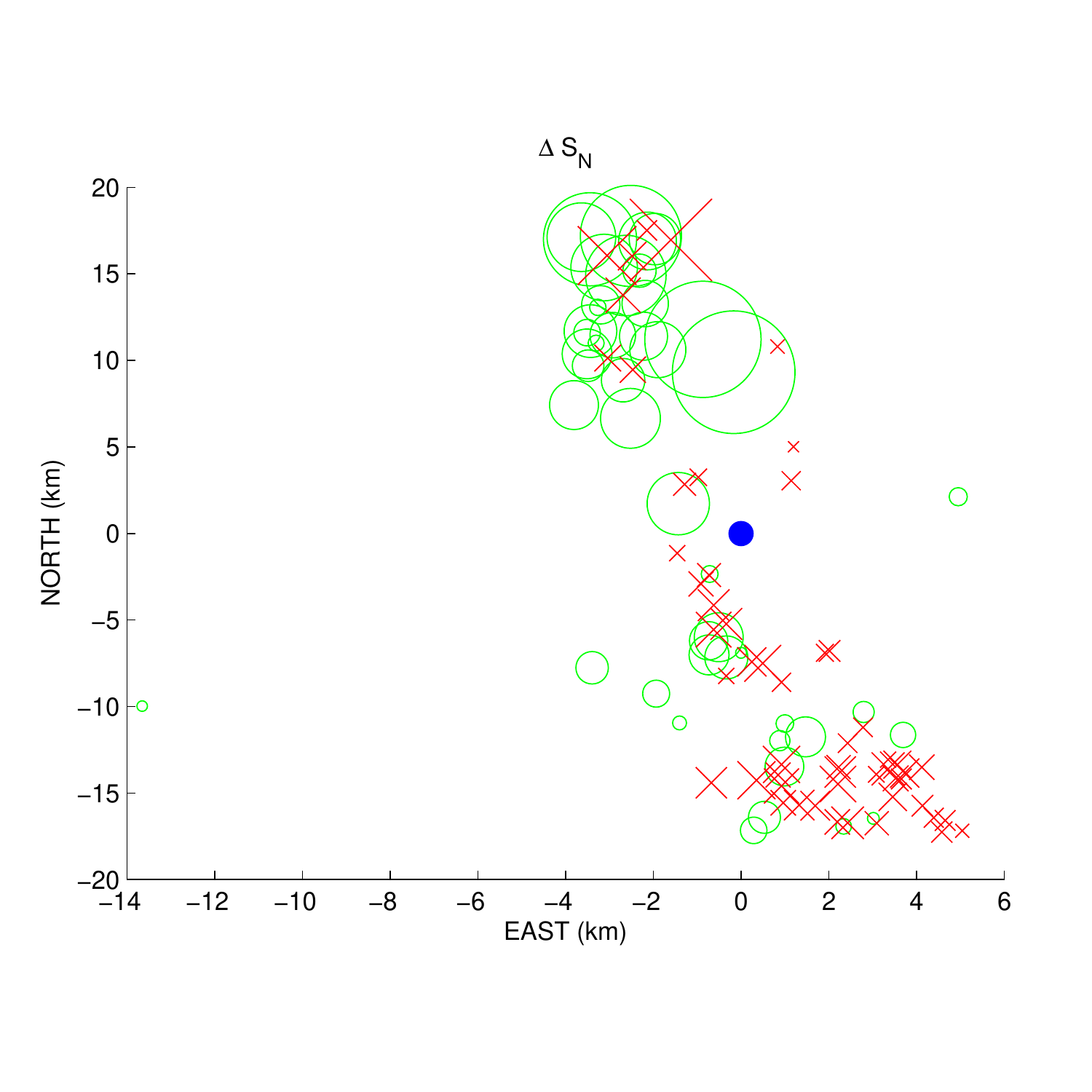} &
\includegraphics[width=.5\textwidth]{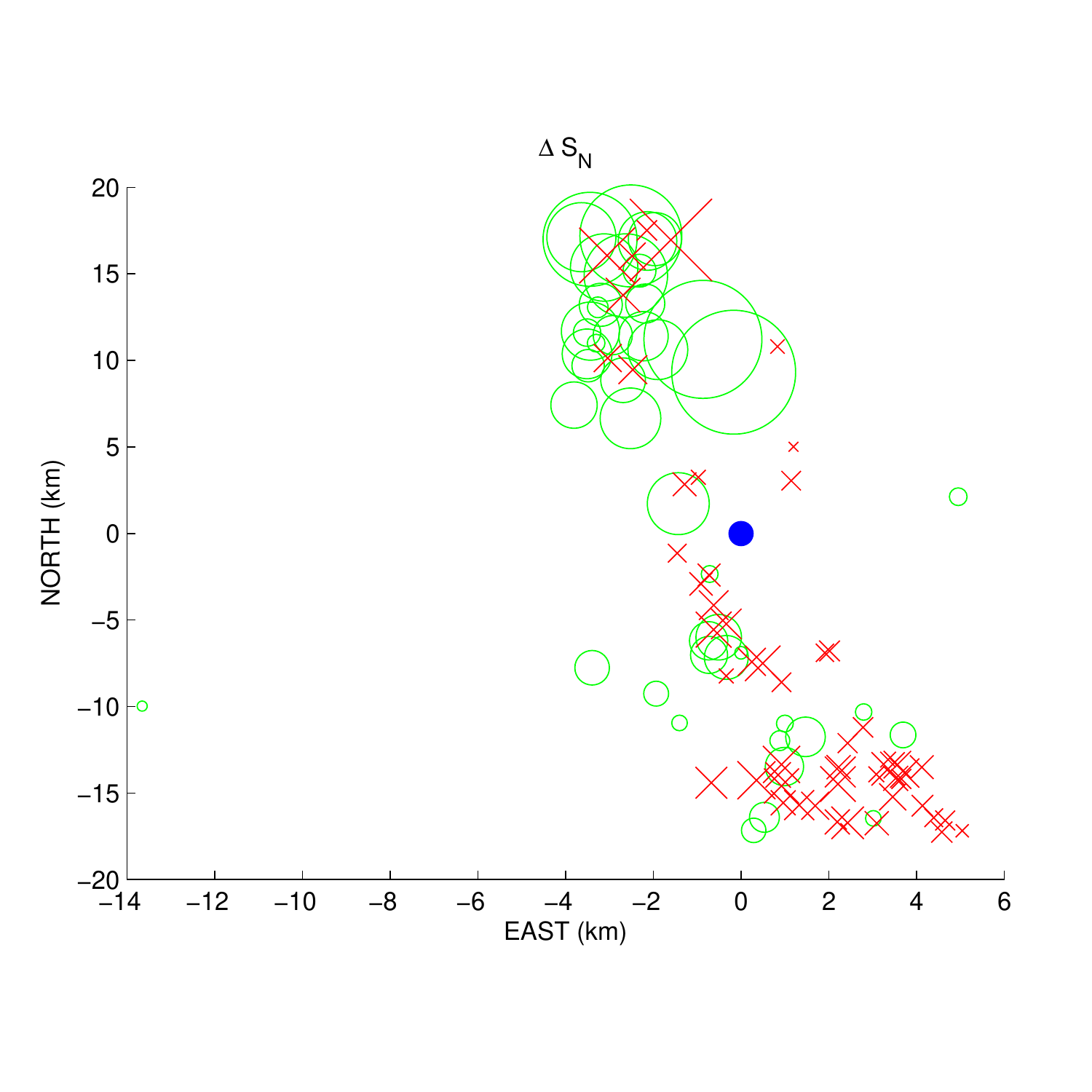} 
\end{tabular}
\end{center}
\vspace{-10 mm}
\caption{Unclamping (green) for the local dataset of Landers aftershocks, homogeneous vs layered Earth.}
\label{fig:dnormaftershockslocal}
\end{figure}
\begin{figure}[htbp]
\begin{center}
\begin{tabular}{cc}
\includegraphics[width=.5\textwidth]{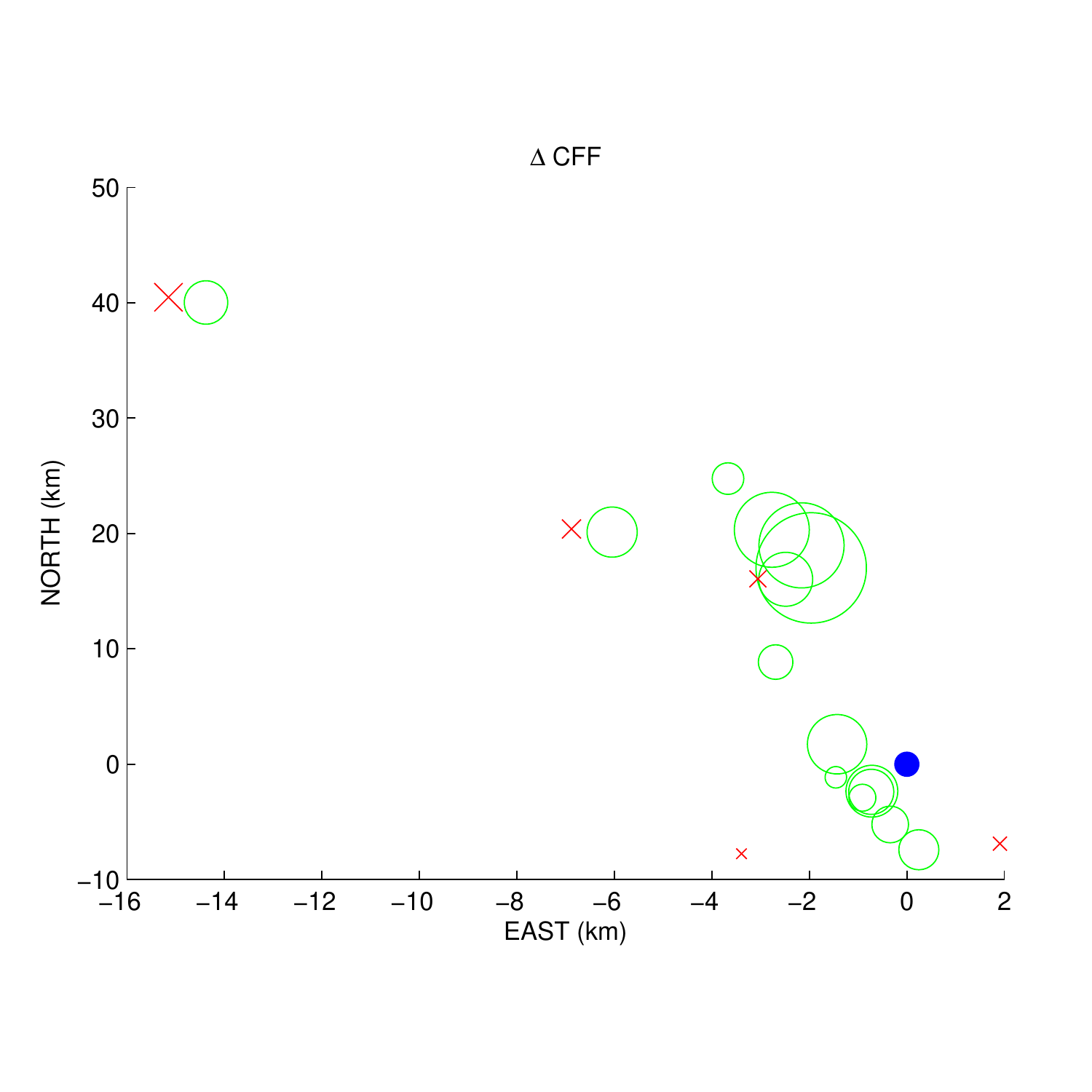} &
\includegraphics[width=.5\textwidth]{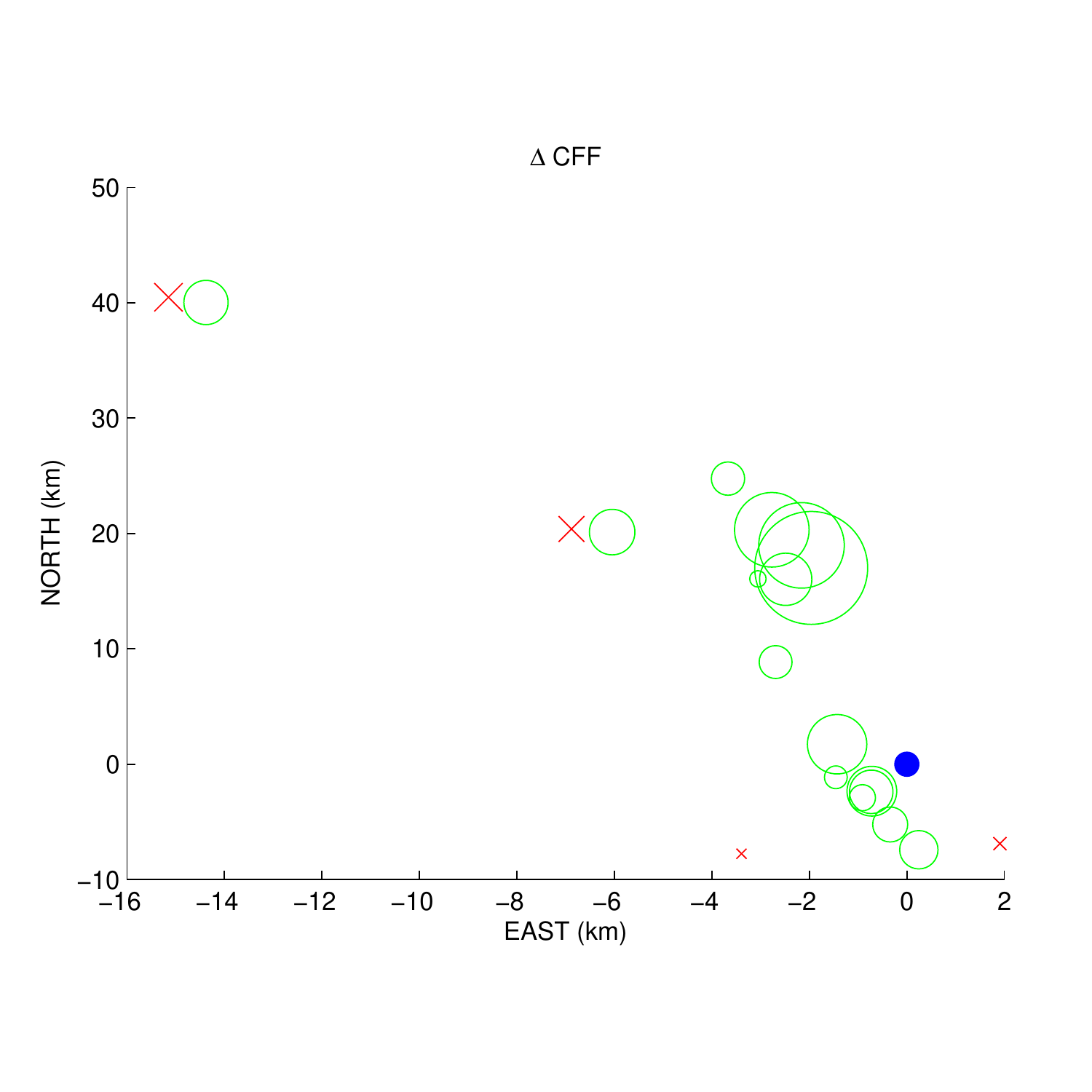} 
\end{tabular}
\end{center}
\vspace{-10 mm}
\caption{Coulomb transfer for the decimated regional dataset of Landers aftershocks (20 events), homogeneous vs layered Earth.}
\label{fig:dcffaftershocks}
\end{figure}

\begin{figure}[htbp]
\begin{center}
\begin{tabular}{cc}
\includegraphics[width=.5\textwidth]{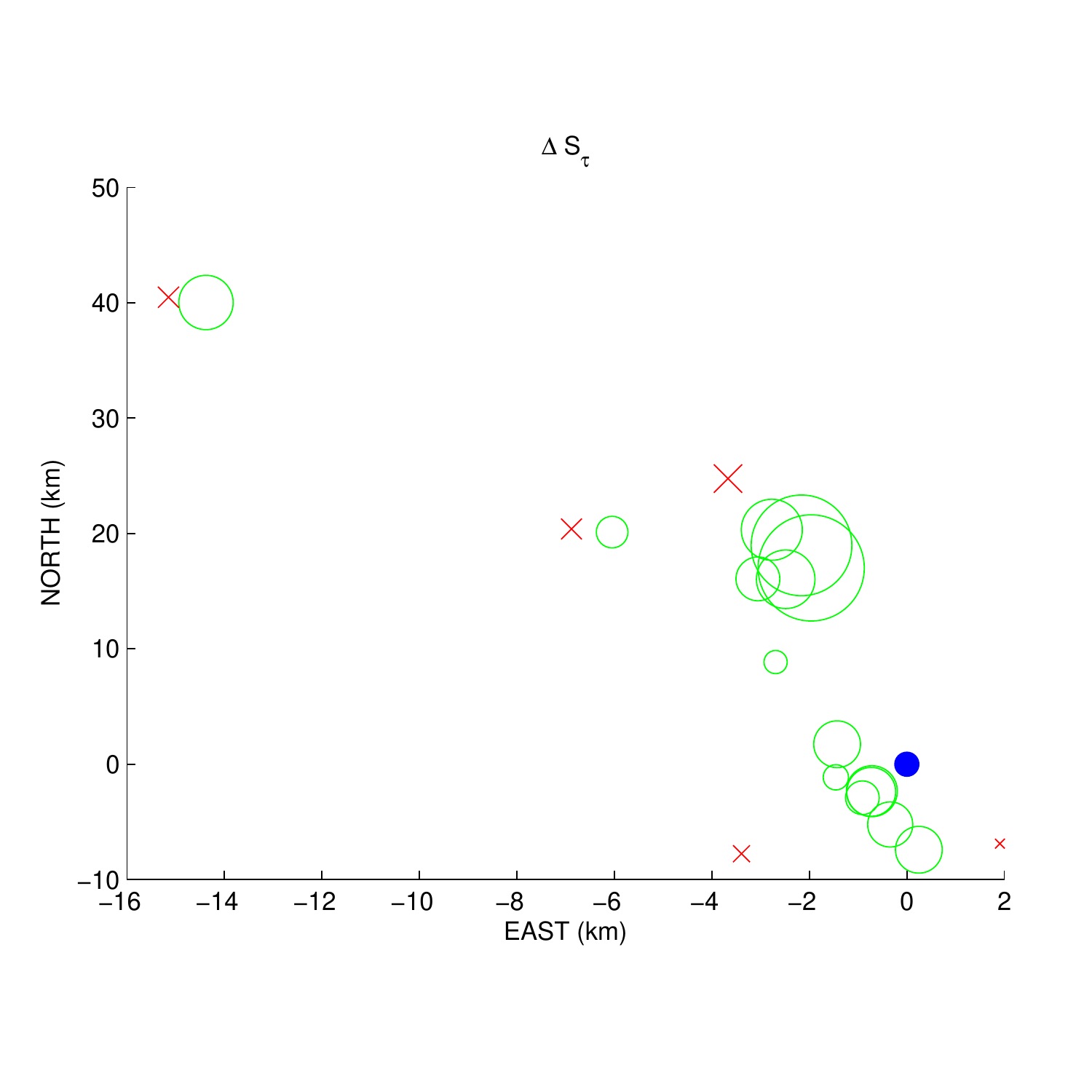} &
\includegraphics[width=.5\textwidth]{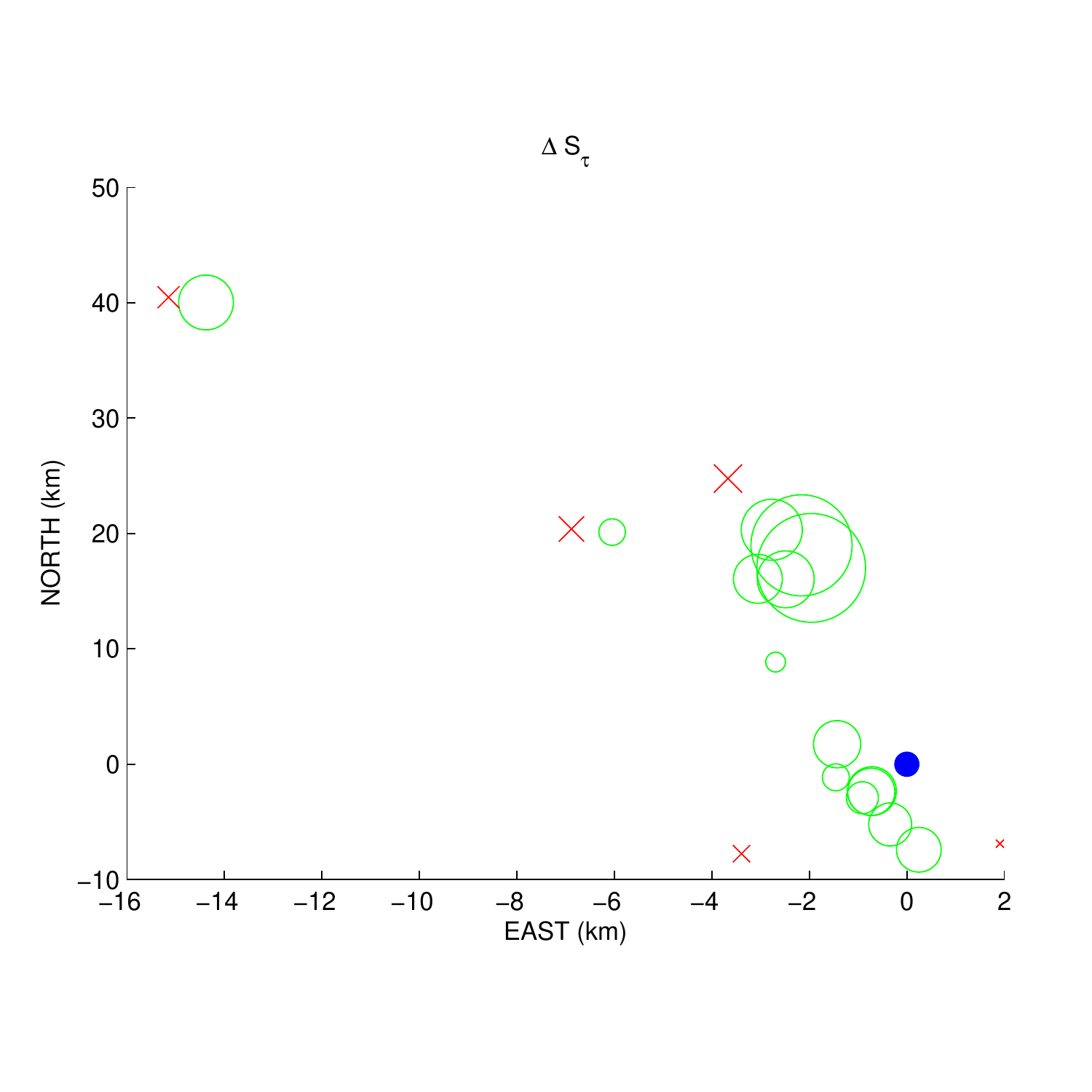} 
\end{tabular}
\end{center}
\vspace{-10 mm}
\caption{Advancement in shear for the decimated regional dataset of Landers aftershocks, homogeneous vs layered Earth.}
\label{fig:dshearaftershocks}
\end{figure}

\begin{figure}[htbp]
\begin{center}
\begin{tabular}{cc}
\includegraphics[width=.5\textwidth]{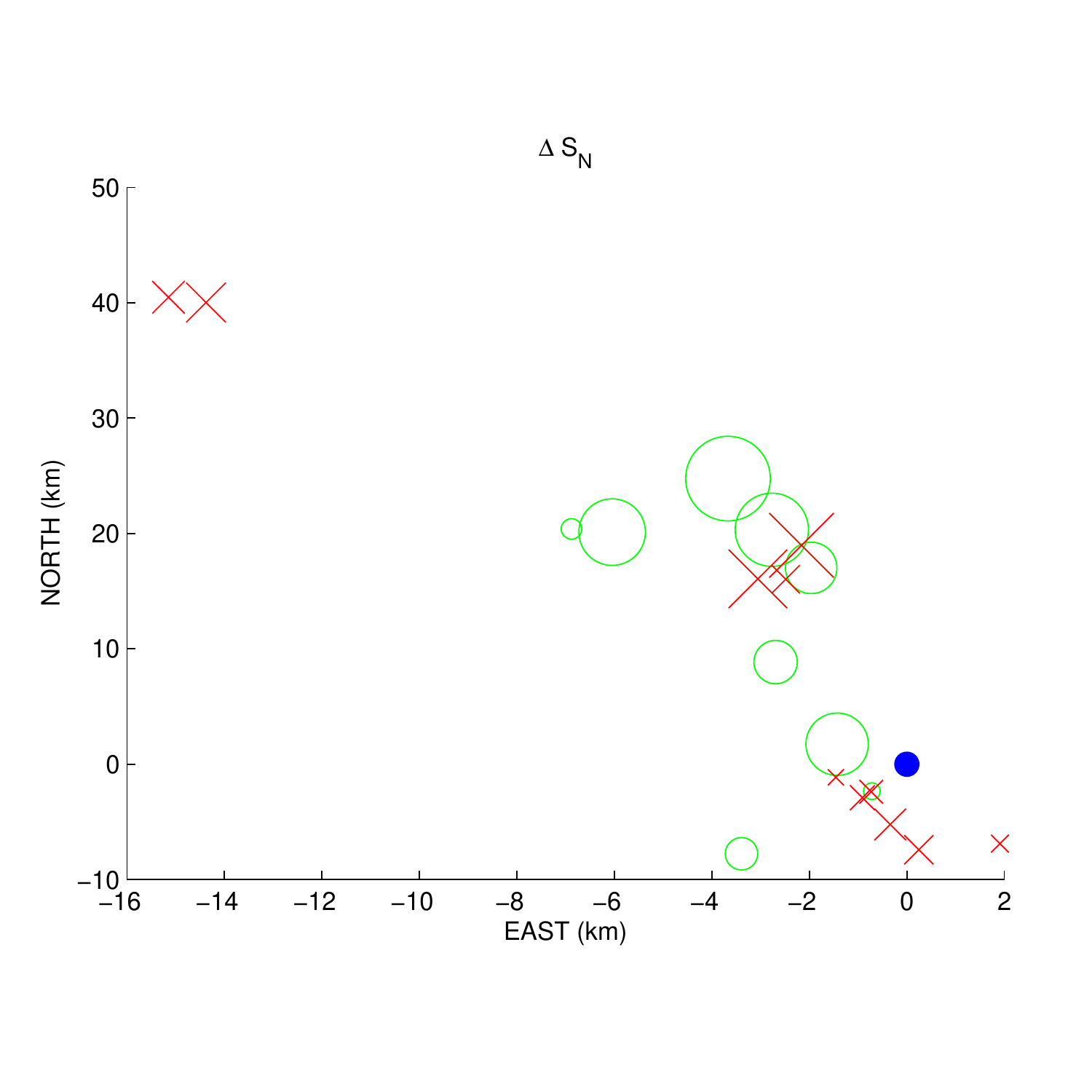} &
\includegraphics[width=.5\textwidth]{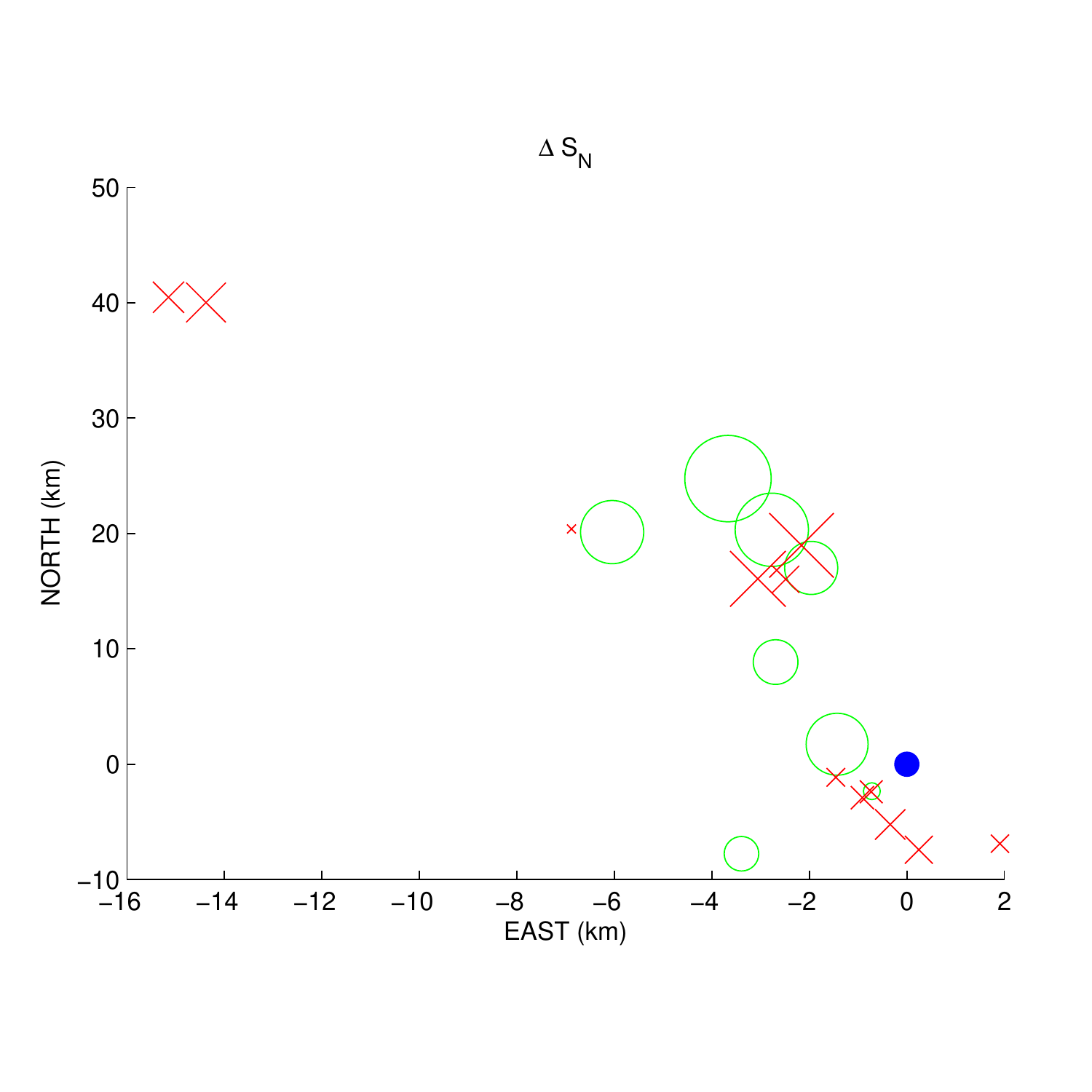} 
\end{tabular}
\end{center}
\vspace{-10 mm}
\caption{Unclamping (green) for the decimated regional dataset of Landers aftershocks, homogeneous vs layered Earth.}
\label{fig:dnormaftershocks}
\end{figure}

\begin{figure}[htbp]
\begin{center}
\begin{tabular}{cc}
\includegraphics[width=.5\textwidth]{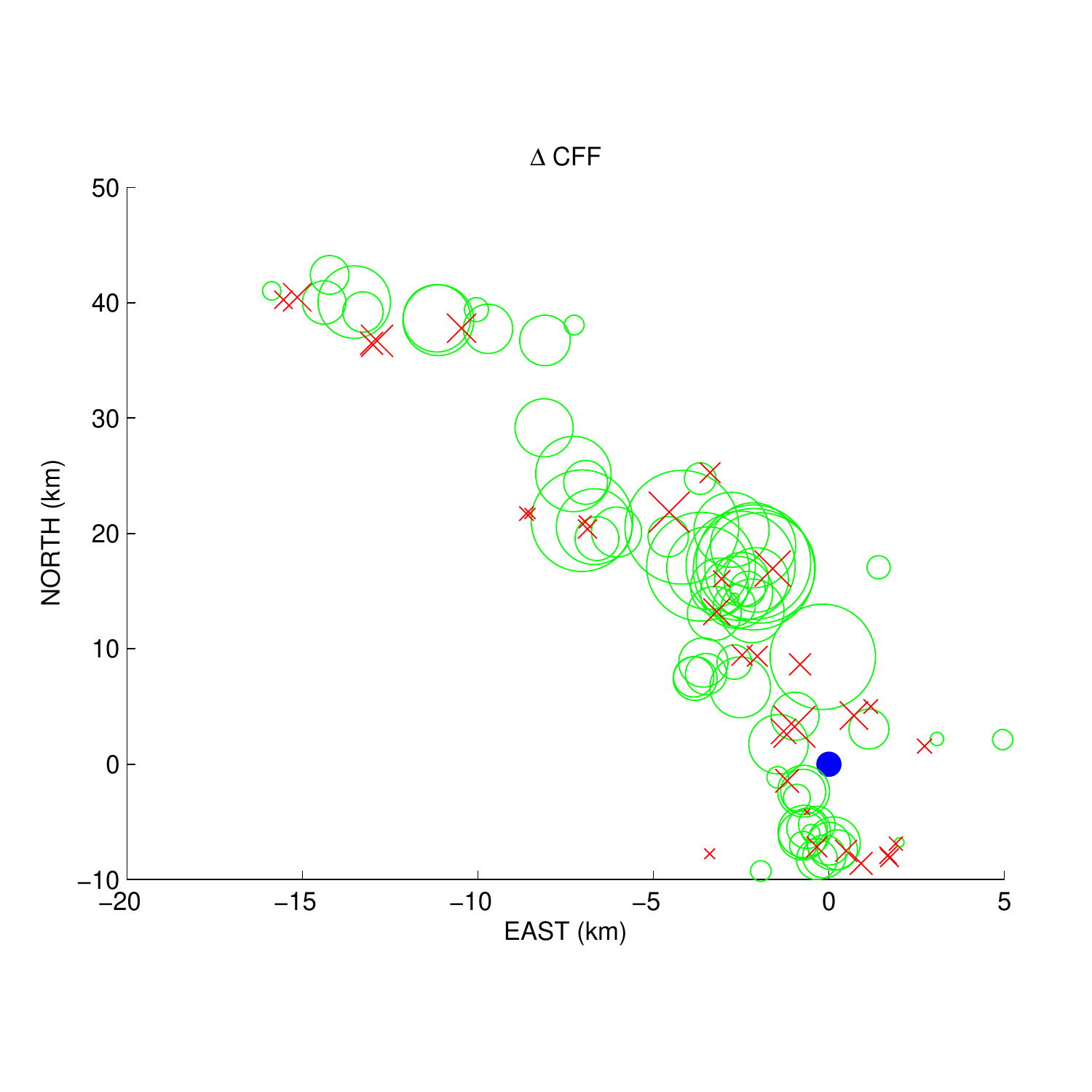} &
\includegraphics[width=.5\textwidth]{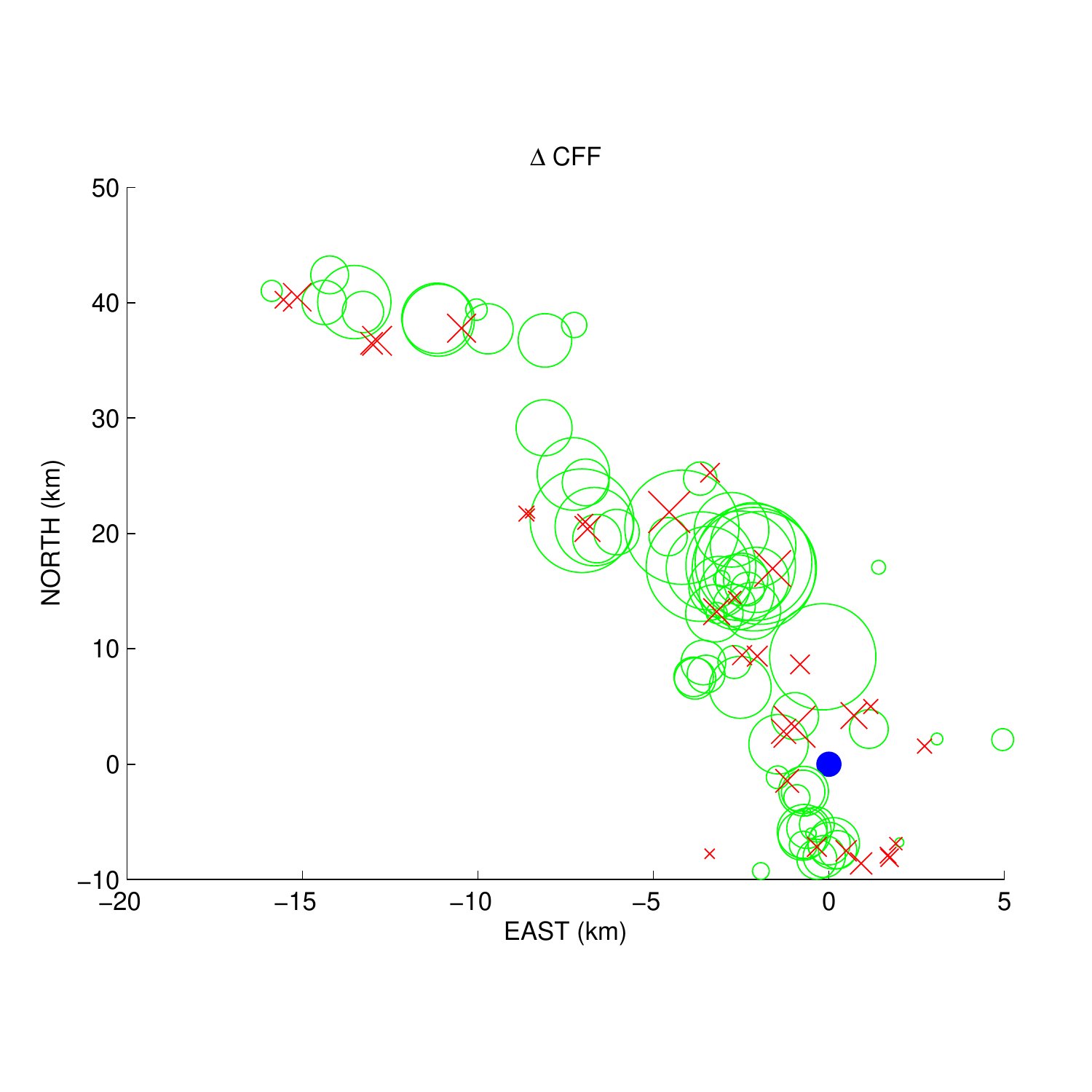} 
\end{tabular}
\end{center}
\vspace{-10 mm}
\caption{Coulomb transfer for Landers aftershocks, homogeneous vs layered Earth.}
\label{fig:dcffspreadaftershocks}
\end{figure}

\begin{figure}[htbp]
\begin{center}
\begin{tabular}{cc}
\includegraphics[width=.5\textwidth]{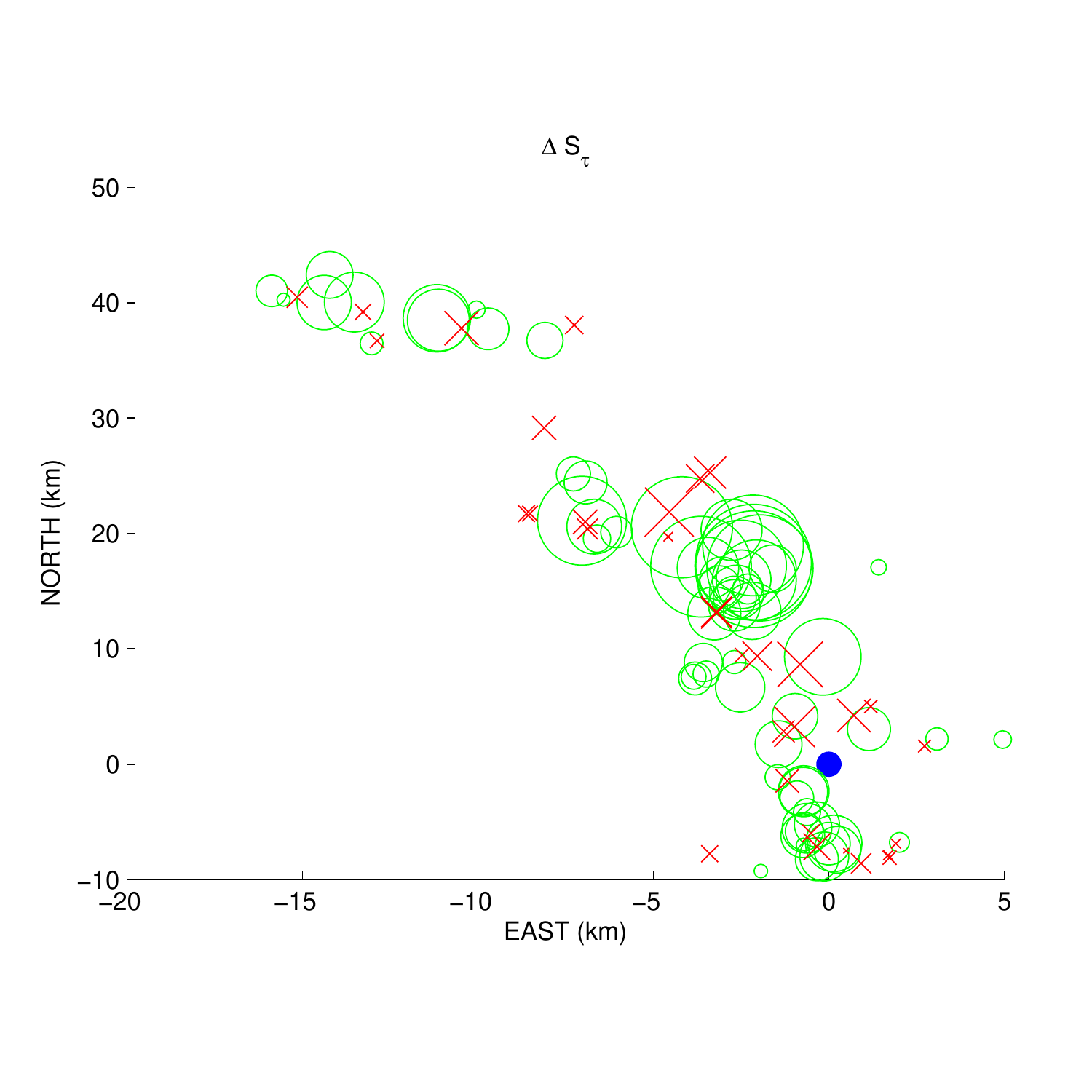} &
\includegraphics[width=.5\textwidth]{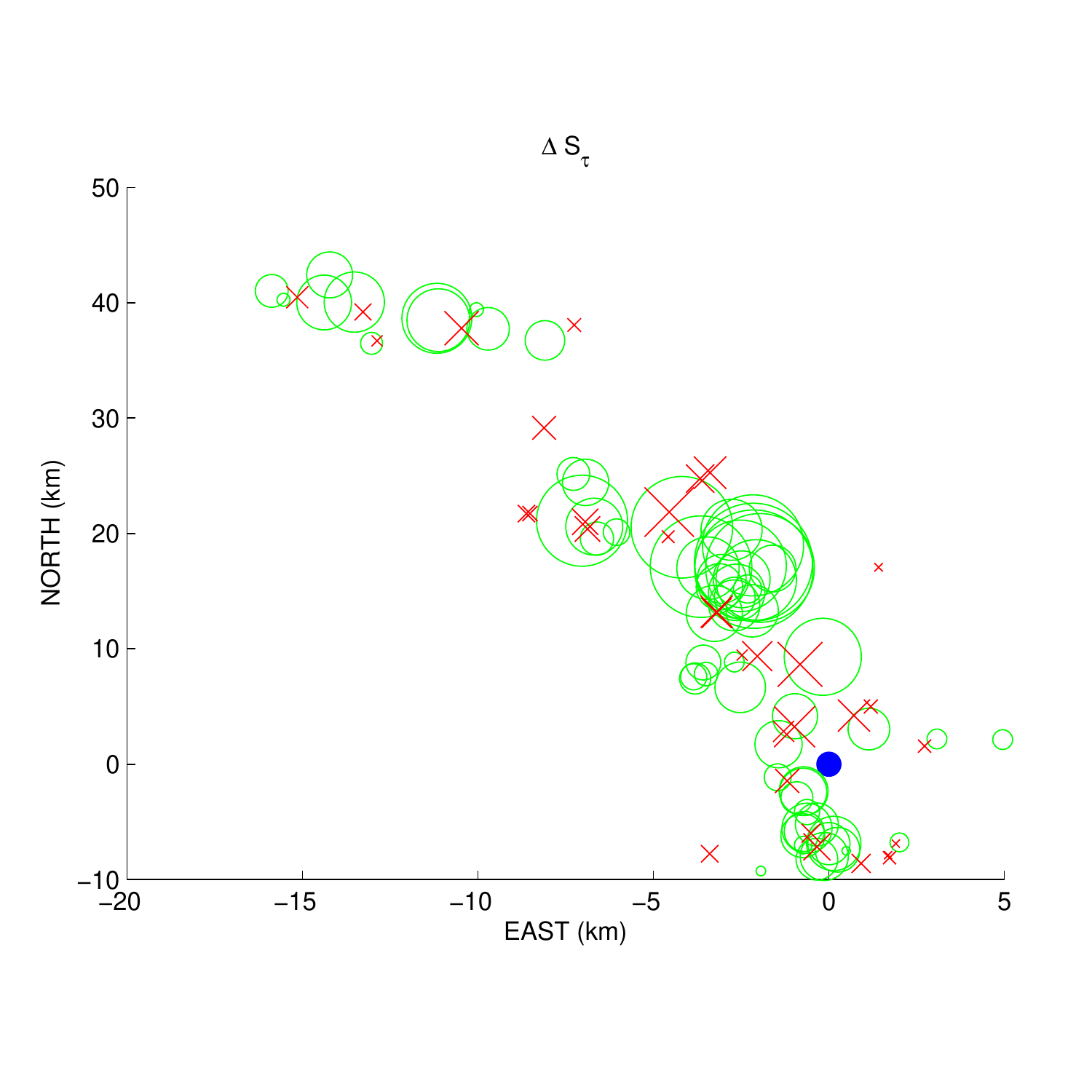} 
\end{tabular}
\end{center}
\vspace{-10 mm}
\caption{Advancement in shear for Landers aftershocks, homogeneous vs layered Earth.}
\label{fig:dshearspreadaftershocks}
\end{figure}

\begin{figure}[htbp]
\begin{center}
\begin{tabular}{cc}
\includegraphics[width=.5\textwidth]{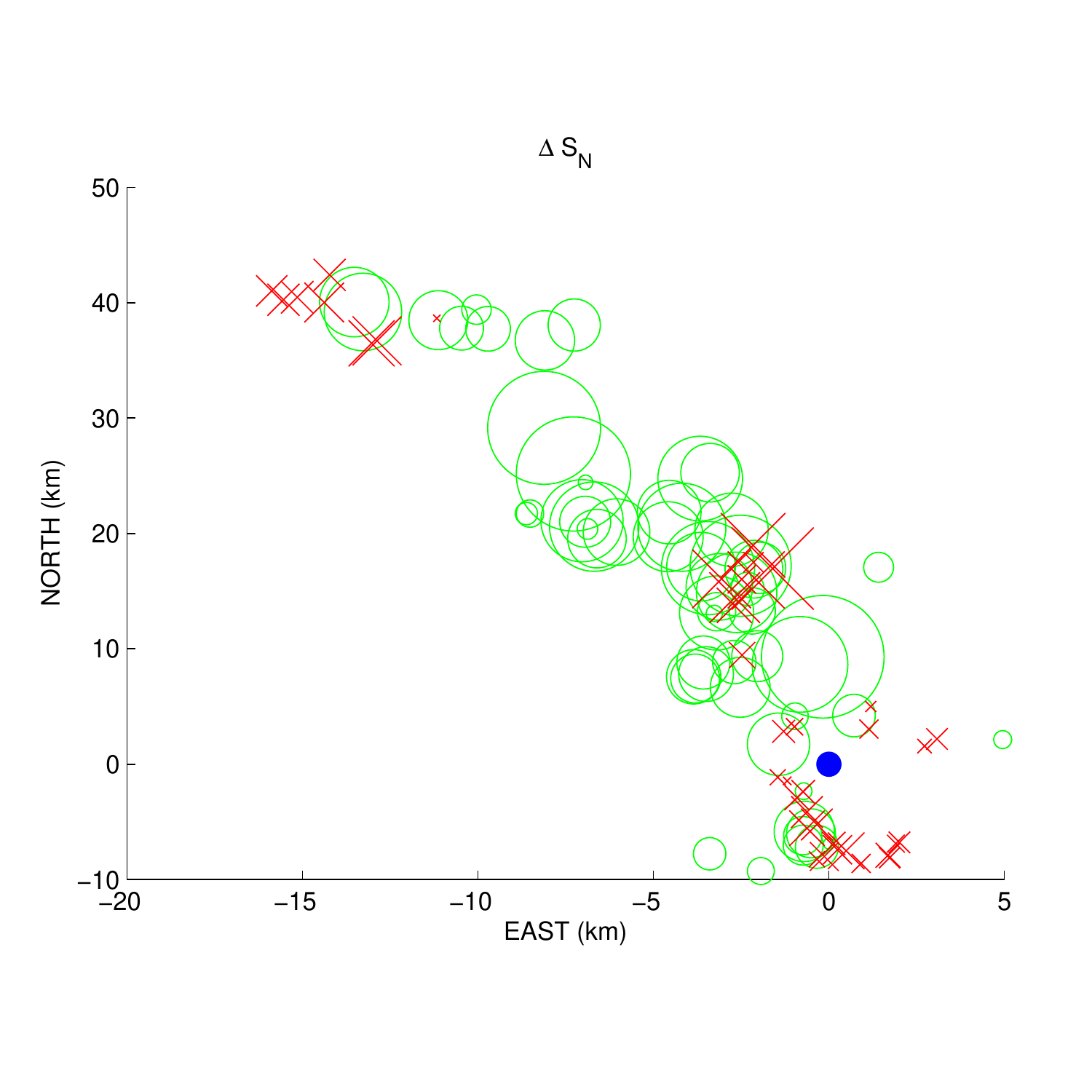} &
\includegraphics[width=.5\textwidth]{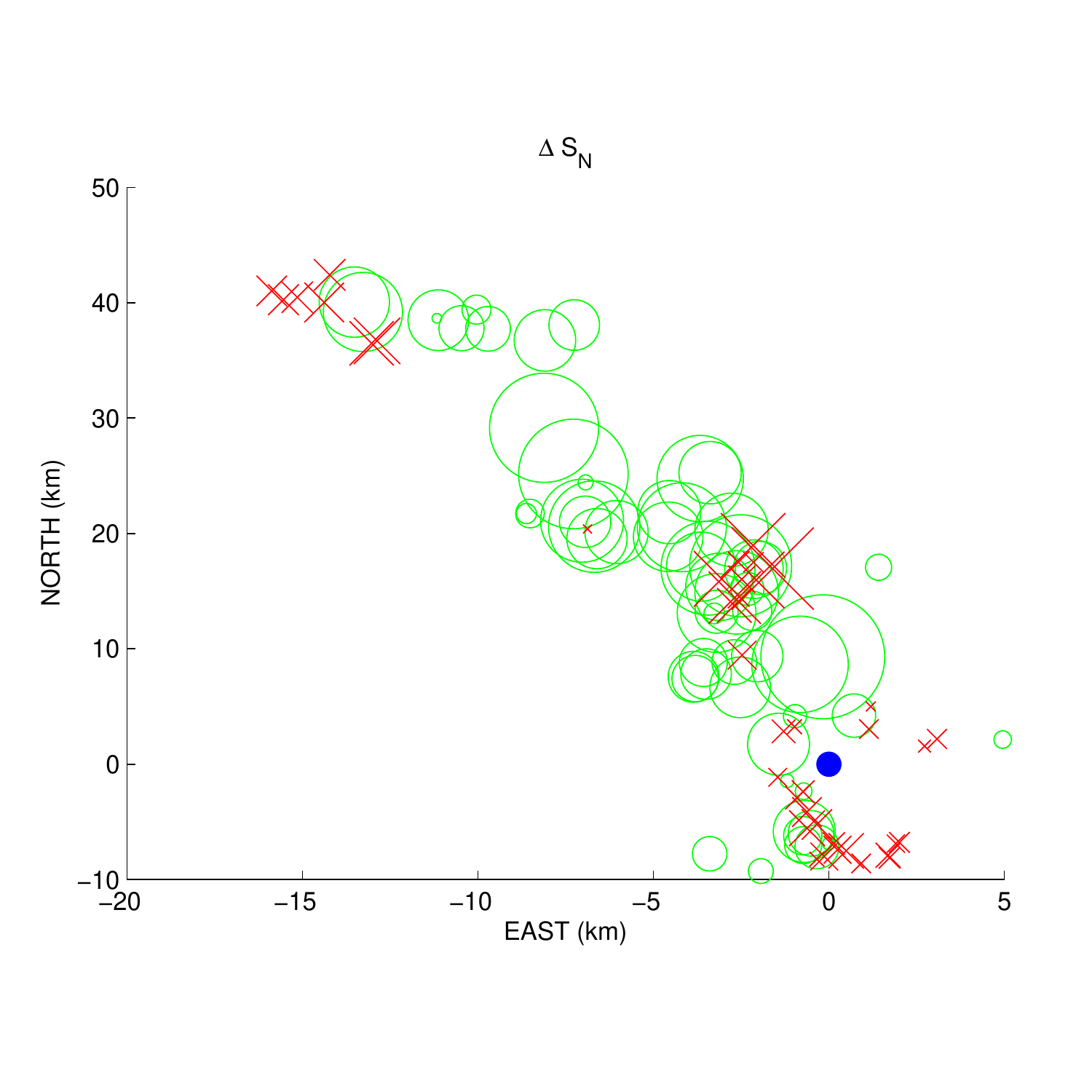} 
\end{tabular}
\end{center}
\vspace{-10 mm}
\caption{Unclamping (green) for Landers aftershocks, homogeneous vs layered Earth.}
\label{fig:dnormspreadaftershocks}
\end{figure}


\begin{figure}[htbp]
\begin{center}
\begin{tabular}{ccc}
\includegraphics[width=.33\textwidth]{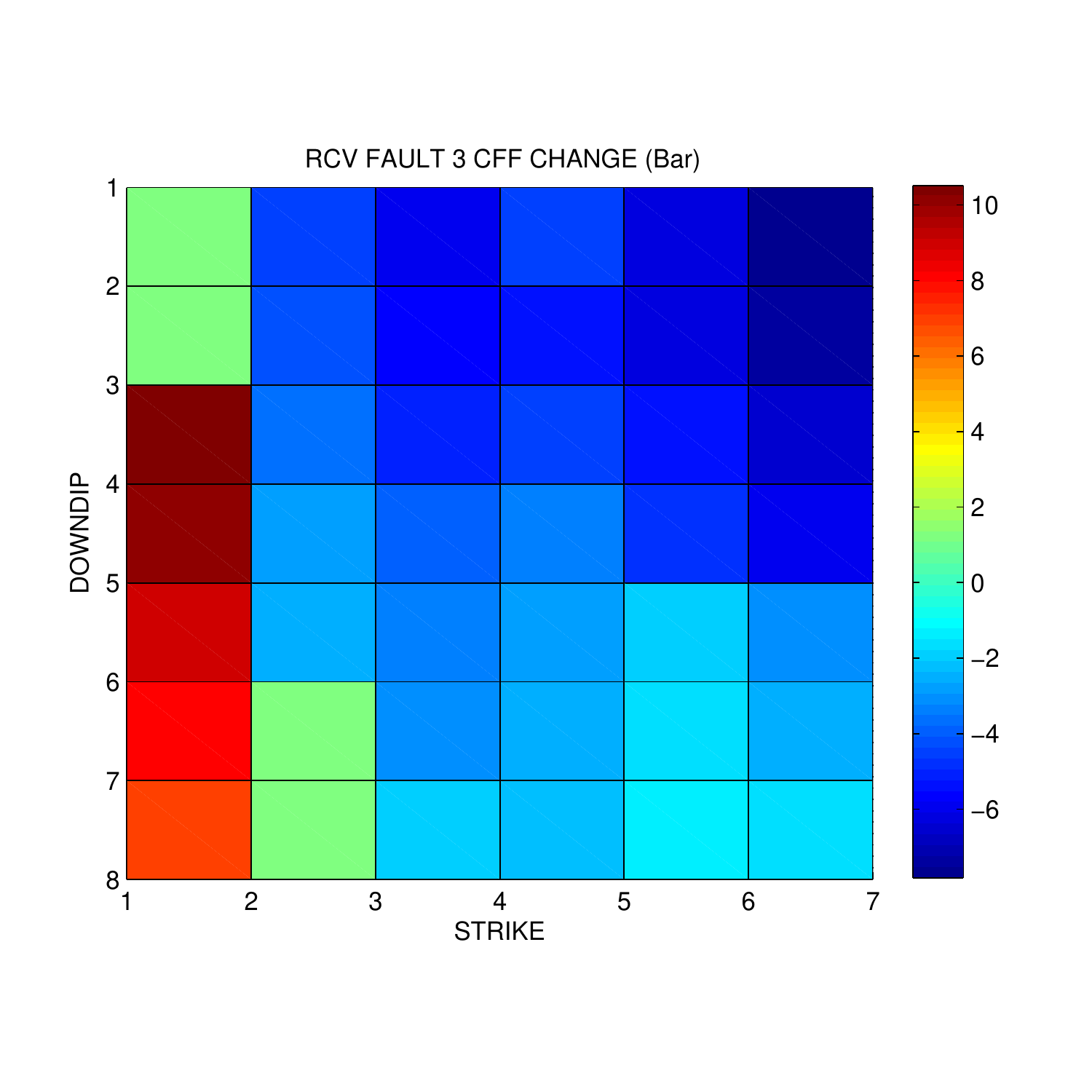} &
\includegraphics[width=.33\textwidth]{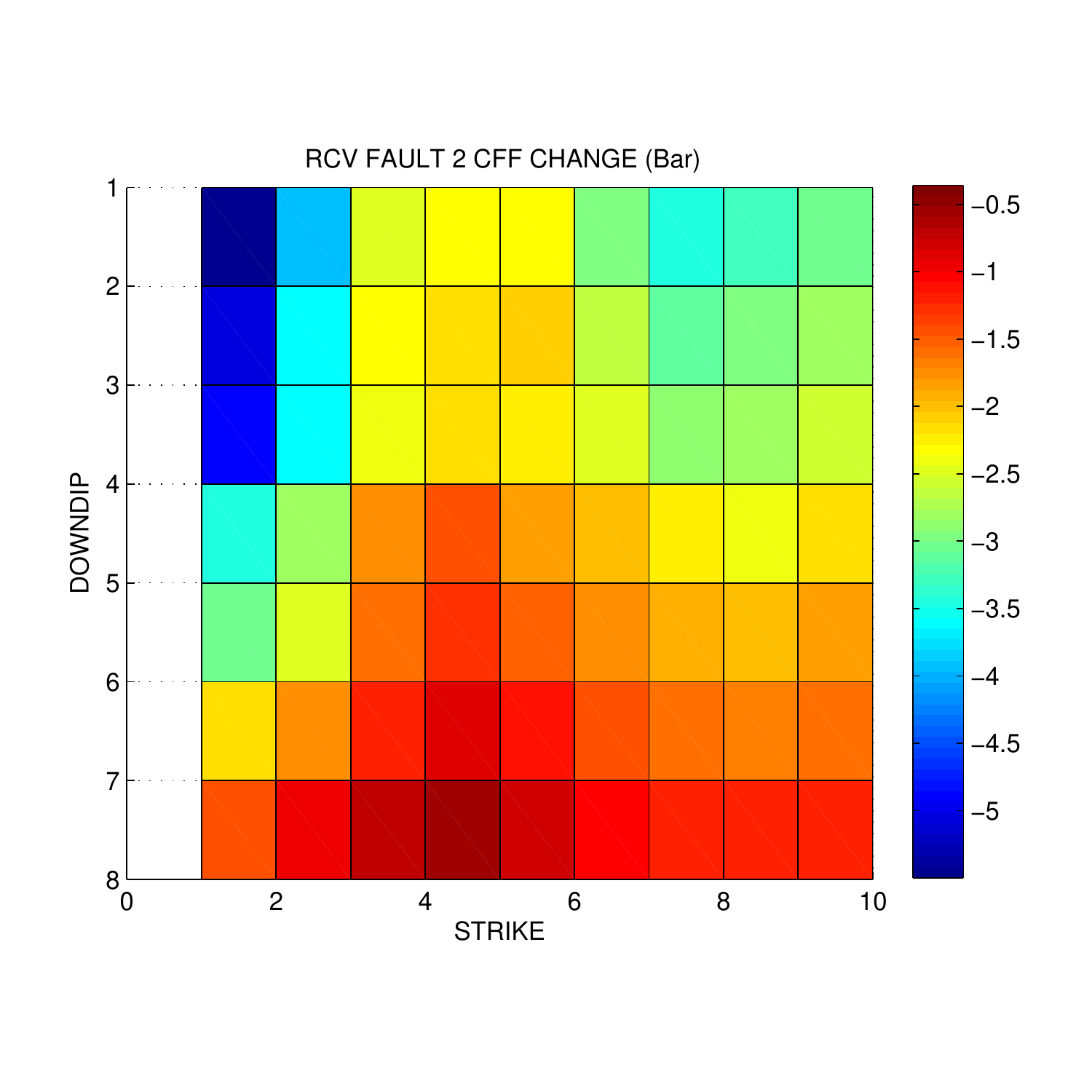} &
\includegraphics[width=.33\textwidth]{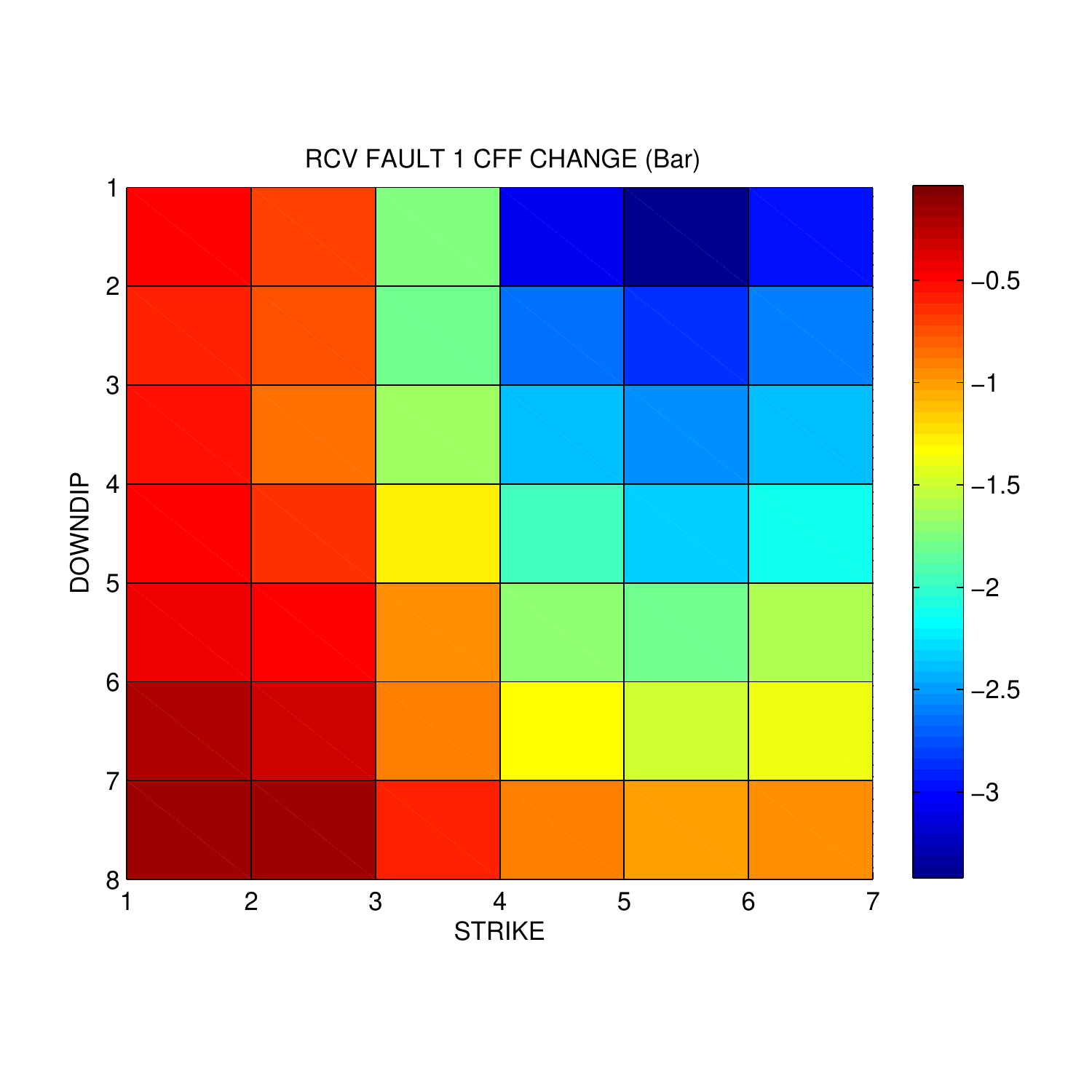}\\ 
\includegraphics[width=.33\textwidth]{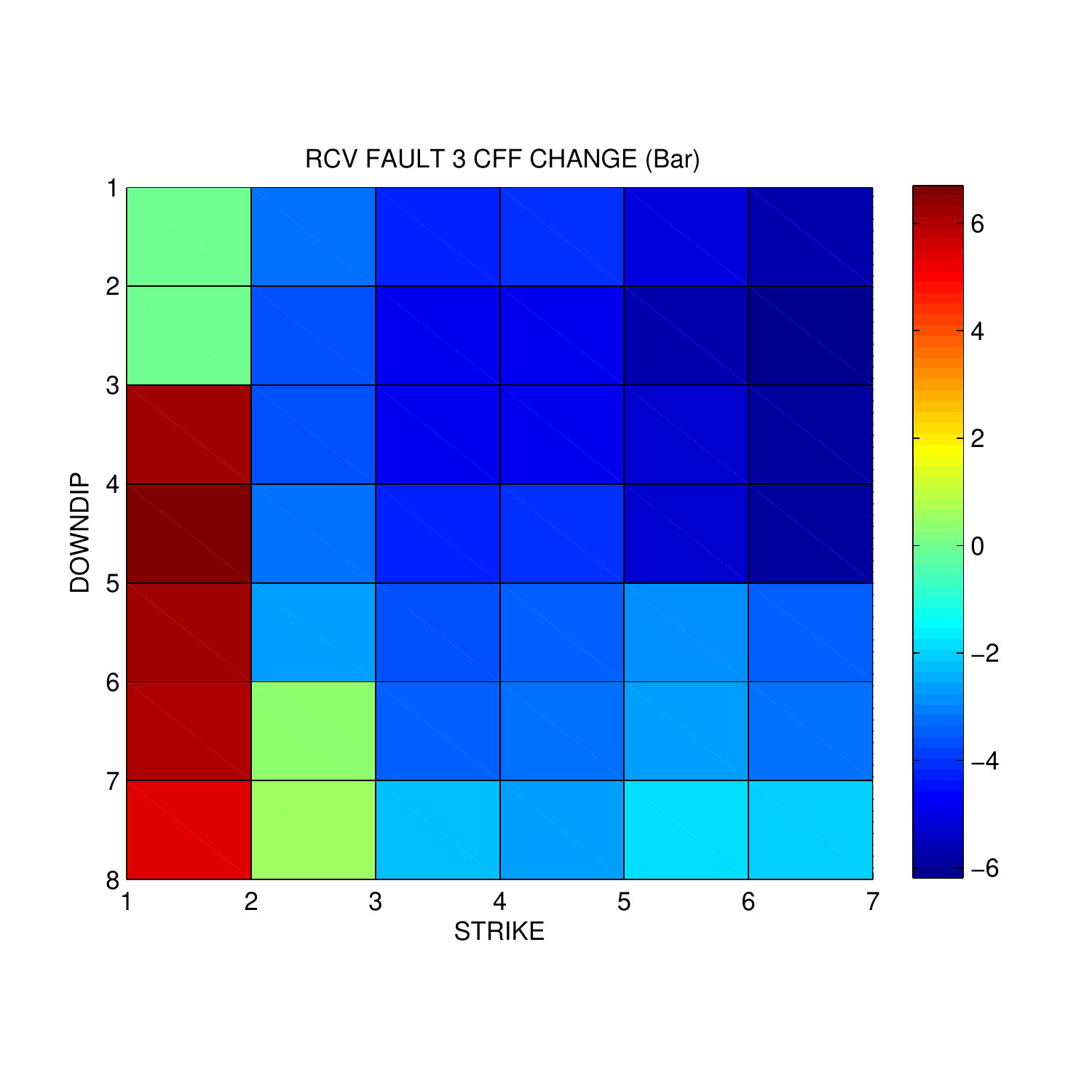} & 
\includegraphics[width=.33\textwidth]{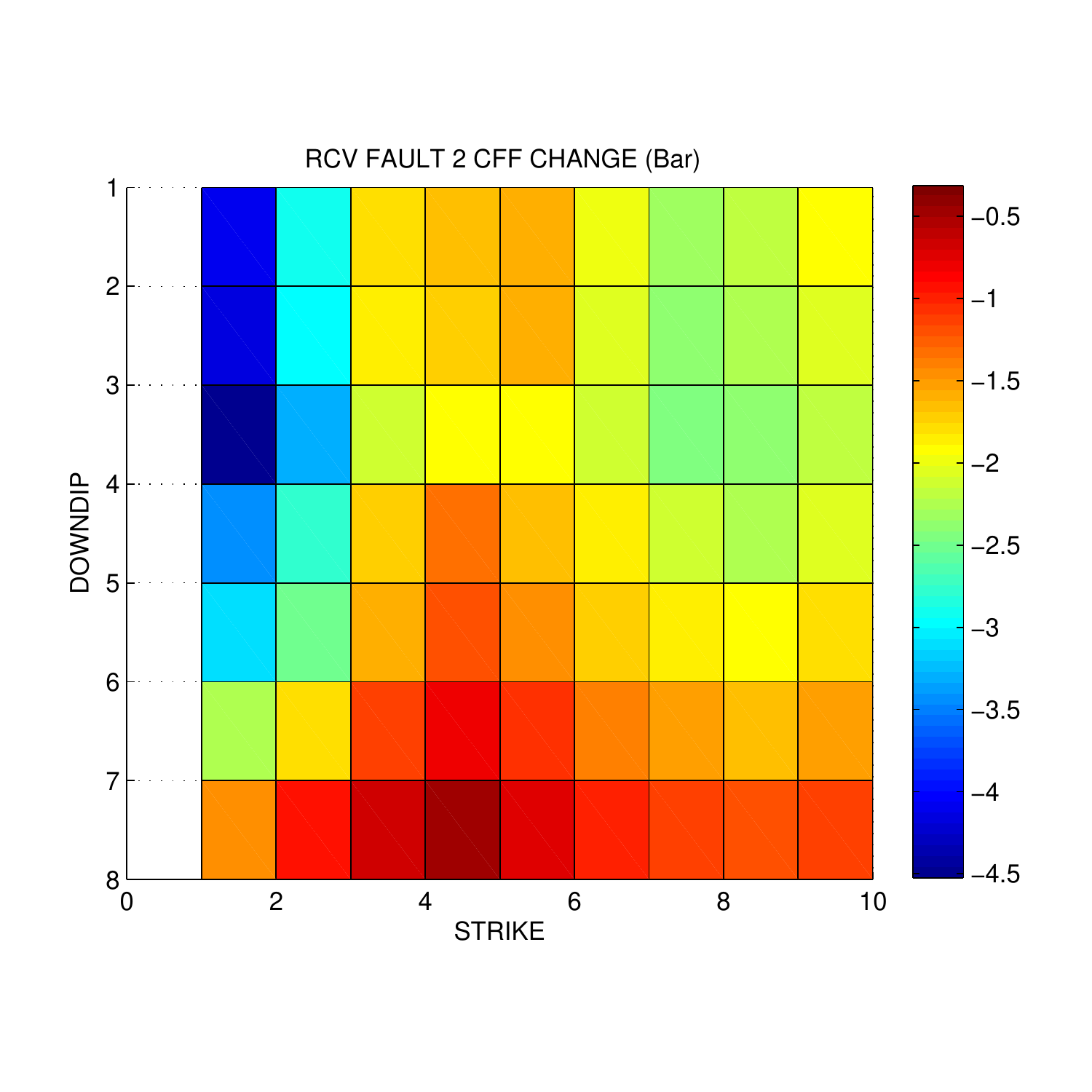} &
\includegraphics[width=.33\textwidth]{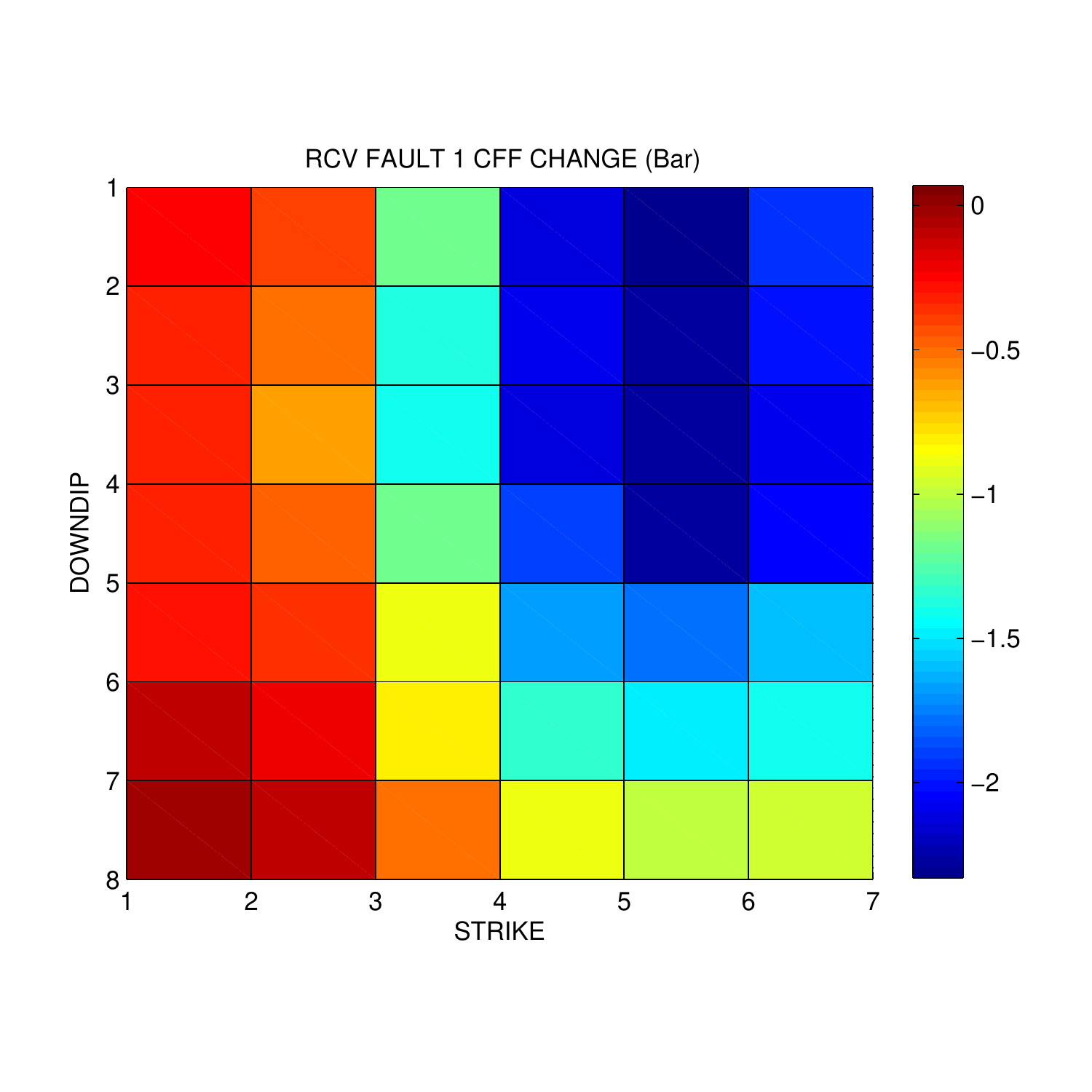} 
\end{tabular}
\end{center}
\vspace{-10 mm}
\caption{Coulomb transfer, Hector Mine faults 3,2,1, $\mu=0.7$, homogeneous vs layered Earth.}
\label{fig:dcffhm7}
\end{figure}
\begin{figure}[htbp]
\begin{center}
\begin{tabular}{ccc}
\includegraphics[width=.33\textwidth]{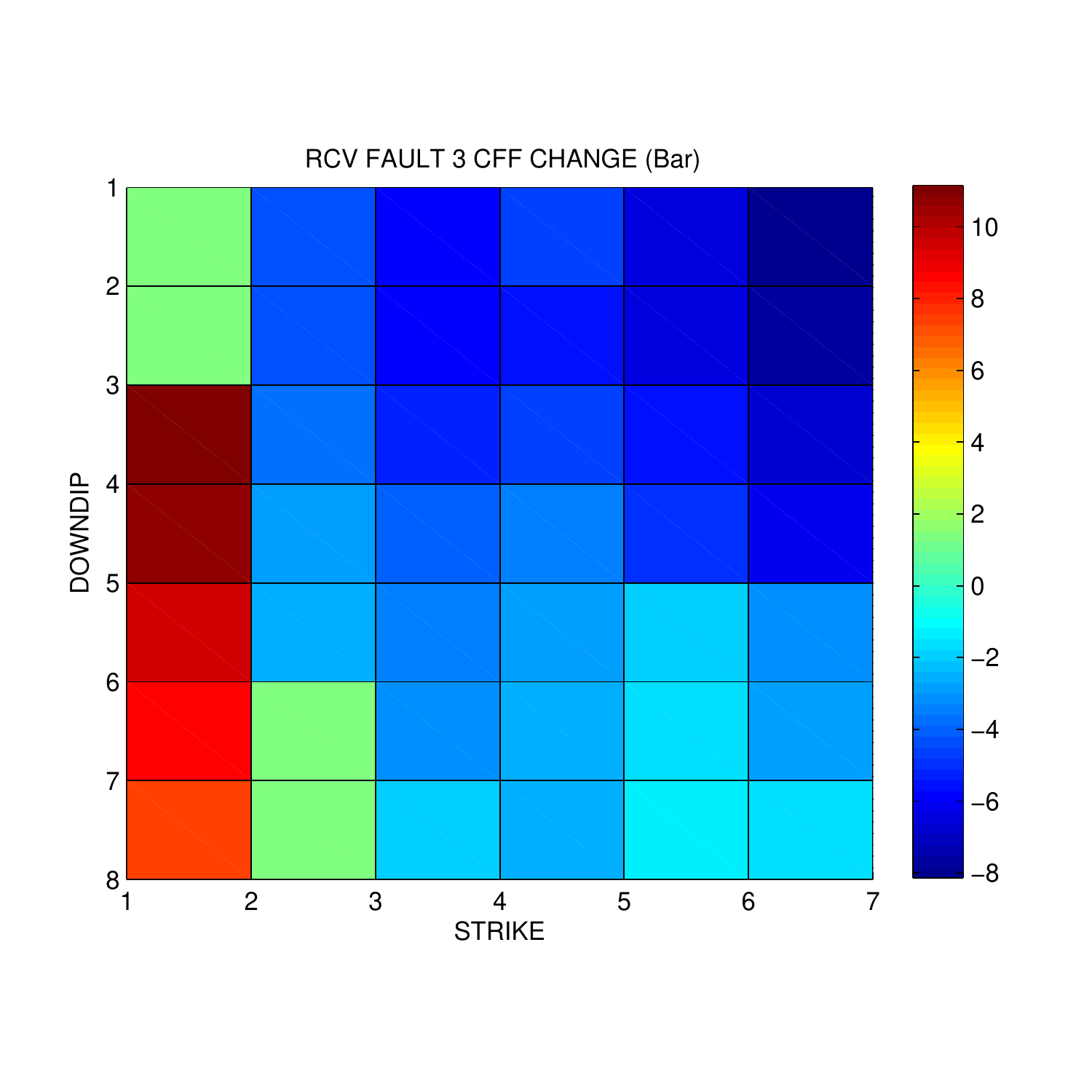} &
\includegraphics[width=.33\textwidth]{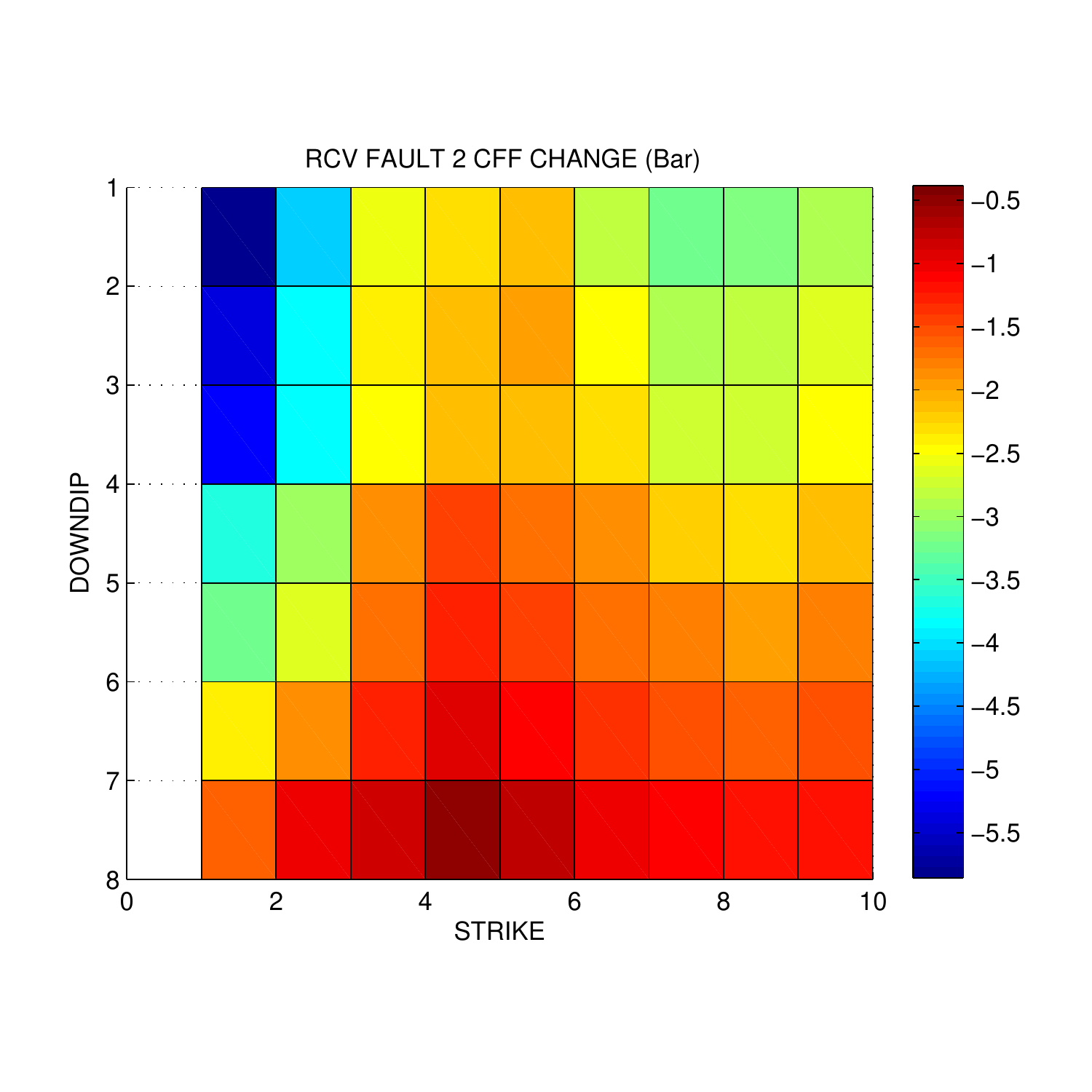} &
\includegraphics[width=.33\textwidth]{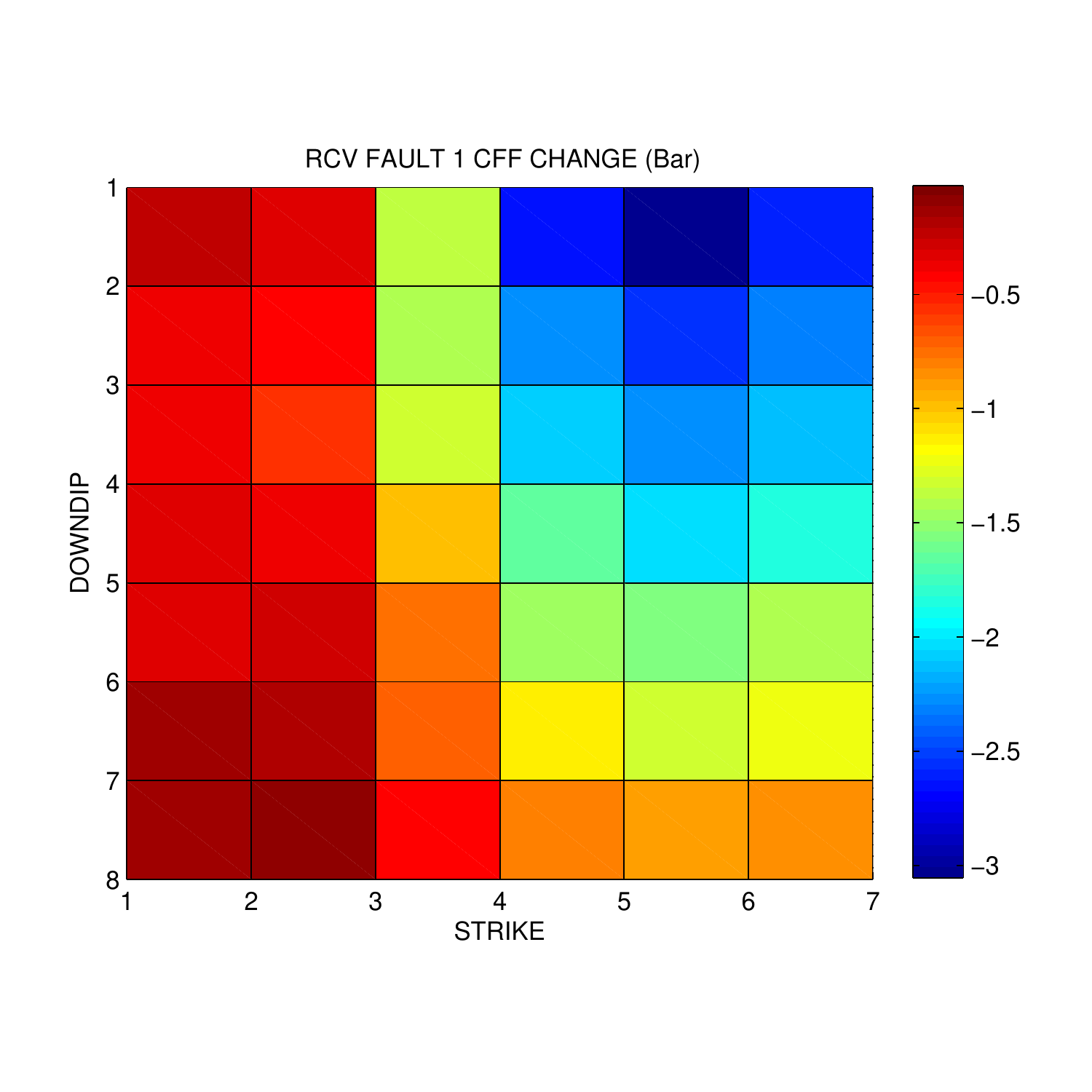}\\ 
\includegraphics[width=.33\textwidth]{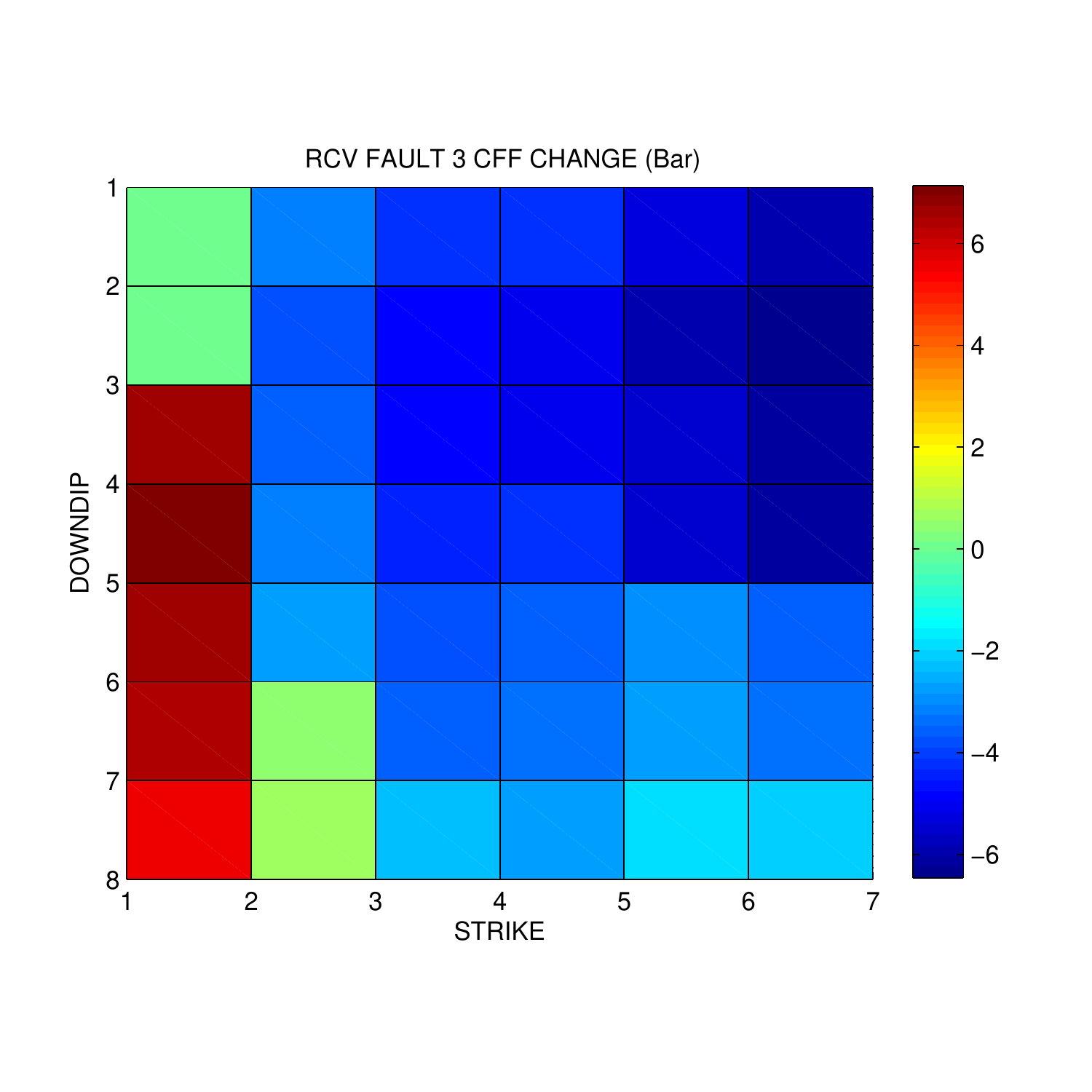} & 
\includegraphics[width=.33\textwidth]{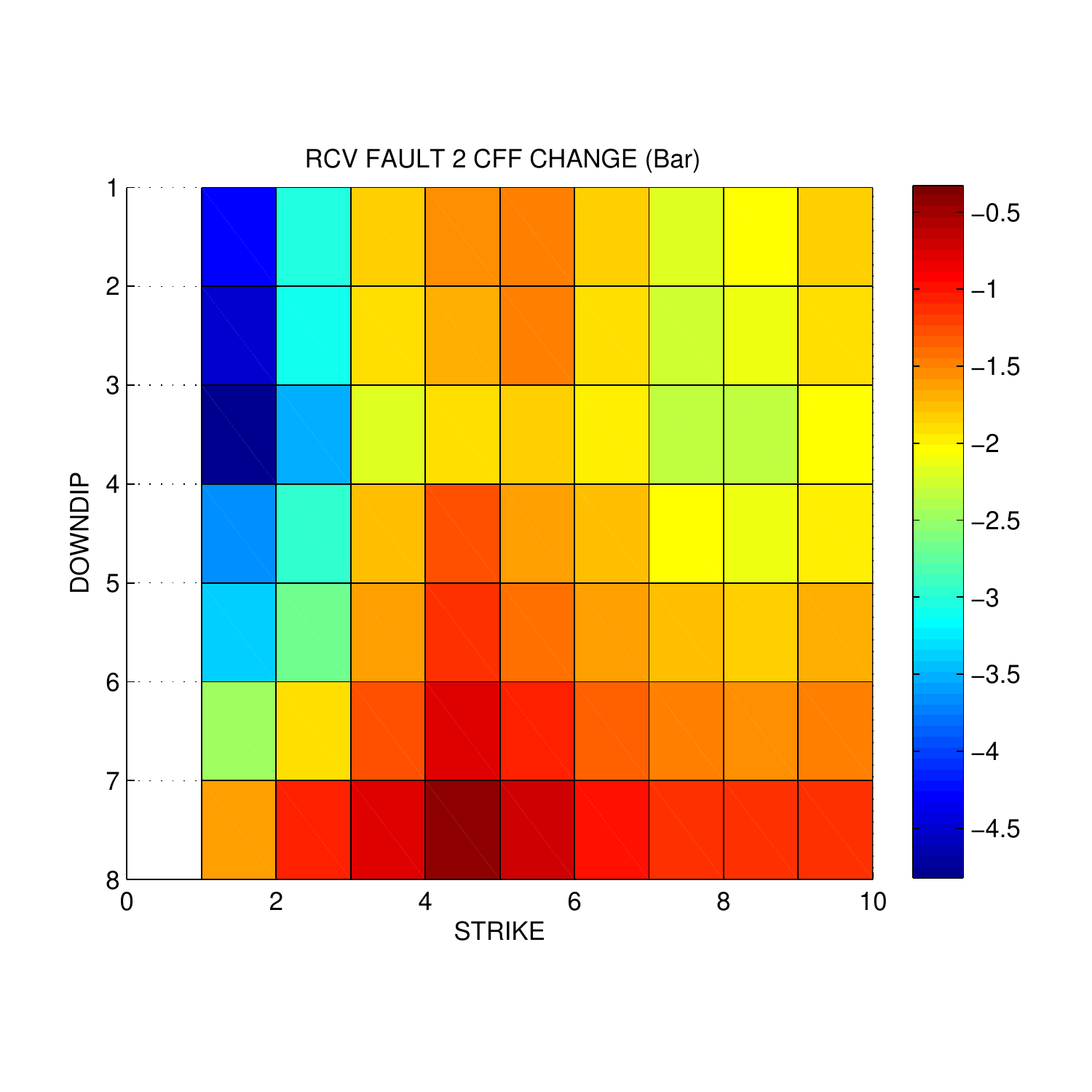} &
\includegraphics[width=.33\textwidth]{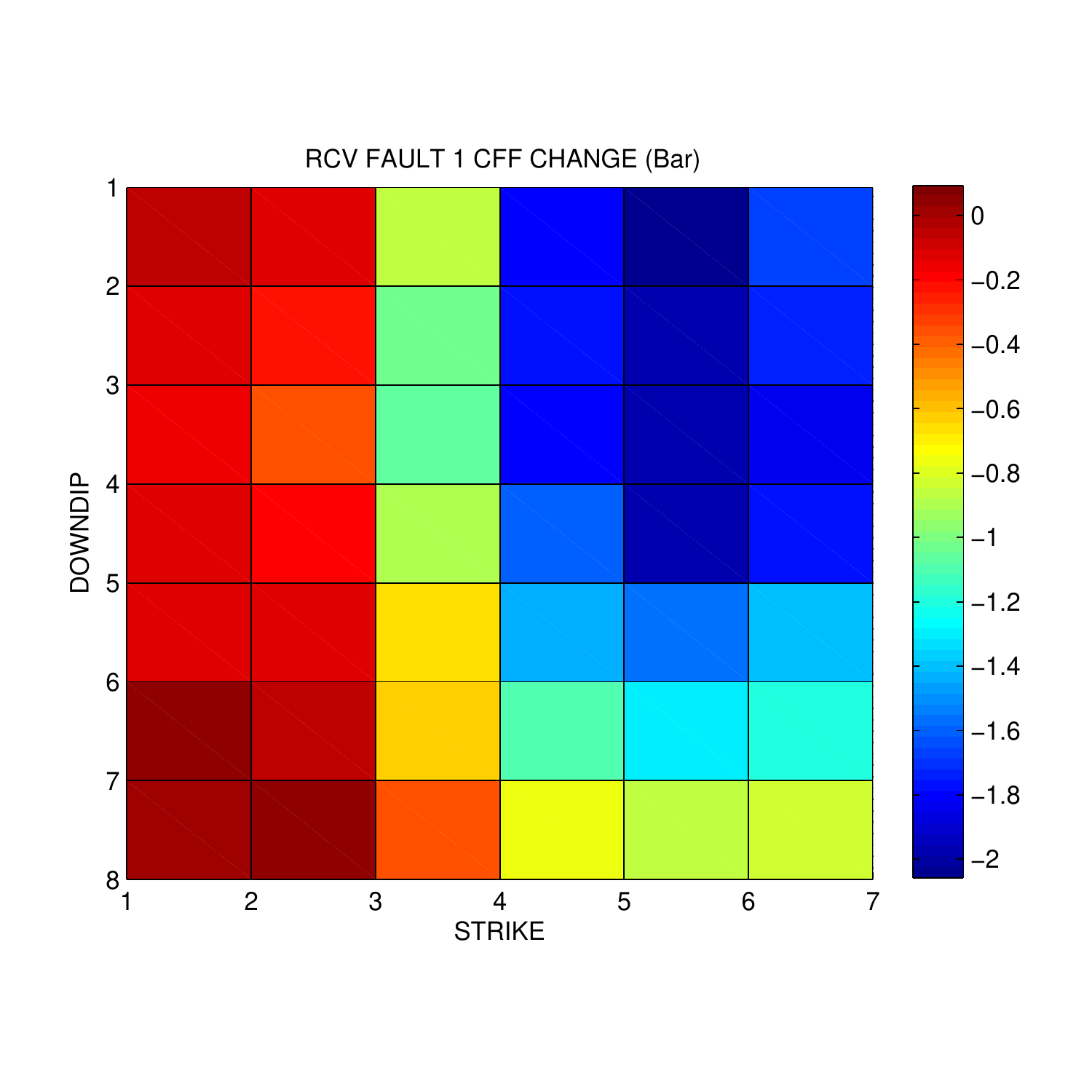} 
\end{tabular}
\end{center}
\vspace{-10 mm}
\caption{Coulomb transfer, Hector Mine faults 3,2,1, $\mu=0.8$, homogeneous vs layered Earth.}
\label{fig:dcffhm8}
\end{figure}

\begin{figure}[htbp]
\begin{center}
\begin{tabular}{ccc}
\includegraphics[width=.33\textwidth]{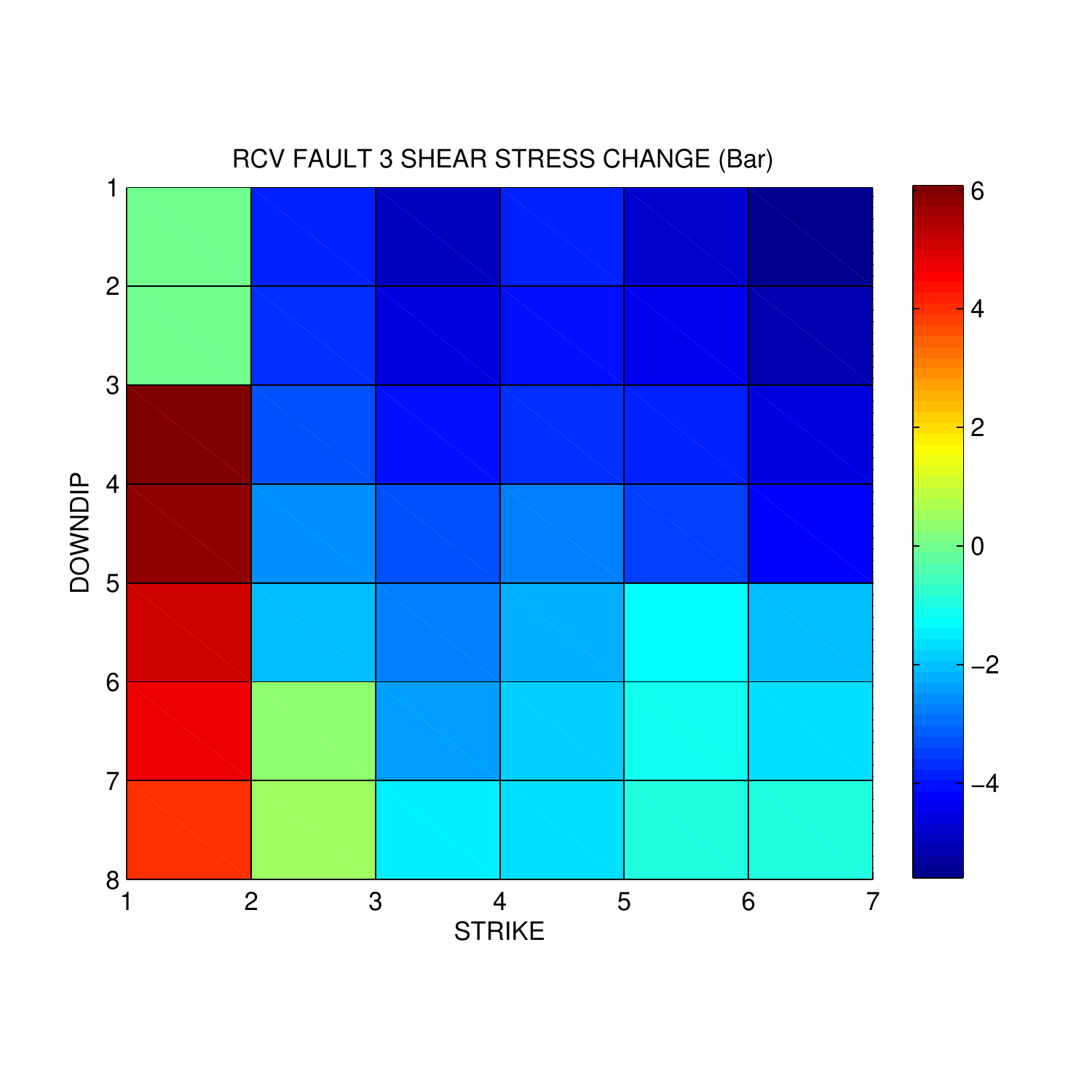} &
\includegraphics[width=.33\textwidth]{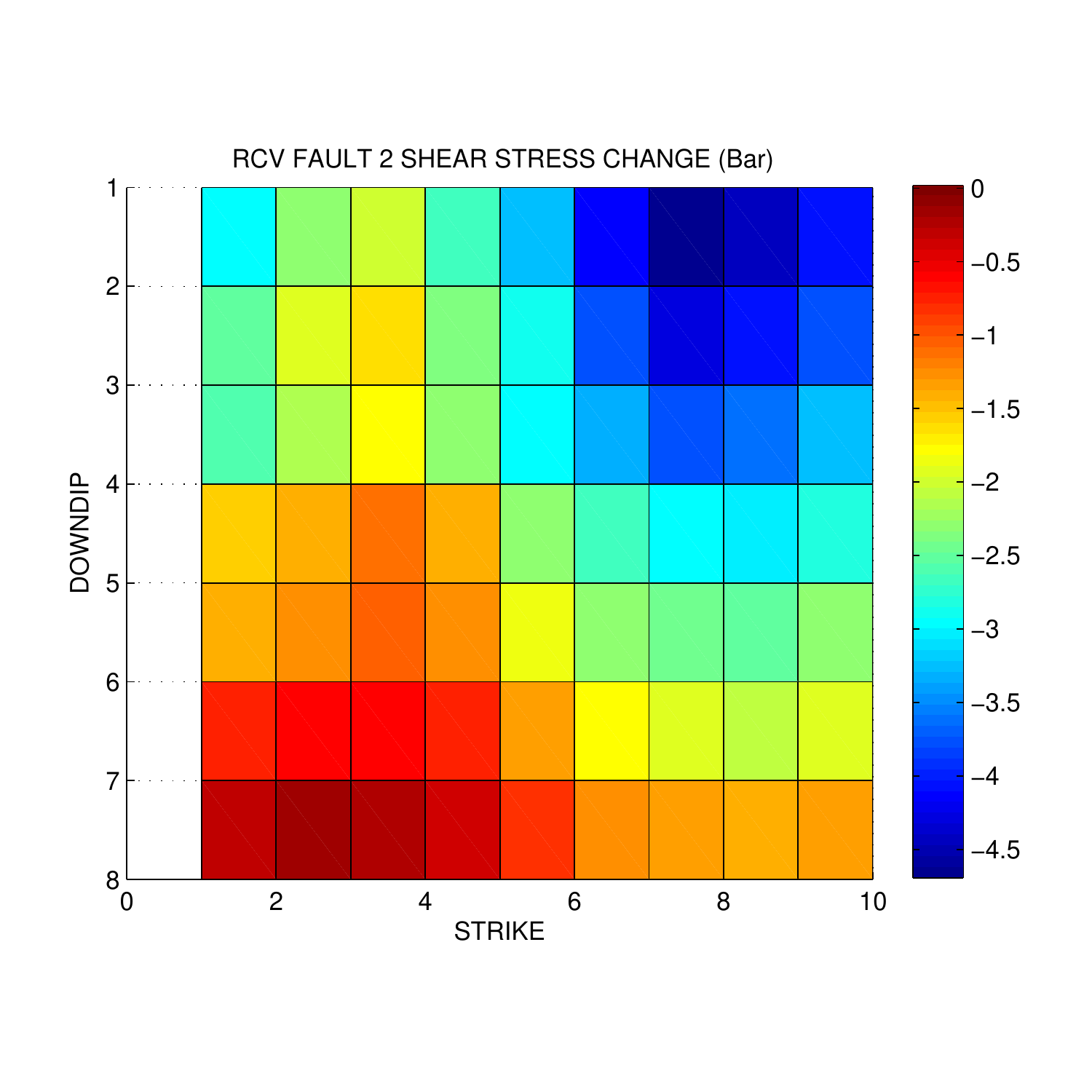} &
\includegraphics[width=.33\textwidth]{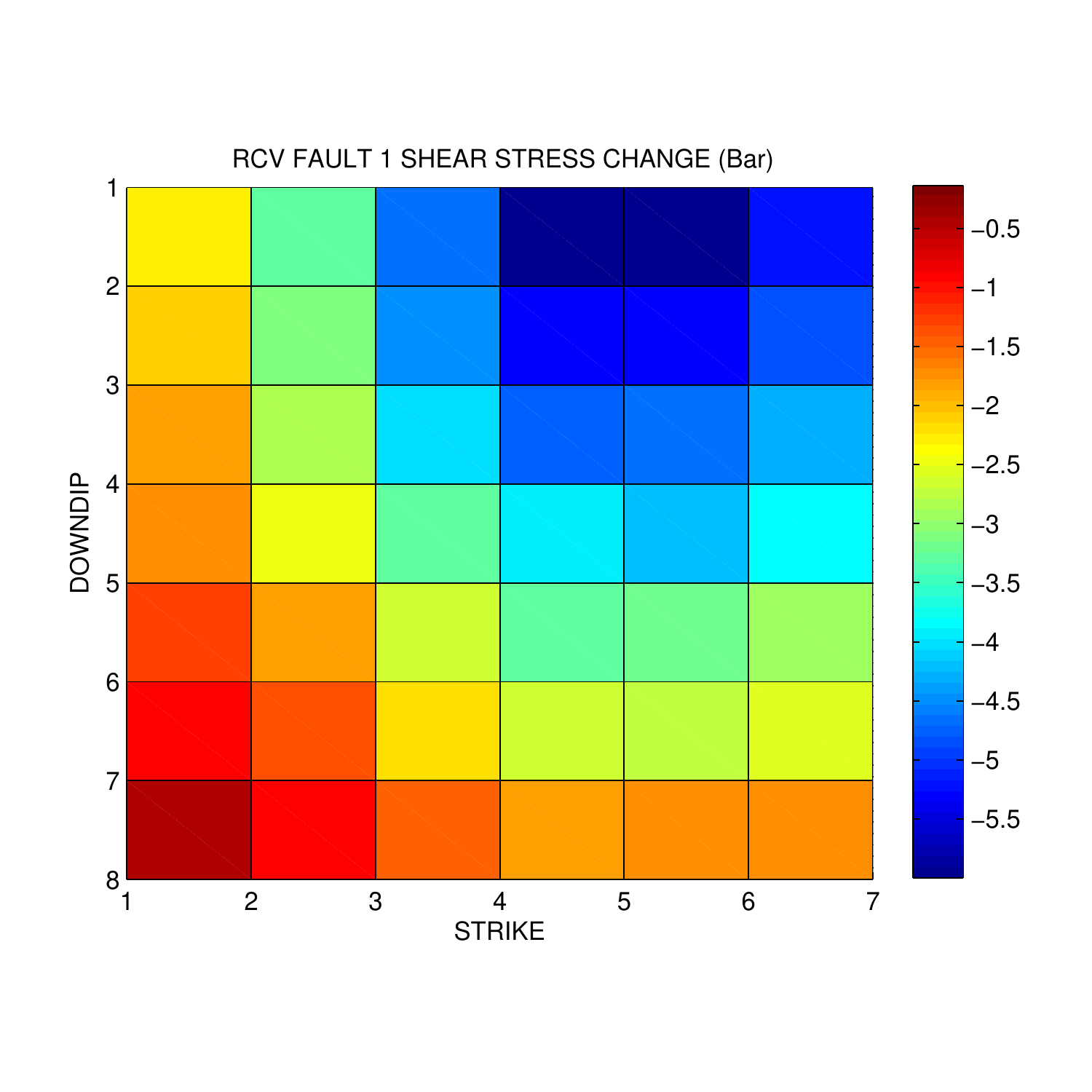} \\
\includegraphics[width=.33\textwidth]{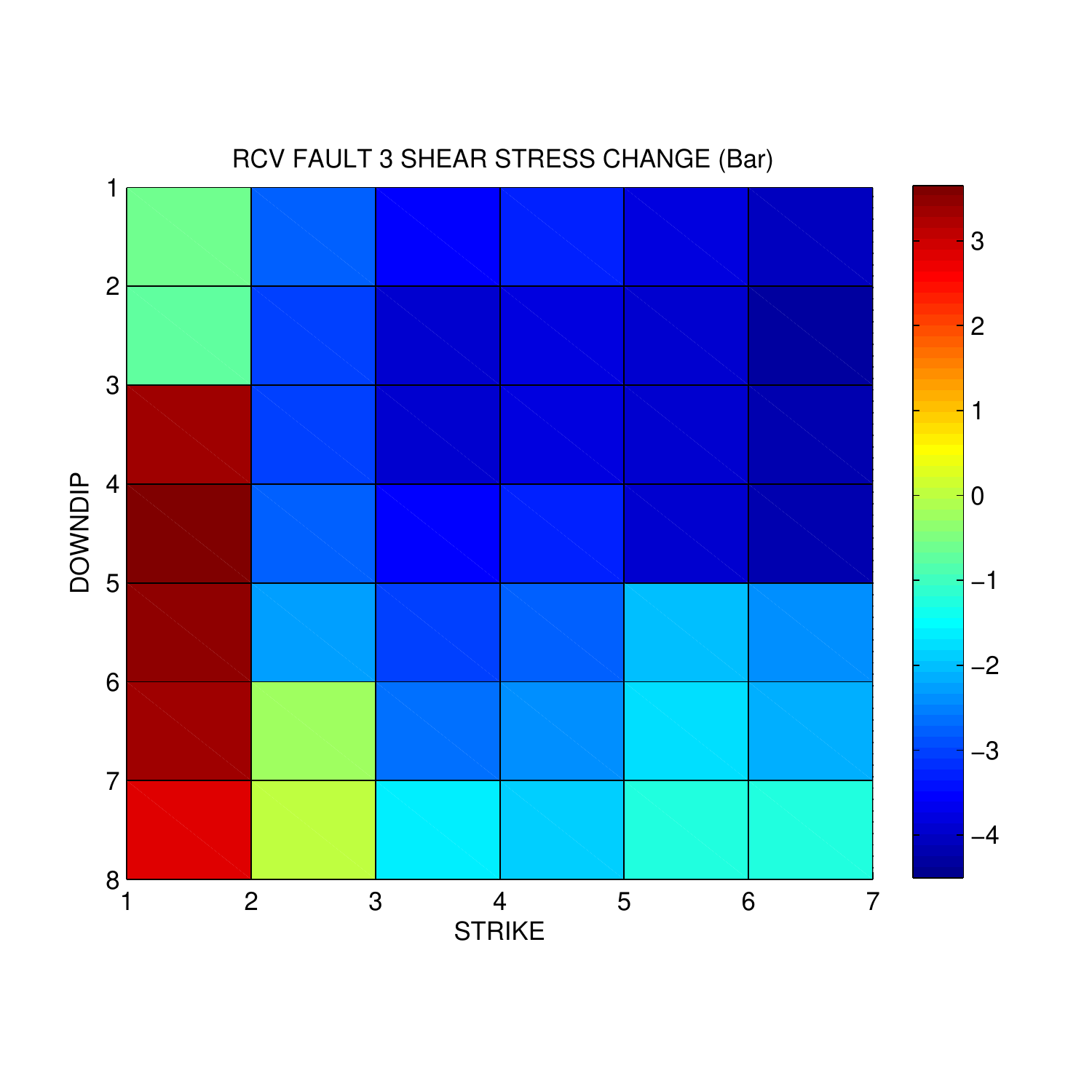} &
\includegraphics[width=.33\textwidth]{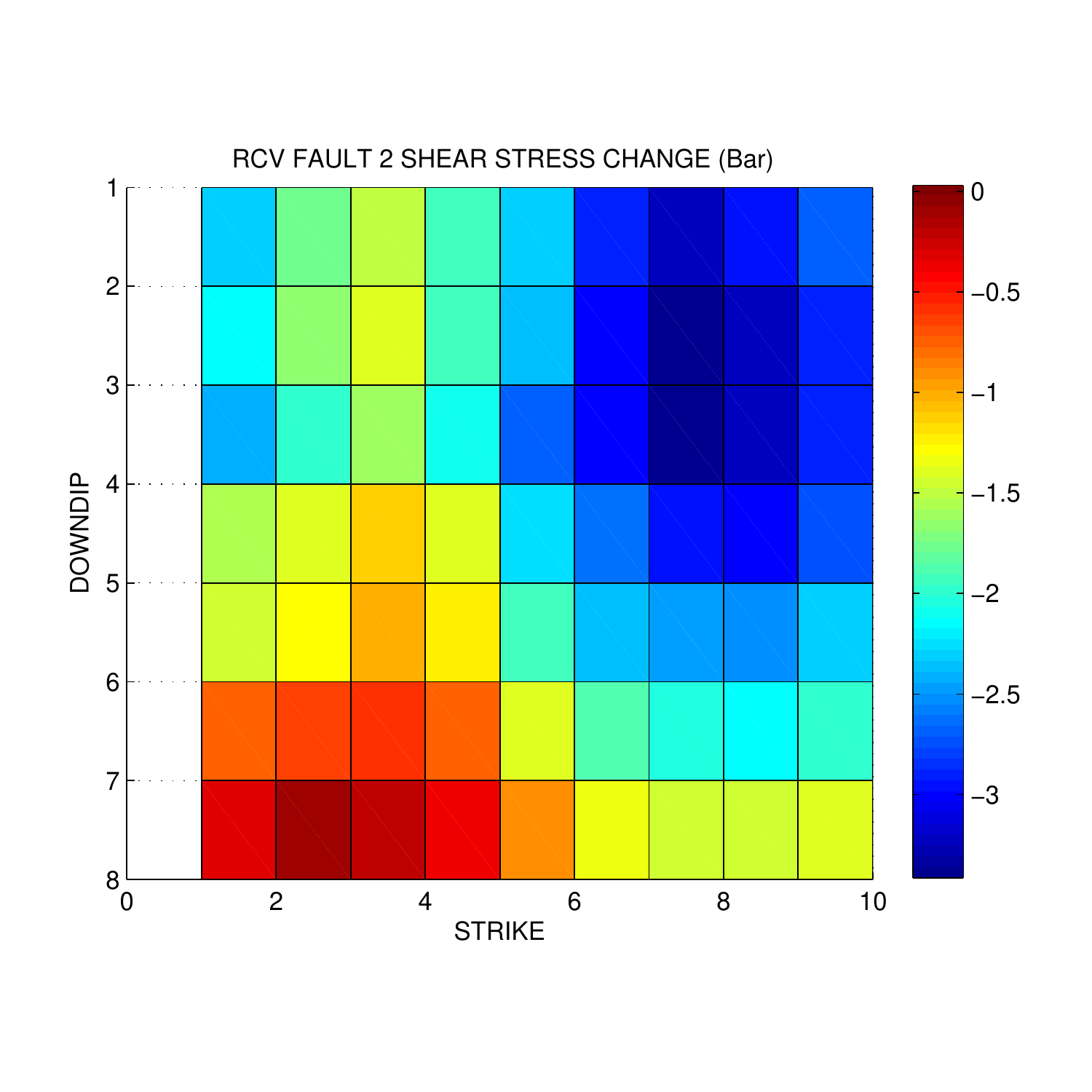} &
\includegraphics[width=.33\textwidth]{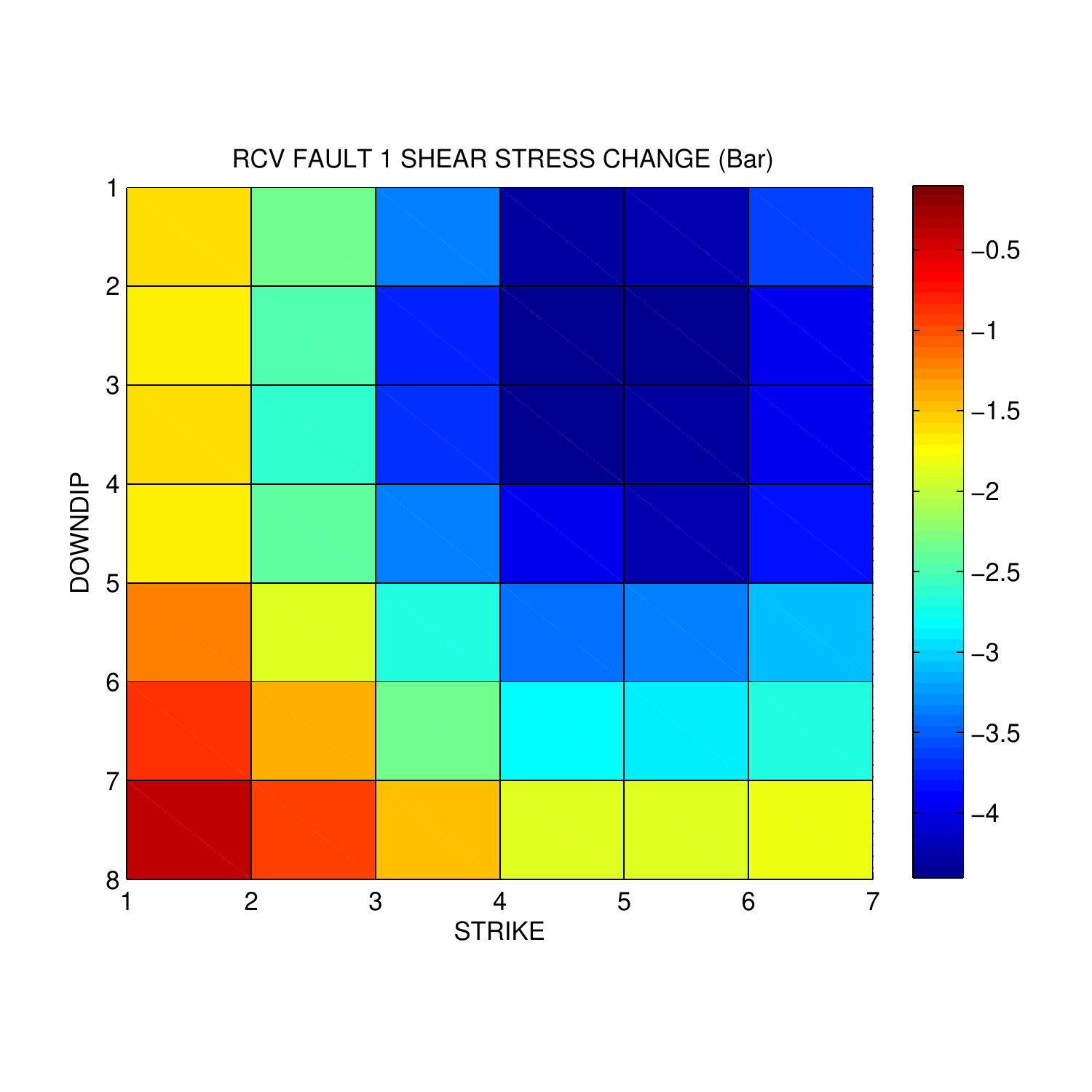} 
\end{tabular}
\end{center}
\vspace{-10 mm}
\caption{Shear Stress Transfer, Hector Mine faults 3,2,1, $\mu=0.7$, homogeneous vs layered Earth.}
\label{fig:dshearhm7}
\end{figure}

\begin{figure}[htbp]
\begin{center}
\begin{tabular}{ccc}
\includegraphics[width=.33\textwidth]{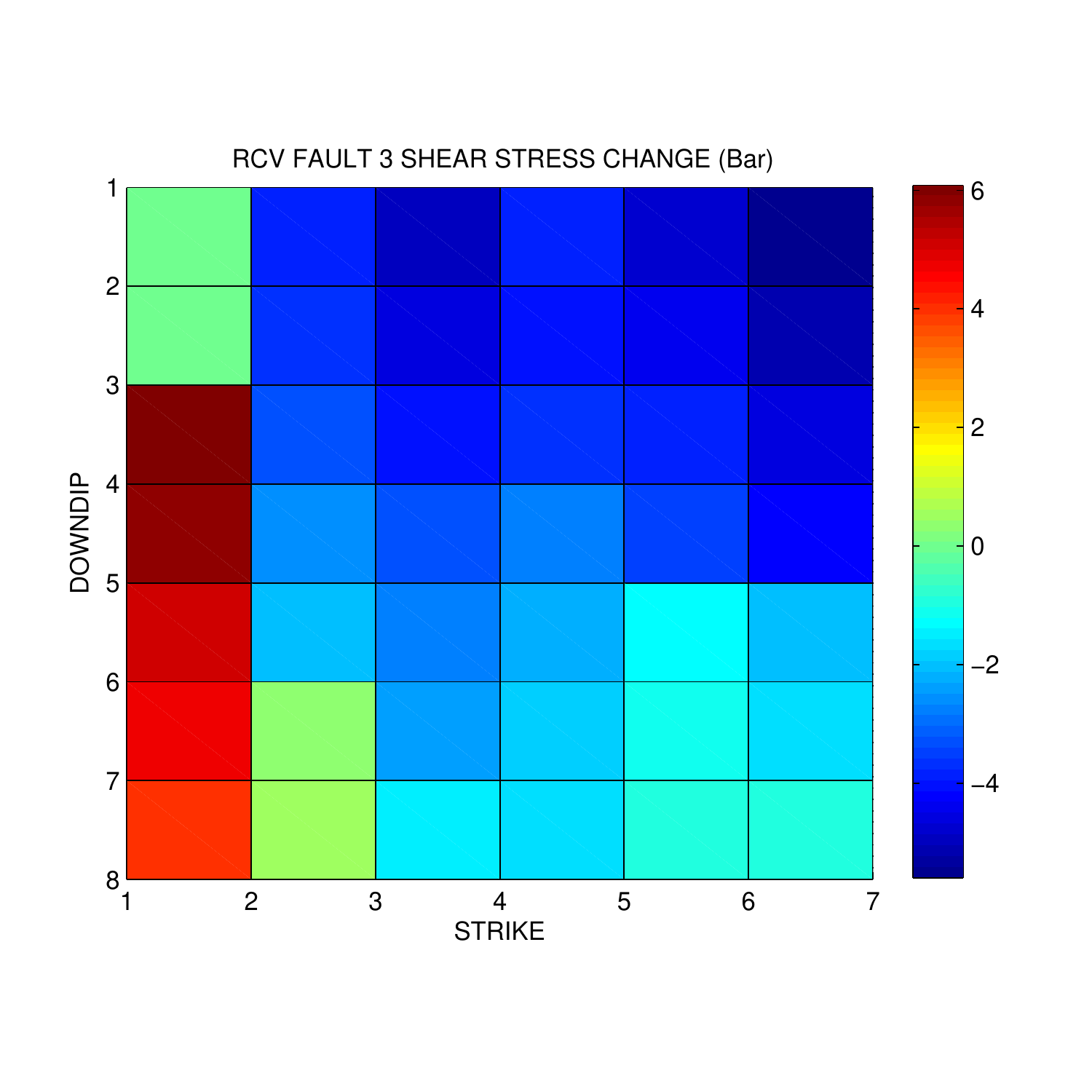} &
\includegraphics[width=.33\textwidth]{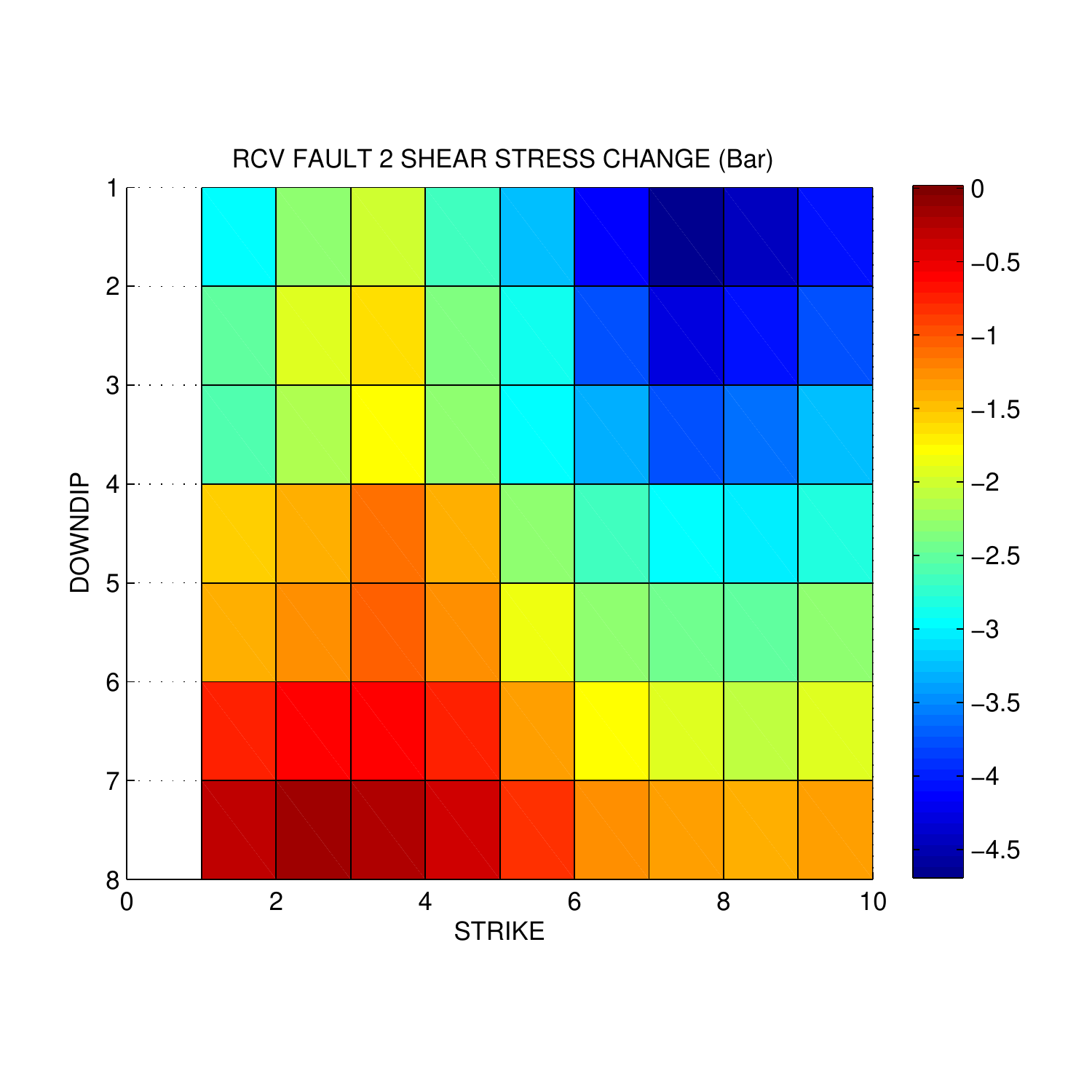} &
\includegraphics[width=.33\textwidth]{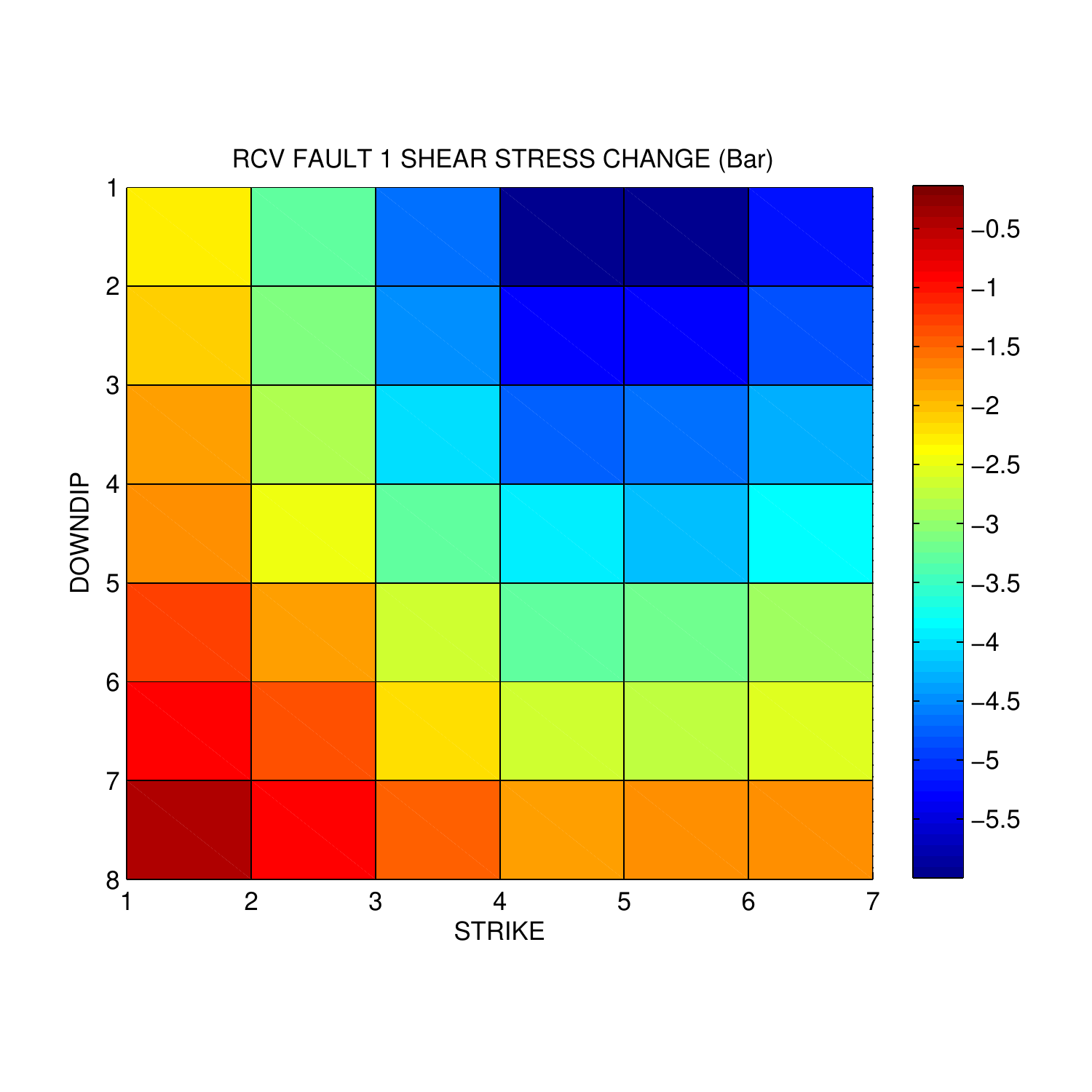} \\
\includegraphics[width=.33\textwidth]{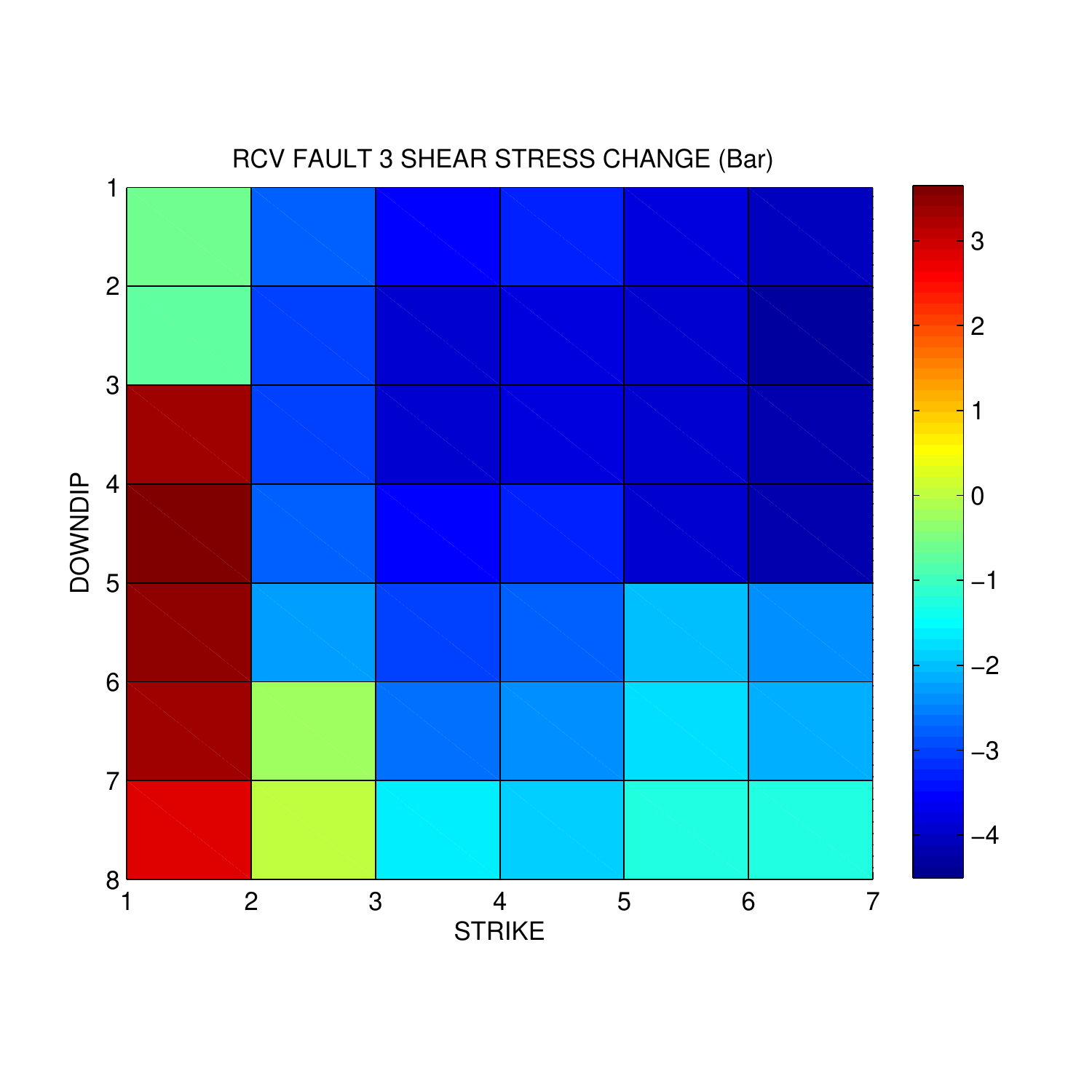} &
\includegraphics[width=.33\textwidth]{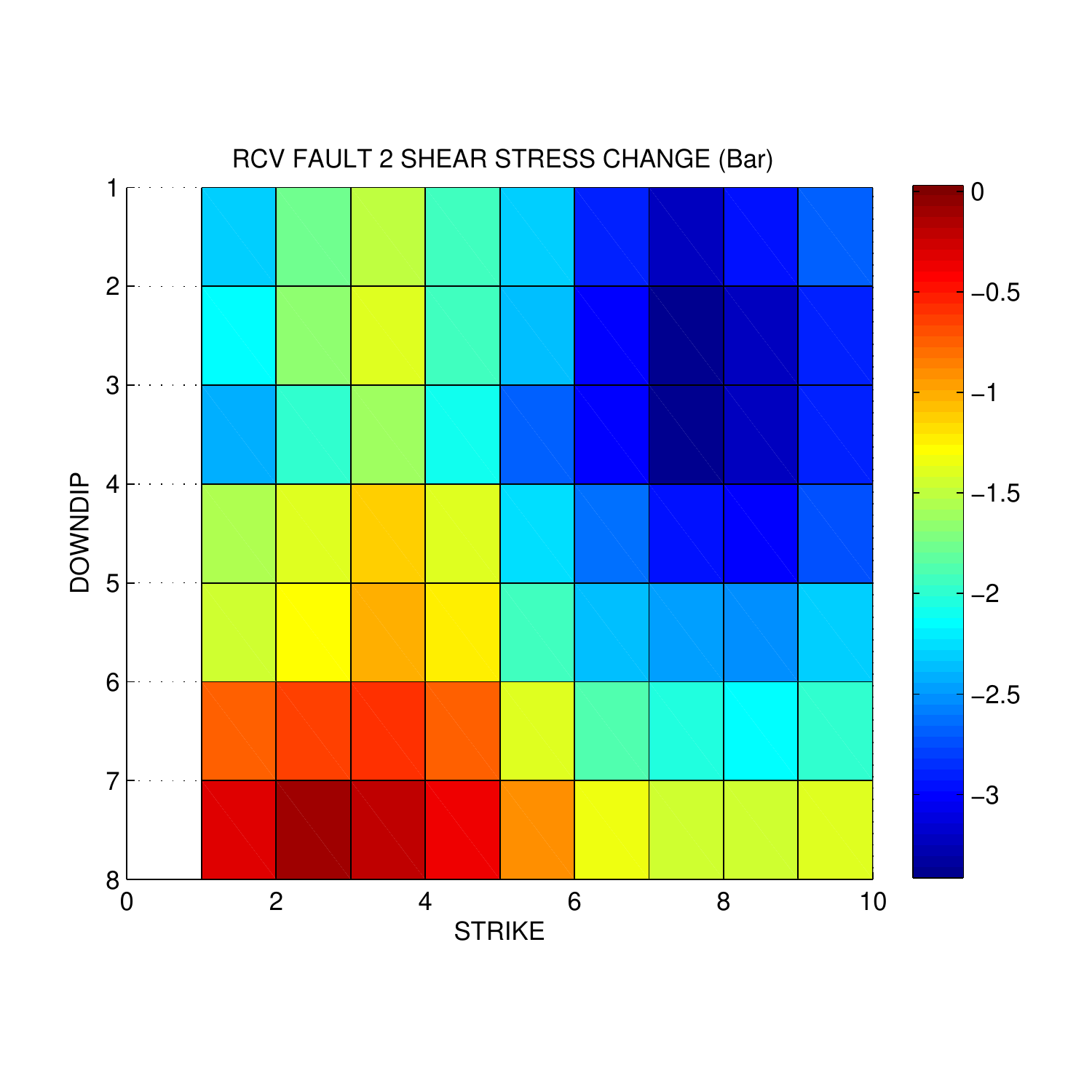} &
\includegraphics[width=.33\textwidth]{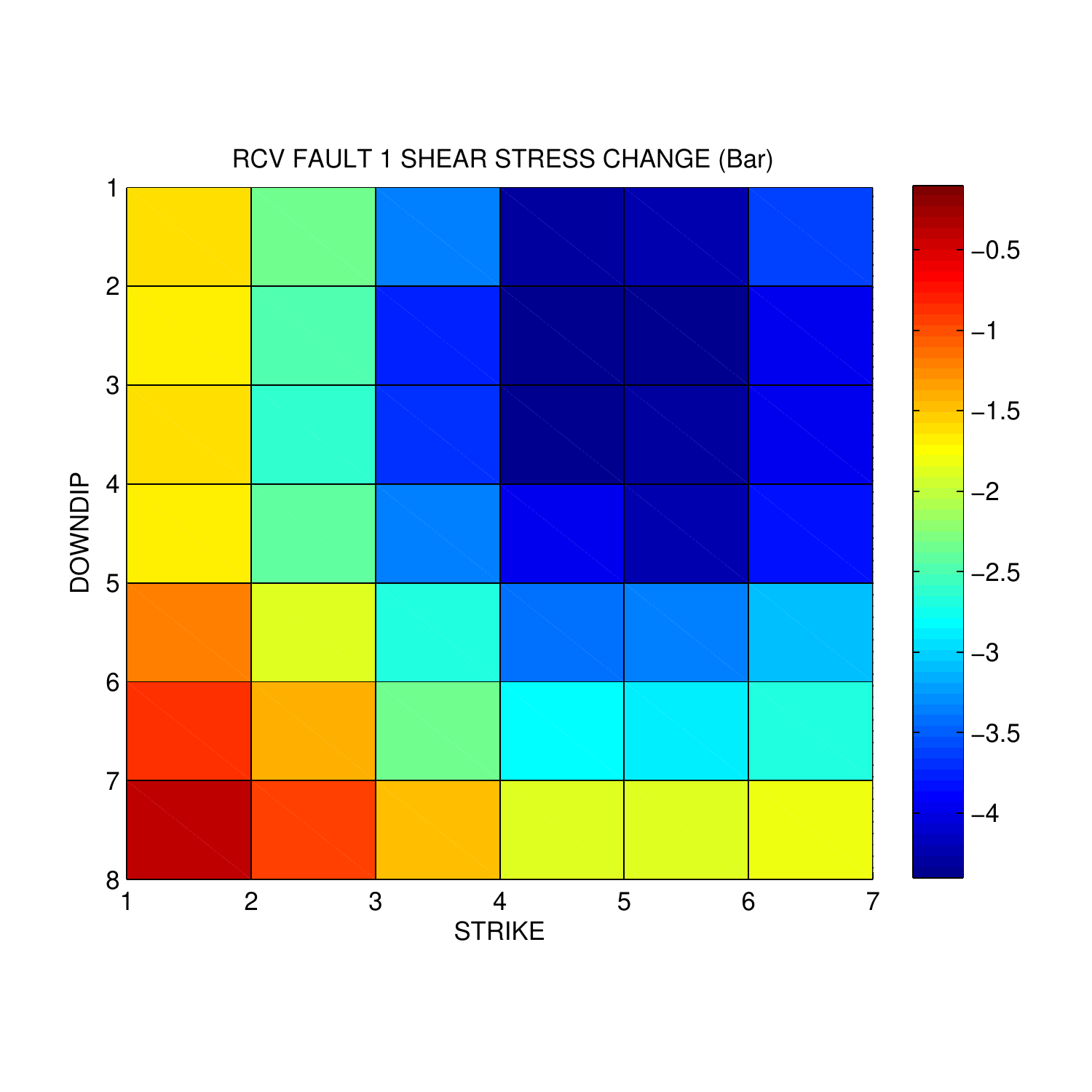} 
\end{tabular}
\end{center}
\vspace{-10 mm}
\caption{Shear Stress Transfer, Hector Mine faults 3,2,1, $\mu=0.8$, homogeneous vs layered Earth.}
\label{fig:dshearhm8}
\end{figure}

\begin{figure}[htbp]
\begin{center}
\begin{tabular}{ccc}

\includegraphics[width=.33\textwidth]{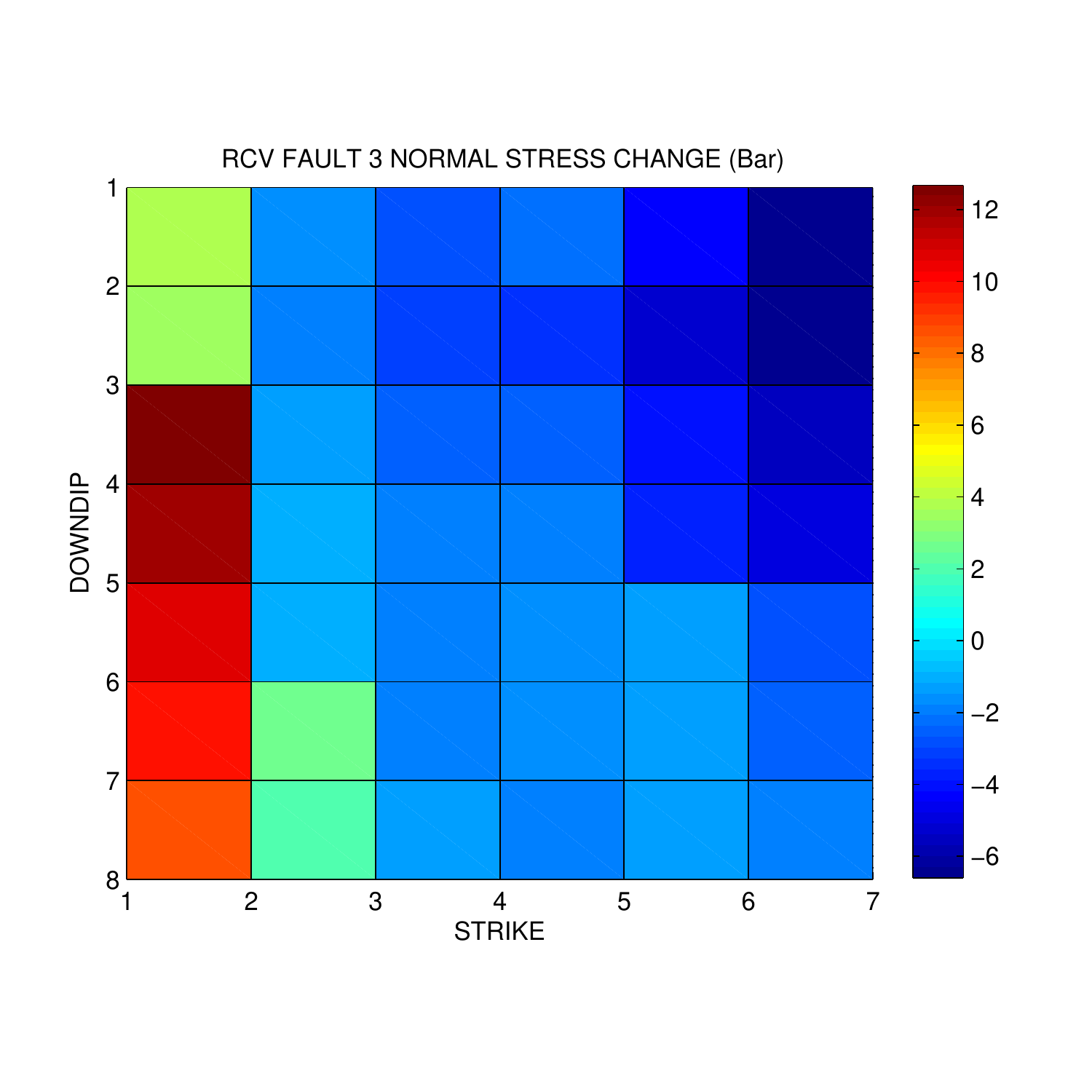} &
\includegraphics[width=.33\textwidth]{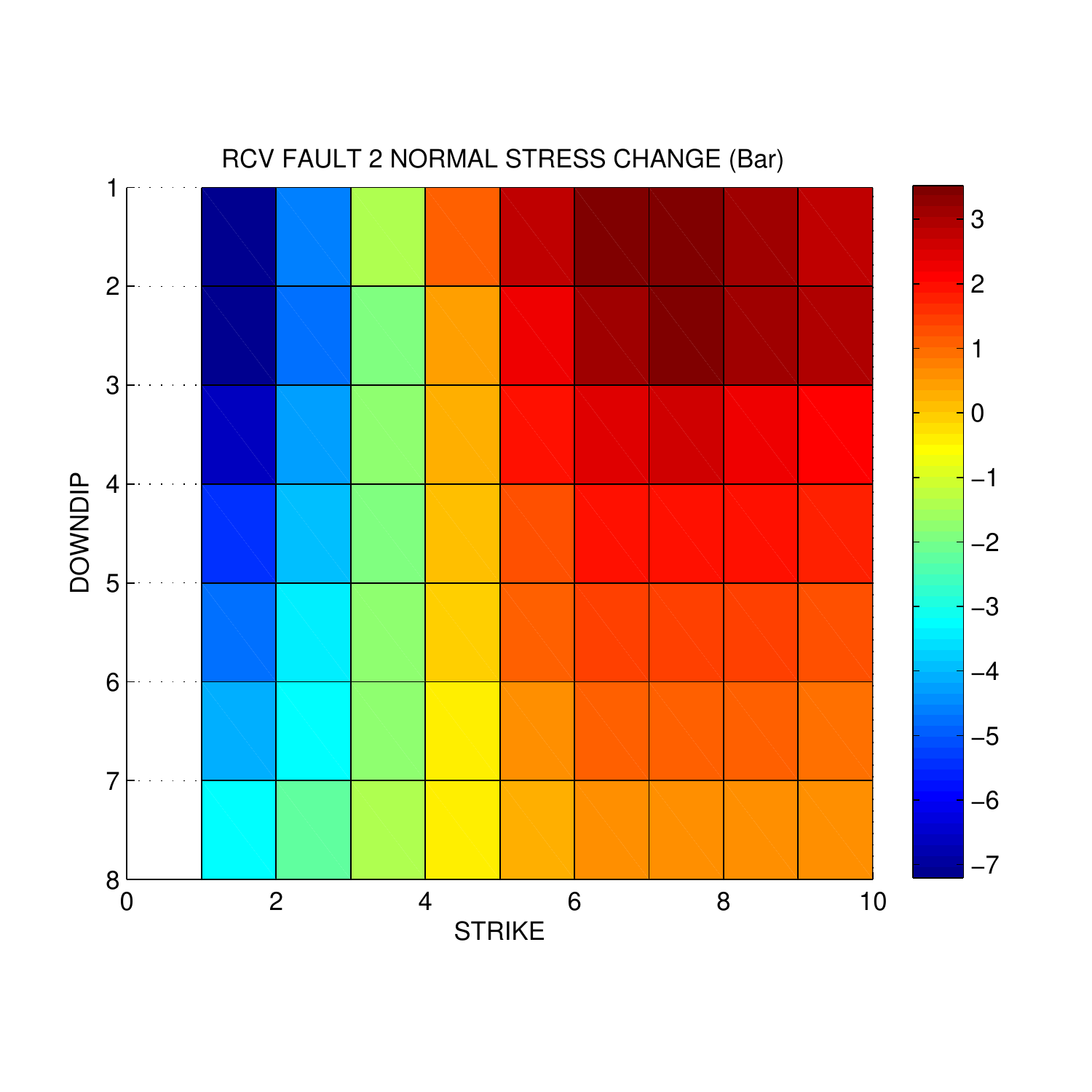} &
\includegraphics[width=.33\textwidth]{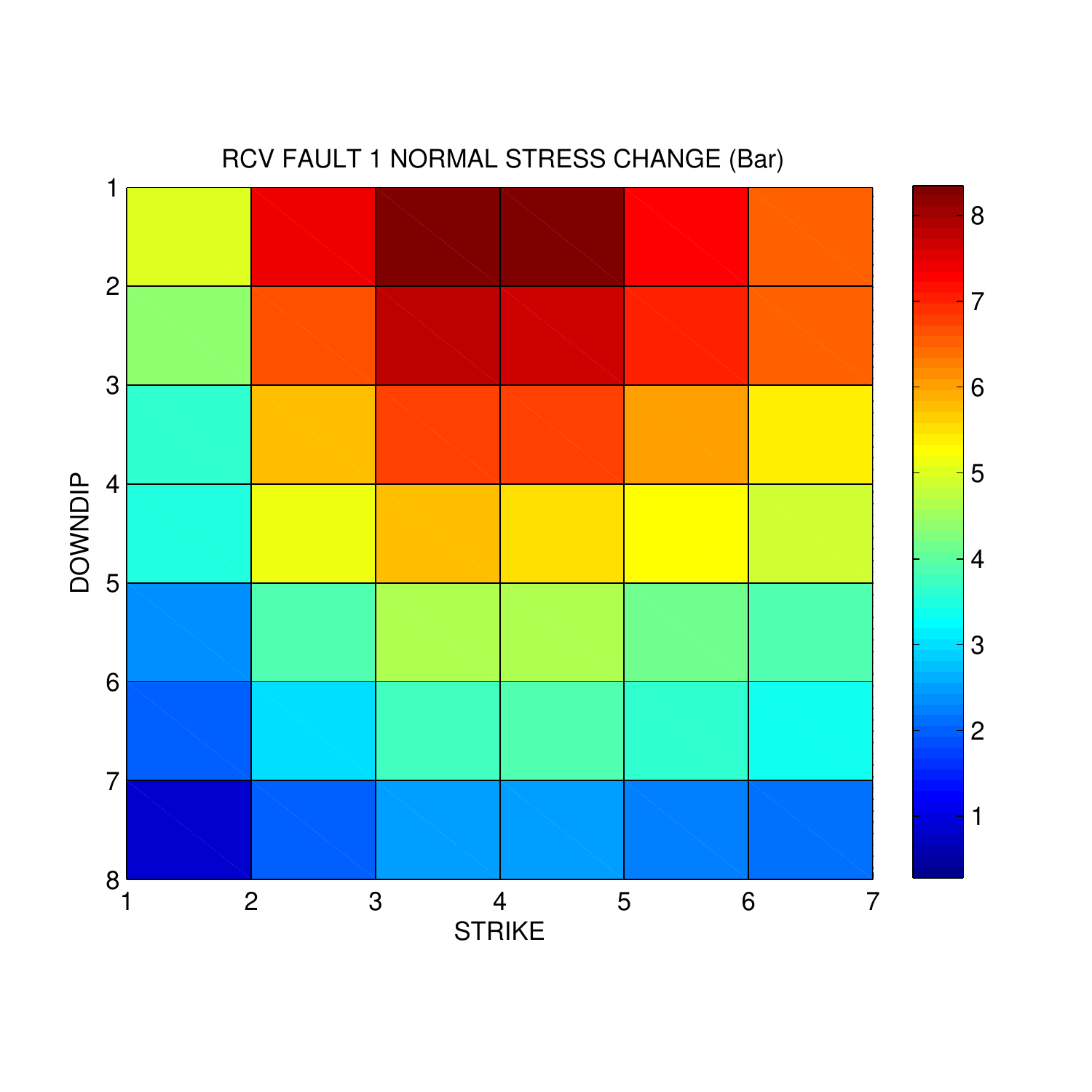} \\
\includegraphics[width=.33\textwidth]{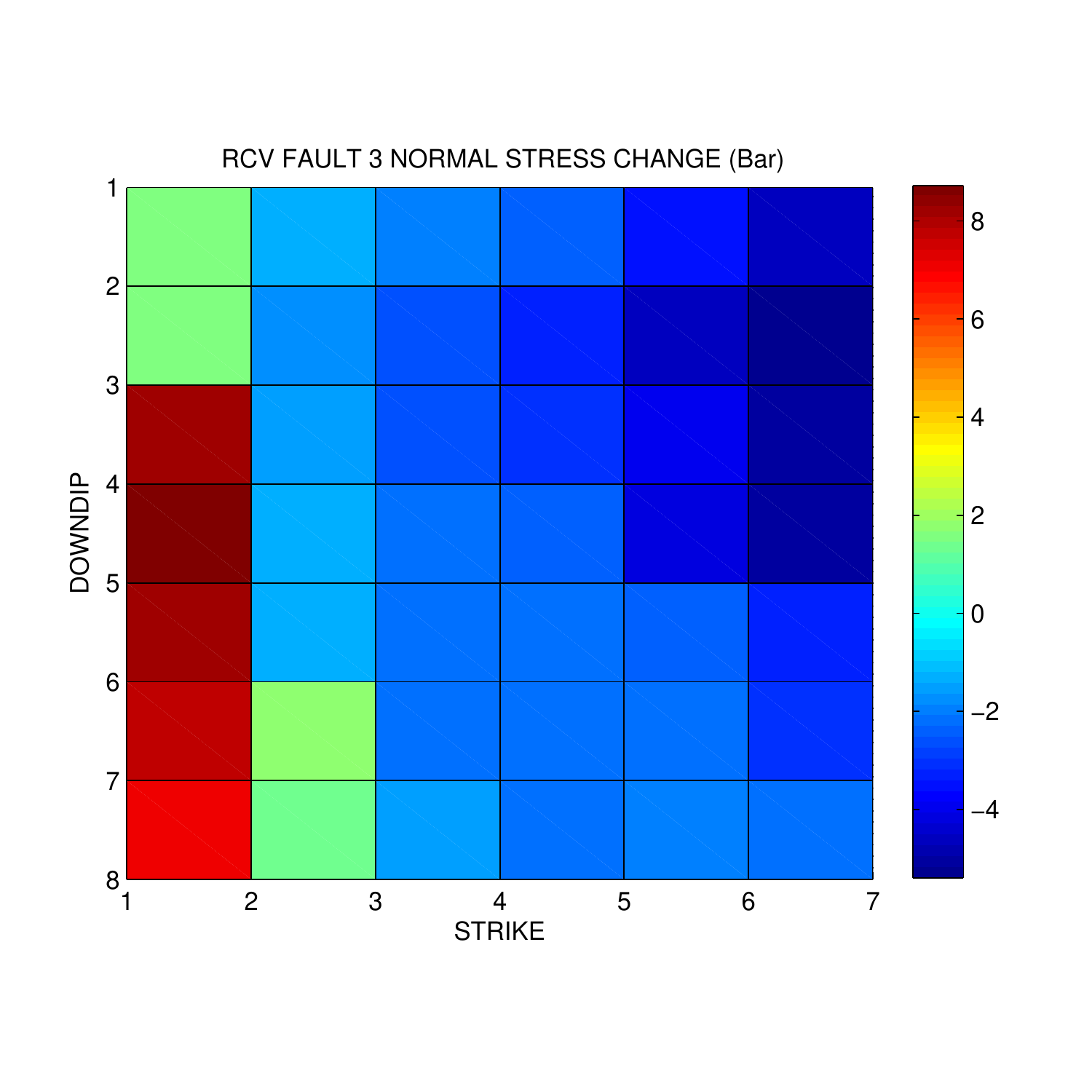} &
\includegraphics[width=.33\textwidth]{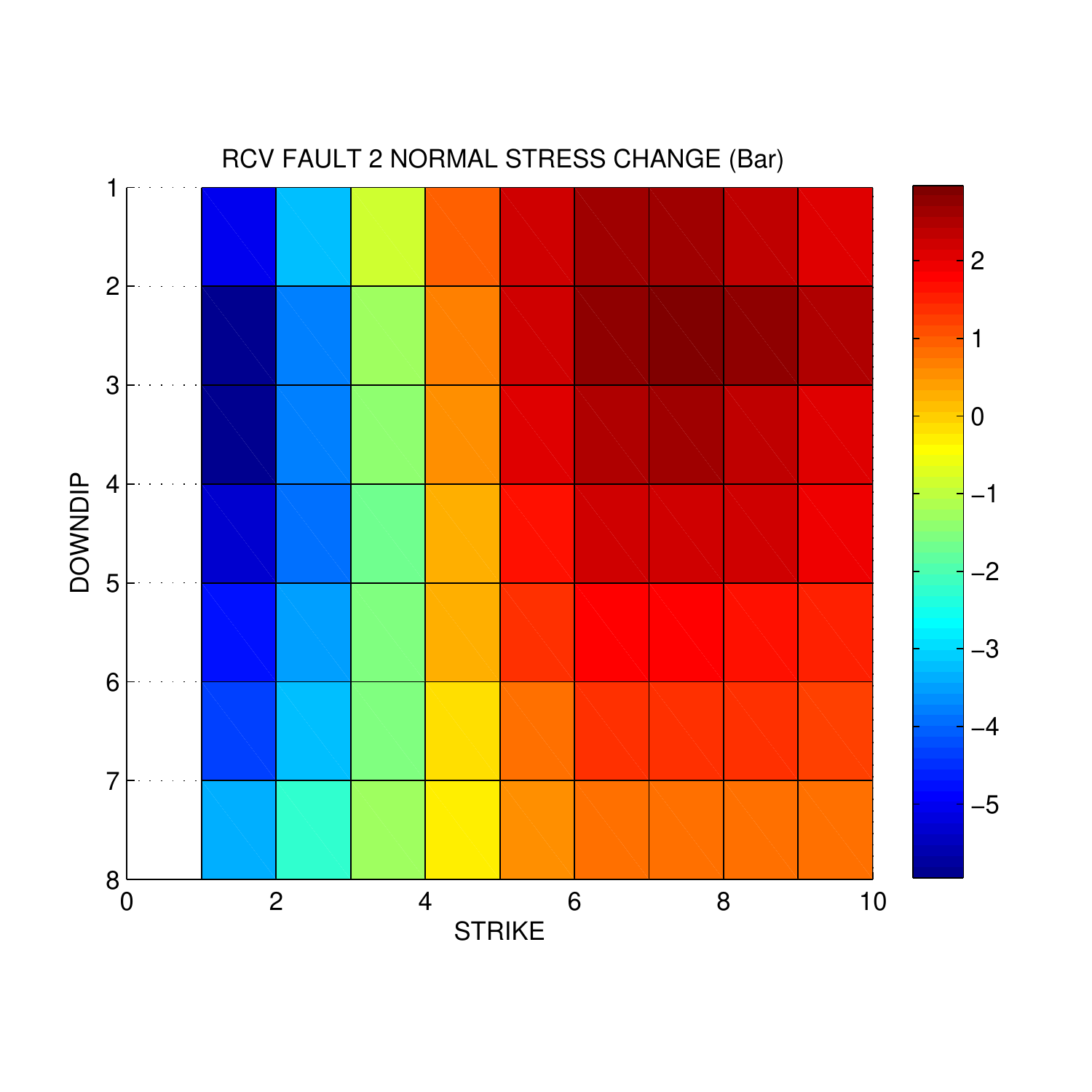} &
\includegraphics[width=.33\textwidth]{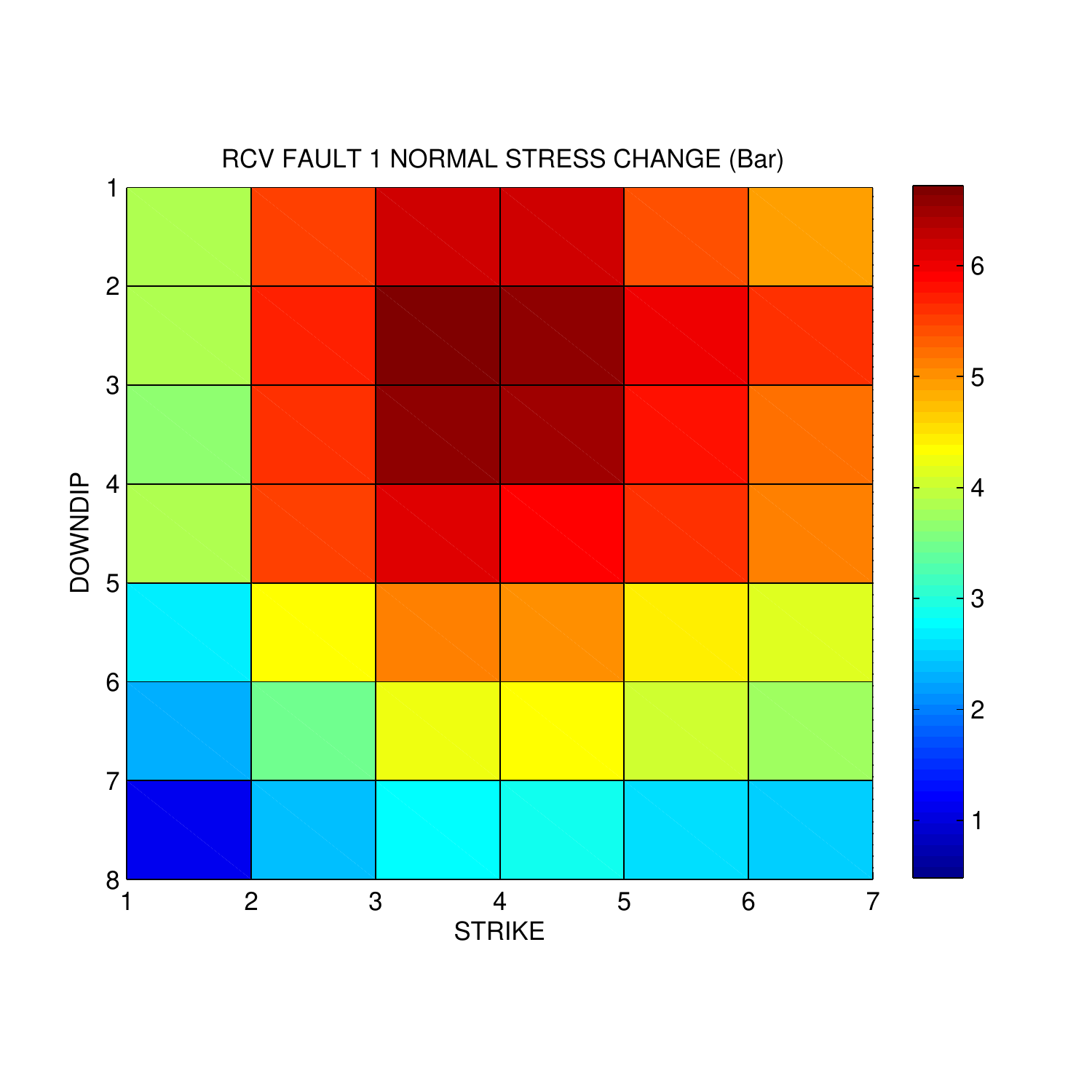} 
\end{tabular}
\end{center}
\vspace{-10 mm}
\caption{(Un)clamping, Hector Mine faults 3,2,1, $\mu=0.7$, homogeneous vs layered Earth.}
\label{fig:dnormhm7}
\end{figure}

\begin{figure}[htbp]
\begin{center}
\begin{tabular}{ccc}
\includegraphics[width=.33\textwidth]{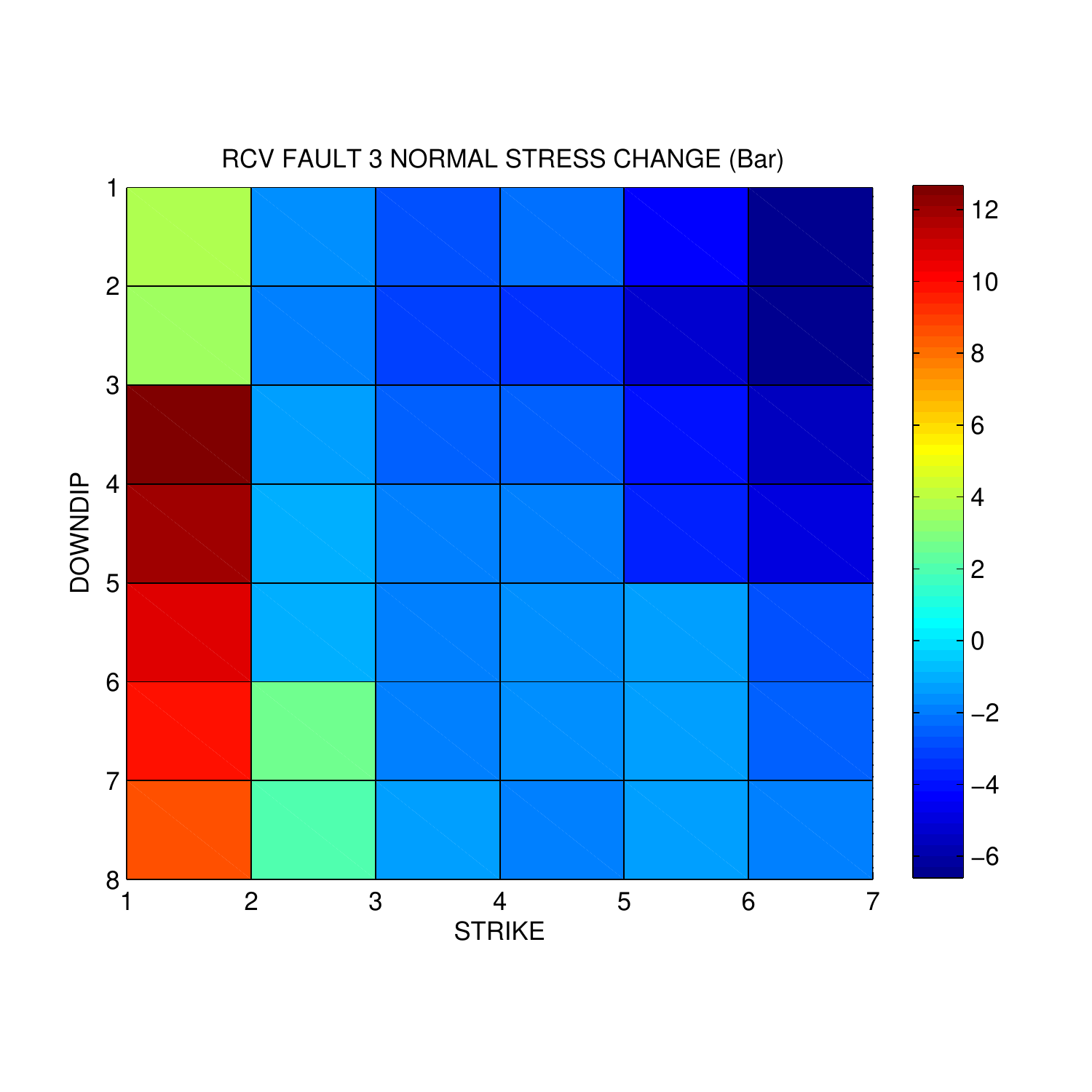} &
\includegraphics[width=.33\textwidth]{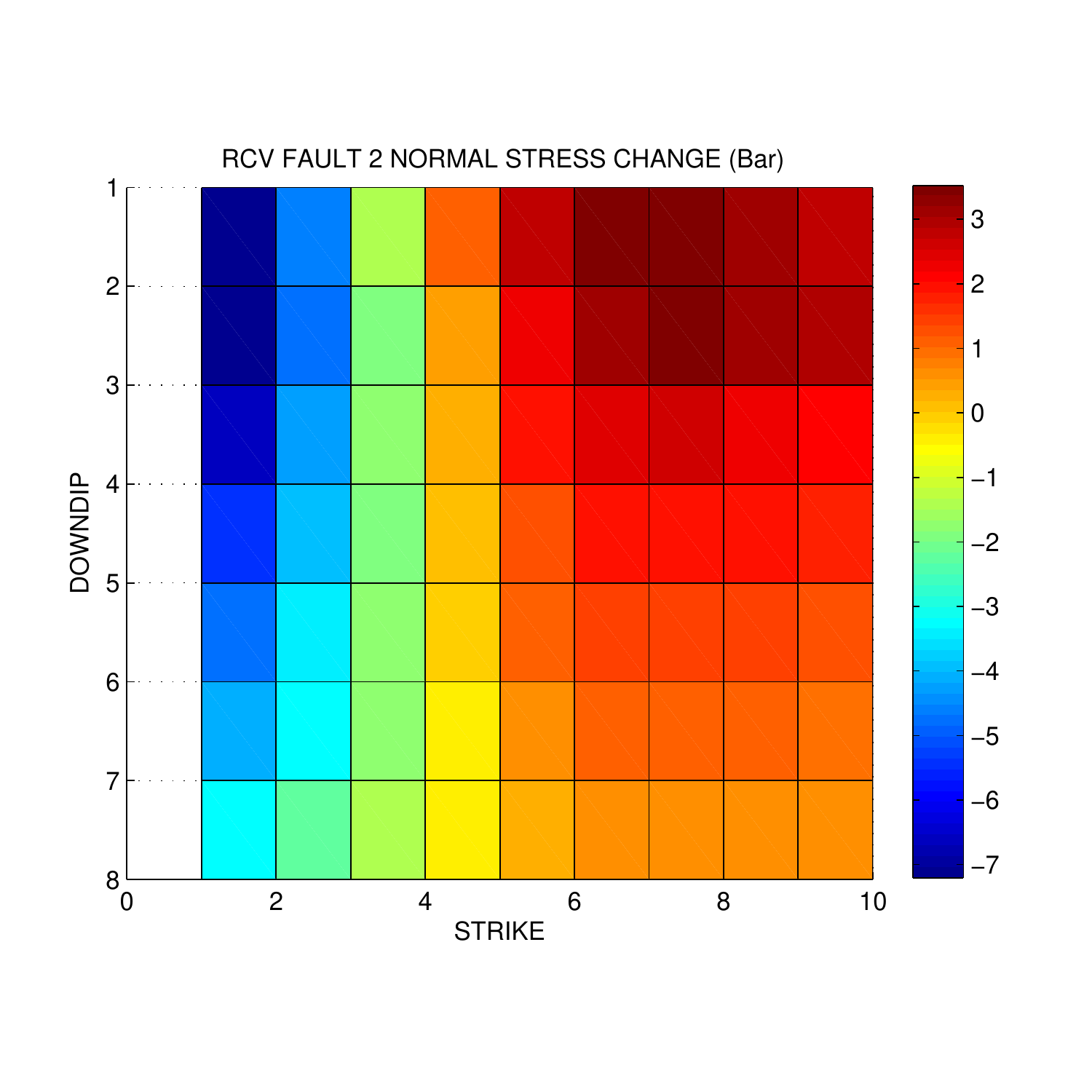} &
\includegraphics[width=.33\textwidth]{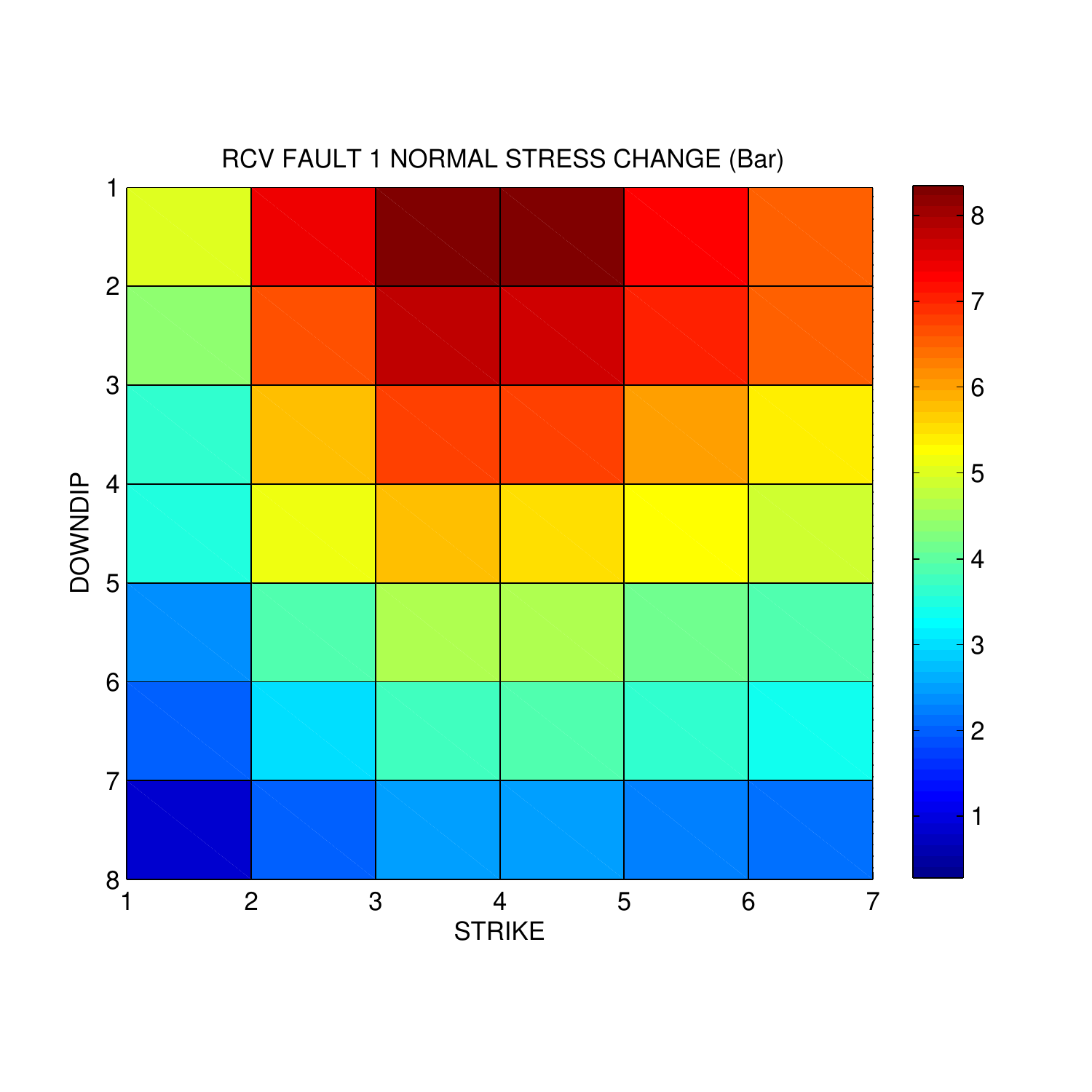} \\ 
\includegraphics[width=.33\textwidth]{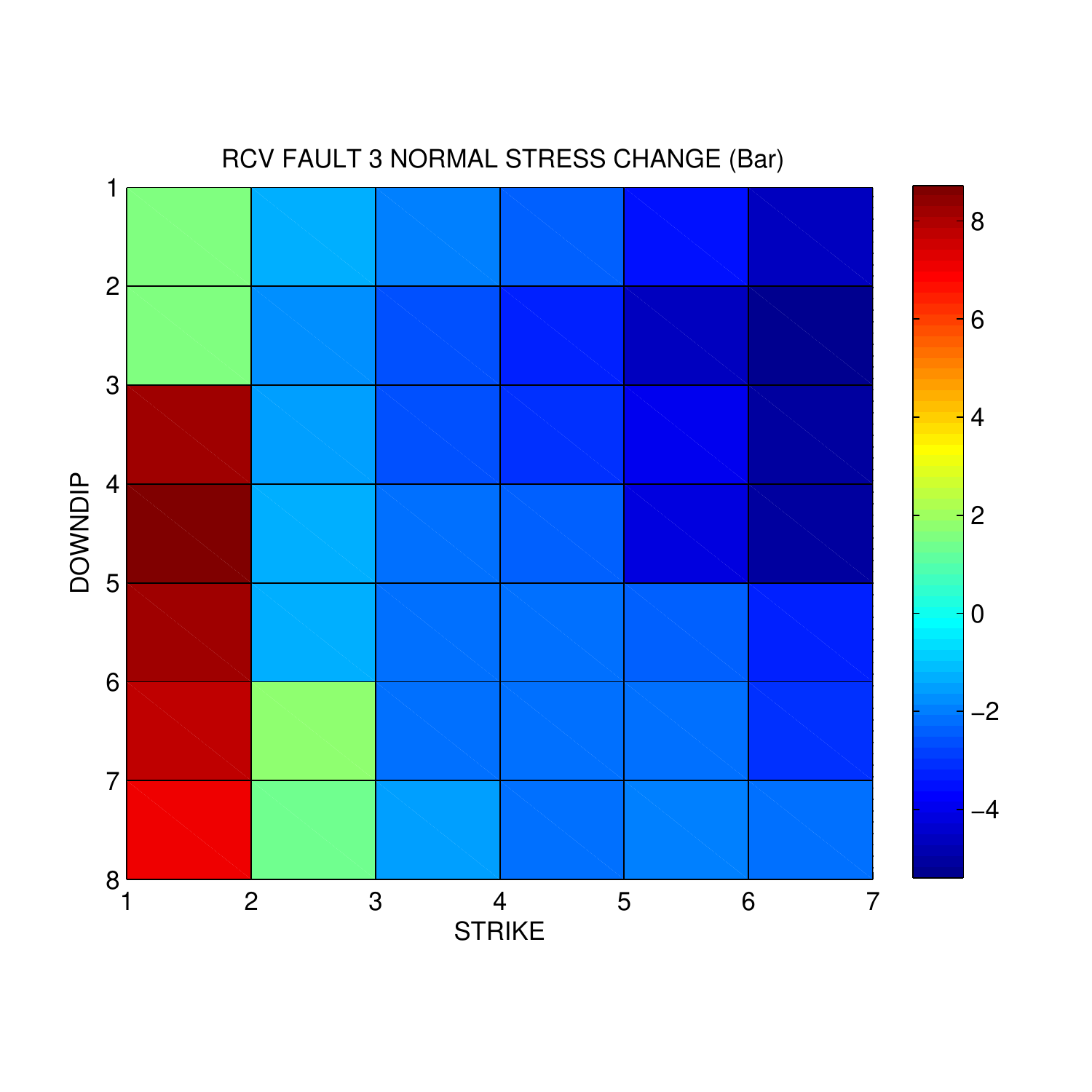} & 
\includegraphics[width=.33\textwidth]{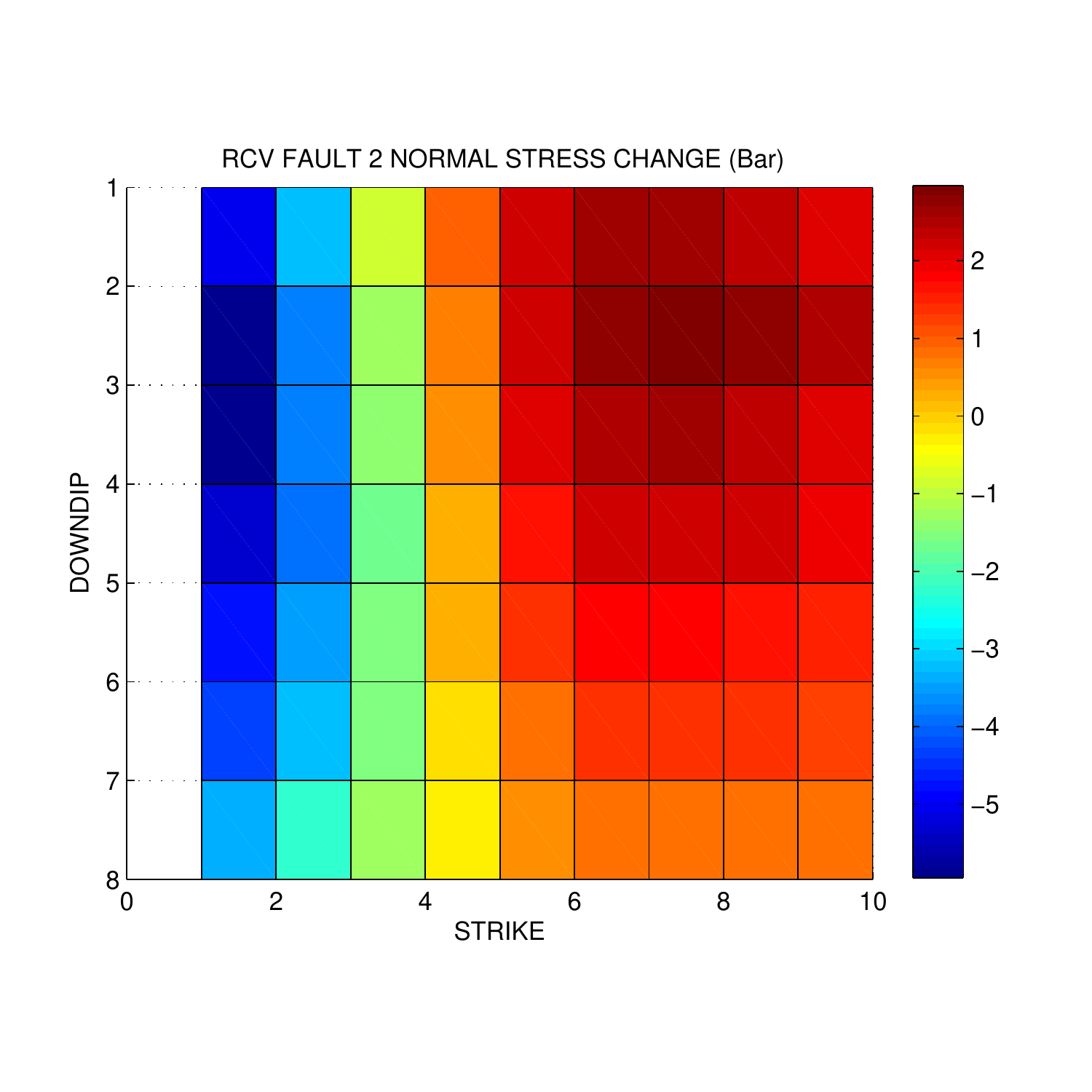} &
\includegraphics[width=.33\textwidth]{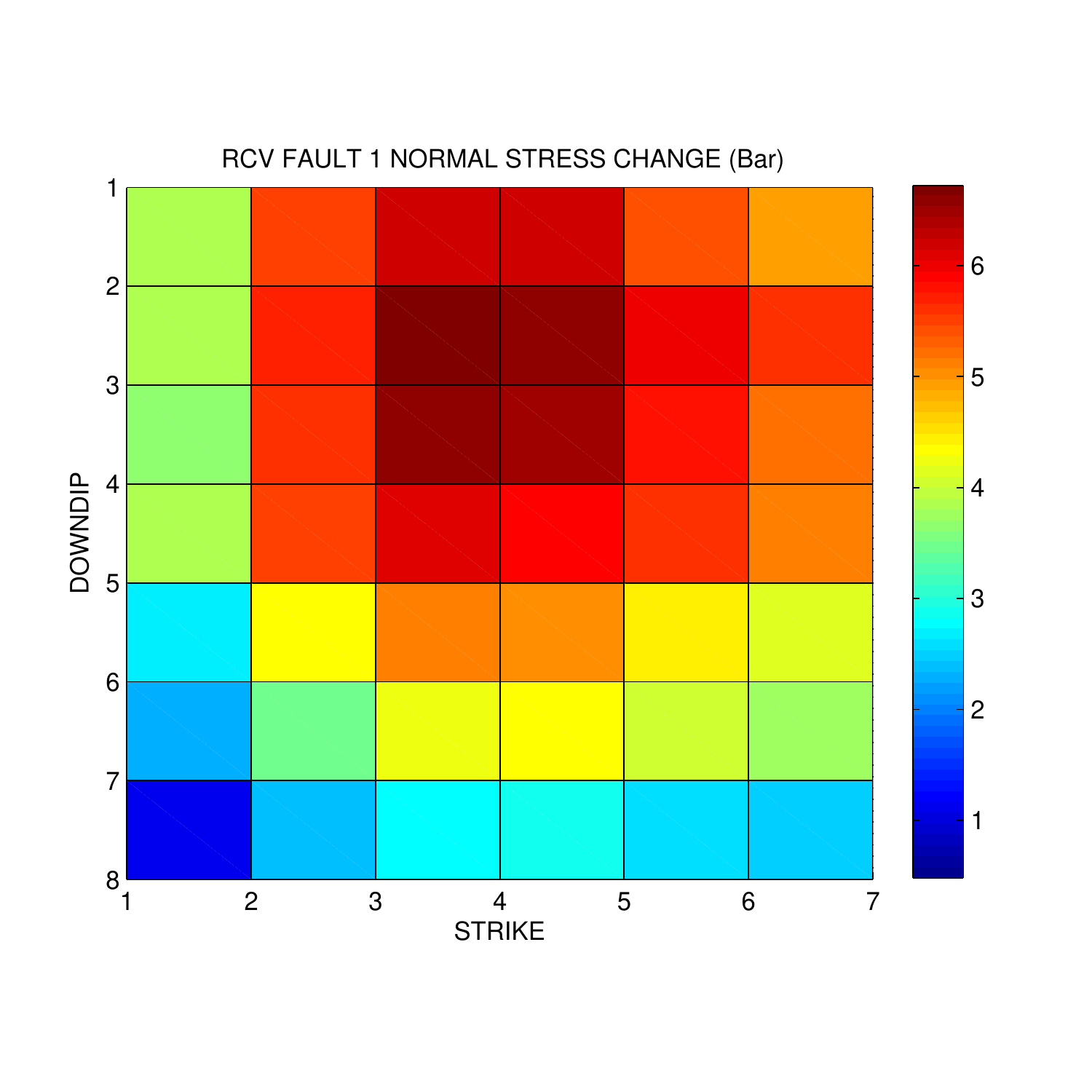} 
\end{tabular}
\end{center}
\vspace{-10 mm}
\caption{(Un)clamping, Hector Mine faults 3,2,1, $\mu=0.8$, homogeneous vs layered Earth.}
\label{fig:dnormhm8}
\end{figure}

\section{Slip Inversion}

As was demonstrated in the previous section, the \emph{estimated} stress transfer is not always such as to encourage slip, hence some constraints in (\ref{eq:INV}) will be violated. Including all constraints may render the problem inconsistent and the solution infeasible for the optimization algorithm (Simplex implemented in {\tt MATLAB linprog}). Therefore we conduct two experiments: invert using all the constraints, or use only \emph{consistent} constraints obtained from direct modeling, allowing us to assess the impact of dataset size on the accuracy.

\begin{figure}[htbp]
\begin{center}
\begin{tabular}{ccc}
\includegraphics[width=.33\textwidth]{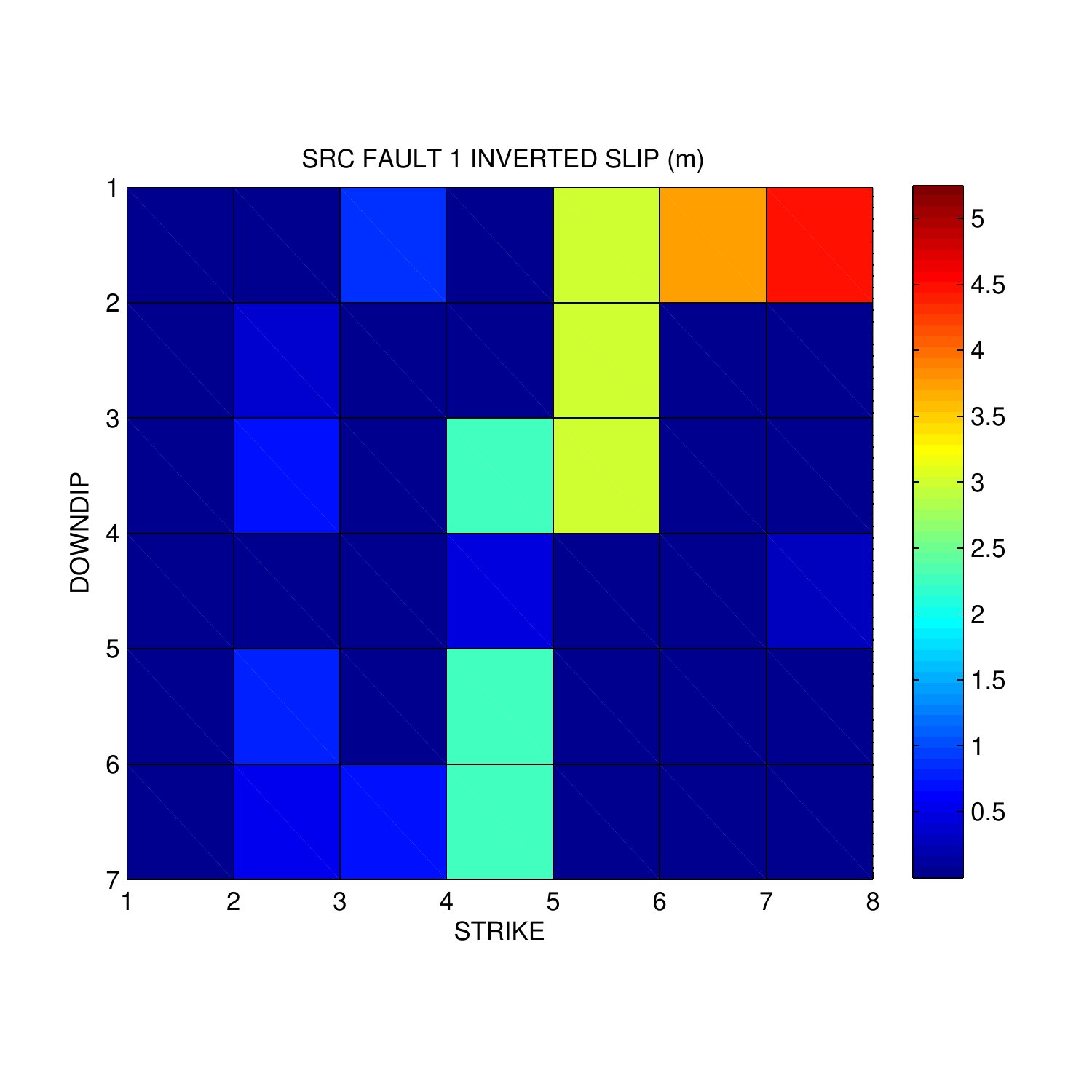} &
\includegraphics[width=.33\textwidth]{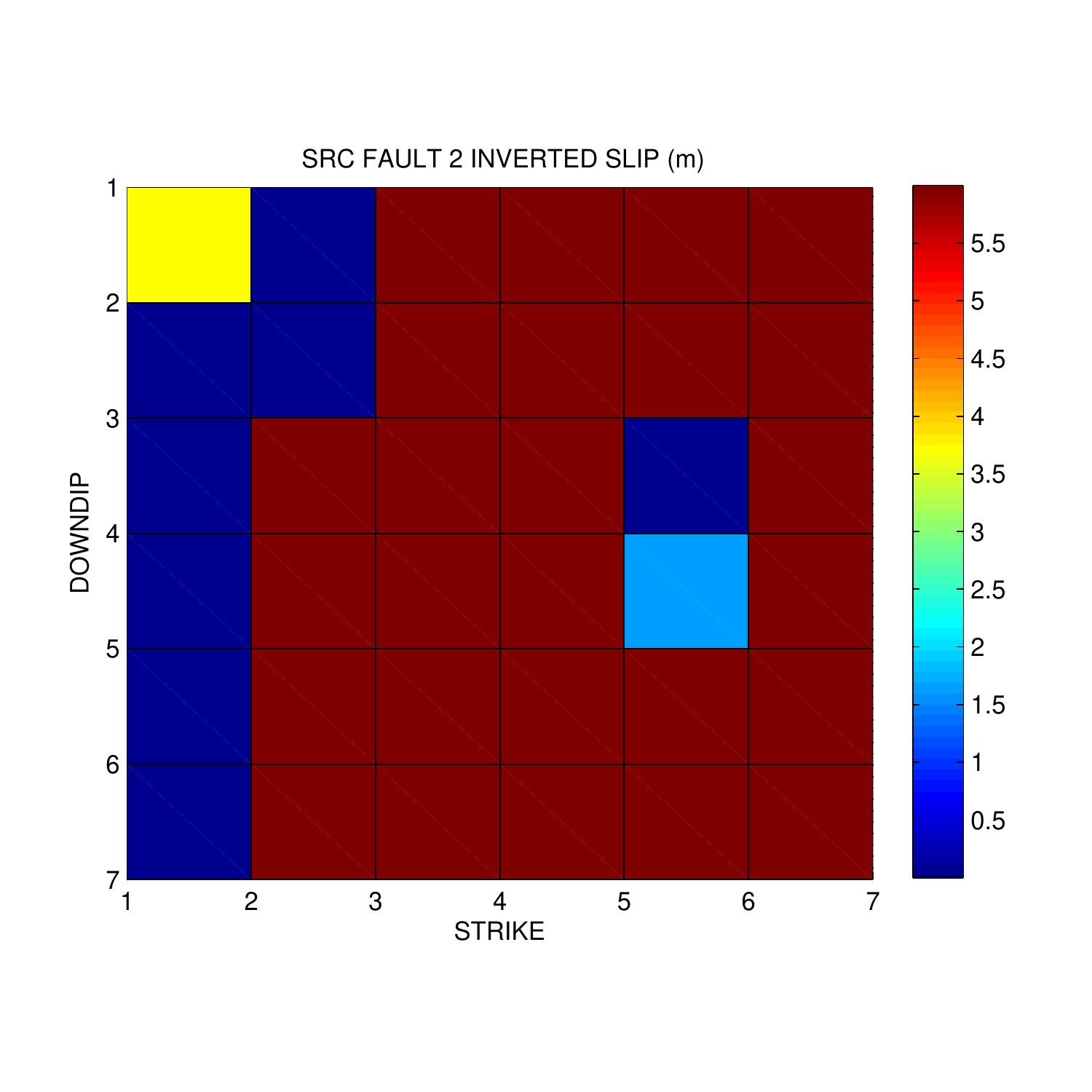} &
\includegraphics[width=.33\textwidth]{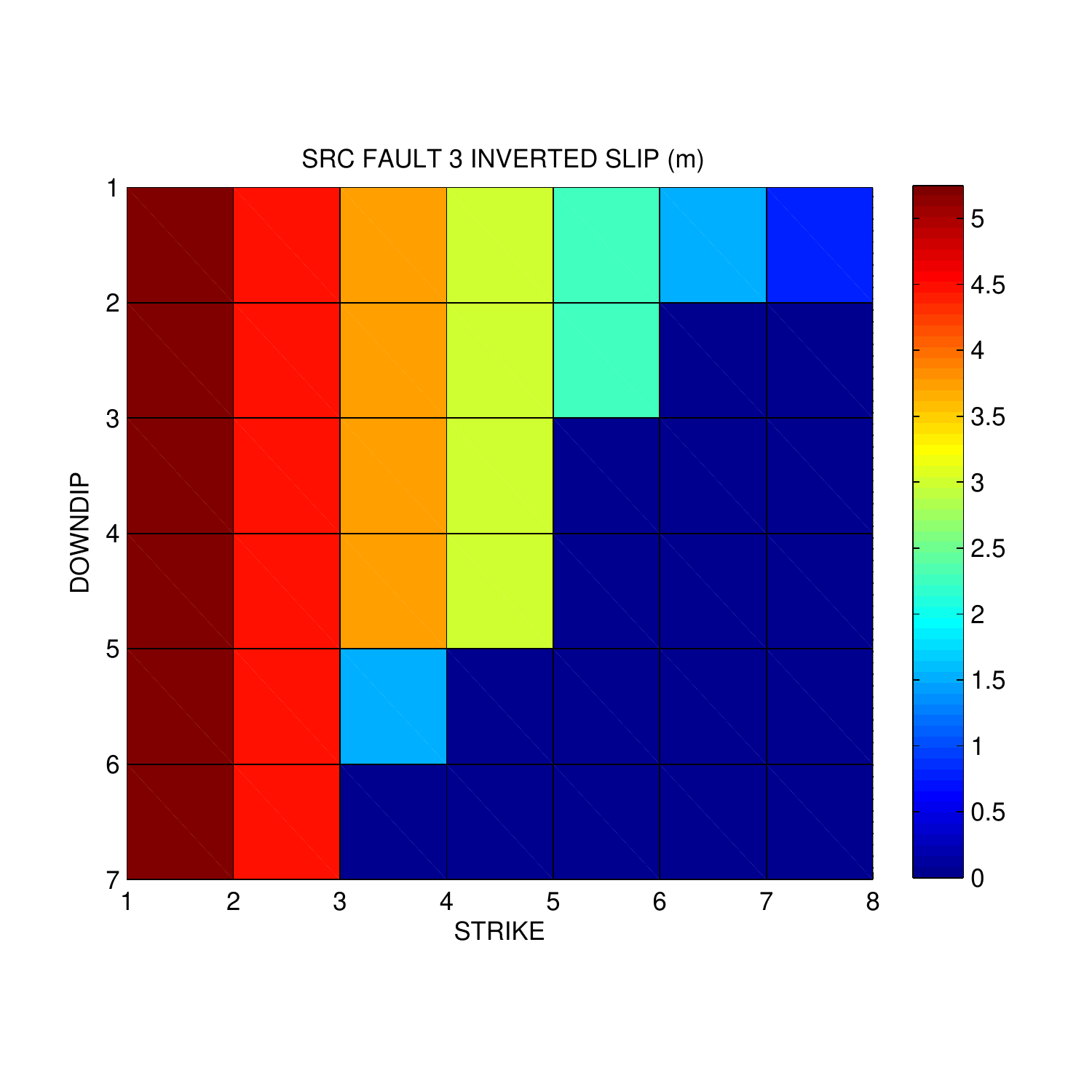}\\ 
\includegraphics[width=.33\textwidth]{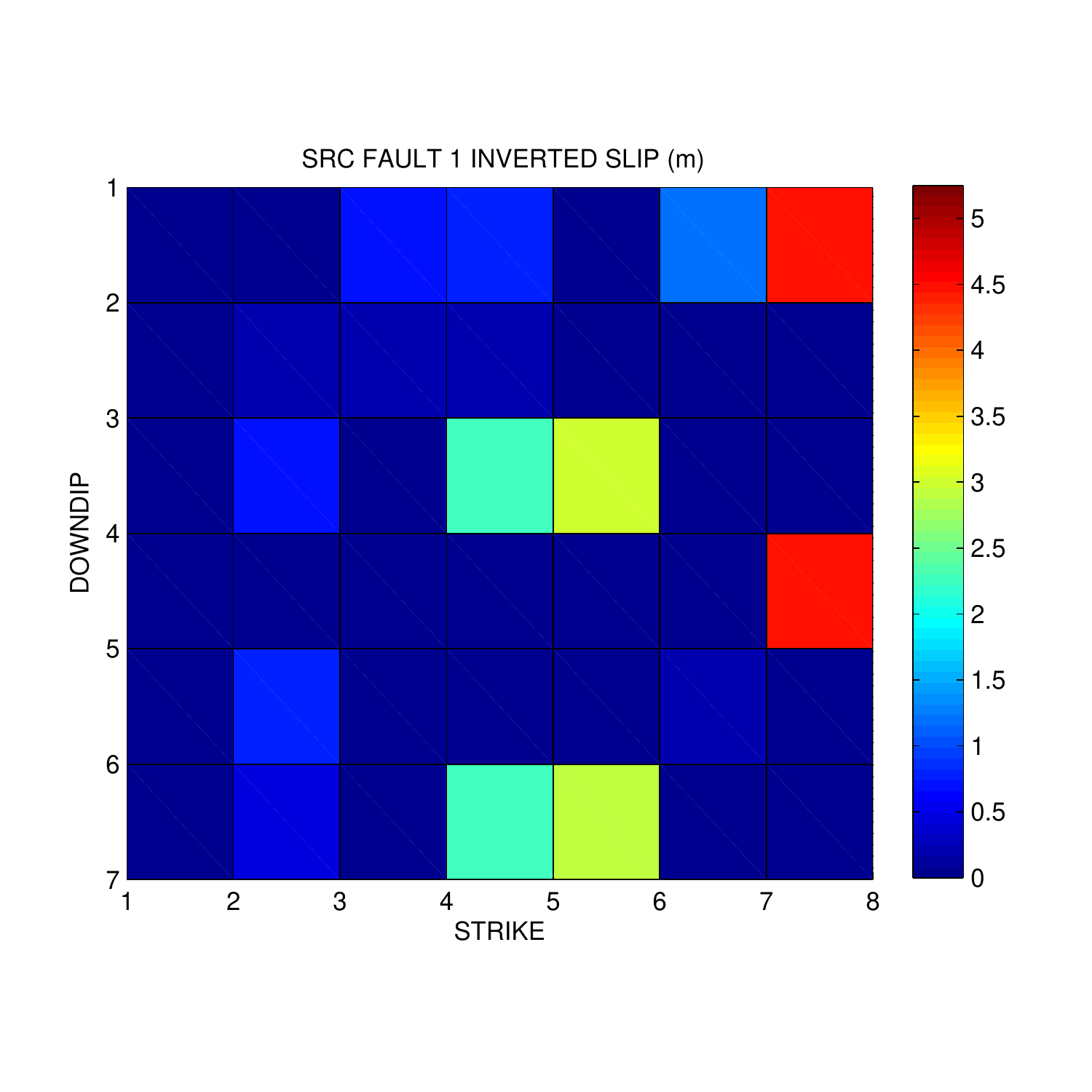} &
\includegraphics[width=.33\textwidth]{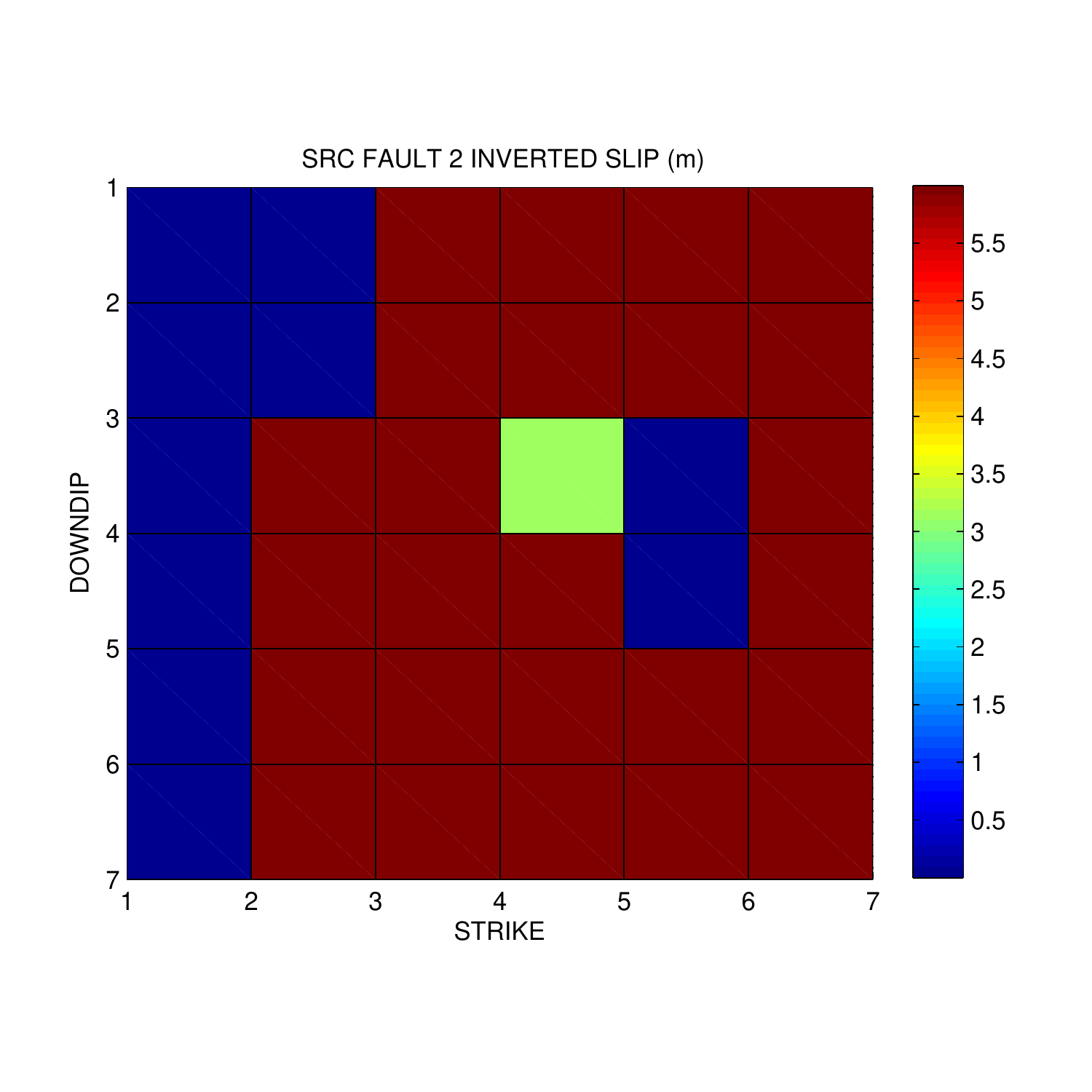} &
\includegraphics[width=.33\textwidth]{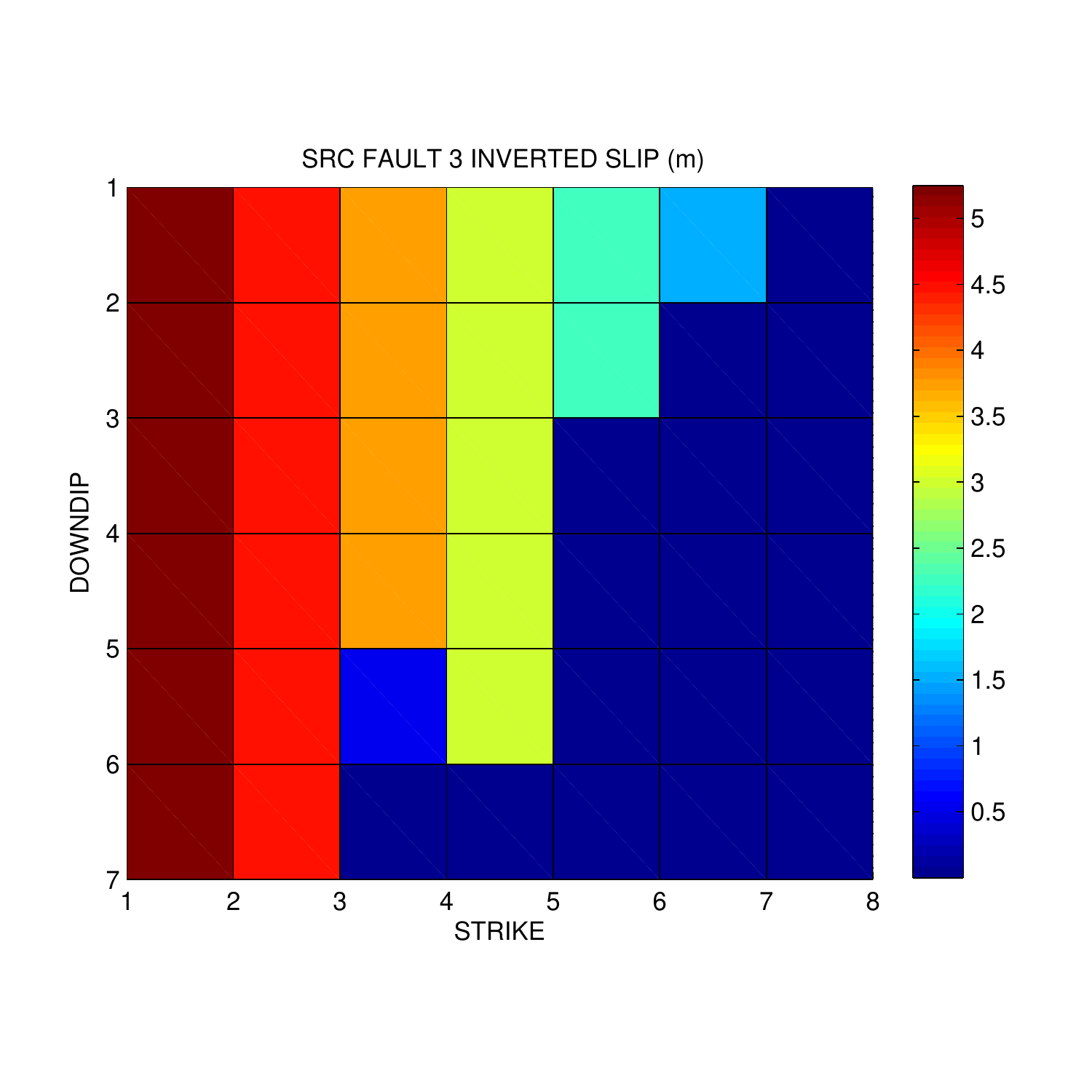}\\ 
\end{tabular}
\end{center}
\vspace{-10 mm}
\caption{Inverted Landers slips from the local dataset, only \emph{consistent} constraints, faults 1-3, $\mu=0.7$, homogeneous vs layered Earth.}
\label{fig:invertedlocal}
\end{figure}

\begin{figure}[htbp]
\begin{center}
\begin{tabular}{ccc}
\includegraphics[width=.33\textwidth]{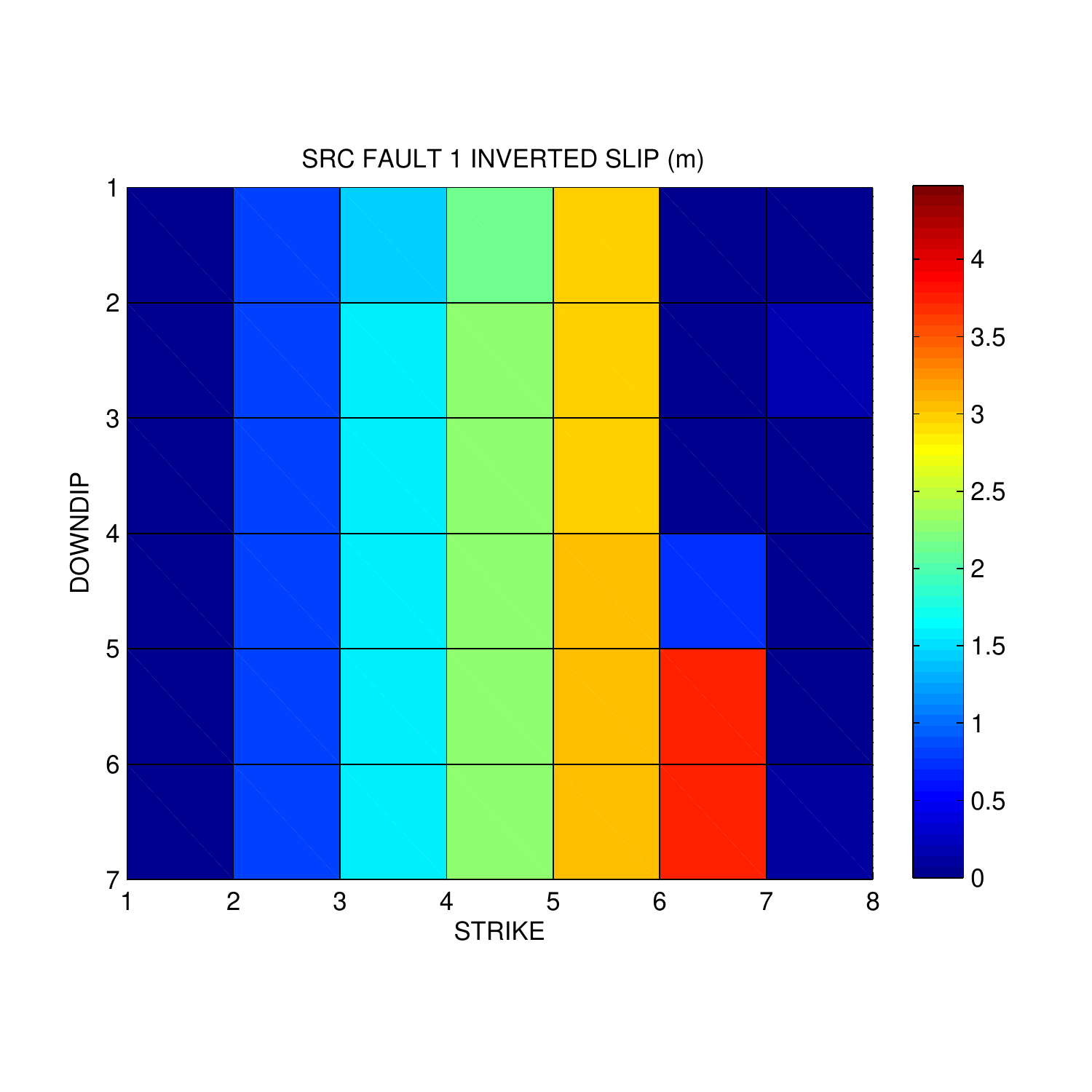} &
\includegraphics[width=.33\textwidth]{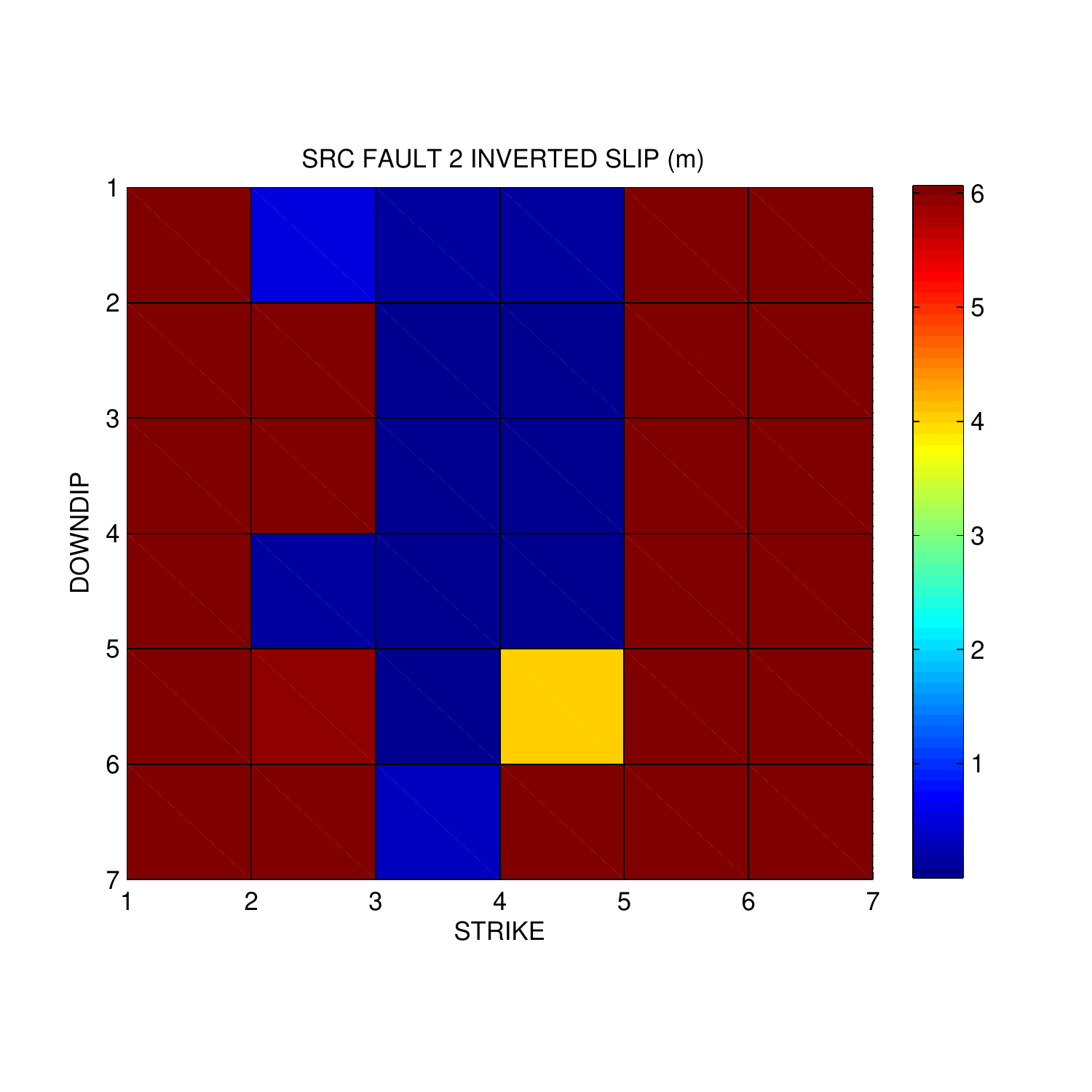} &
\includegraphics[width=.33\textwidth]{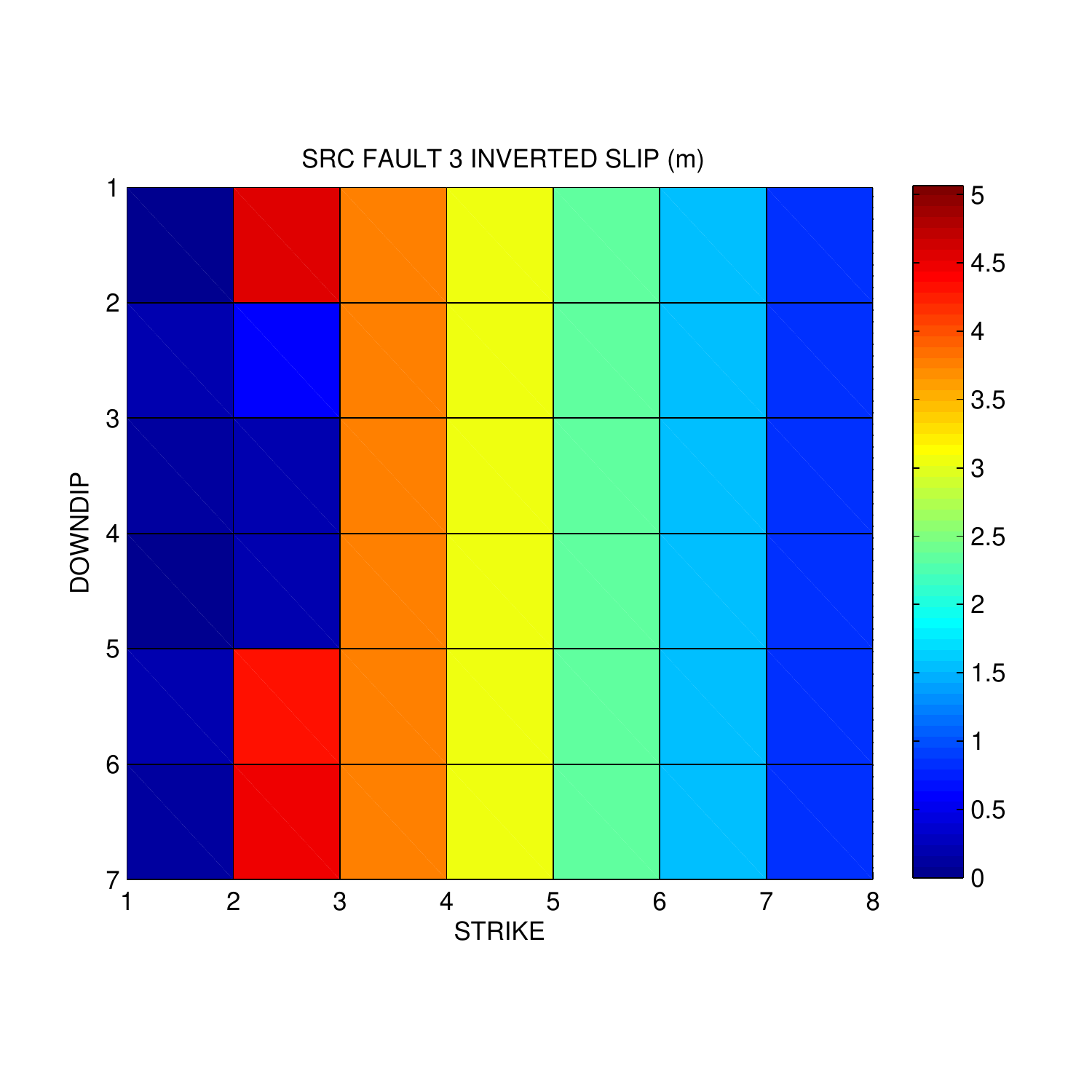}\\ 
\includegraphics[width=.33\textwidth]{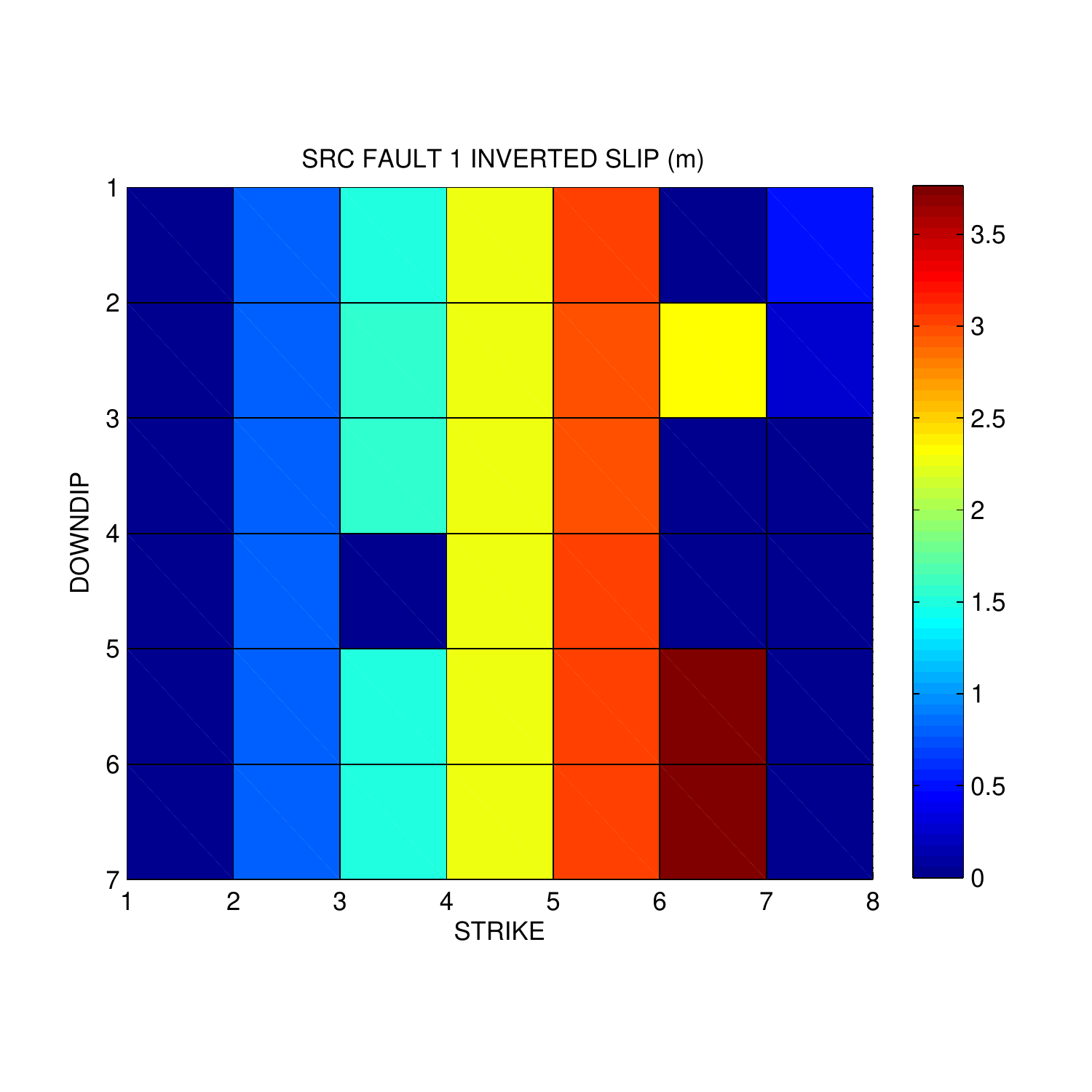} &
\includegraphics[width=.33\textwidth]{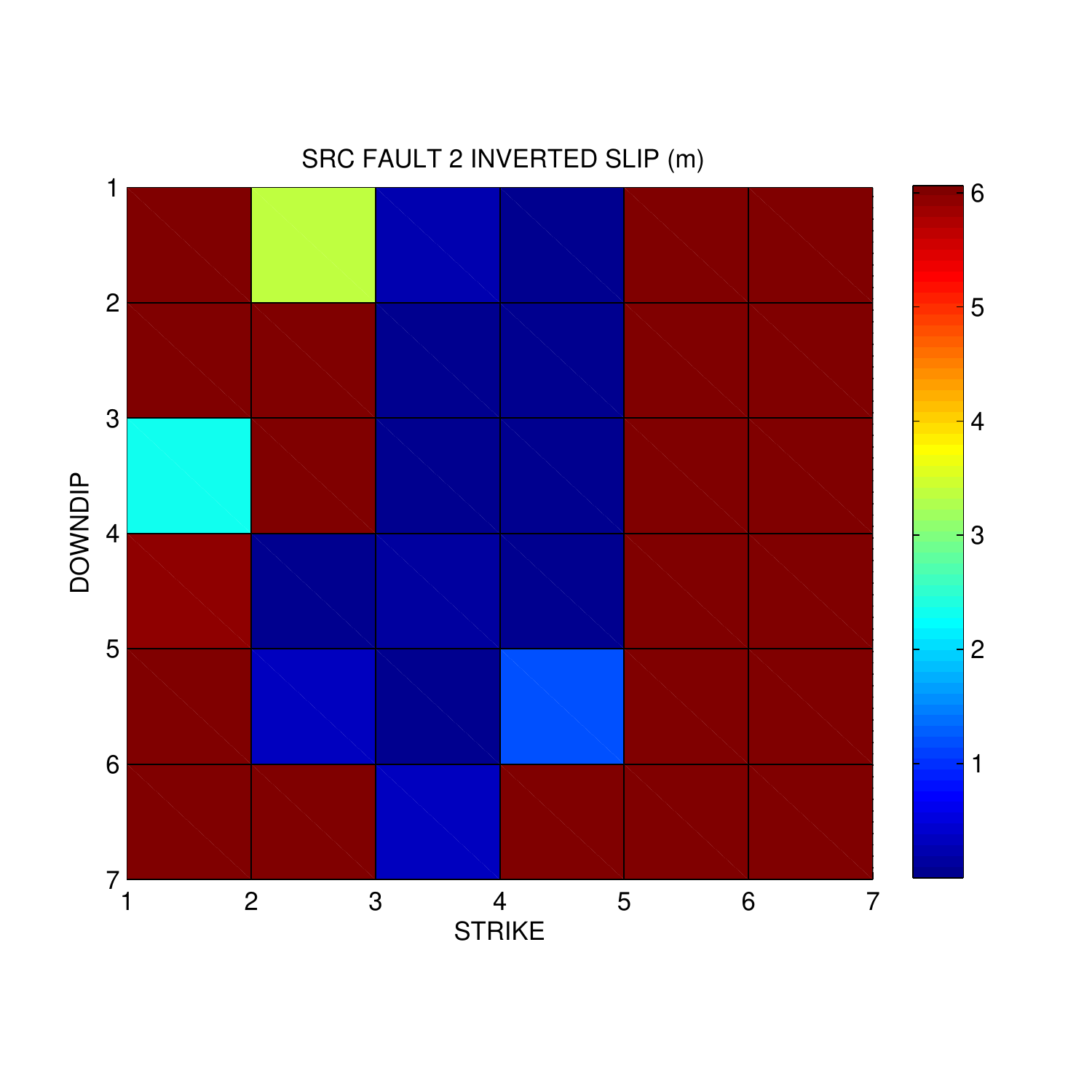} &
\includegraphics[width=.33\textwidth]{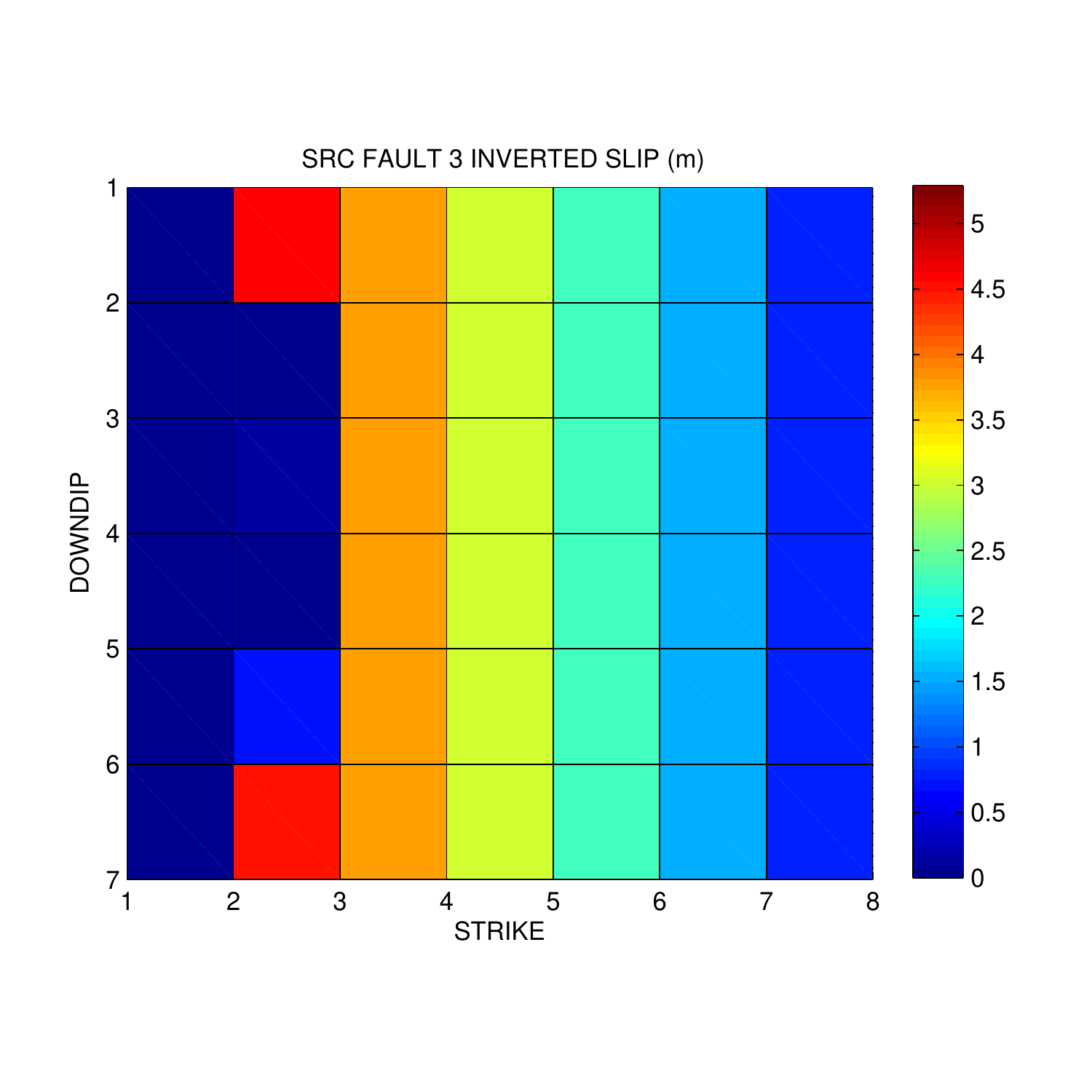}\\ 
\end{tabular}
\end{center}
\vspace{-10 mm}
\caption{Inverted Landers slips from the local dataset, \emph{all} constraints, faults 1-3, $\mu=0.7$, homogeneous vs layered Earth.}
\label{fig:invertedlocalall}
\end{figure}

\begin{figure}[htbp]
\begin{center}
\begin{tabular}{ccc}
\includegraphics[width=.33\textwidth]{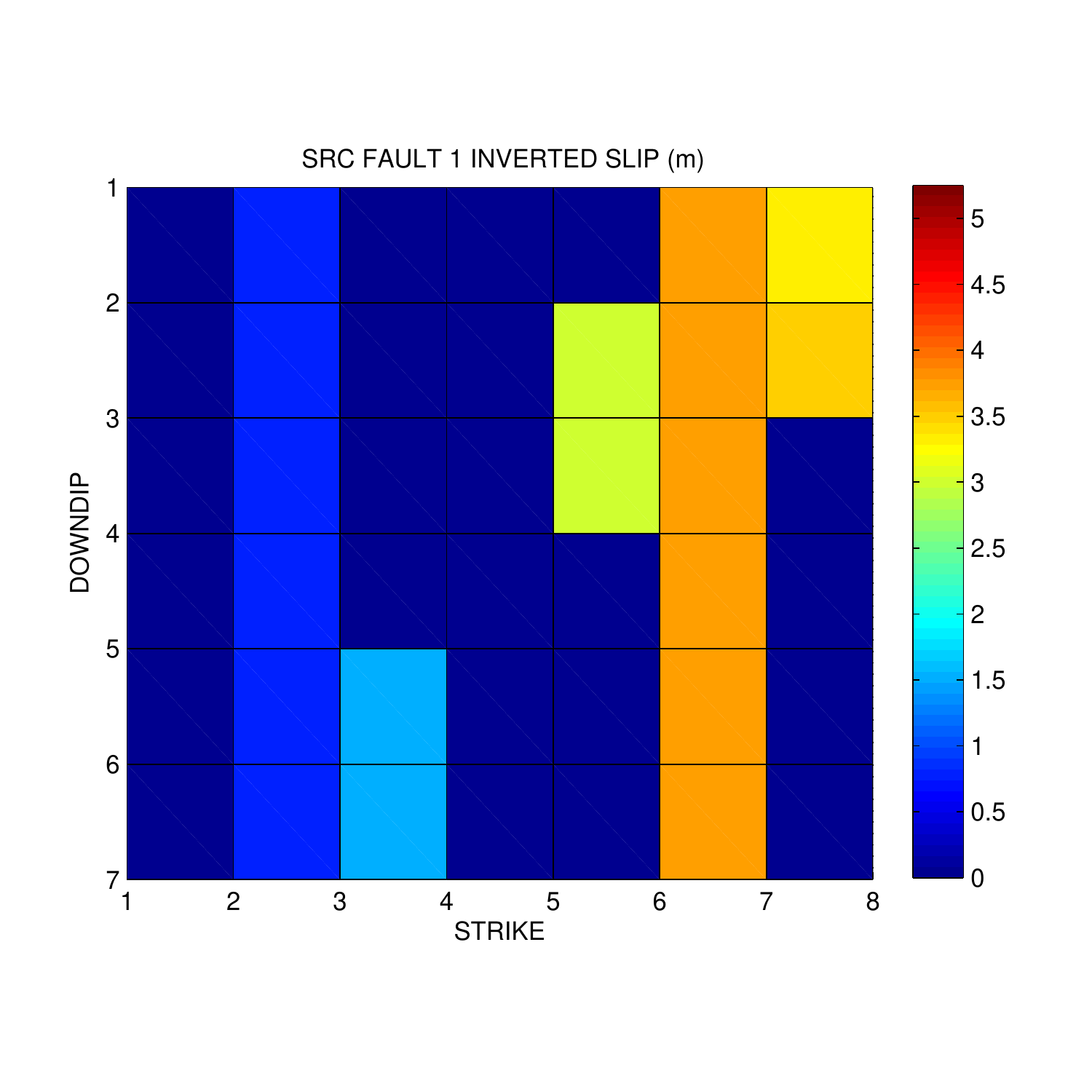} &
\includegraphics[width=.33\textwidth]{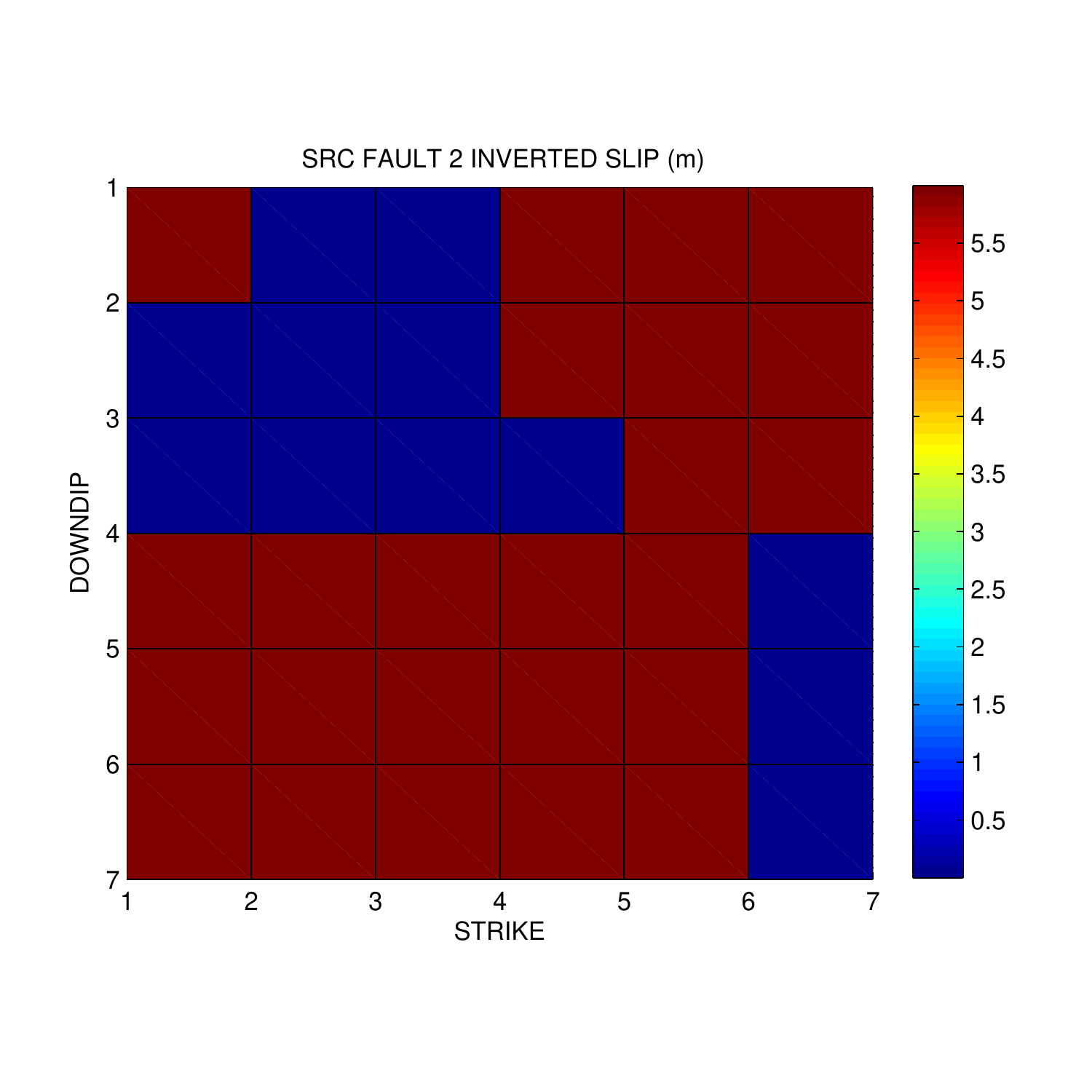} &
\includegraphics[width=.33\textwidth]{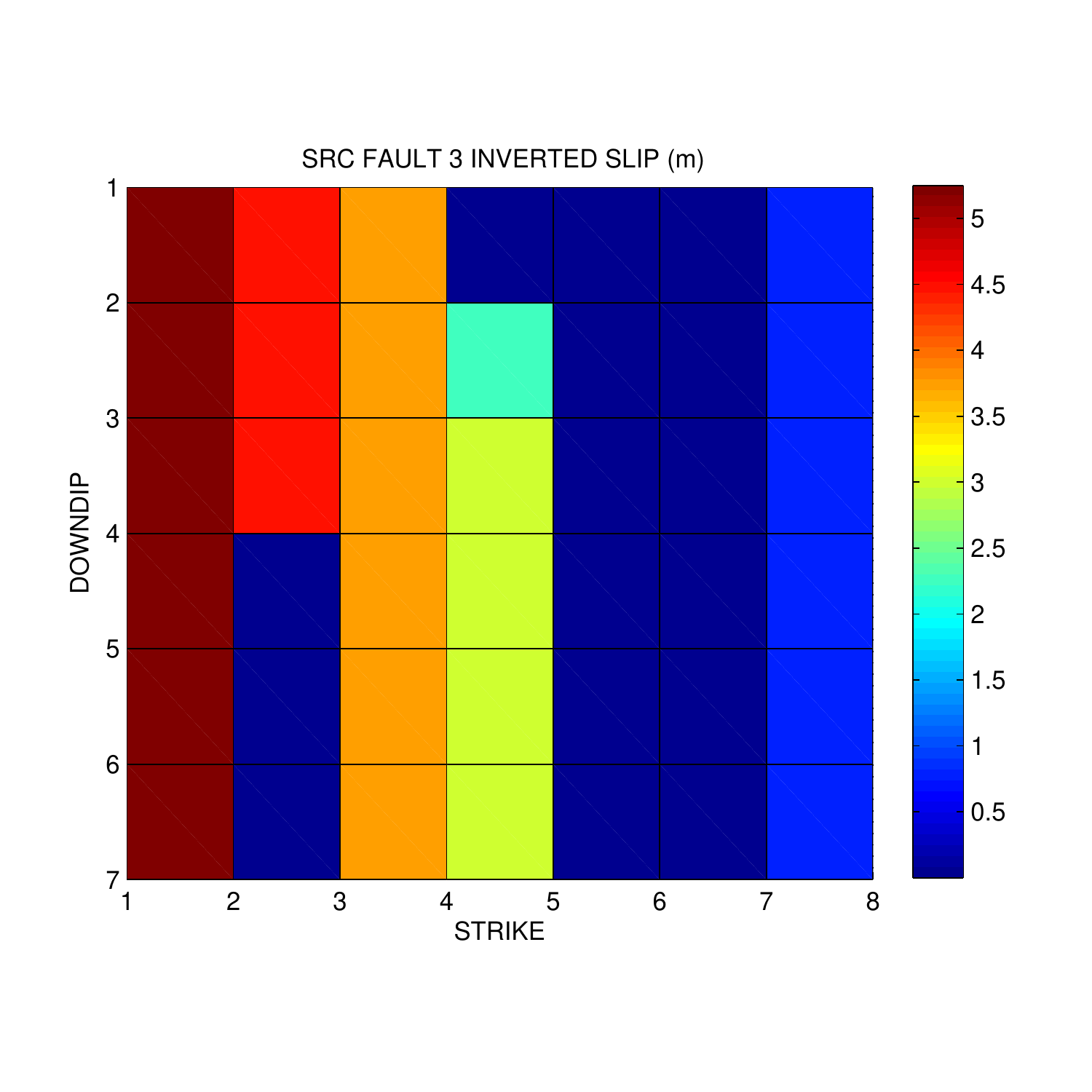}\\ 
\includegraphics[width=.33\textwidth]{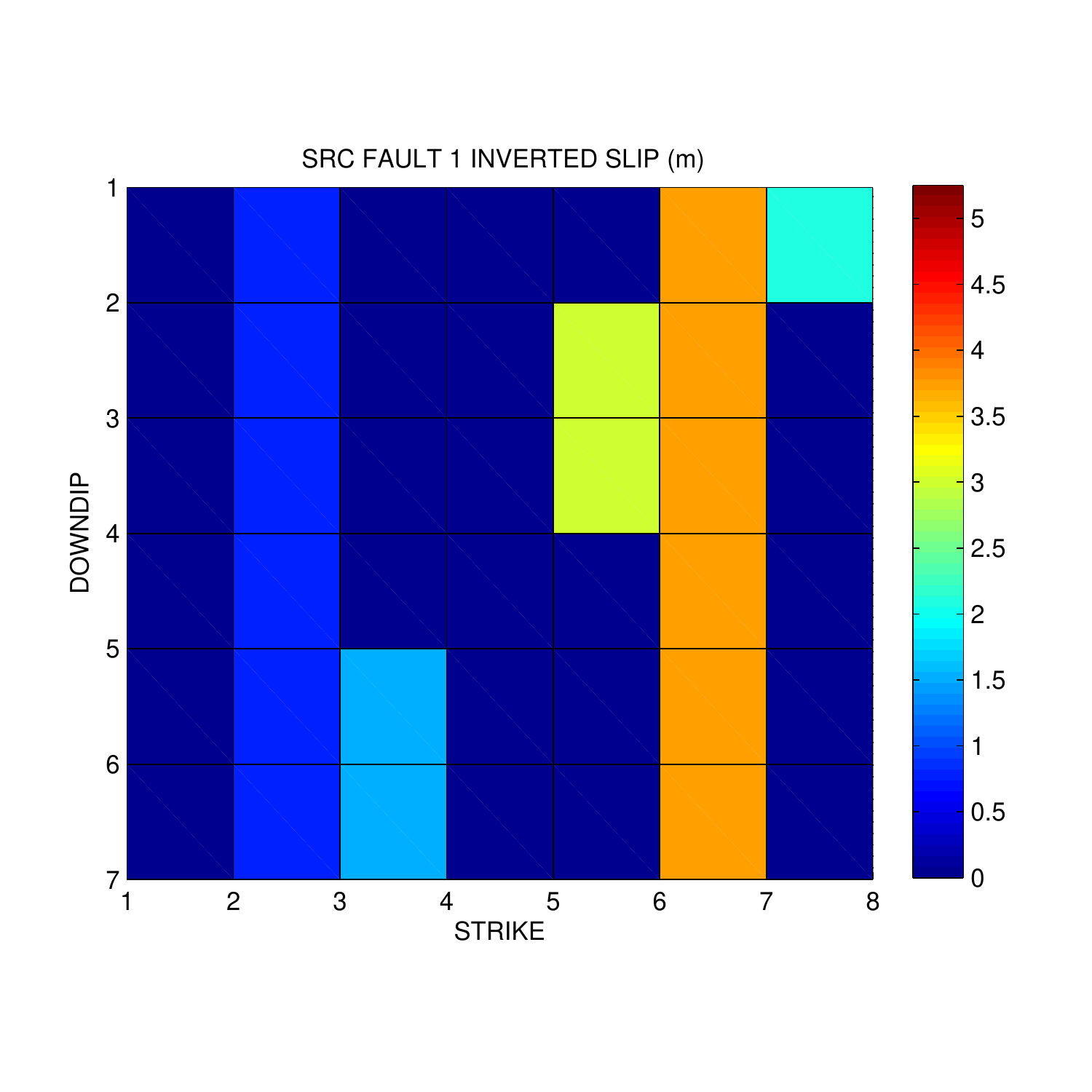} &
\includegraphics[width=.33\textwidth]{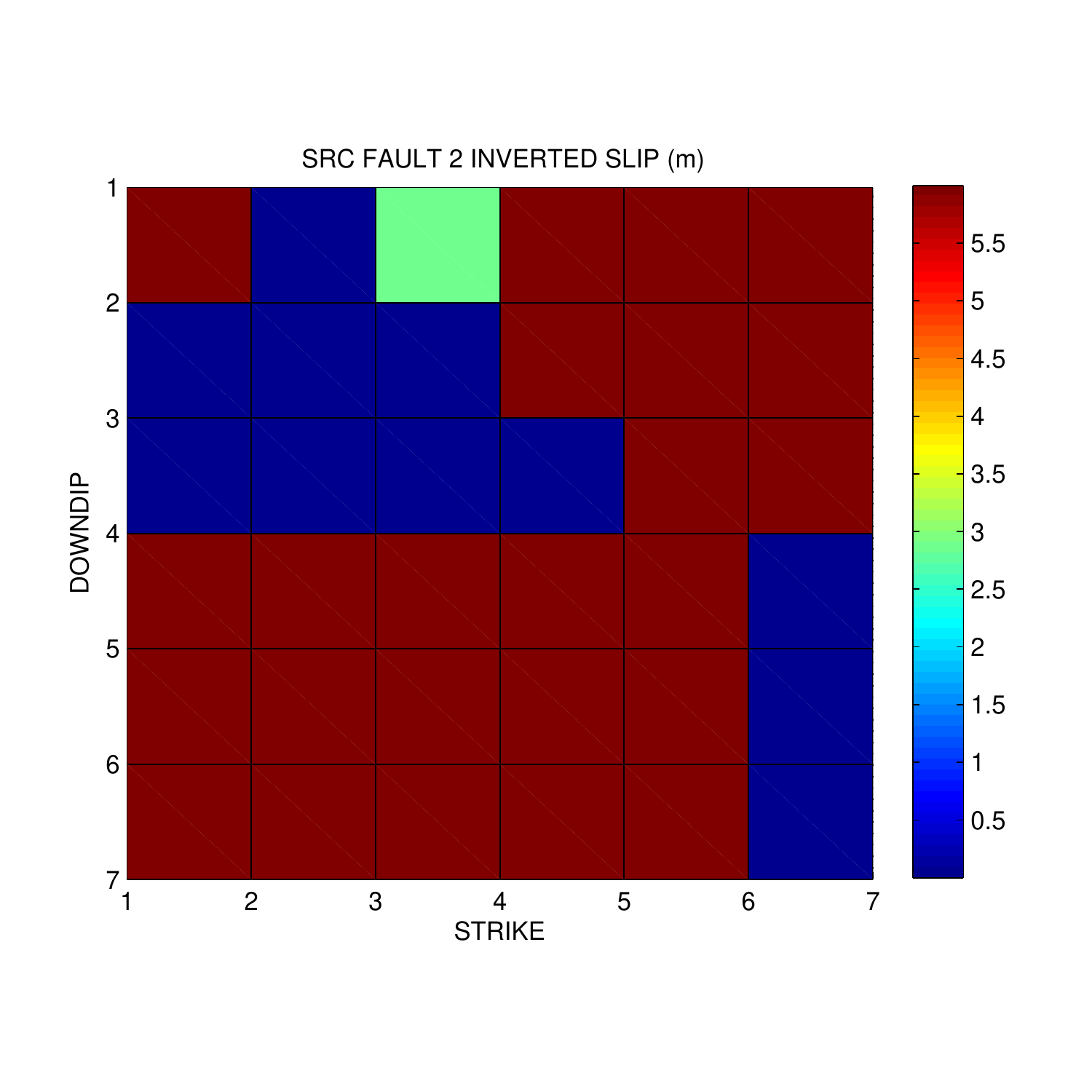} &
\includegraphics[width=.33\textwidth]{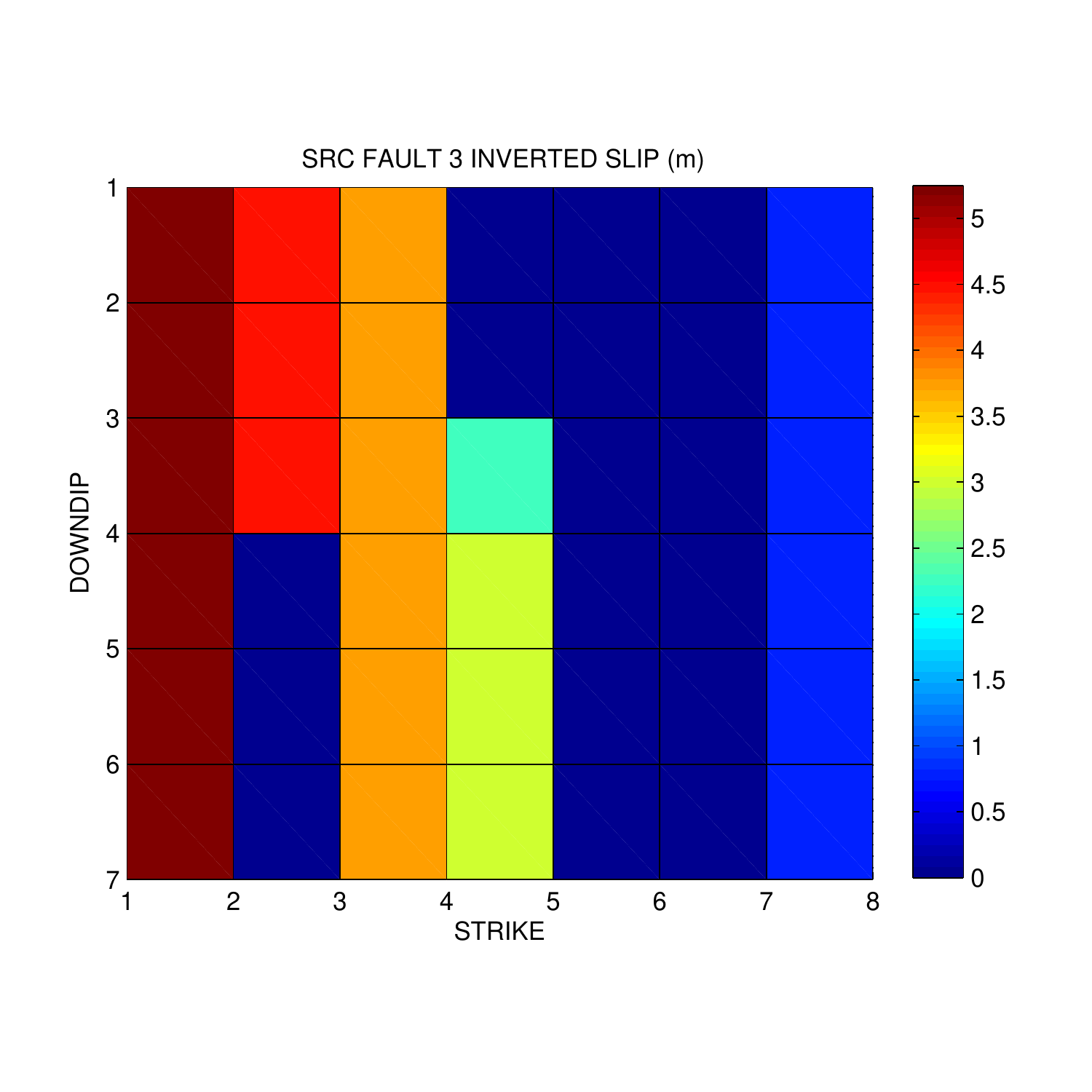}\\ 
\end{tabular}
\end{center}
\vspace{-10 mm}
\caption{Inverted Landers slips from the decimated (20 faults) regional dataset, only \emph{consistent} constraints, faults 1-3, $\mu=0.7$, homogeneous vs layered Earth.}
\label{fig:inverted}
\end{figure}

\begin{figure}[htbp]
\begin{center}
\begin{tabular}{ccc}
\includegraphics[width=.33\textwidth]{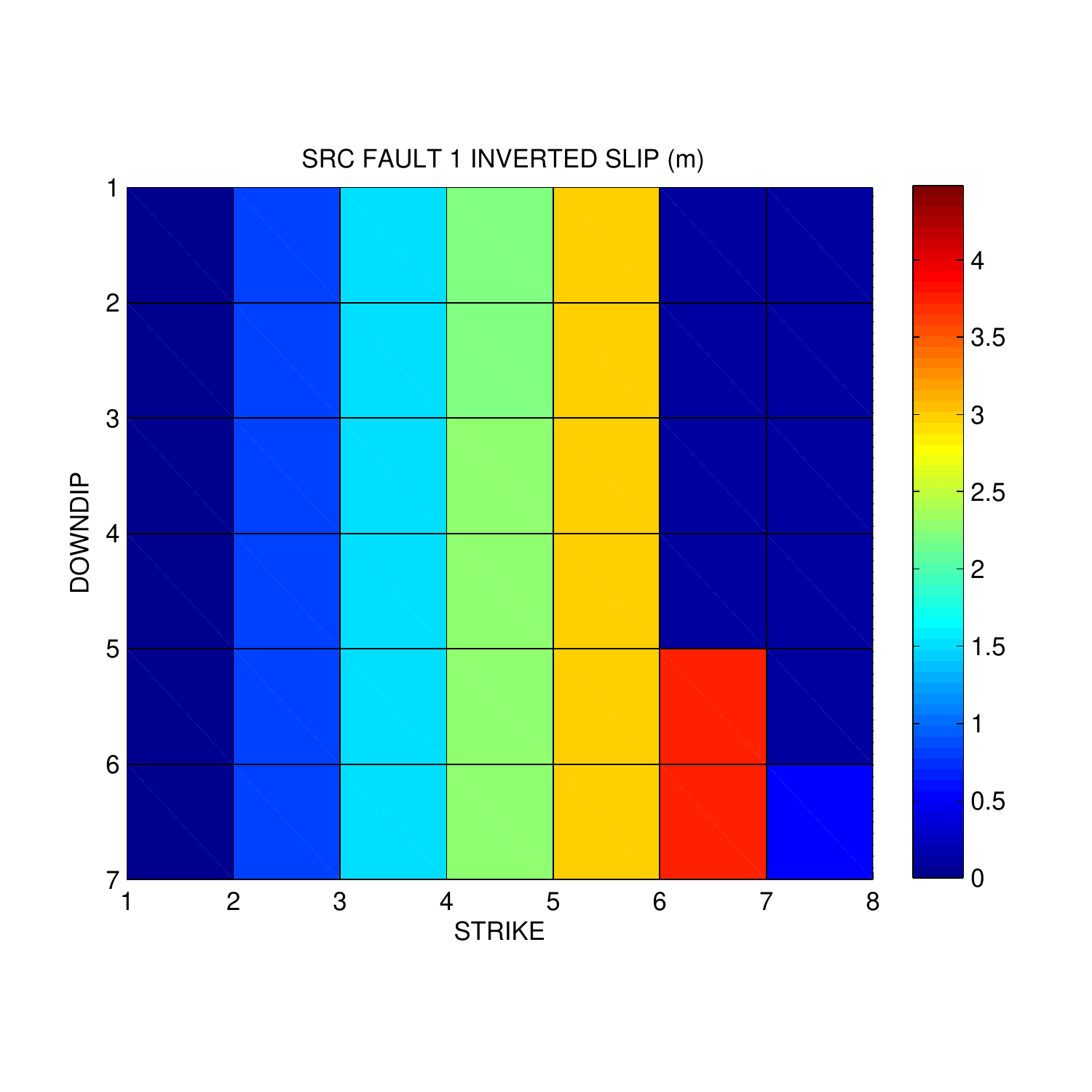} &
\includegraphics[width=.33\textwidth]{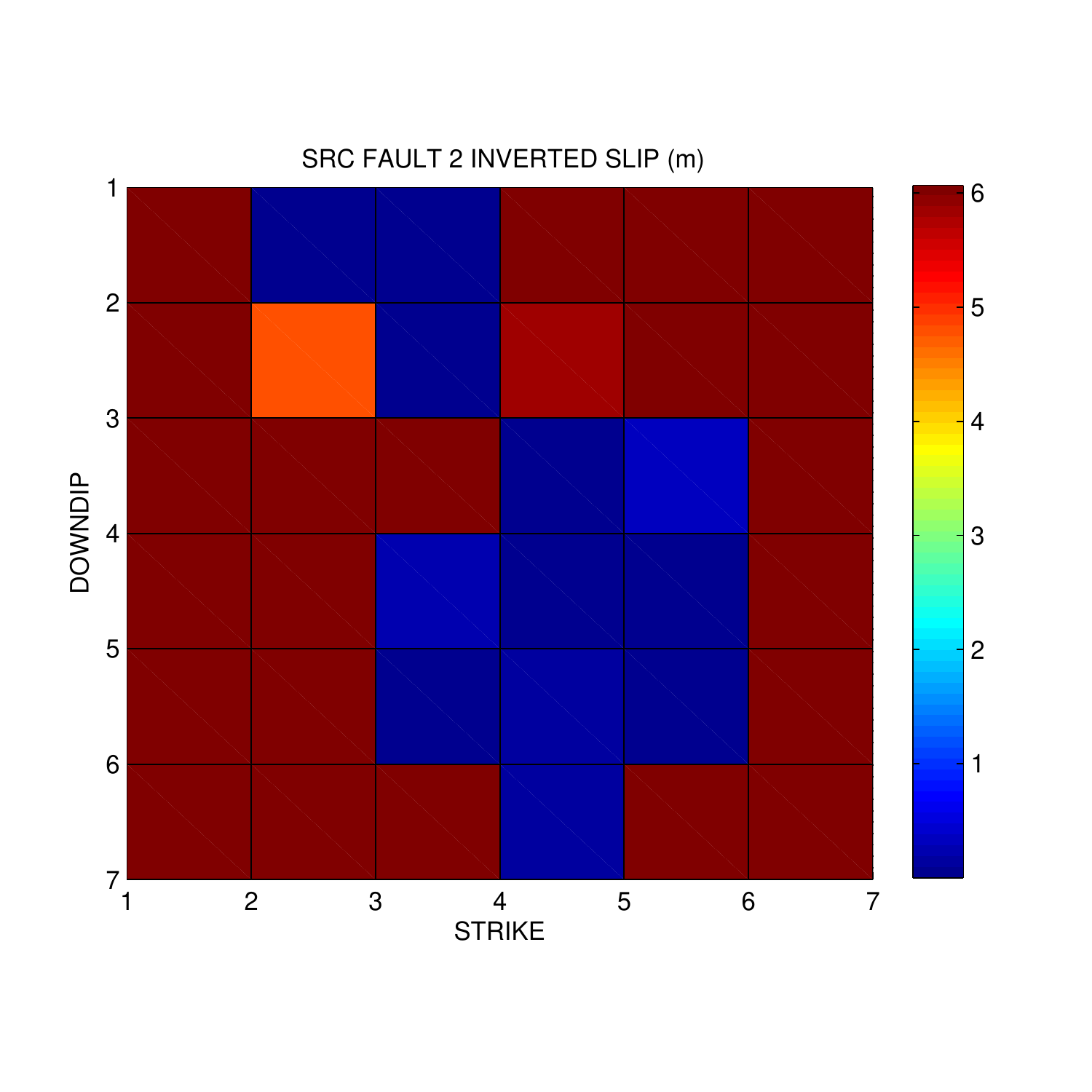} &
\includegraphics[width=.33\textwidth]{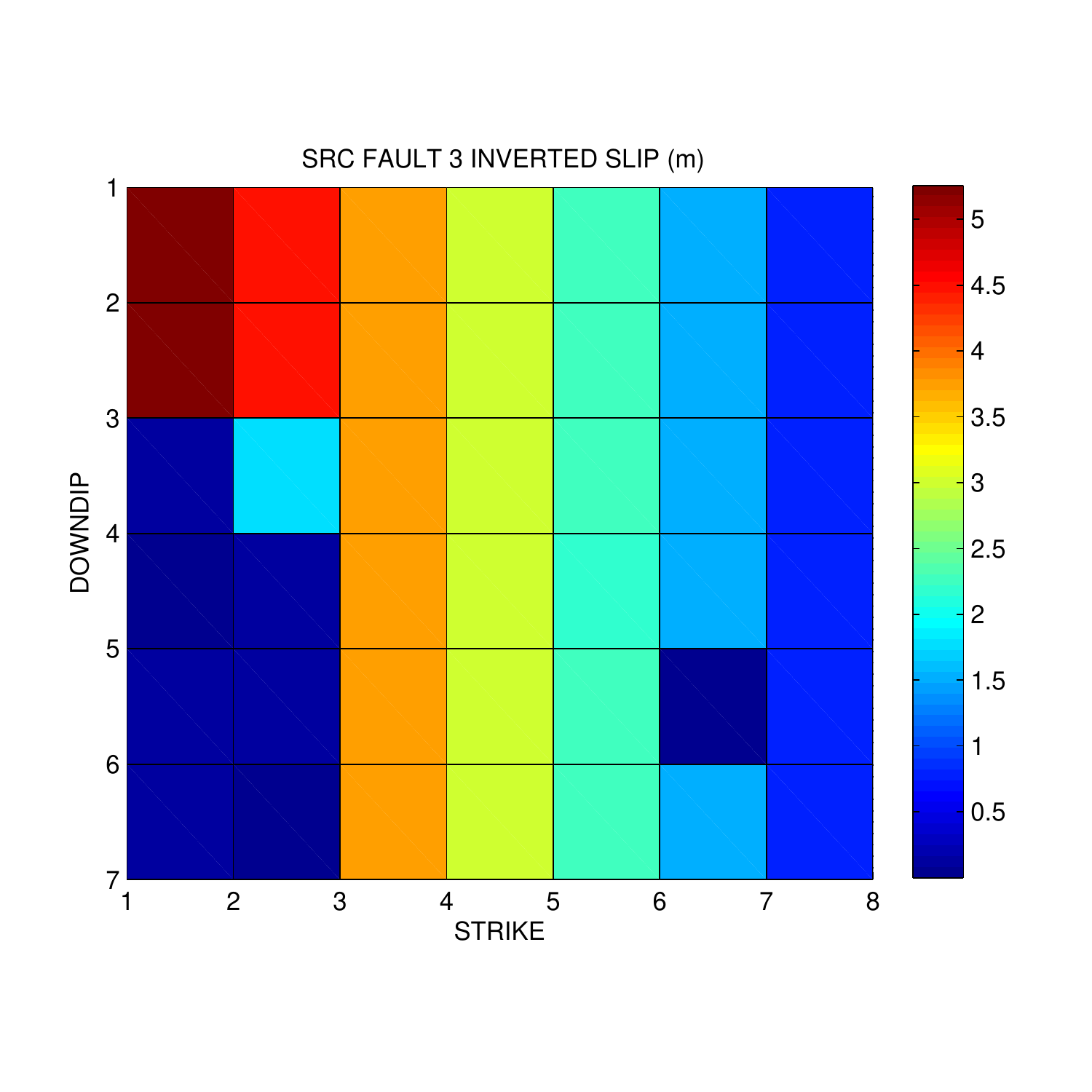}\\ 
\includegraphics[width=.33\textwidth]{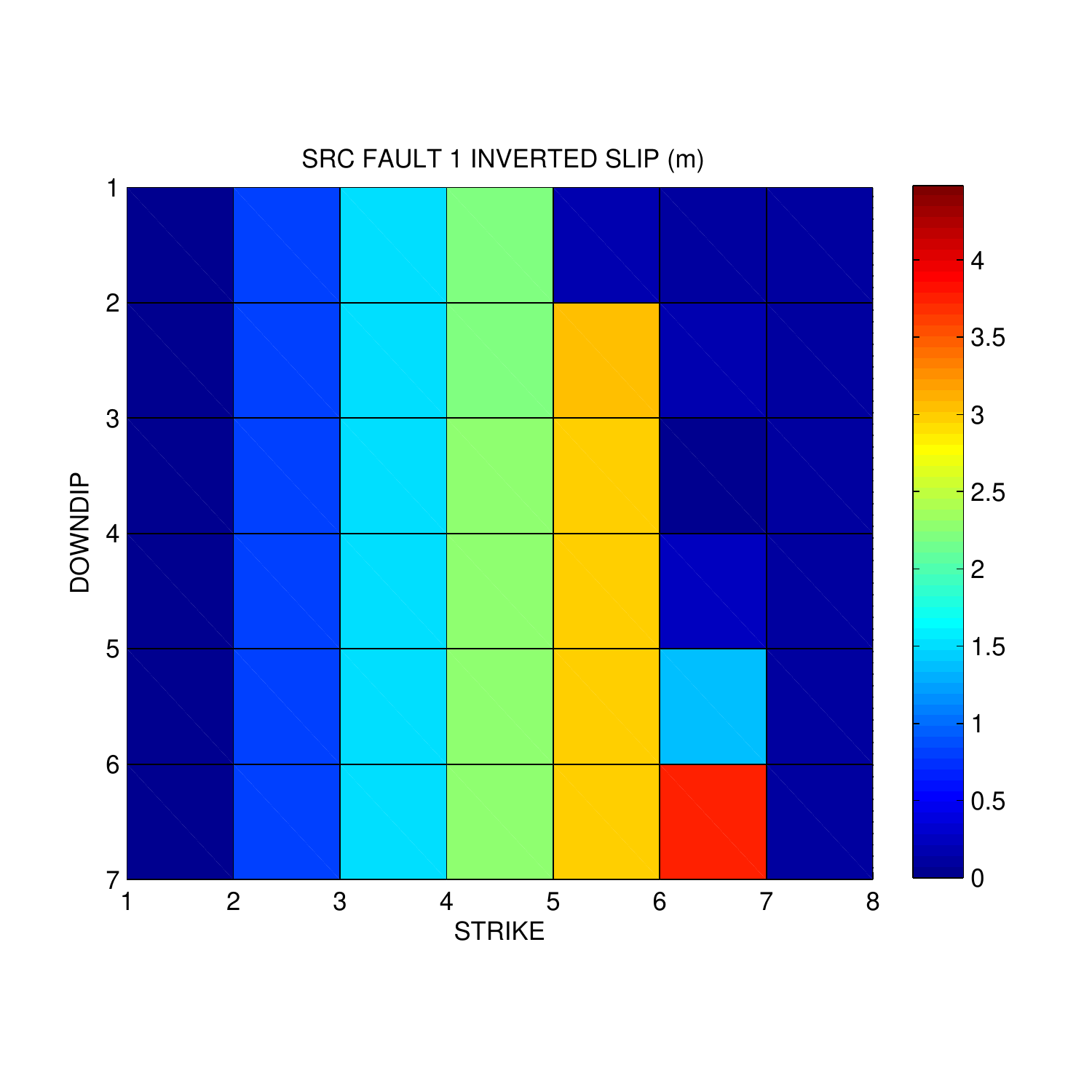} &
\includegraphics[width=.33\textwidth]{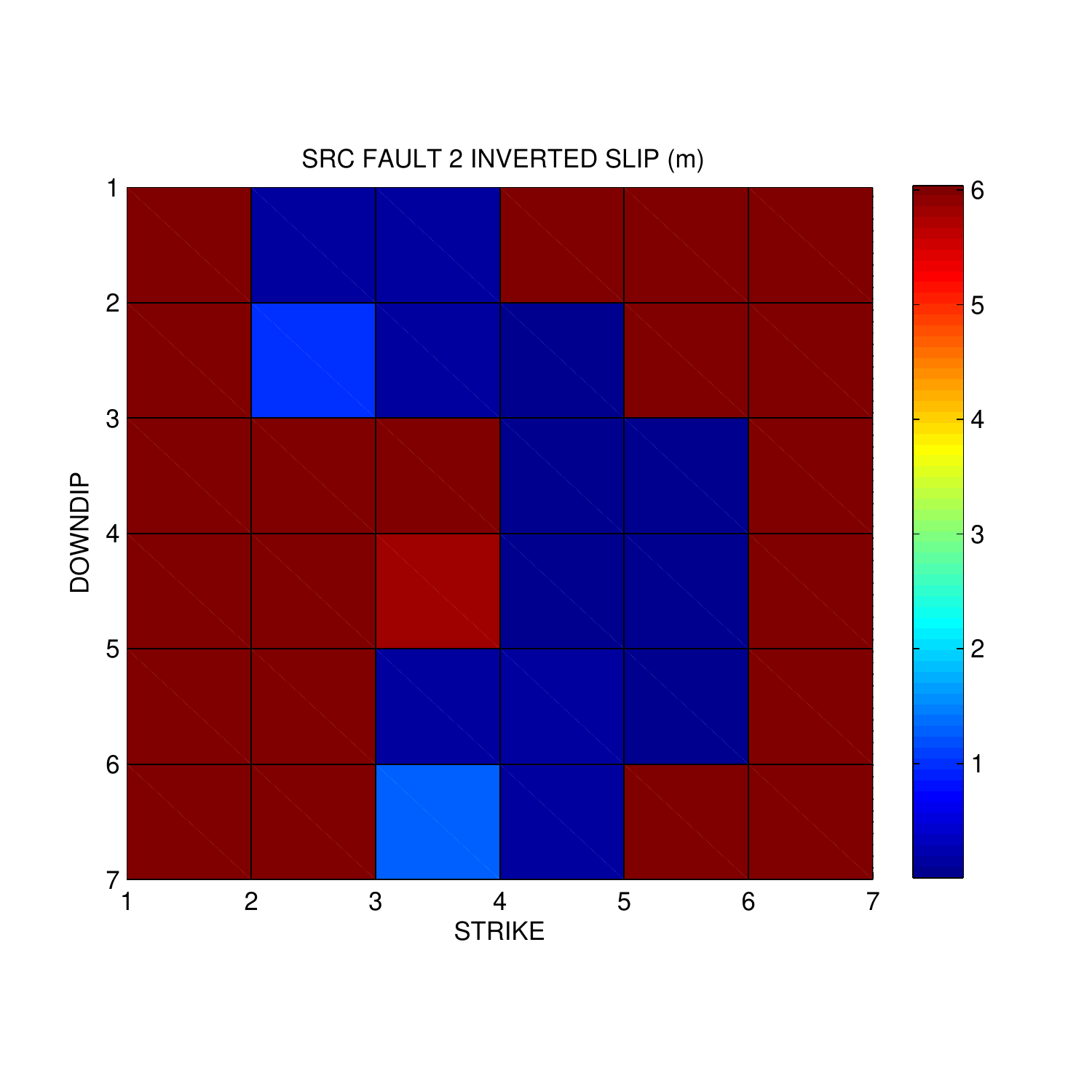} &
\includegraphics[width=.33\textwidth]{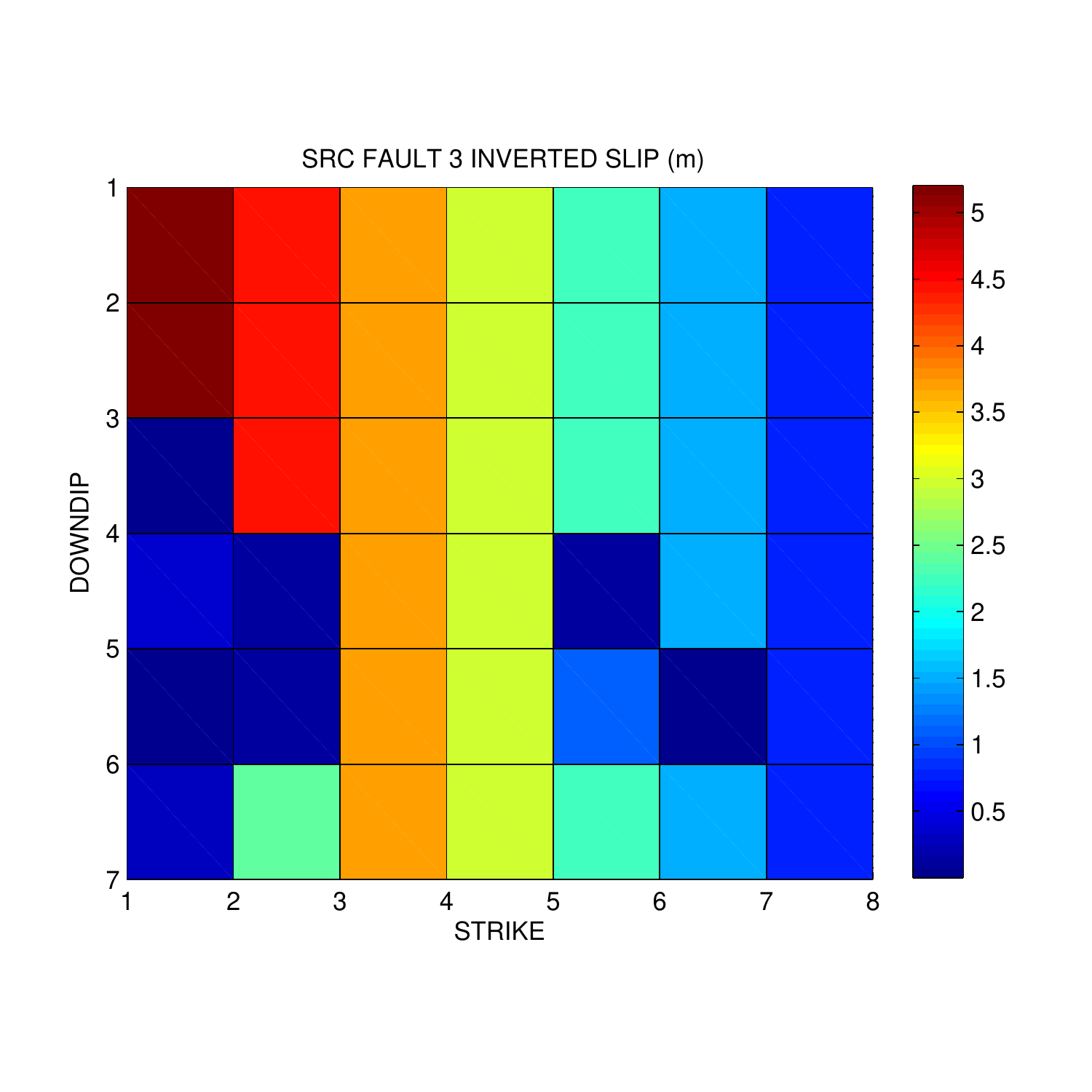}\\ 
\end{tabular}
\end{center}
\vspace{-10 mm}
\caption{Inverted Landers slips from the decimated (20 faults) regional dataset, \emph{all} constraints, faults 1-3, $\mu=0.7$, homogeneous vs layered Earth.}
\label{fig:invertedall}
\end{figure}

\begin{figure}[htbp]
\begin{center}
\begin{tabular}{ccc}
\includegraphics[width=.33\textwidth]{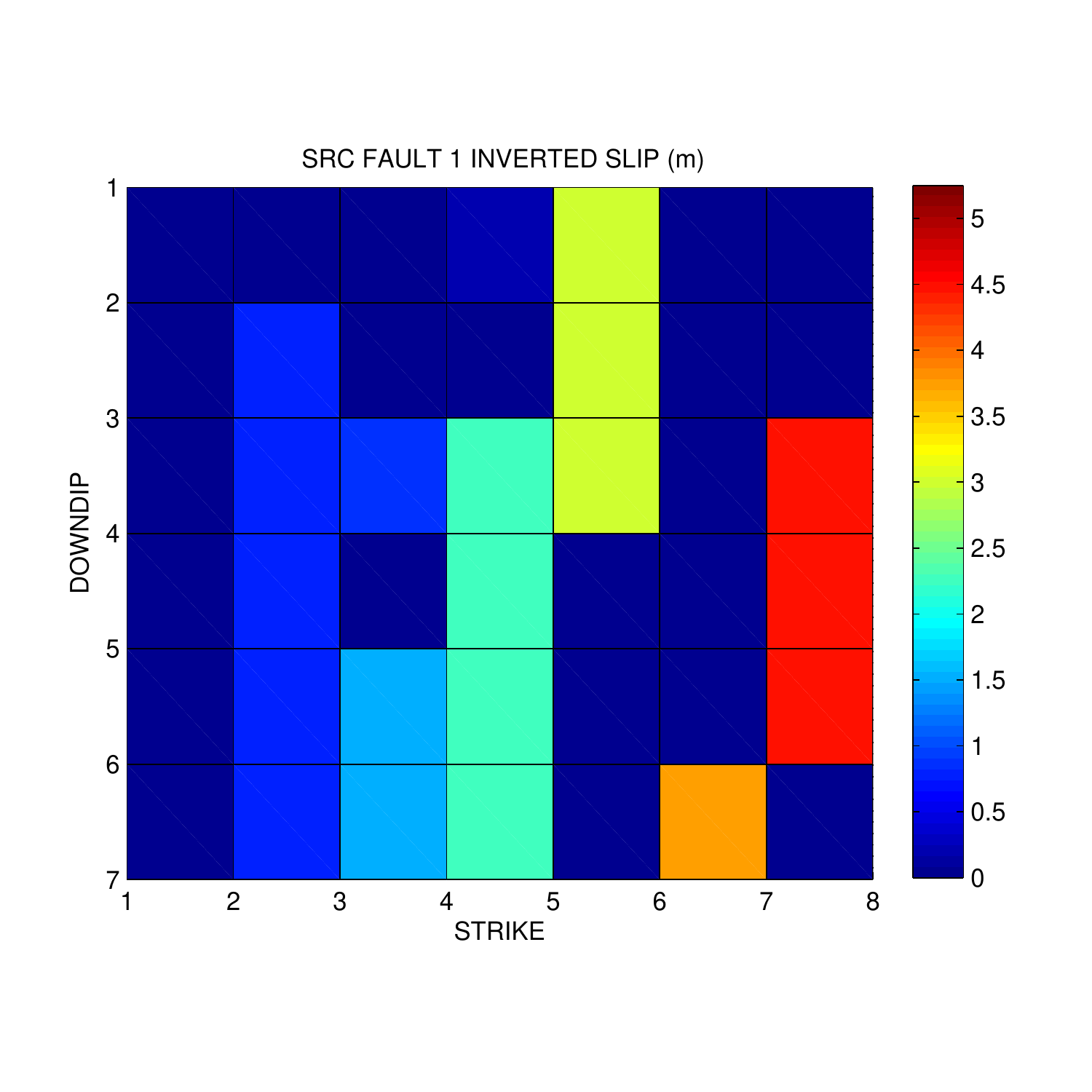} &
\includegraphics[width=.33\textwidth]{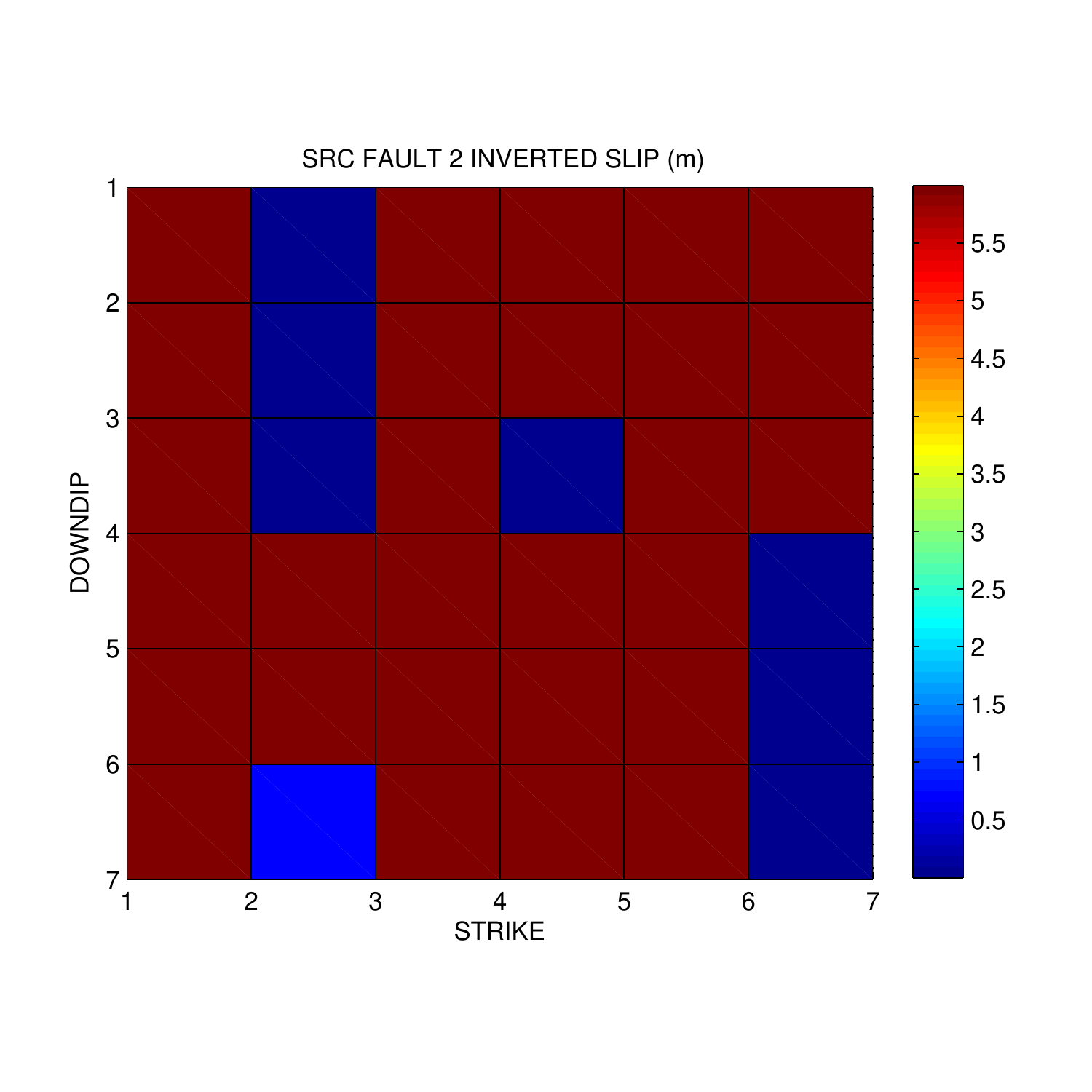} &
\includegraphics[width=.33\textwidth]{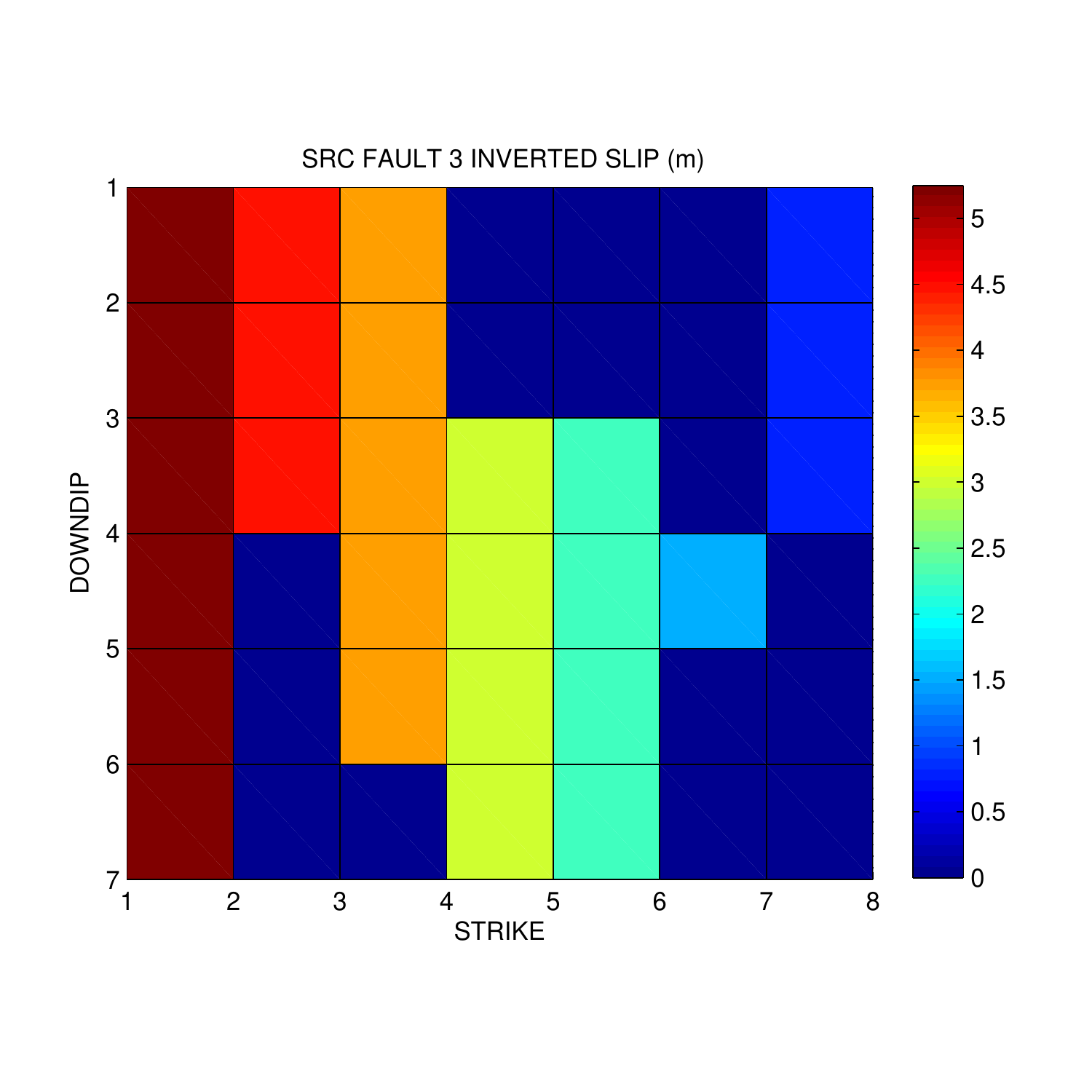}\\ 
\includegraphics[width=.33\textwidth]{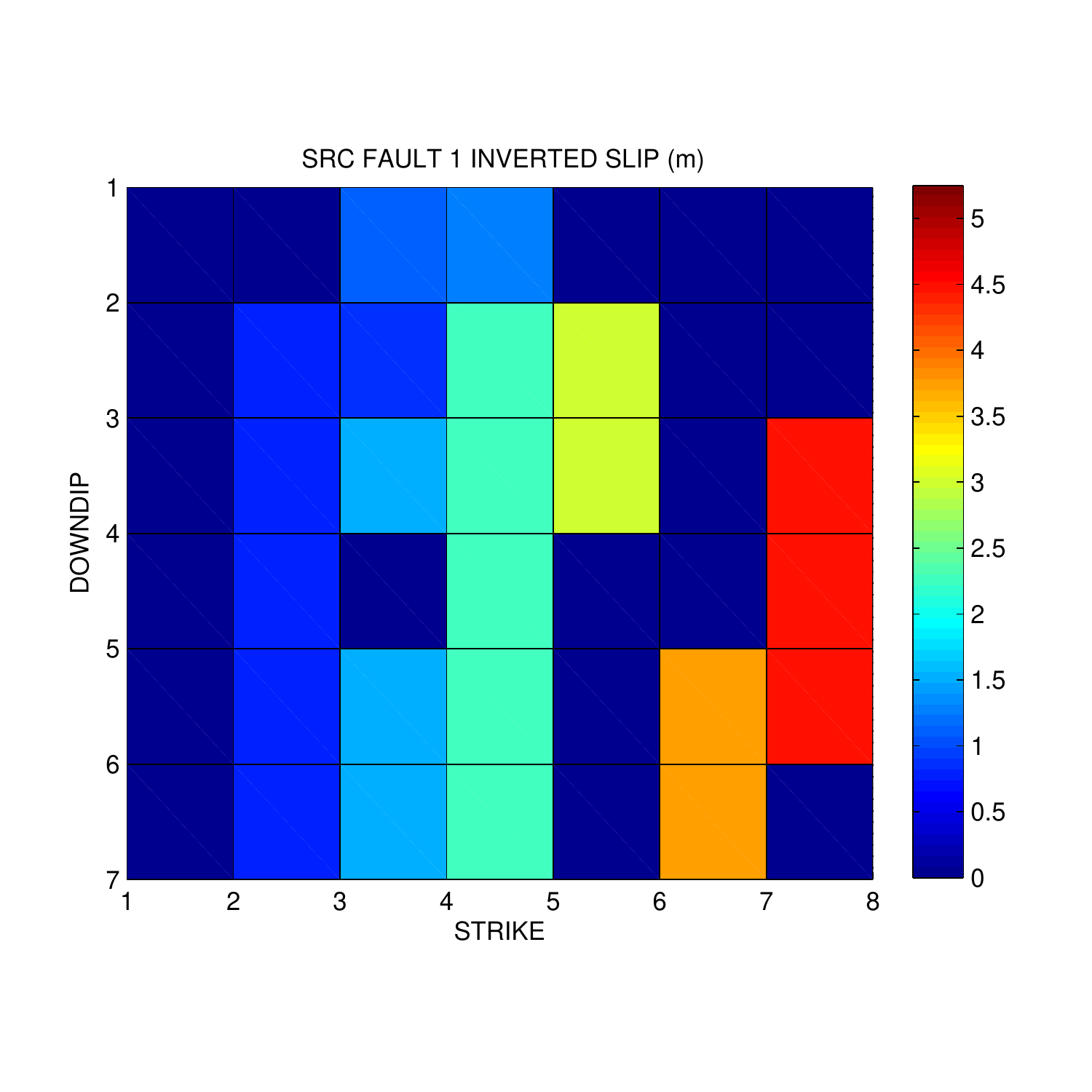} &
\includegraphics[width=.33\textwidth]{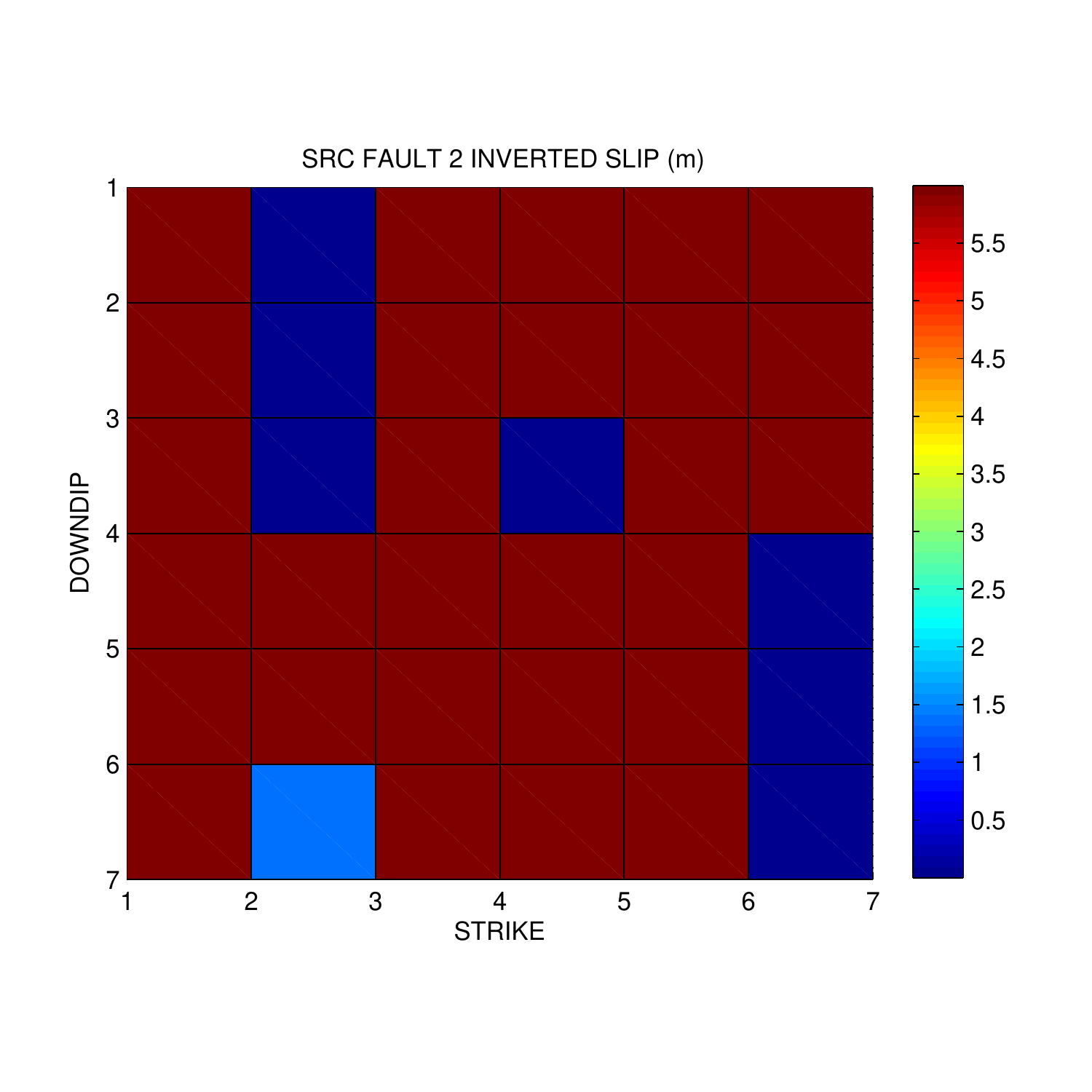} &
\includegraphics[width=.33\textwidth]{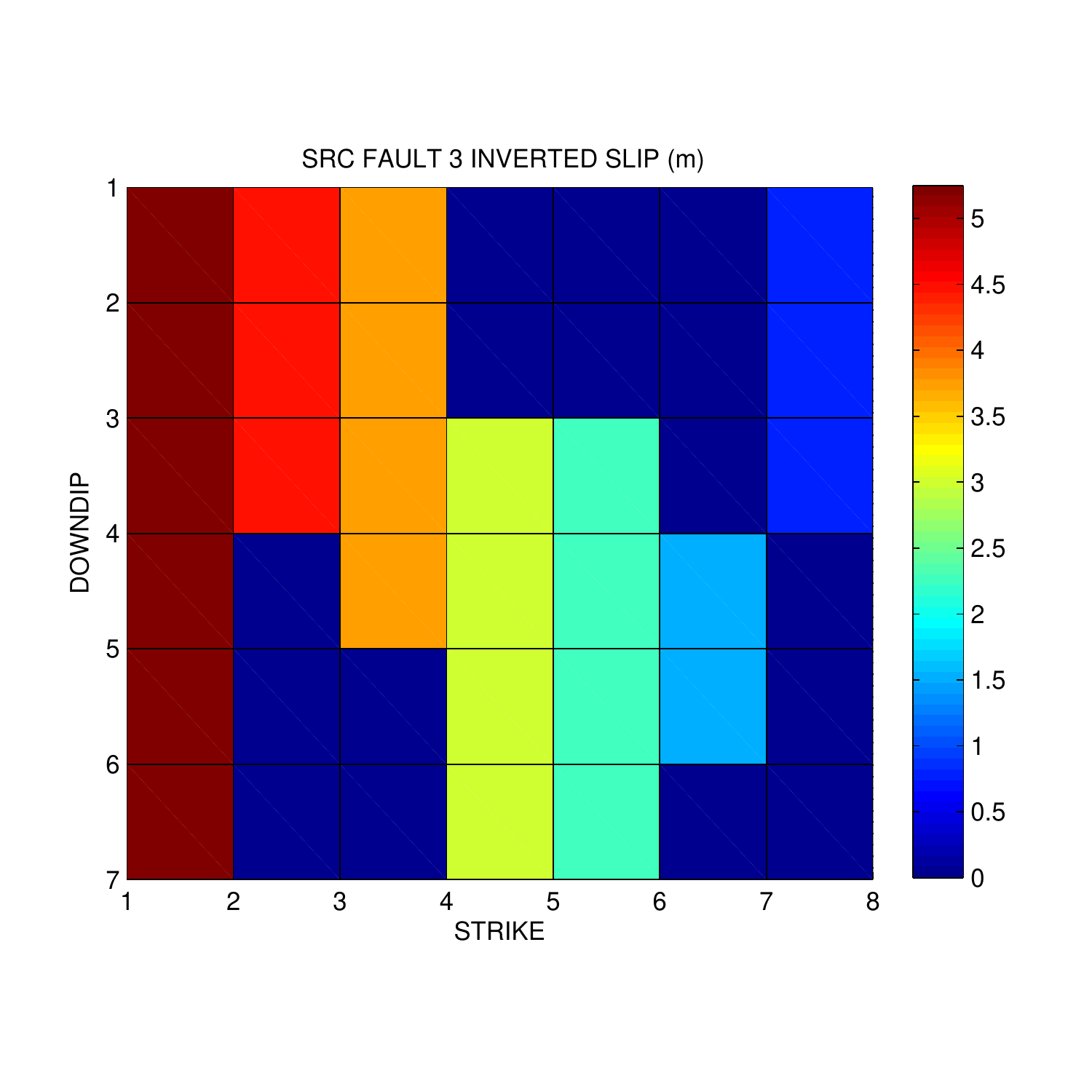}\\ 
\end{tabular}
\end{center}
\vspace{-10 mm}
\caption{Inverted Landers slips using the regional dataset, only \emph{consistent} constraints, faults 1-3, $\mu=0.7$, homogeneous vs layered Earth.}
\label{fig:invertedspread}
\end{figure}

\begin{figure}[htbp]
\begin{center}
\begin{tabular}{ccc}
\includegraphics[width=.33\textwidth]{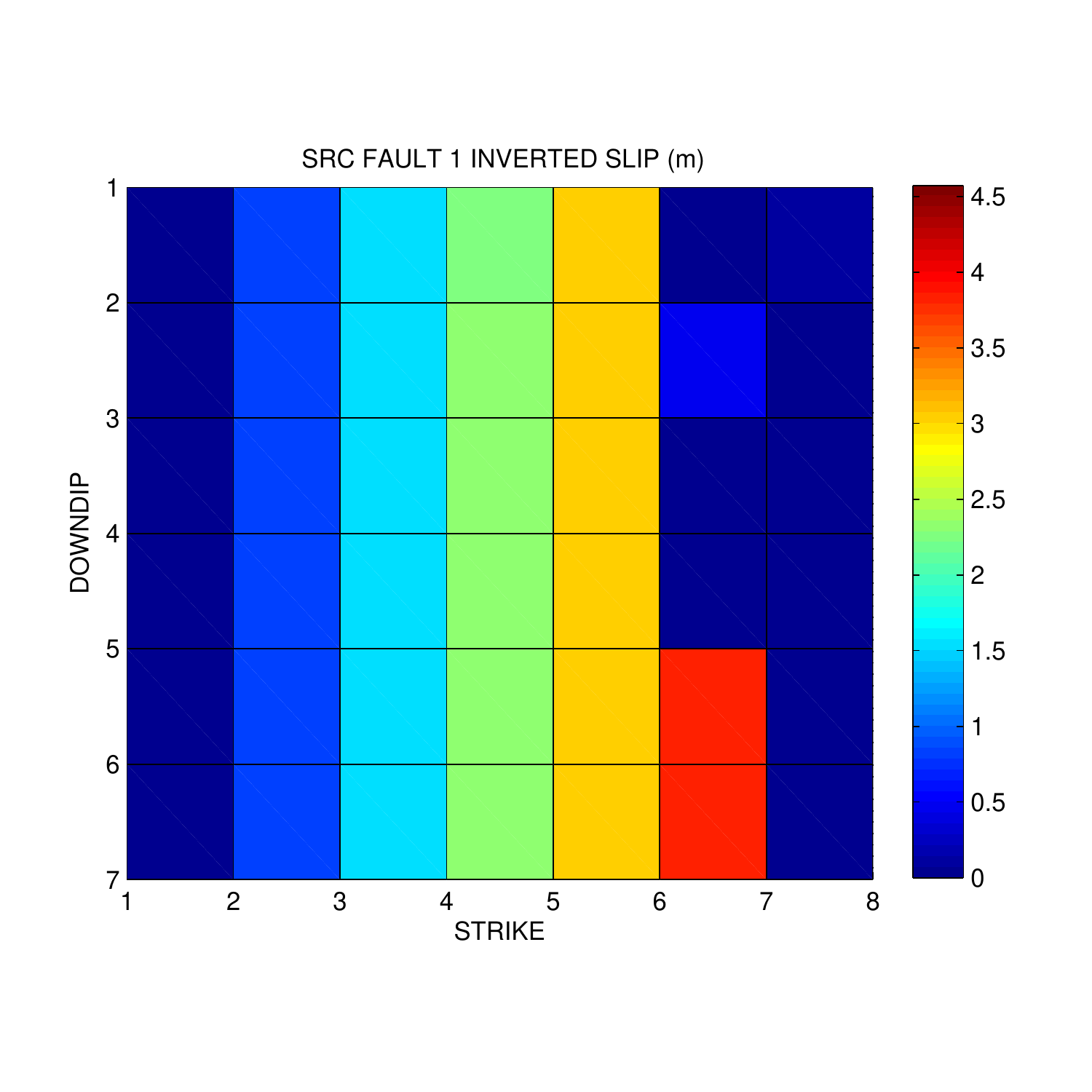} &
\includegraphics[width=.33\textwidth]{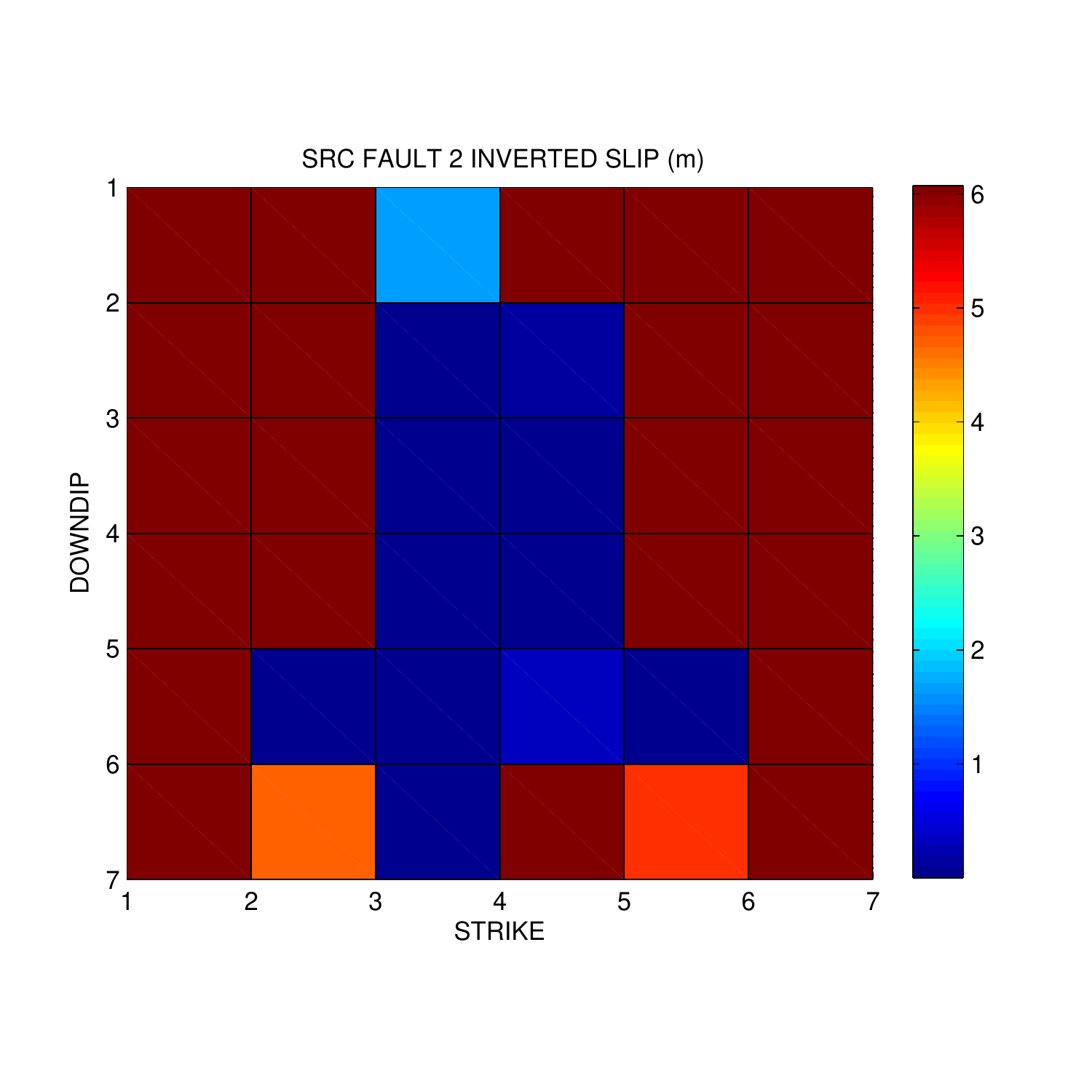} &
\includegraphics[width=.33\textwidth]{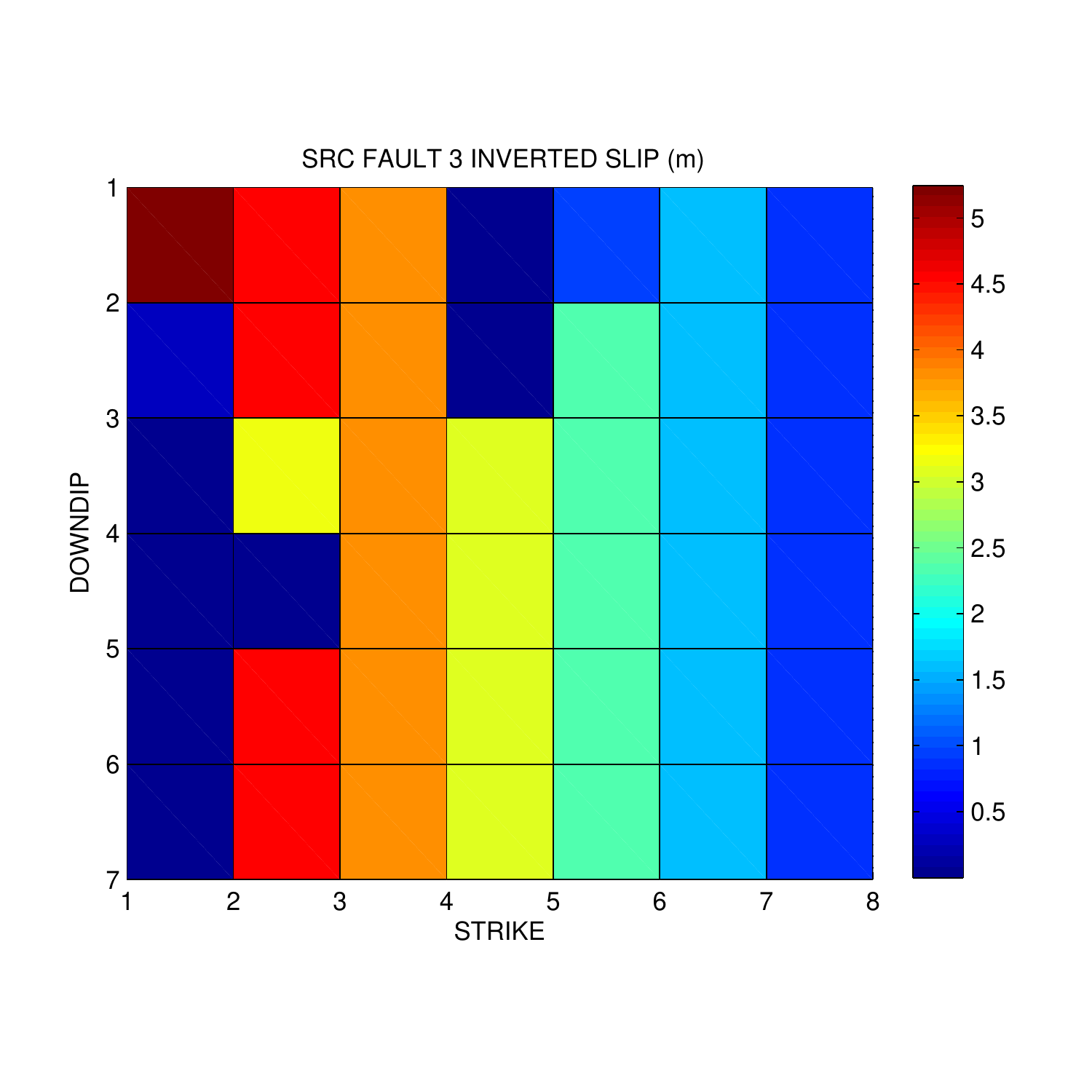}\\ 
\includegraphics[width=.33\textwidth]{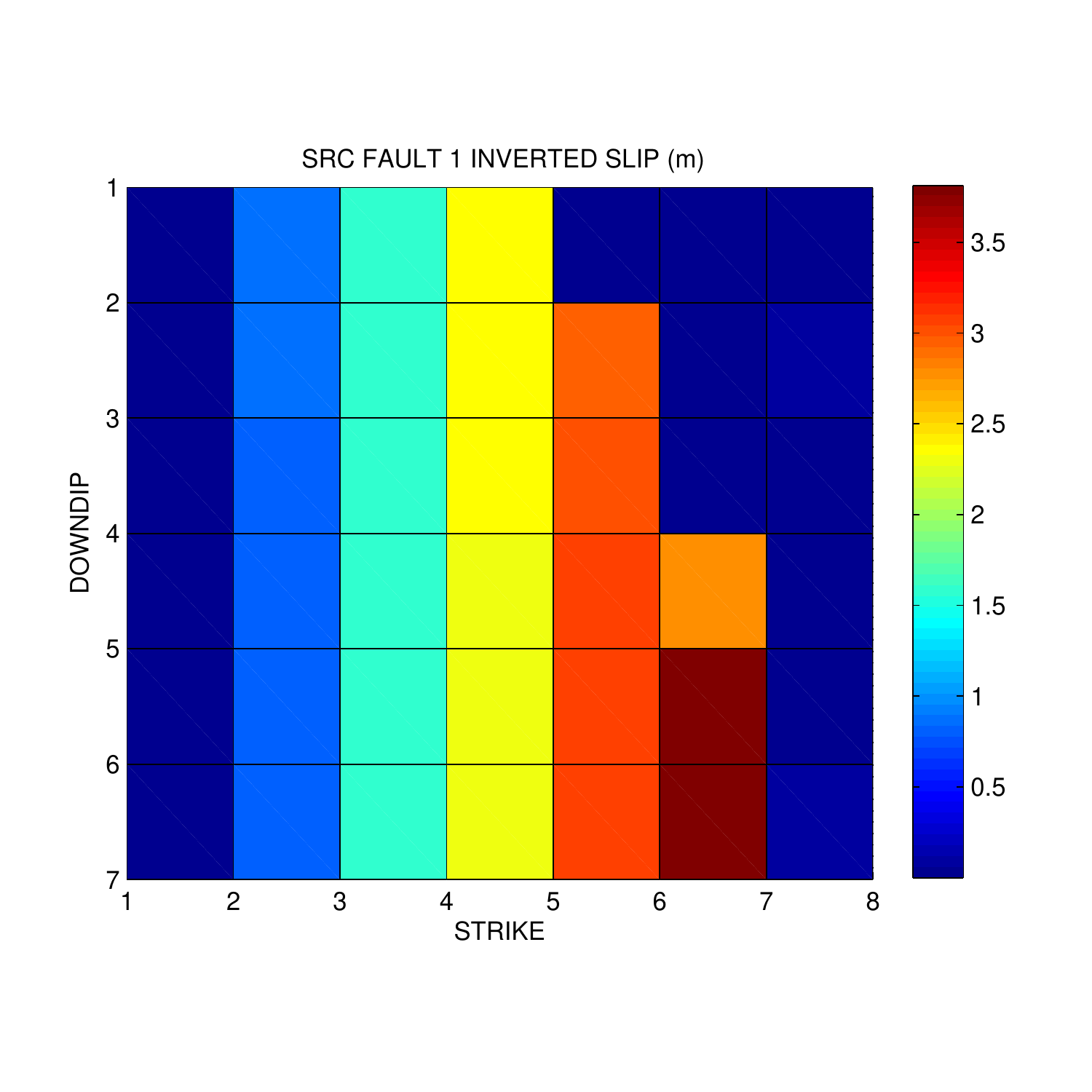} &
\includegraphics[width=.33\textwidth]{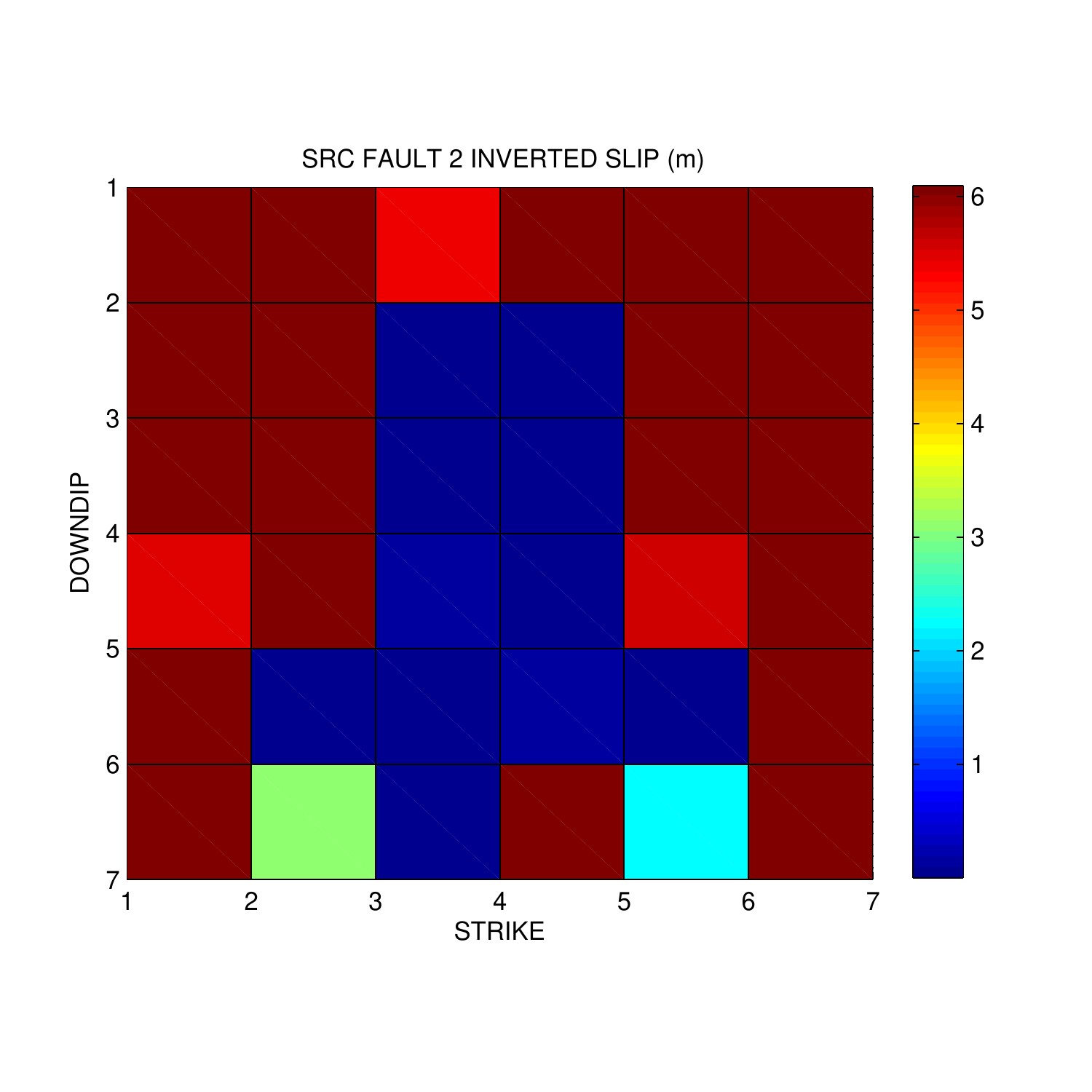} &
\includegraphics[width=.33\textwidth]{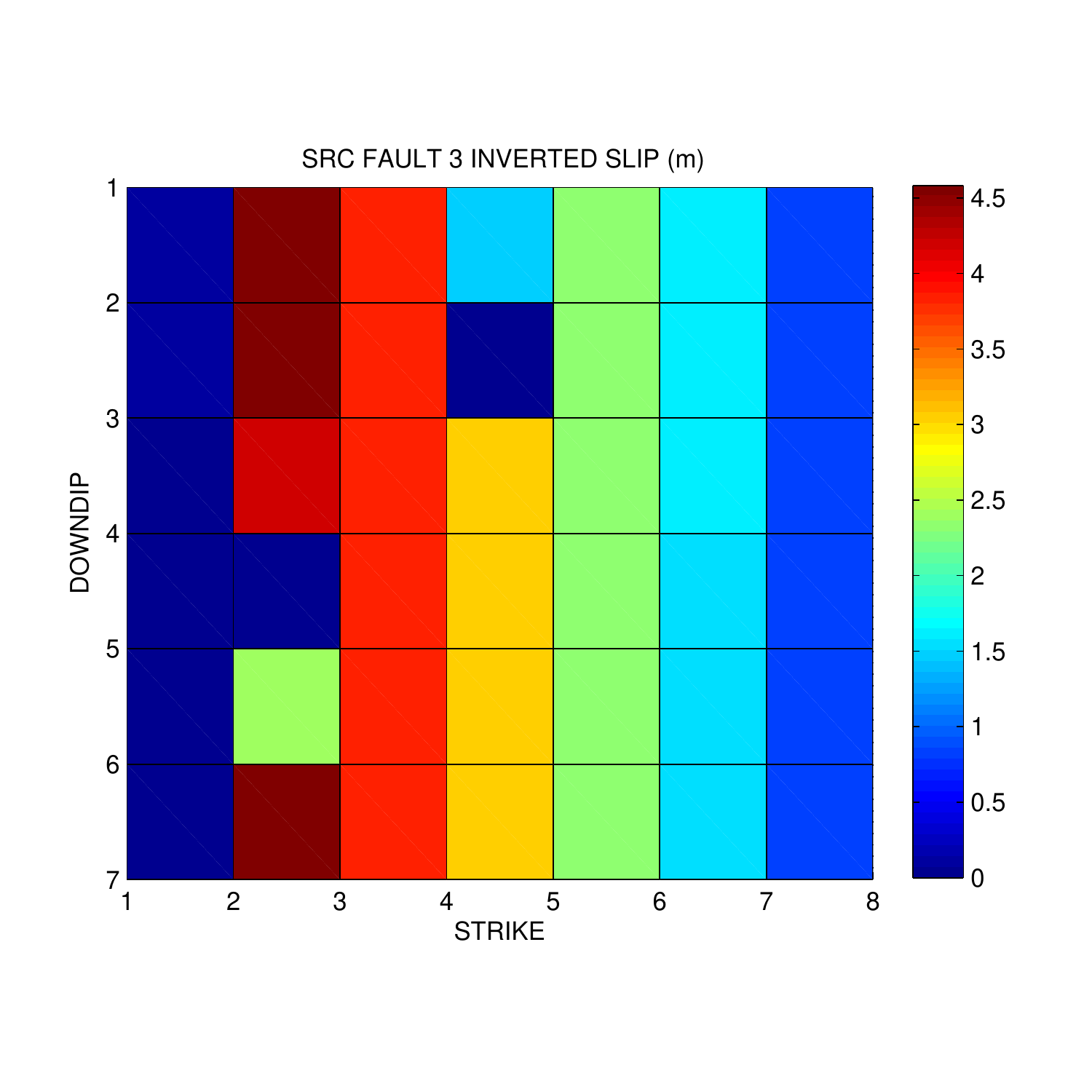}\\ 
\end{tabular}
\end{center}
\vspace{-10 mm}
\caption{Inverted Landers slips using the regional dataset, \emph{all} constraints, faults 1-3, $\mu=0.7$, homogeneous vs layered Earth.}
\label{fig:invertedspreadall}
\end{figure}

\begin{figure}[htbp]
\begin{center}
\begin{tabular}{ccc}
\includegraphics[width=.33\textwidth]{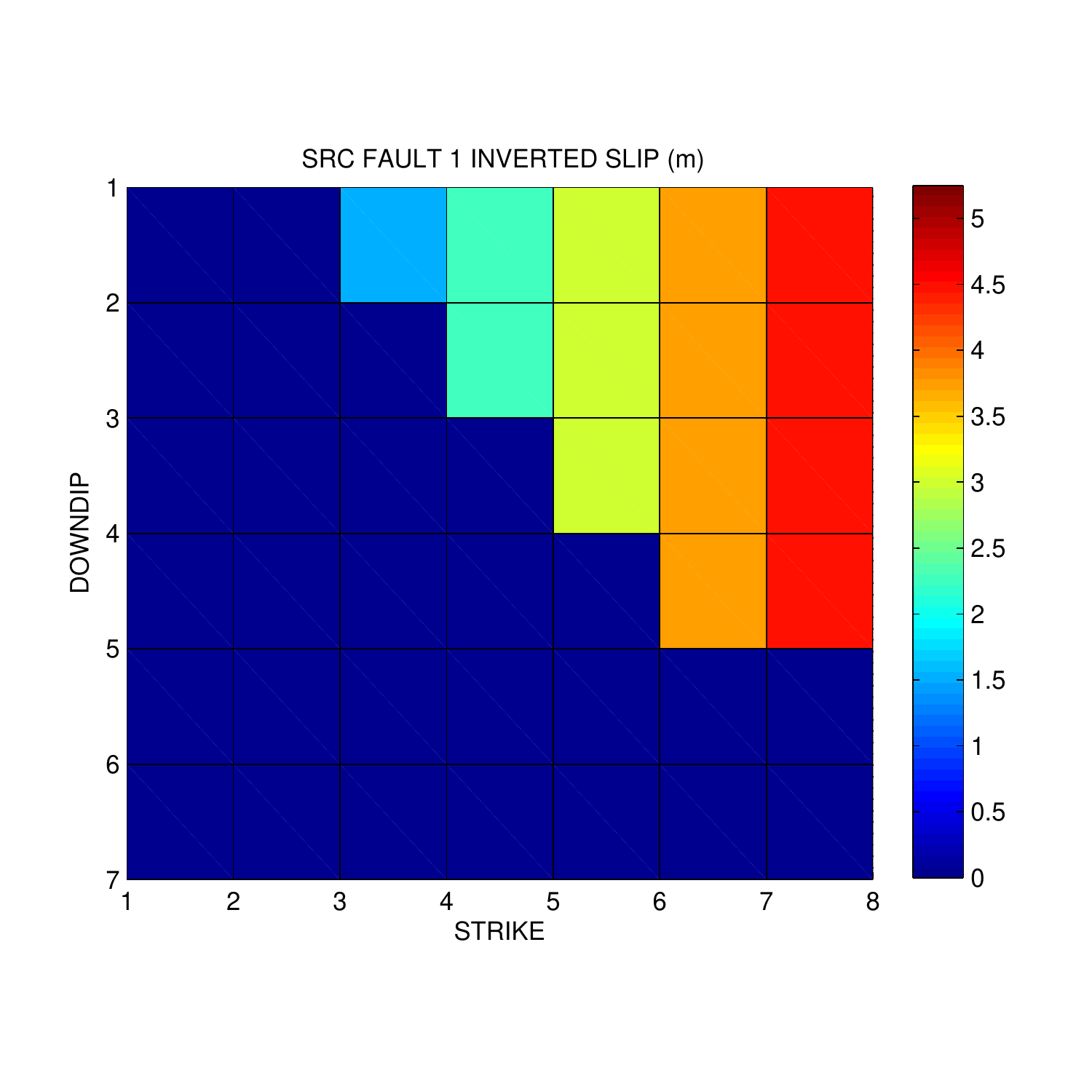} &
\includegraphics[width=.33\textwidth]{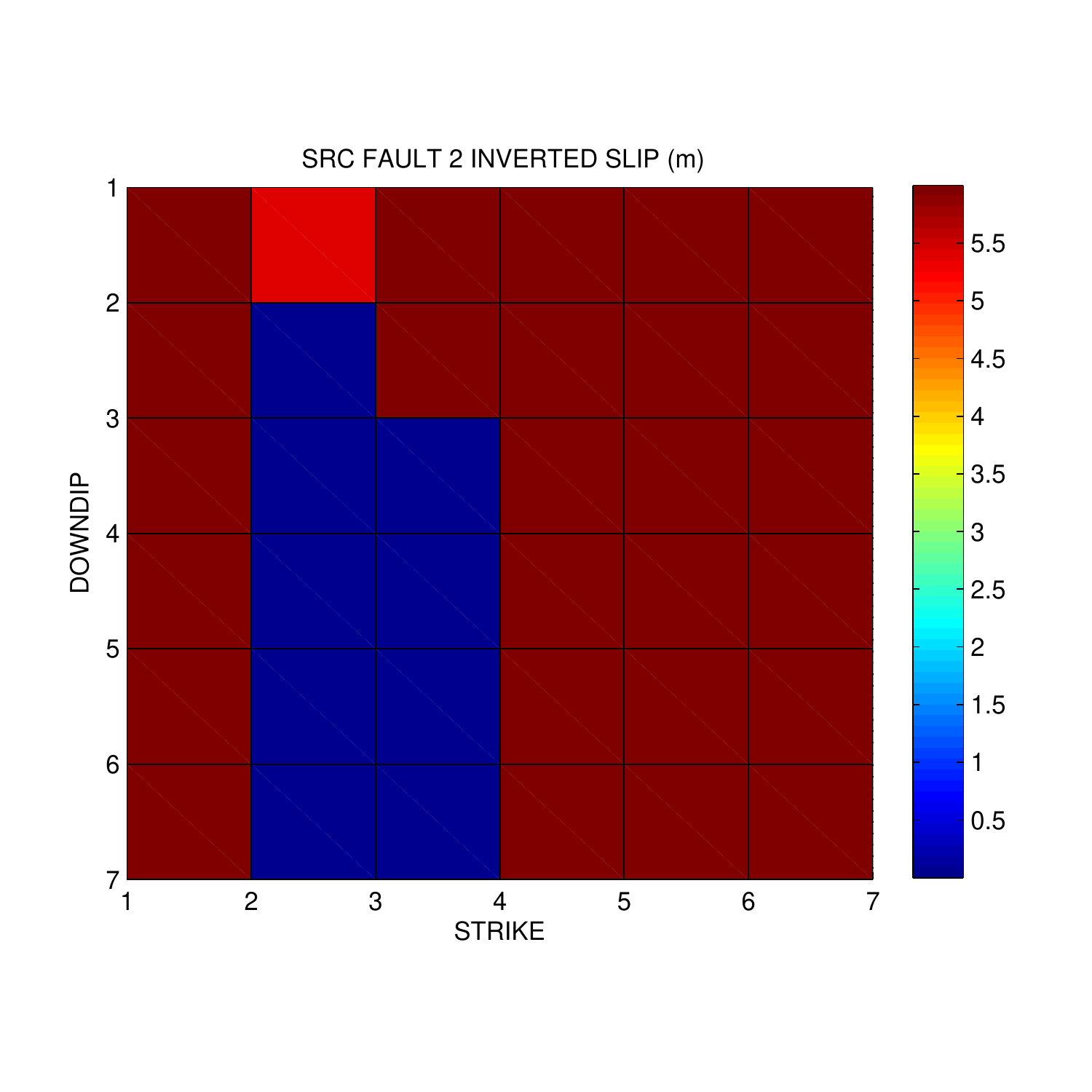} &
\includegraphics[width=.33\textwidth]{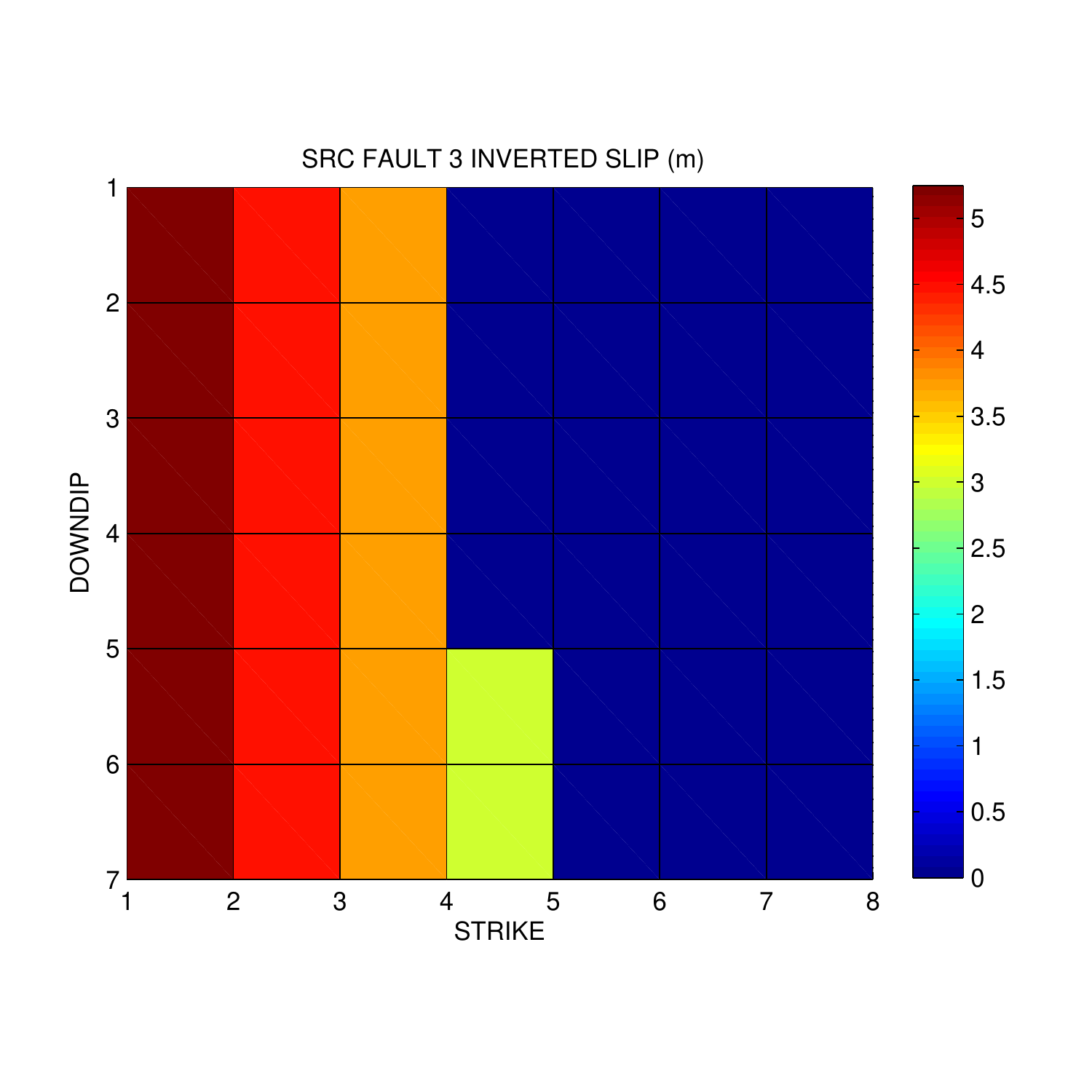}\\ 
\includegraphics[width=.33\textwidth]{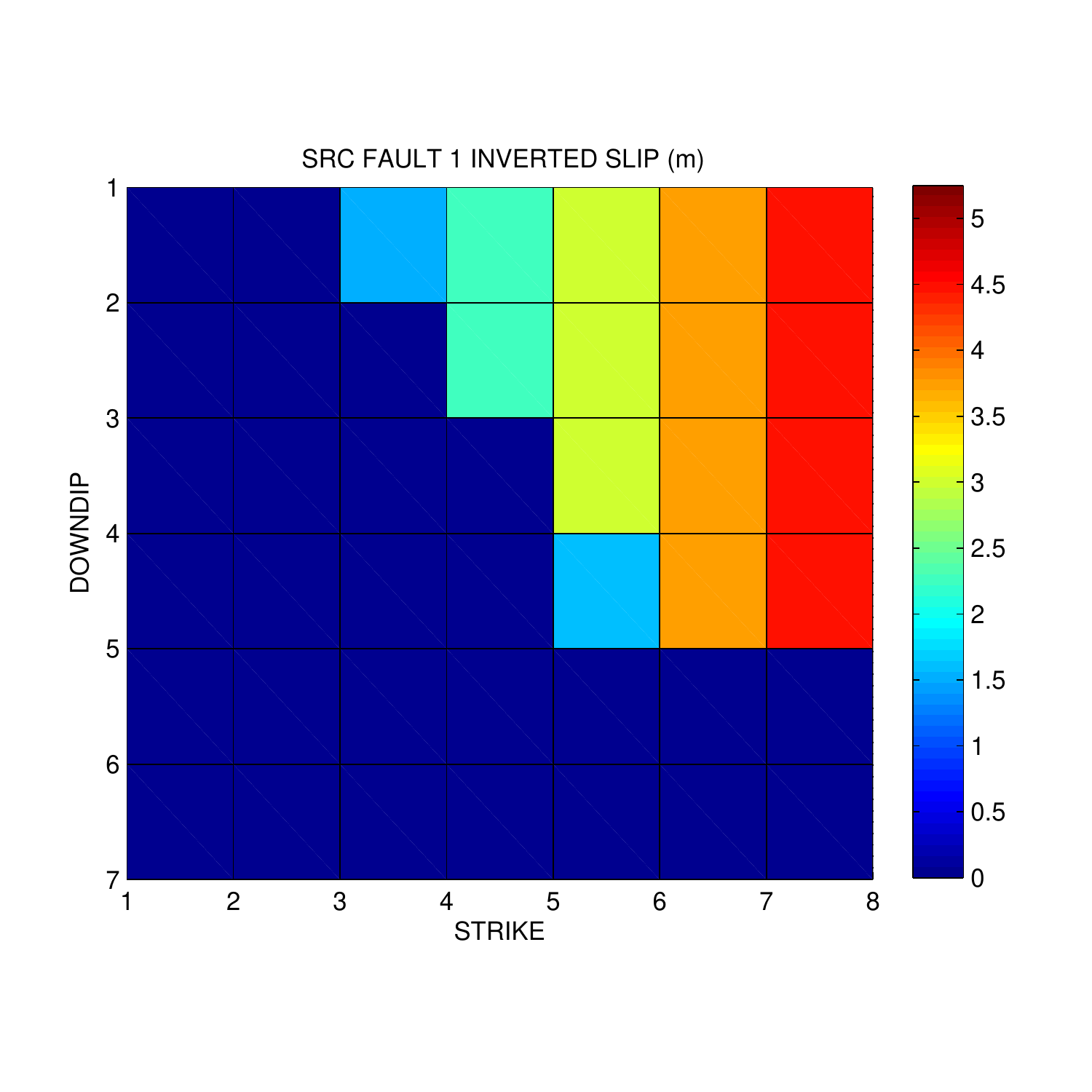} &
\includegraphics[width=.33\textwidth]{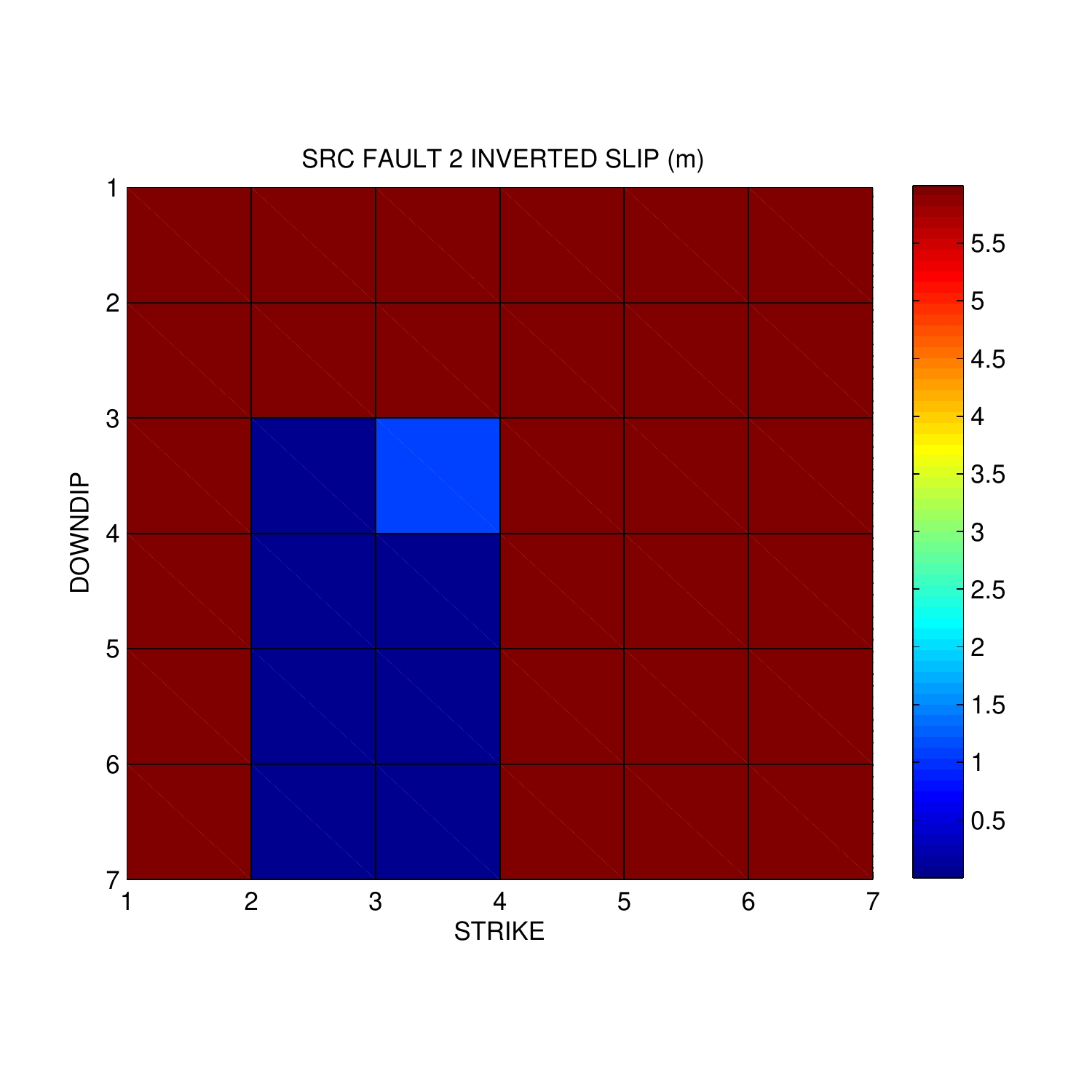} &
\includegraphics[width=.33\textwidth]{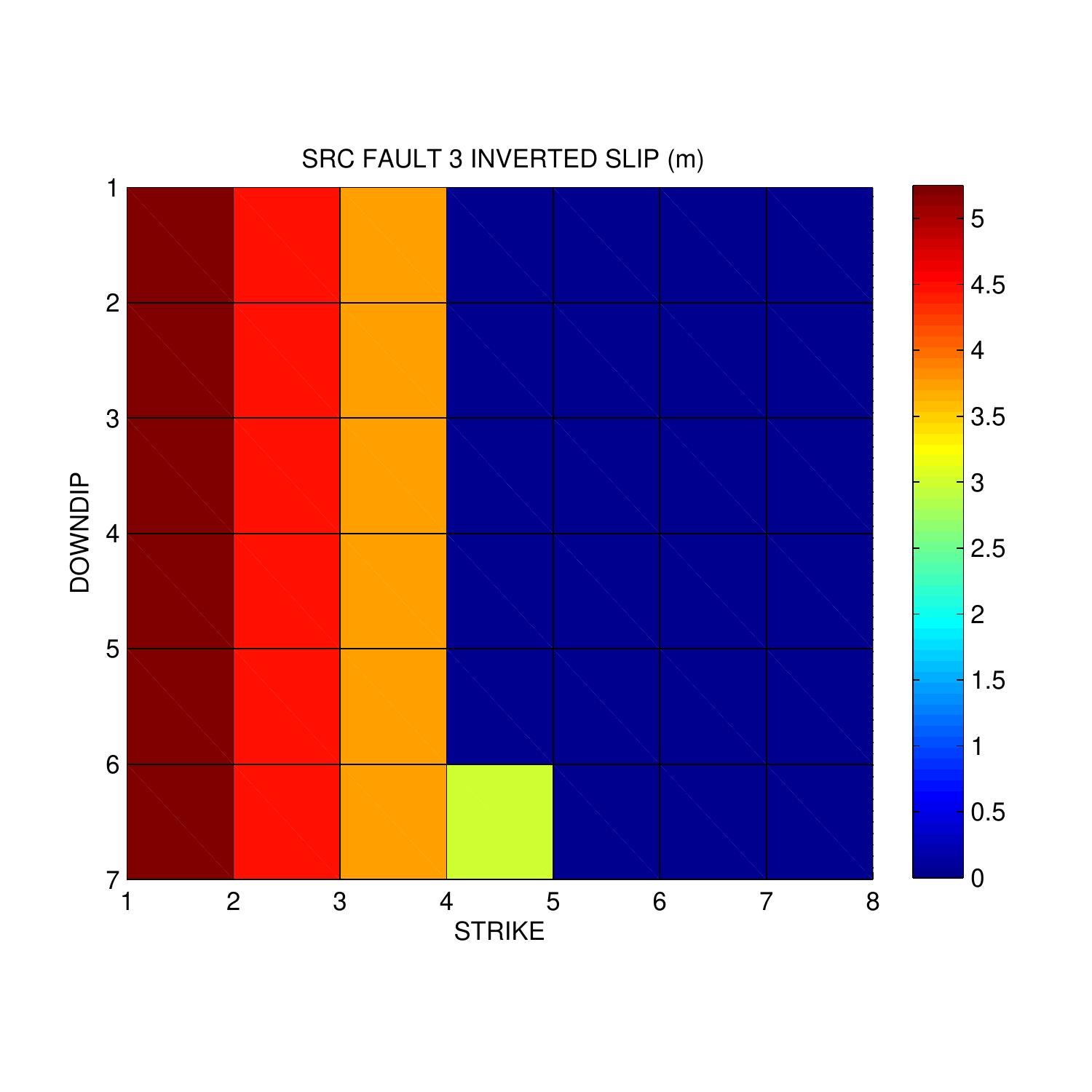}\\ 
\end{tabular}
\end{center}
\vspace{-10 mm}
\caption{Inverted Landers slips using the Hector Mine event only, un-clamping constraints, faults 1-3, $\mu=0.8$, homogeneous vs layered Earth. }
\label{fig:invertedhm}
\end{figure}

\begin{figure}[htbp]
\begin{center}
\begin{tabular}{cc}
\includegraphics[width=.33\textwidth]{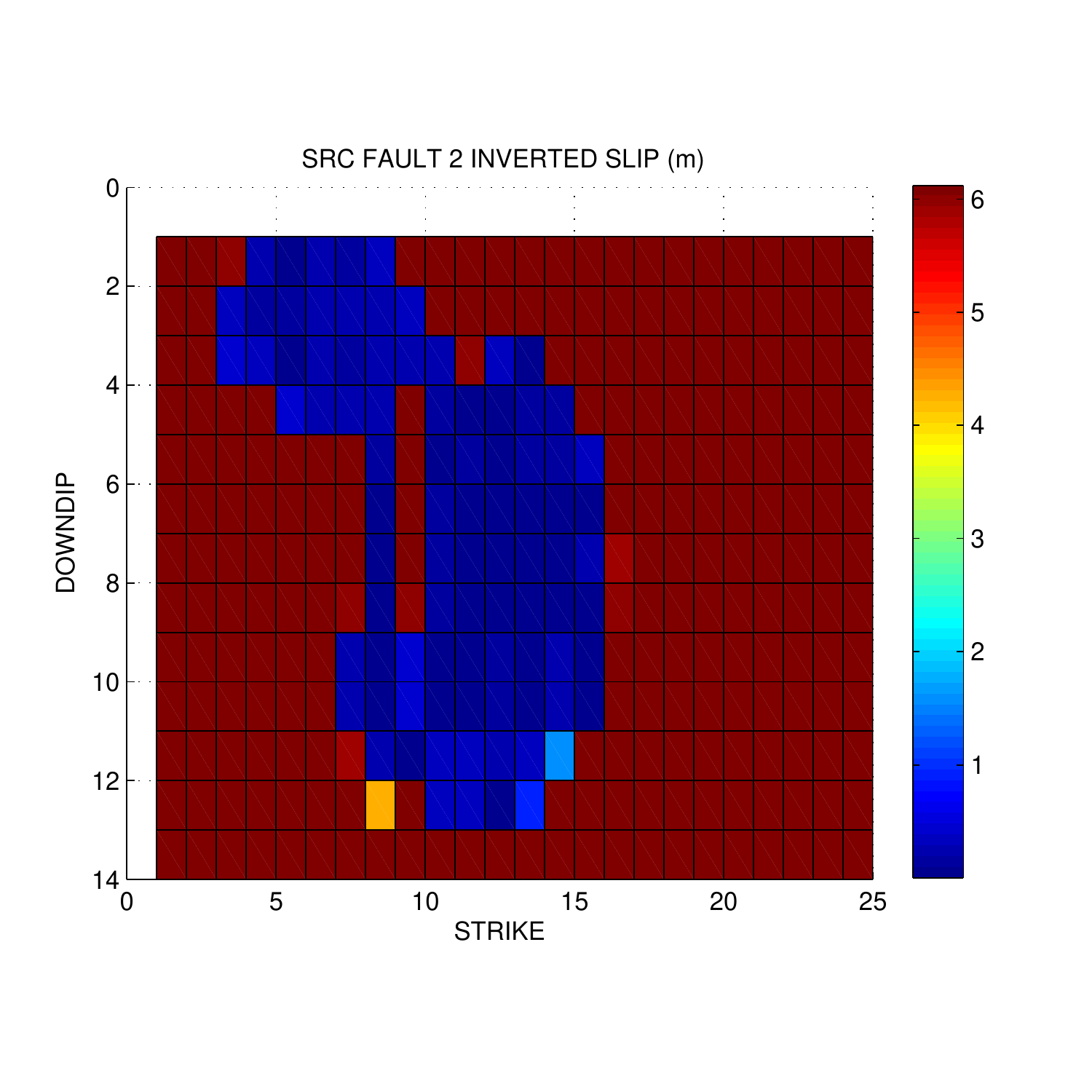} &
\includegraphics[width=.33\textwidth]{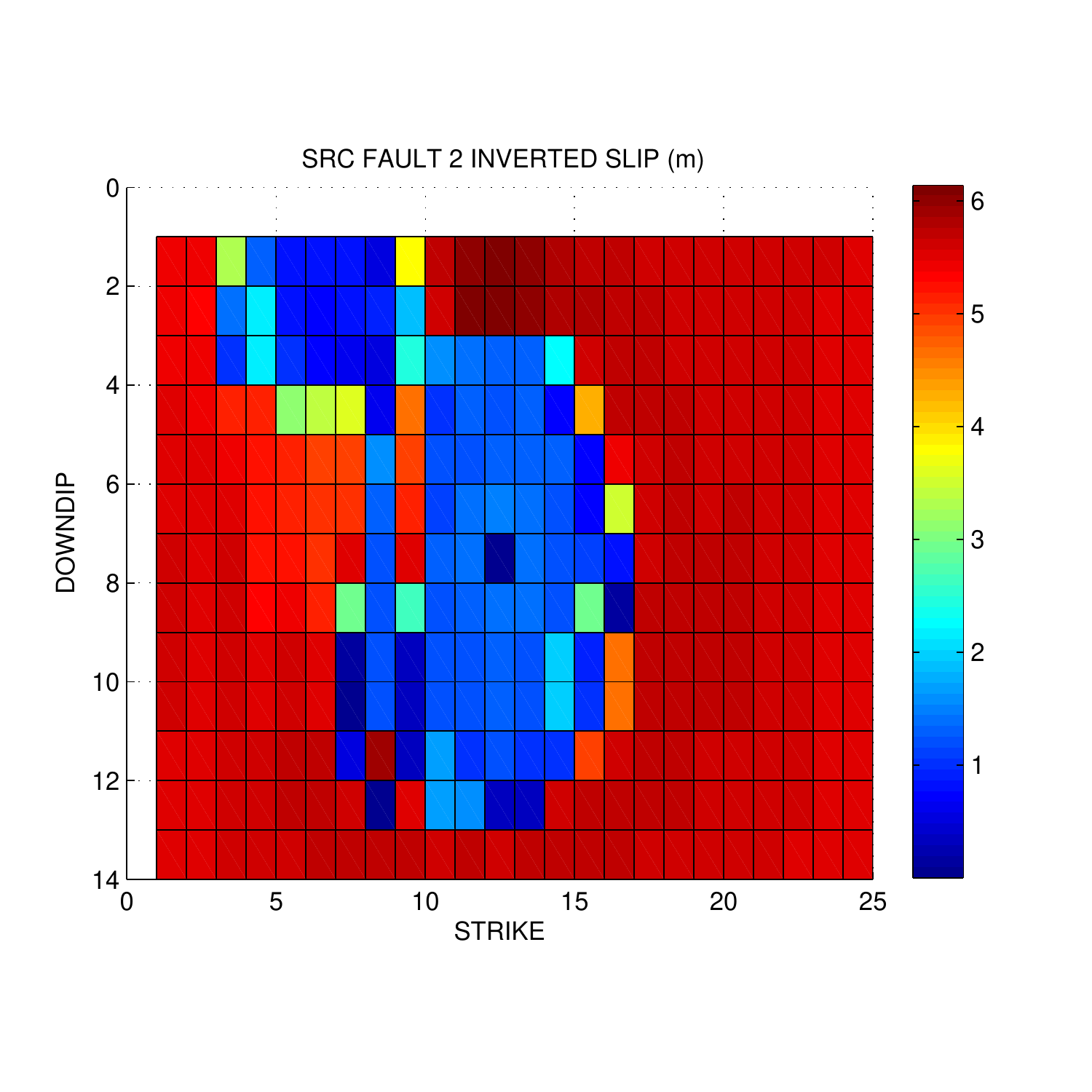} 
\end{tabular}
\end{center}
\vspace{-10 mm}
\caption{Inverted Landers slips from the decimated (20 faults) regional dataset, \emph{all} constraints, high-resolution fault grid, fault 2, $\mu=0.7$, homogeneous vs layered Earth.}
\label{fig:invertedhrsmall}
\end{figure}

Fig~\ref{fig:invertedlocal},\ref{fig:invertedlocalall} demonstrate inversion results for the (three idealized) Landers faults using the local dataset (i.e., events clustered around Landers). Despite the events being clustered around source fault 1, inversion indicates the most significant slips on source fault 2 (i.e., Homestead Valley) consistent with our earlier estimate that this part of the rupture accounts for more that half of the total Landers moment. This is confirmed by the inversion of the \emph{decimated} regional dataset consisting of only 20 faults: fig~\ref{fig:inverted},\ref{fig:invertedall} again indicate the greatest moment density on the source fault 2. Including the entire regional dataset produces qualitatively similar results of fig~\ref{fig:invertedspread},\ref{fig:invertedspreadall}. Of course, this is an obvious consequence of the \emph{equality constraint} in (\ref{eq:INV}) for the total moment on a source fault. However, another, more subtle, feature of all these plots is that fault 2 is predicted to have small slips roughly in the interior of the southern half of the fault, with the slips increasing between faults 1 and 2. Taking advantage of the spatial distribution of Hector Mine, we may expect inversion of Hector Mine to produce a ``more accurate'' prediction of slips on fault 2. Fig~\ref{fig:invertedhm} demonstrate the result of solving for Landers slips using only \emph{un-clamping constraints} for Hector Mine. The results appears to be in a qualitative agreement with the estimated Landers slips (see Figure 5 of \cite{WALD}), correctly identifying the low-slip area 7-10 {\tt km} from the southern tip of the Homestead Valley fault, starting at a depth of $\approx$7 {\tt km}. While this result should not be over-interpreted, it may indicate inverse stress transfer as a viable mechanism for slip inversion. Unlike the traditional methods of inverting subsurface slips from the observed surface displacements (e.g., GPS or INSAR, \cite{PS}), inverse problem (\ref{eq:INV}) estimates the slips from the \emph{assumption that the source fault triggered a given set of events}. Such inversion is inevitably qualitative by nature but still may prove useful as e.g. an additional constraint or regularization for a more traditional inversion method. Further study of inverse stress transfer in applications to large-scale triggering problems may help assess its usefulness. For example, fig~\ref{fig:invertedhrsmall} shows the result of inverting the slips on Landers fault 2 using the decimated regional dataset with all constraints, on a high-resolution fault grid. The result is biased by the inconsistent constraints but appears not unreasonable in the context of known Landers slips (\cite{WALD}). A similar experiment could not be carried out for Hector Mine as the receiver because of the numerical cost of solving this problem using {\tt MATLAB}. An obvious extension of this work would be to use high-resolution fault grids and fast compiled solvers.


\section{Conclusions and Perspectives}

Results of our study can be summed up as follows:
\begin{itemize}
\item Elastic heterogeneity appears to have a significant qualitative and quantitative impact on stress transfer, even when average moduli remain constant.
\item Stress transfer is sensitive to the spatial distribution of slips on the source faults, and may provide a mechanism for qualitative slip inversion.
\item Application to Landers and 100 random ``aftershocks'', both in a neighbourhood of Landers and spread around the entire rupture, indicates that, with a few exceptions, Landers appears to have encouraged slips for these events.
\item Application to Hector Mine indicates un-clamping as the dominant slip encouraging effect of Landers, in agreement with earlier works.
\item Inverse stress transfer may be a useful technique for qualitative slip inversion, however, further study is required to ascertain the extent of its applicability and stability to data uncertainty
\end{itemize}

\section{Appendix A. Reproducibility}

The computational toolkit developed for this study consists of the following components:

\begin{itemize}
\item Elastostatic Green's Tensor Computation Module {\tt elastostatic\_green.m} 
\item Surface Displacement Modeling Module {\tt model\_surface\_displacements.m}
\item Direct and Inverse Stress Transfer Analysis Module {\tt transfer\_stress.m}
\item Auxiliary functions: {\tt interp\_indices.m, NN\_interp\_indices.m, ROT.m}
\end{itemize}

The Green's tensor computation module reads inputs from three files: 
\begin{itemize}
\item control parameter file (specified by the \verb+$ELASTIC_GT_CTRL+ environment variable in the batch mode or user-specified input);
\item elastic Earth model file (\verb+$ELASTIC_GT_MODEL+ or user input);
\item source faults (\verb+$ELASTIC_GT_FAULTS+ or user input).
\end{itemize}

Following are sample input files used for the computation of the Green's tensor on the regional Landers/Hector Mine grid for the layered Earth model (see above):
\begin{verbatim}
controls.dat:
nx   sigma    ny     nz    lx    ly    lz     
256   1.5     80    20    128    40    20

input_hmodel.dat:
z   mu  nu
2  21  0.2
5  23  0.2
8  32  0.2
12 36  0.2
15 38  0.2
18 40  0.2
40 45  0.2

Landers.dat:
x      y      z length width     strike   dip   rake  mag
0      0      9   30     14        -5       90    180  6.8035
22    -3.9    9   25     14        -26      90    180  7.1269
38.5  -11.7   9   30     14        -26      90    180  6.9209
\end{verbatim}

The stress transfer module requires receiver module definition (\verb+$ELASTIC_GT_RECEIVERS+ or user input) similar to the source fault file, but excluding the magnitude column and including friction and Skempton columns:
\begin{verbatim}
x    y        z     length  width     strike   dip   rake  mu   B
50   5.85     10    26       16       -35      77    180   0.7   0.5
44   13.65    10    38       16       -15      77    180   0.7   0.5
16.5  23.4    10    26       16       -35      77    180   0.7   0.5
\end{verbatim}

Constraints of inverse problem (\ref{eq:INV}) can be limited to consistent ones by uncommenting operator {\tt continue} on line 823 of {\tt transfer\_stress.m}.

\bibliographystyle{plain}
\bibliography{p}
\end{document}